\documentclass[11pt]{article}
\usepackage[letterpaper,includefoot,top=1in,bottom=1in,left=1in,right=1in]{geometry}
\usepackage{multirow}
\usepackage{float}
\usepackage[centertags,sumlimits]{amsmath}
\usepackage{amsfonts}
\usepackage{amssymb,graphics,psfrag}
\usepackage{array,epsfig,multirow,stmaryrd,graphicx}
\usepackage[all,cmtip]{xy}
\usepackage{mathrsfs}
\usepackage{color}
\usepackage{arydshln}  
\usepackage[bookmarks,breaklinks]{hyperref}
\usepackage{multirow}
\usepackage{eufrak}
\usepackage{verbatim}
\numberwithin{equation}{section}
\numberwithin{table}{section}
\usepackage{caption}
\usepackage{subcaption}
\usepackage{float}

\usepackage{bbm, dsfont}

\newcommand{\pid}{(2 \pi)^3} 

\DeclareMathAlphabet{\mathbbold}{U}{bbold}{m}{n}

\newcommand{\pa}{\partial}

\definecolor{rojo}{rgb}{0.8,0.1,0}

\newcommand{\ci}[1]{\overset{\circ}{ #1}}

\newcommand{\id}{\mathbbm{1}}

\newcommand{\beq}{\begin{equation}}
\newcommand{\eeq}{\end{equation}}
\newcommand{\be}{\begin{equation}}
\newcommand{\ee}{\end{equation}}
\newcommand{\bea}{\begin{eqnarray}}
\newcommand{\eea}{\end{eqnarray}}
\newcommand{\ben}{\begin{eqnarray*}}
\newcommand{\een}{\end{eqnarray*}}
\newcommand{\ba}{\begin{aligned}}
\newcommand{\ea}{\end{aligned}}
\newcommand{\bt}{\begin{tabular}}
\newcommand{\et}{\end{tabular}}
\newcommand{\bc}{\begin{center}}
\newcommand{\ec}{\end{center}}

\newcommand{\cO}{\mathcal{O}}

\newcommand{\cC}{\mathcal{C}}
\newcommand{\cD}{\mathcal{D}}

\newcommand{\cN}{\mathcal{N}}

\newcommand{\cA}{\mathcal{A}}

\newcommand{\cM}{\mathcal M}

\newcommand{\ds}{\displaystyle}

\DeclareMathOperator{\SO}{\mathit{SO}}

\newcommand{\dd}{d}

\newcommand{\bbZ}{\mathbb{Z}}

\newcommand{\bbP}{\mathbb{P}}

\newcommand{\IP}{\mathbb{P}}

\newcommand{\nn}{\nonumber}

\newcommand{\cref}{{\bf [check ref]}}

\newcommand{\ra}{\rightarrow}

\def\id{\operatorname{id}}
\def\ord{\operatorname{ord}}

\usepackage{amsmath}
\graphicspath{{./Pictures/}}

\entrymodifiers={+!!<0pt,\fontdimen22\textfont2>}

\begin{document}

\baselineskip=16pt
\setlength{\parskip}{6pt}

\begin{titlepage}
\begin{flushright}
\parbox[t]{1.4in}{BONN-TH-2013-21}
\end{flushright}

\begin{center}


{\Large \bf Landscaping with fluxes and the $E_8$ Yukawa Point in F-theory}
%


\begin{center}
   \normalsize \bf{Nana Cabo Bizet}$^{a,b}$ \footnote{\texttt{nana@fisica.ugto.mx}},
   \normalsize \bf{Albrecht Klemm}$^c$ \footnote{\texttt{aklemm@th.physik.uni-bonn.de}}
   \normalsize \bf{Daniel Vieira Lopes}$^d$ \footnote{\texttt{dvieiralopes@unb.br}},
   \end{center}
\vskip 0.0cm
 \emph{$^{a}$ Departamento de F\'{\i}sica, Divisi\'on de Ciencias e Ingenier\'{\i}as, Campus Le\'on, \\ [.1cm]  
 Universidad de Guanajuato,Loma del Bosque 103, CP 37150,  Le\'on, Guanajuato, M\'exico} 
\\[0.1cm]
 \emph{$^{b}$ Centro de Aplicaciones Tecnol\'ogicas y Desarrollo Nuclear  \\[.1cm]
 Calle 30 \# 502 e/5ta y 7ma Avenida 11400, Playa, La Habana, Cuba.} 
\\[0.1cm]

\emph{$^{c}$ Bethe Center for Theoretical Physics, Universit\"at  Bonn, Nussallee 12, D-53115 Bonn}
\\[0.1cm]
 \emph{$^{d}$ CIFMC - Universidade de Bras\'ilia, \\ [.1cm]
          Campus Universitario Darcy Ribeiro, Bras\'ilia - CEP 0910-900}  \\[0.25cm]
\vspace*{0.05cm}
\end{center}
\vskip 0.0cm

\begin{center} {\bf ABSTRACT } \end{center}
Integrality  in the Hodge theory of Calabi-Yau fourfolds  is essential 
to find the vacuum structure  and the anomaly cancellation mechanism of 
four dimensional F-theory compactifications. We use the Griffiths-Frobenius geometry 
and homological mirror symmetry to fix the integral monodromy basis in the 
primitive horizontal subspace of Calabi-Yau fourfolds.  The Gamma class
and supersymmetric localization calculations in the 2d gauged
linear sigma model on the hemisphere are used to check and extend this method.  
The result allows us to study the superpotential and the Weil-Petersson metric and an 
associated tt* structure  over the full complex moduli space of compact fourfolds 
for the first time.  We show that integral fluxes can drive the theory to N=1 supersymmetric 
vacua at orbifold points and  argue that fluxes can be chosen that fix the complex moduli of F-theory compactifications at gauge 
enhancements including such with U(1) factors. Given the mechanism it is natural to 
start with the most generic complex structure families of elliptic Calabi-Yau 4-fold 
fibrations over a given base. We classify these families in toric ambient spaces and 
among them the ones with heterotic duals. The method  also applies to the creating 
of matter and  Yukawa structures in F-theory. We construct two SU(5) models in F-theory with 
a Yukawa point that have a point on the base with an $E_8$-type singularity on the fiber and 
explore their embeddings in the global models. The explicit resolution of the singularity
introduce a higher dimensional fiber and leads to novel features.

\hfill \today
\end{titlepage}

\section{Introduction}
The exceptional gauge groups that can arise in F-theory provide 
an attractive framework for constructing GUT models with 
promising phenomenological properties. This is a common feature 
with the heterotic string, where the $E_8\times E_8$ gauge group 
in 10d is one of two ways to fulfill worldsheet modularity 
as well as anomaly cancellation in supergravity. The worldsheet 
conformal field theory description of the heterotic string gives 
a computational framework to make quantitative predictions at least 
in a region of the coupling space. While this is absent in F-theory 
important moduli dependent terms of the  $\mathcal{N}=1$ effective action from F-theory are 
governed by holomorphic quantities such as the flux superpotential and the gauge 
kinetic terms, which are exactly computable using the variation 
of the mixed Hodge structures and are technically obtained  by a Griffiths 
residuum calculation on the F-theory manifold or its mirror. This is particularly 
effective for fourfolds $M$ where $H_3(M)$  is usually small or zero~\cite{Klemm:1996ts} 
and the only relevant data are $G_4$ fluxes on $H_4(M)$.   F-theory 
phenomenology is related to singular fourfolds, typically with a singular 
elliptic fibre, and there are two useful ways to learn about the structure of the 
singularity either by analyzing the homology  of the smooth space 
after blowing it up or after deforming it in the complex structure moduli space. 
The former method is at first glance more effective, but since it is only leading 
order in the local parameters  it misses the subleading effects that do play 
an important  physical role. The deformations 
close to the singularity  are described by the deformation of the  limiting 
mixed Hodge with the corresponding residue integrals. This formalism gives not 
only amplitudes such as the superpotential exactly 
as a period integral, but more fundamentally the exact $\mathcal{N}=1$ coordinates. The point 
of view that it is more natural to consider the complex deformations has appeared 
in many works, see e.g.~\cite{Grassi:2013kha}, and 
reconciles  e.g. the higher codimension singularity analysis in comparison 
with the simpler group theory expectation from the heterotic string. 
Calculation of local complex residua also organize the corrections to 
Yukawa couplings with many cancellations among them, see 
e.g.~\cite{Cecotti:2009zf}. While the strength of F-theory should be in the exact geometrical 
calculation of the holomorphic terms,  there are not many results along this line, 
simply because it seems a daunting task to analyze the usually huge numbers of periods on  
phenomenological interesting fourfolds. However using constraints from 
the  Griffiths-Frobenius structure, which leads to an analog of special
geometry for Calabi-Yau fourfolds, and  constraints  from homological 
mirror symmetry, we solve the problem of reconstructing the periods  in 
an integer monodromy basis  from the topological data.  Another reason for the interest in the  integral basis is that it allows us to 
define the $tt^*$  structure for fourfolds, which  is expected to control non-holomorphic 
corrections beyond the Weil-Petersson  metric.  
In order to exemplify the method we calculate 
first the  explicit basis and the monodromies of the periods over 
the primitive horizontal subspace in  nine one-parameter families including
the sextic in $\mathbb{P}^5$  and the intersections of five quadrics in $\mathbb{P}^9$ 
and study the periods at the degenerate  points. For the first time one gets hence 
the exact information about the ${\cal N}=1$ coordinates the superpotential and the Weil-Petersson 
metric in the entire complex moduli space of a Calabi-Yau fourfold. 
Like the seminal calculation for the quintic
 in $\mathbb{P}^4$, these examples have 
universal features,  which generalize to multimoduli 
Calabi-Yau fourfolds. In particular they have beside the large complex 
structure singularity, the conifold singularity, which is the most generic singularity,  
and orbifold singularities, which are the model for smooth geometries with additional 
discrete symmetries. Part of the results can be checked  using the $\hat \Gamma$ 
class and  the hemi-sphere partition function obtained recently by Hori and Romo 
using supersymmetric localization and match perfectly. These latter formalisms are 
useful to extend our results to hypersurfaces and complete intersections in toric ambient spaces 
with many moduli and to the much more general Calabi-Yau spaces embedded into 
Grassmannians and  flag-varieties described by non-abelian 2d $\sigma$-models.

Of course the above approaches to singularities are related by mirror symmetry 
and one outcome of the analysis is that, once the leading order analysis in 
the A-model resolutions is known, the structures mentioned above are rigid 
enough so that the exact expression follow relatively swiftly.  F-theory fourfolds with  a realistic rank of the 
gauge group  are within the reach of the techniques, developed in the paper.  
Explicite heterotic/F-theory moduli identifications ala Friedman, Morgan Witten   
allows to find good ${\cal N}=1$ coordinates, moduli stabilizing superpotentials  
as well as the uncorrected scalar metric and the $tt^*$ structure also for 
bundle moduli in ${\cal N}=1$ heterotic string compactifications in the stable
degeneration limit.

The  main question we address here with these new techniques is whether integer 
flux  superpotentials stabilize F-theory moduli  at phenomenological 
viable compactifications.  Using induced actions of global symmetries 
on the complex  moduli space and the notation of sub-monodromy systems 
we argue that these superpotentials can drive F-theory to the loci in the moduli space with gauge symmetry 
enhancements  and matter structures  including the heterotic string limit.  
We show explicitly in  section~\ref{orbifoldpoints}  that further moduli can be stabilized at orbifold
points and contrary to the few known attractor points in threefolds, fourfold 
orbifold points can be  attractors  with a {\sl vanishing} scalar potential. 
The notion of sub-monodromy systems is very useful in the degeneration limit 
and we can give an exact description of the moduli of the central
fibre and the heterotic gauge bundle in the stable degeneration limit.     

The superpotentials that we calculate are linear in the periods, which have by 
definition modular properties. The reason is that for the theory to make sense 
in the deformation family of fourfolds over the vacuum manifold, they have to 
transform with integral monodromies that respect the intersection pairing in  
homology, which we called $Q$, or $\eta$ when represented as matrix. 
Otherwise the flux quantization condition and  the metric of the scalars 
would not be uniquely defined. These transformation properties makes the 
periods  meromorphic modular forms w.r.t. to the discrete symmetry group 
$\Gamma$ generated by the monodromies. The latter are more fundamental, 
because  the  Riemann Hilbert problem is solved in this context, meaning that the 
monodromies determine the periods as those  meromorphic sections over the complex 
moduli that have reasonable behaviour at the boundaries of the moduli space. 
If non-perturbative effects  respect $\Gamma$, then the corresponding superpotential
will be modular and hence reconstructable from a very restricted set of meromorphic sections.  A 
striking example of such a non-perturbative superpotential with modular properties was 
given in~\cite{Donagi:1996yf} and as we explain in section \ref{E8BPSstates} it 
is simply the period on the mirror manifold.              

By the same principle F-theory periods encode very valuable information about the non-perturbative corrections 
to the perturbative theories that arises in degeneration  limits.  This implies  for the heterotic 
string~\cite{Grimm:2009ef}\cite{Jockers:2009ti}\cite{Grimm:2009sy} that small instantons as 
well as the associated heterotic 5-brane moduli can be identified and $e^{1/g_s}$ 
effects  can be calculated. Moreover the  open/closed duality between the F-theory at genus zero 
closed string sector and the Type II disk open string string sector~\cite{Aganagic:2001nx}\cite{Aganagic:2000gs} that has been 
computationally established in those  non-compact limits,  allows~\cite{Jockers:2009ti} to study 
D-instanton corrections in orientifold compactifications. The relations between fourfold 
periods  and type II $D_5$-brane open string superpotentials has been intensively 
studied~\cite{Grimm:2009ef}\cite{Grimm:2008dq}\cite{Alim:2009bx}\cite{Alim:2011rp}.
Moreover the fourfold  superpotential also captures important  aspects of the vacuum 
structure of  manifolds with $SU(3)$ structure, see e.g.~\cite{Grimm:2010gk}.

The paper is divided roughly into two parts and organized as follows:   
Until section five we deal  with  properties of F-theory  that are useful to analyze the complex 
structure moduli, the flat coordinates and the integral structure on the 
primitive horizontal subspace, including the  modular properties of the 
amplitudes. In section~\ref{elementsofF} we give a conceptual overview over 
challenges in F-theory model building on compact Calabi-Yau manifolds.  
The geometrization of the Type IIB axio-dilaton in elliptic fibrations, its 
dependence on the moduli of the fourfold as well as a short discussion of 
the non-perturbative  limits can be found in~\ref{Ftheorymap}, 
while the basic properties of its compactification are spelled out in 
section~\ref{compactification}.  The section~\ref{fluxquantization} gives 
those topological and geometrical properties of fourfolds in particular its 
Hodge structure and its integer structures that are used  throughout 
the paper.  The construction of Batyrev and Borisov is set up in section 
\ref{batyrev} together with a discussion of implementation of twisted 
fibration structures in section \ref{fibretwistingdata}. Beside the aspect 
of  mirror symmetry that turns out to be essential for F-theory,  other  
main advantages of the construction is that it gives a good model of the moduli 
space ${\cal M}_{cs}$ described  in section~\ref{complexmoduli} and the points of 
maximal unipotent monodromy, see section \ref{mumpoints}. Using 
the model of the moduli space developed in section~\ref{complexmoduli} 
we study the induced  group actions on it in  section \ref{groupactions} 
and give a local  account of the  behaviour of the periods near orbifold 
points. In preparation of the discussion of the sub-monodromy 
systems that extend this local picture, we discuss in section~\ref{transitions}
the transitions that are described geometrically by embedding polyhedra 
into each other.  In section~\ref{gaussmaninS} we define the notion of the 
sub-monodromy system using the Picard-Fuchs ideal and its ${\cal D}$-module 
and  argue that the embedding of polyhedra defines  sub-modromy 
systems on the locus ${\cal S}\subset {\cal M}_{cs}$. We discuss  
the relevance of the degenerations  of Calabi-Yau manifolds and its description using limiting mixed Hodge structure 
to the sub-monodromy systems, the fixing of an integral basis and 
the approach of gluing amplitudes from degenerate components in 
stable limits. We  correct some statements in the physics literature concerning the 
application of Schmids nilpotent orbit approximation, which is 
incompatible with the $\hat \Gamma$-classes. In section~\ref{effectiveaction} 
we review the terms that are calculable exactly and approximatively from 
the holomorphic--  and the real geometrical structures of 4-folds respectively. 
\ref{integralmonodromie} --- \ref{minimizingW}  are the main sections of the first part 
where we fix the integral monodromy basis and draw conclusions 
for the ${\cal N}=1$ vacua structure in F-theory.  In 
section~\ref{frobeniusstructure} we use the Frobenius 
structure, the Griffiths transversality and the real structure 
of the Calabi-Yau n-folds to define the real  Griffiths-Frobenius 
structure, which in particular includes special geometry and 
the $tt^*$ structure.   It remains incompletely specified until we 
find the integral basis. We review this procedure for 3-fold 
stressing the general aspects in~\ref{threefoldbasis}. 
In section~\ref{fourfoldbasis} we find the integral basis for nine 
fourfolds  and all its monodromy- and connections matrices as well 
as the flat coordinates throughout ${\cal M}_{cs}$ by explicit 
calculation.  In section~\ref{HMS} we relate the calculation 
to homological mirror symmetry  and check the occurrence on the 
$\hat \Gamma$ class at the maximal unipotent degeneration. 
In section~\ref{hemisphere} we check and extend the results 
of~\ref{fourfoldbasis} using the hemisphere partition function  
of Hori and Romo, with boundary data  specified by  Chan-Paton 
factors and matrix factorizations. For complete intersection 
and hypersurfaces in toric varieties one gets a rational $K$-theory 
basis of twisted line bundles on the mirror, whose intersection is 
simply defined by the Chern character map and  the Hirzebruch-Riemann-Roch 
index theorem. Continuation of  Barnes integral representation 
of  the HPF relates the rational basis to our integral basis  and 
the intersection forms for the two basis match perfectly. We also 
generalize  the HPF to local CY and relate the HPF to the Picard-Fuchs ideal.  
We discuss the general features for minimizing the superpotential in 
section~\ref{minimizingW}. In section~\ref{orbifoldpoints} we use the 
integral basis to establish that the superpotential drives the 
moduli to the orbifold point with a vanishing scalar potential.  We analyze 
the superpotential at the most generic singularity, the node in 
section~\ref{nodalpoints}, which has very different features depending 
on the dimension $n$ of the CY mod 4. Given the superpotential 
in the integral basis we come back to the sub-monodromy system in 
section \ref{submonodromy} and formulate a general conjecture based on the examples discussed 
before.   In section~\ref{elliptictdata} we classify the clean sheet models starting with
the heterotic basis in section~\ref{heteroticbasis} and a  small statistic of  the topological data  of 
the more general 4319 toric almost Fano basis in section~\ref{toricfanobases} and the 
generic global fibres in the complete intersection class~\ref{globalfiber}. 
For the embeddings of polyhedra we get a sub-monodromy system and a superpotential  
that restricts the moduli to subloci ${\cal S}$ . This will be used in section~\ref{compactCY} 
to construct first non-abelian gauge groups in the models discussed in 
section~\ref{elliptictdata} by restricting polyhedra to reflexive subpolyhedra. 
We start in sections~\ref{tatem} and~\ref{stabledegenerationlimit} 
by  enforcing the stable degeneration limit and identify the sub-monodromy 
system that governs the moduli space, metric and flat coordinates 
of the central fibre as well as of the one that governs the same data on the  
heterotic gauge bundle moduli in section~\ref{E8BPSstates}.  We explain the 
occurrence of the modularity in the partition functions for the gauge degree 
of freedom, which can be interpreted in terms of $[p,q]$-strings. In 
section~\ref{addU1} we extend that to models with additional 
$U(1)'s$.  In particular we use Nagells algorithm to bring the 
corresponding normal forms of the fibres into Weierstrass form and 
relate the gauge groups to vanishing of the coefficients of the normal 
forms for the $E_7$ fibre and the $E_6$ fibre in sections \ref{E7u1} and
\ref{E6u12} respectively.  In section~\ref{extremalbundle} we construct 
series of Calabi-Yau $n$-folds that go  back to an observation 
of Pinkham regarding the realization of  Arnolds strange dualities in K3's. 
The cases hold the current records for the topological data for Calabi-Yau n-folds w.r.t to 
 Euler number and number of complex deformations for all n. As such they 
are another logical starting point for the {\sl landscaping} with fluxes, but 
with an $E_6,E_7,E_8,\ldots$ gauge group unhiggsable by complex 
structure deformations.      

The second part of the paper is devoted to study the F-theory models with $SU(5)$ gauge group and codimension 3 singularities (Yukawa points) expected to be of $E_8$ type i.e. on the studied locus the vanishing order of the coefficients in the Weierstrass form describing the elliptically fibered
fourfold is the one of a codimension 1 $E_8$. In the work of Esole and Yau \cite{Esole:2011sm} small resolutions of $SU(5)$ GUT models were studied, to obtain singularities with  codimension higher than 1 which resolutions differ from the expectations obtained by following Kodaira
classification. The study realized by Marsano and Schafer-Nameki \cite{Marsano:2011hv} considers the codimension 3 $E_6$ kind singularity, to show that even if after resolving a different structure arises, the physics is as expected. Motivated by the fact that the Yukawa couplings giving rise to the neutrino mixing matrix 
were obtained by Heckman, Tavanfar and Vafa \cite{Heckman:2009mn} to arise from an $E_8$ singularity, and by the previously mentioned works 
we explore the resolution of codimension 3 singularities expected to be of $E_8$ type, to find results that differ from the 
expectations, see sec \ref{case1co3}. In section \ref{SU5andE8yukawa}  we present those models. Section \ref{embedE8} is devoted to explore the realization of the codimension 3 $E_8$ singularities in the global toric models constructed in sec \ref{compactCY}. In section \ref{su5models} we describe the models with $SU(5)$ gauge group which carry a codimension 1 $SU(5)$ singularity, and its resolution. Those models present further singularities in higher codimension, and in section \ref{co3spectral} we propose a way to study the expected codimension 3 $E_8$ enhancement based on the F-theory/Heterotic duality via the spectral cover construction and the group decomposition $E_8\rightarrow SU(5)\times SU(5)_{\perp}$. We choose two different local models which give an $E_8$ type vanishing order for the coefficients in the Weierstrass form of the elliptic fibration. The first of those is Case 1, which contains matter representations 5 and 10. This model  is studied in section \ref{case1}. In section \ref{case1resol} we perform its resolution, by blowing-up ten times with $\mathbb{P}^1$  and  $\mathbb{P}^2$'s in the ambient space.  In section \ref{case1co1} we describe the codimension 1 locus obtained after the resolution. In section \ref{case1co2}
we describe the codimension 2 locus, the matter curves 10 and 5 obtained
in the resolution with its respective gauge group charges and its weights.
Section \ref{case1co3}   is devoted to the analysis of the codimension 3 locus, which turns out to differ from
the expected $E_8$. In the appendix \ref{appBcase1} we give the details of this blowup, and in appendix \ref{Appendix C} we describe how the charges are computed.  In section \ref{case2} we study the resolution of an $SU(5)$ model with a codimension 3 $E_8$ singularity in which there is a certain symmetry  among the two basis coordinates which vanishing (above the $SU(5)$ locus) gives the singularity locus, we denote
this Case 2. We follow identical steps as for Case 1, and in sec \ref{case2resol} we perform the resolution by eight blow-ups, in sec \ref{case2co1}, \ref{case2co2} and \ref{case2co3} we describe the blow-up obtained codimension 1, 2 and 3 locus respectively. In sec \ref{case2pqsym} we comment on the implications of the symmetry present in Case 2.  Appendix \ref{appBcase2} contains the details of the Case 2 resolutions. Finally we give our conclusions in section \ref{conclusions}.

\section{Elements of F-theory phenomenology} 
 \label{elementsofF}
In F-theory~\cite{Vafa:1996xn} the relevant data are described 
in terms of the geometry of a Calabi-Yau n-fold $M_n$, elliptically fibred ${\cal E}_F\rightarrow B_{n-1}$ over a 
base $B_{n-1}$ and a choice of $G$ fluxes. By definition $M_n$ 
has an unique holomorphic $(n,0)$-form $\Omega$ and an $(1,1)$ K\"ahler form $J$. 
For compactifications to eight dimensions  $M_2$ is hence an elliptic K3 manifold, 
$B_1=\mathbb{P}^1$ and F-theory, or equivalently Type IIB on $B_1$ 
with varying axion $C_0$ and dilaton $\phi$ background encoded in the complex structure $\tau$ 
of ${\cal E}_F$ as $\tau=C_0 + i e^{-\phi}$, with $e^{\phi}=g_{IIB}$ the type IIB string coupling, 
yields a more general 8d theory then the $E_8\times E_8$ 
heterotic string\footnote{The heterotic $SO(32)$~\cite{Morrison:1996pp,Aspinwall:1996nk} and 
the CHL~\cite{Kachru:1997bz,Bershadsky:1998vn,Berglund:1998va} string can be accommodated 
in the F-theory geometry.} on $T^2$. 
In fact it contains the latter in the stable degeneration 
limit, in which the $K3$ decomposes into two half $K3$'s, i.e. rational elliptic 
surfaces describable as the 9-fold blow of $\mathbb{P}^2$ called ${\rm dP}_9$, 
with the heterotic torus $T^2$ as the central fibre. In short the degenerate 
geometry is given by $\lim_{het} K3 ={\rm dP}_9 \cup_{T^2} {\rm dP}_9$.

Lower dimensional F-theory/heterotic duals can be obtained  by fibering 
the dual 8d compactification data K3/($T_2$ + gauge bundles) over a common 
base $\bar B_{n-2}$, i.e. in compactifications to 4d $B_{3}$ becomes a rational 
fibration $\mathbb{P}^1\rightarrow \bar B_2$ over a basis $\bar B_2$. The F-theory K3 is 
realized as an elliptic fibration over 
the above $\mathbb{P}^1$, so that $M_4$ becomes an elliptic fibration 
over $B_3$ as well as a  $K3$ fibration over $\bar B_2$, while the 
heterotic Calabi-Yau 3-fold $Z_3$ is an elliptic fibration 
${\cal E}_{het}\rightarrow \bar B_{2}$. This construction still has a 
heterotic limit $\lim_{het} M_4 =P_4 \cup_{Z_3} P_4$, where $P_4$ 
has the fibration structure of $M_4$ with $K3$ replaced by the half 
$K3$, and  the matching of the  compactification data on both sides is 
well understood~\cite{Friedman:1997yq}. This includes a dictionary between  
heterotic moduli and the moduli of $M_n$. In particular the bundle moduli are 
mapped to the complex structure moduli of $M_n$  and the choice of the  
intermediate Jacobian $J_{M_n}=H_3(M_n,\mathbb{R})/H_3(M_n,\mathbb{Z})$. For $n=4$ 
one always has $H_{3,0}(M_4)=0$ and frequently $H_{2,1}(M_4)=0$ 
hence $H_{3}(M_4)=0$, so the intermediate Jacobian becomes either trivial or an 
abelian variety\footnote{In the latter case one can twist it~\cite{Friedman:1997yq} by a 
$G_4$ flux in $H^{2,2}(M_4)$, so that this part of the moduli is described 
by Deligne cohomology of $M_4$.}.  This very geometrical description of  F-theory 
in terms of the fourfold geometry allows to determine the holomorphic terms in the  
four dimensional $\mathcal{N}=1$ effective action. The  easiest example is the 
superpotential, which makes  (heterotic) moduli stabilization possible\footnote{Unfortunately the adiabatic argument requires 
the elliptic fibration structure and large volumes on the heterotic side, which prevents a direct F-theory description of heterotic orbifolds.}. 
The latter is determined by the fourfold periods  and includes 
in particular non-perturbative corrections to the heterotic string and the orientifold theories 
that arise in the limits of F-theory and have been identified 
in~\cite{Grimm:2009ef}-\cite{Grimm:2010gk}  .

Using the M-theory/F-theory lift, it was explained 
in~\cite{Gukov:1999ya} that a (half) integer quantized 
$[x]=[G_4- c_2[M_4]/2]\in H_4(M_4,\mathbb{Z})$, primitive ($J\wedge G_4=0$) flux $G_4=d C_3$ in 
$H_4(M_4,\mathbb{Z})$, which fulfills the tadpole condition 
(\ref{tadpolecondition}) and is compatible with the fibration 
structure, leads to a superpotential 
\begin{equation} 
 W_{cs}({\underline a})=\int_{M_4} 
\Omega({\underline a}) \wedge \frac{G_4}{2 \pi} \ ,
\label{supo} 
\end{equation} 
which fixes the otherwise unobstructed complex structure 
moduli ${\underline a}$ of $M_4$ so that $G_4$ is of Hodge type $(2,2)$, 
which implies that it is self-dual. For arbitrary $G_4$, i.e. if  the primitivity 
condition $J\wedge G_4=0$ does not hold, one can view it as a condition 
enforced on the K\"ahler  moduli ${\underline t}$ by minimizing 
a superpotential 
\begin{equation} 
W_{ks}({\underline  t}) =\int_{M_4}  J^2({\underline t})\wedge\frac{G_4}{4 \pi}\ .
\label{supo2} 
\end{equation} 
Minimization of (\ref{supo},\ref{supo2}) encompasses geometric as well as 
the bundle moduli of the heterotic string. The splitting of the 
superpotential and more importantly the K\"ahler potential into the 
two types of chiral $\mathcal{N}=1$ fields is an approximation, which will be in 
general invalidated by non-perturbative effects. Nevertheless because 
of the positivity of the individual contributions to the $F$-term 
superpotential the minimization of the terms  coming 
(\ref{supo}) and (\ref{supo2}) can reveal the structure of 
the vacuum manifold even in the presence of certain mixings. 
An example for a non-perturbative superpotential that gives 
additional constraints on the  K\"ahler moduli that depend mildly 
on ${\underline a}$ is the non-perturbative superpotential 
that comes from the $M5$ brane wrapping a $D$ in $M_4$, which has 
to have  Dirac index $\chi_D(G)=h^{00}-h^{10}-h^{30}+n(G_4)=1$ with 
$n(0)=h^{2,0}$ and $n(G_4)\le h^{2,0}$~\cite{Witten:1996bn,Donagi:1996yf,Kallosh:2005gs}. 
This superpotential can be mapped to F-theory for vertical divisors and has 
been identified  with the supersymmetry breaking effect due to gluino 
condensation~\cite{Katz:1996th}. This can eventually  provide a 
mechanism for the KKLT uplift. Even if the splitting is only 
mildly broken, the constraints from (\ref{supo}) that depend 
on the primitive part of $G_4$  and the one of (\ref{supo2}) 
that depend on the non-primitive part of $G_4$ can in 
general not be separated, because an (half) integer basis of 
$H^n(M_n,\mathbb{Z})$ correlates elements in $H_{\rm prim}^n(M_n)$ 
and $H_{\rm notprim}^n(M_n)$ with rational coefficients as 
discussed in section~\ref{fluxquantization}.              

While (\ref{supo}) can stabilize complex moduli, one of the most 
challenging problems in F-theory is to find out whether there are  
$G_4$ fluxes that drive the moduli to the degenerations that one needs to 
create those structures suitable for phenomenology, in particular the heterotic string and
more general gauge  symmetry enhancements at codimension one, matter curves and Yukawa 
points at higher codimension in the base e.t.c. An equally important and challenging problem is to fix  
the other moduli without creating unwanted further structures.  For 
illustration consider the universal  K3, with stringy moduli space 
${\cal M}_{K3}= O(4,20,\mathbb{Z})\backslash {O}(4,20)/(O(4)\times O(20))$. While 
type II compactifications on this geometry have generically no gauge  
symmetry enhancement, one could realize  the K3 in many  
ways algebraically, e.g. as section of the anticanonical bundle in 
those  toric varieties  $\mathbb{P}_{\Delta_3}$, which correspond to the $4319$ 
3d reflexive polyhedra. The corresponding complex families frequently 
have a rank $r$  gauge group, because they live in codimension $r$ in the K3 
moduli space specified by the Noether-Lefschetz divisors~\footnote{The 
Noether-Lefschetz divisor of the fibre is preserved in algebraic realizations of 
$K3$ fibred Calabi-Yau manifolds, a fact that was recently used 
to prove the Yau-Zaslow conjecture relating the elliptic  genus of 
the heterotic strings to IIB counting functions of higher genus curves 
in K3  using modular forms~\cite{MR2669707}.}.
However picking the algebraic representation of the K3 in  
$\mathbb{P}_{\Delta_3}$ is ad hoc and one wishes to have a physical
mechanism, e.g. a potential that starting for the universal K3 explains, why the 
physical model lives on a set of measure zero in the moduli space.   For 
three or four dimensional Calabi-Yau spaces there is no universal 
moduli space  in the sense of ${\cal M}_{K3}$, but there are good 
indications that there are huge connected components of universal moduli 
spaces, which can be connected by transitions. E.g. one can argue that 
all  Calabi-Yau 3-folds realized as anticanonical  bundle  in  
$\mathbb{P}_{\Delta_4}$  are connected~\cite{greenekantor} in one
moduli space. For fourfolds from  the  purely geometric point of
view  this is likely to hold for embeddings of $r$ transversal polynomial 
constraints in in $\mathbb{P}_{\Delta_{5+r}}$, which we review in 
sect.~\ref{batyrev}. However $\mathcal{N}=1$ vacua require in general fluxes 
that have to change a least in certain transitions~\cite{Klemm:1996ts,Intriligator:2012ue} 
to fulfill the tadpole cancellation condition. Therefore one can  have 
discrete connected components.  

The most sensible way to address the flux problem is hence to 
start within a generic member of a family Calabi-Yau of manifolds 
in a particular component with no gauge group. Its convenient to 
fix the base $B_3$ of the family, since transitions by blow ups in the base  
of a fourfold were studied already in~\cite{Klemm:1996ts} and first 
classify all elliptic toric hypersurface Calabi-Yau manifolds, 
which have no gauge group and no tensionless  strings. These we 
call the {\sl clean sheet fibrations}.  We first 
classify all those basis, which are $\mathbb{P}^1$ fibrations over 
a  toric bases $\bar B_2$ in section~\ref{heteroticbasis}. The 
interest in these models is that they have heterotic duals 
with completely higgsed gauge bundle over the same basis $\bar B_2$.
These models  are all members of a class geometries, for which the basis 
of the elliptic fibration is a toric projective space $\mathbb{P}_{\Delta_3}$ associated to 
the before mentioned $4319$ three dimensional reflexive polyhedra. The 
corresponding classification of properties of the clean sheet models is 
done in section~\ref{toricfanobases}. We then search for fluxes, which can drive 
the family to gauge theory limits and create further structures at 
higher codimension. We call this process landscaping by fluxes.  To implement 
it concretely it we use Batyrev's approach to mirror symmetry, which constructs 
mirror pairs of hypersurfaces and complete intersections in 
toric  ambient spaces. The reason for staying in this manifestly  mirror 
symmetric class is that  structures implied by {\sl homological mirror symmetry} 
are very  useful to get a basis for $H_{\rm  prim}^n(M_n)$ and the periods.
The bulk of the investigation focusses on the question, how to pick the part of $G_4$ in 
$H_{\rm prim}^n(M_n)$ that does the landscaping to the desired 
configurations in the complex moduli space. The conditions to find 
an ${\cal N}=1$ supersymmetric flux vacuum are reviewed 
in sect. \ref{minimizingW}. On an ${\cal N}=2$  Calabi-Yau background 
with fluxes they are equivalent to find solutions to the attractor 
equations for supersymmetric black holes~\cite{Curio:2000sc}. The study of the minima of  
flux superpotentials requires to find special points in the period domain of Calabi-Yau spaces a  
problem related in a very fascinating way to  arithmetic and number 
theory~\cite{Moore:1998pn}.

In the second part of the paper we study a degeneration of the 
complex structure, which  is phenomenologically motivated.                     
$SU(5)$ GUT model building with D branes and         
orientifold-branes has perturbatively no ${\bf 5}_{H_u} {\bf 10} 
\, \, \, {\bf 10}$ coupling, which would give the required order $1$ 
Yukawa coupling for the top quark, a fact which is 
remedied in $F$ theory when one starts with an 
$E_{k\ge 6}$ symmetry enhancement in complex 
codimension three over the base. It has been further 
argued that a hierarchy in the Cabibbo-Kobayashi-Maskawa 
(CKM) quark mixing matrix leads naturally to $k\ge 7$, while 
the corresponding  hierarchy in Pontecorvo-Maki-Nakagawa-Sakata 
(PMNS) matrix for the lepton sector and an 
phenomenological acceptable $\mu$ terms favors 
$k=8$\cite{Beasley:2008dc,Beasley:2008kw,Heckman:2009mn}.  

Gauge symmetry enhancement in the 4d part of the 7-brane 
with gauge group $G$ occurs at divisors $D_g$ in $B_3$ 
wrapped by the 7-brane --- for the decoupling scenario 
chosen to be the above $\bar B_2$ --- over which the elliptic fibre 
degenerates  to ${\cal E}_{C_{\hat g}}$: A collection of vanishing  
$\mathbb{P}^1$, whose intersection is given by the negative 
of Cartan-matrix $C_{\hat g}$ of the corresponding affine Lie algebra 
$\hat g$ associated to $G$. This yields an elliptic singularity 
of the Kodaira type labelled by ${\hat g}$ (also the $\hat{}$ 
is usually dropped). These singularities can be 
classified on complex co-dimension one loci over the base by 
the vanishing order of the Weierstrass data \footnote{There are some cases in which one needs
as an extra condition the factorization of a polynomial to cause the singularity, those are
the split cases \cite{Bershadsky:1996nh}.}  of the elliptic fibration 
$M_4$ at $D_g$ as made explicit below. In order to capture in 
addition the monodromy data for elliptic singularities in 
codimension 2 in the base, which acts as an {\sl outer automorphism} on the 
$\mathbb{P}^1$ in ${\cal E}_{C_{\hat g}}$, which correspond to the simple 
roots, to yield non-simply laced Lie groups, one needs information encoded, for simple fibre 
types  in the Tate  form or analogous normal forms of the elliptic fibre.        

The matter spectrum and its representations is 
determined from enhancements of the elliptic fibre 
singularity ${\cal E}_{C_{\hat g}}\rightarrow 
{\cal E}_{\hat C''}$ over a co-dimension two (matter)-curve 
$\Sigma_M=D_g\cap D_{g'}$, or $\Sigma_M\subset D_g$ if $D_g$ 
is not smooth, in the base. The chiral spectrum is determined by 
the gauge symmetry breaking  $G_4$-flux $G_4^b$, in particular 
the chiral index is given by  
\begin{equation} 
\chi_+-\chi_-=\int_{\Sigma_M} i^*(G_4^b)\  .
\label{chirality} 
\end{equation}

The tree level Yukawa couplings are  related to a further 
enhancement of the elliptic fibre ${\cal E}_{\hat C_{g''}}\rightarrow 
{\cal E}_{\hat C'''}$ over a co-dimension three (Yukawa)-point 
$P=D_g\cap D_{g'} \cap D_{g'''}$.

Commonly it was assumed in the phenomenological analysis of 
F-theory that $C''$ and $C'''$ are given by the group 
theory expectations and that this can be confirmed by the 
Tate data of $M_4$. However a direct resolution of the singularities 
at the claimed codimension two and three  where the enhancements 
to the $E_6$ exceptional group was claimed shows a picture which 
contradicts these expectations \cite{Esole:2011sm}. 

We analyze this mismatch and the physical 
consequences of the actual resolution for the $E_8$ point.
First we discuss the $SU(5)$ model, which embedded into 
F-theory has attractive phenomenological features. 
We then describe the complex structure to obtain 
the $E_8$ point and proceed with a detailed analysis 
of the  resolutions  process in increasing codimension.

In such general case, the singularities enhancements can be 
resolved via small resolutions as done in~\cite{Esole:2011sm} and further 
explored in~\cite{Marsano:2011hv}.

\section{F-theory and Calabi-Yau $n$-fold geometry}
In this section we review the general construction of F-theory 
and prepare for the compactifications with gauge symmetry 
enhancements over divisors in the base using Kodaira's and 
Tate's algorithms. We then argue that this gauge symmetry 
enhancement is induced by the $F$-terms of $G_4$ fluxes 
and describe the realization of that mechanism for toric 
Calabi-Yau manifolds using general arguments about the symmetries 
in the complex moduli space, the constraints of Griffiths transversality,
the  Frobenius structure and some general properties of the period map.

\subsection{The F-theory `map` and its perturbative limits} 
\label{Ftheorymap}  
Every family of elliptic curves can be written in the Weierstrass form, which 
reads in affine complex coordinates $x,y$ as 
\begin{equation}\label{Weierstrass}
y^2 - 4 x^3 + f x + g =0 ,
\end{equation} 
where $\Delta=f^3-27 g^2$ is the discriminant of the curve. 
The complex structure of the elliptic curve $\tau$ is related 
to $f$ and $g$ by 
\be 
j(q)=12^3 \frac{f^3}{\Delta},
\label{j-function}
\ee  
where $q=\exp(2 \pi i \tau)$ and $j(q)=(12 E_4)^3/(E_4^3-E_6^2)=\frac{1}{q}+744+196884 q+\ldots$. 
For $k\in 2 \mathbb{N}_+$  
\be 
E_k=\frac{1}{2\zeta(k)}\sum_{n,m\in \mathbb{Z} \atop (n,m)\neq (0,0)} \frac{1}{(m \tau +n)^k}=1+\frac{(2 \pi i)^k}{(k-1)! \zeta(k)}\sum_{n=1}^\infty \sigma_{k-1}(n)q^n\, ,
\label{eisensteinseries}
\ee 
are normalized (and regularized by the second equal sign for $k=2$) Eisenstein series, with 
$\sigma_k(n)$ the sum of the $k$-th power of the positive divisors of $n$ and $\zeta(k)=\sum_{r\ge 0}1/r^k$, which equals 
$-(2 \pi i)^k B_k/(2 k (k-1)!)$ for $k\in 2 \mathbb{N}_+$, with  
$\sum_{k=0}^\infty B_k x^k/k!\, :=x/(e^x-1)$.                     

For a family of elliptic curves $f$ and $g$ depend on one complex modulus, 
but in a fibration they depend on the coordinates $\underline{{\bf u}}$ of 
the base $B_{n-1}$ and of the complex moduli $\underline{{\bf a}}$ of $M_n$, so
that (\ref{j-function}) yields a `map` from $(\underline{{\bf u}},\underline{{\bf a}})$ to the 
axio-dilaton $\tau$. The complex moduli of $M_4$, typically in the order of thousands for compact 
fourfolds~\cite{Klemm:1996ts} describe the complex moduli of $B_{n-1}$, 
but mostly the profile of $j$ over $B_{n-1}$. Like $j$ is compactified 
in $\mathbb{P}^1$ with special points,  the full complex moduli space parametrized by $\underline{{\bf a}} \in {\cal M}_{cs}$ is compact 
with normal crossing divisors.  Eq. (\ref{j-function}) defines the `map` up to a ${\rm PSL}(2,\mathbb{Z})$ action on $\tau$, 
as $j(\tau)$ is invariant under $\tau \mapsto \tau_\gamma=\frac{a \tau + b}{c \tau+ d}$ with 
$\gamma={\tiny \left(\begin{array}{cc} a&b \\ c &d \end{array}\right)\in {\rm SL}(2,\mathbb{Z})}$, 
which follows from the definition of $j$ and the first identity 
in (\ref{eisensteinseries}) that obviously implies 
$E_k( \tau_\gamma)=(c \tau +d)^{k} E_k(\tau)$ for\footnote{For $k=2$ 
the second equality is taken as a regularization description 
for the sum in the first terms, which breaks the modular 
transformations and lead to almost modular forms.}  $k>2$. This 
corresponds to an ambiguity in the choice of a type IIB duality 
frame. One cannot chose globally a weak coupling 
duality frame, as $\tau$ undergoes monodromies over paths in 
$B_{n-1}$ around $\Delta$, which generate a finite index subgroup 
$\Gamma_M\in {\rm PSL}(2,\mathbb{Z})$, which does not preserve 
the weak coupling choice $\tau\sim i \infty$ ($1/j\sim q=0$). This means that F-theory 
necessarily contains non-perturbative physics. 

The closest one gets to a perturbative description is to consider limits in $\underline{{\bf a}}$, referred to as weak 
coupling limits, where the profile of $j$ is such that $1/j\sim0$ 
almost everywhere over $B_{n-1}$. Depending on the global limiting profile, 
near the points where this fails, some of the complex structure moduli 
$\underline{\bf a}$ can be understood as moduli of seven 
branes in orientifolds~\cite{Sen:1996vd} or as the moduli of the heterotic 
string~\cite{Friedman:1997yq}.  By analyzing these 
perturbative limits and identifying  carefully the map between the F-theory 
parameters and the parameters of the perturbative theory one can learn 
about the non-perturbative corrections. For example by distinguishing between 
the brane and the bulk moduli one can learn about the disk-instanton corrections of the 
brane~\cite{Grimm:2009ef}\cite{Grimm:2008dq}\cite{Alim:2009bx}\cite{Alim:2011rp}        
Non-perturbative corrections in the heterotic string have been identified in~\cite{Jockers:2009ti}\cite{Grimm:2009sy}. 
By a  tentative identification of the dilaton, D-Instanton  corrections can be 
isolated ~\cite{Jockers:2009ti}.

Picking the correct field basis and the   integer flux basis is essential for the  {\sl anomaly} 
cancellation mechanisms in the perturbative theory in four dimension. An interesting  
example is the central charge formula for the $D$-branes (\ref{Abasis})  that directly  
reflects the  properties of the integrals basis. It is almost the one that governs the anomaly 
inflow for the flat branes~\cite{ Minasian:1997mm}, but the replacement of the  square 
root of the A-roof genus by the $\hat \Gamma$ class induces  corrections that affect 
the anomaly  inflow mechanism in perturbative limits of global F-theory compactifications.          
Many  of the four dimensional  conditions  will be traced back to~(\ref{tadpolecondition1}) 
and one would expect all local limits fulfilling the global tadpole cancellation to 
be consistent four dimensional theories.   That the integral basis  is crucial in the analysis can be also seen from 
the chirality index (\ref{chirality}) depends on the integrality of the cohomology 
of the  flux basis and the dual cycles.  On the other hand understanding the 4d  and 
3d anomaly cancellations mechanism better, give new  methods to fix the integral 
basis and insights in the geometry of fourfolds.         

The map $(\underline{{\bf u}},\underline{{\bf a}})$  to $j$ will have ramifications points, 
which lead to monodromies acting the cohomology 
of the fibre along paths in $(\underline{{\bf u}},\underline{{\bf a}})$ . The Galois group of the map
is crucial for the global fibration structure and the physical spectrum. In particular it determines 
the number of sections of the elliptic fibration and the question, which $\Gamma_M$ and which subgroup 
$\Gamma={\rm PSL}(2,\mathbb{Z})/\Gamma_M$ is realized in the theory~\cite{Bershadsky:1998vn,Berglund:1998va}.

\subsection{Compactifying the Weierstrass model}  
\label{compactification} 
In compactifying (\ref{Weierstrass}) over $B_{n-1}$ to $M_n$ the triviality of the canonical 
bundle $K_{M_n}=0$ requires, at least for a suitable choice 
of a birrational model~\cite{Ka1,Grassi1}, 
\be 
K_{B_{n-1}}= -\sum a_g [D_g]\, , 
\ee 
where $a_g>0$ depends on the Kodaira type of the 
singular fibre given in table A.1. 

One convenient method to construct compact algebraic Calabi-Yau 
fourfolds with tunable gauge theory enhancements is 
to construct first an ambient space as a fibration of a 
projective toric variety $\mathbb{P}_{\Delta^{F}}$  over a
projective toric variety $B_{n-1}=\mathbb{P}_{\Delta^{B}}$ and 
as a second step the algebraic Calabi-Yau manifolds in 
this ambient space.

A special role in the construction of smooth elliptic fibrations is 
played by Fano varieties $X$, which are smooth projective varieties with 
ample anticanonical divisor $-K_X$. 
By Kleinman's criterium~\cite{MR0206009} a divisor is ample if it is in the 
interior of the cone spanned by the numerically effective divisors 
or equivalently its intersection with all numerically effective curves is positive. 
Physically it is sensible to include also smooth projective varieties 
as basis $B_{n-1}$ for which $-K_{B_{n-1}}$ intersects only 
semi-positive on the numerically effective curves. 
This is a generalization of the notion in~\cite{Friedman:1997yq}, 
where it was used to include the resolution of del 
Pezzo surfaces with ADE singularities. The resolution 
introduce rational $-2$ curves $C$ as exceptional divisors, 
for which the adjunction formula $(K+C)\cdot C=2g-2$ 
implies $K\cdot C=0$. In the 2d  toric case these are 
only $A_k$ singularities and correspond to the points on the edges 
of the 2d polytopes in Fig. 1. As in~\cite{Friedman:1997yq} 
we call the corresponding varieties {\sl almost} 
Fano varieties even though following notion of 
pseudo ampleness~\cite{MR0206009} pseudo Fano might be 
more appropriate.  

If $B_{n-1}$ is a toric almost Fano basis in this sense  
one can construct a smooth elliptic fibration. This is 
because $[\Delta]=-12 K_{B_{n-1}}$, $[f]=- 4 K_{B_{n-1}}$ and 
$[g]=- 6 K_{B_{n-1}}$ and by the semipositivity the 
generic discriminant $\Delta$ vanishes at a divisor 
$D$ in $B_{n-1}$ at most with  order one and since 
$[f]$ and $[g]$ can then not vanish both at $D$ one gets according 
to table A.1 at most an $I_1$ singularity in codimension one. 
The existence of the positive support function $\phi$,  
described in section~\ref{batyrev}, guarantees that the 
basis $B_{n-1}=\mathbb{P}_{\Delta^{B}}$, where $\Delta^{B}$ 
is reflexive, are  {\sl toric almost Fano varieties}.

One gets different fibre types by constructing
the generic smooth fibre as the anticanonical divisor 
in a 2d toric almost Fano variety corresponding to a 
reflexive polyhedron. This leads to the generic 
$E_6,E_7$ and $E_8$ fibre types and to non-generic cases 
of $E_{n<6}$ fibre types. $E_n$ fibres give rise to models with $k=9-n$ 
sections and $U(1)^k$ global gauge symmetry.  A framework to 
describe the compact Calabi-Yau manifolds is again as the anticanonical divisors  
(or suitable complete intersections) in toric ambient spaces 
related to reflexive polyhedra discussed in section~\ref{batyrev}. 

The corresponding F-theory compactifications have no non-abelian gauge 
group or equivalently within a given fibre type the 
most generic Higgs bundle over the base. Then in further 
steps one  enforces by specialization the complex structure
singular fibres along gauge divisors, the matter curves 
and the Yukawa points.

\subsection{General global properties of Calabi-Yau fourfolds and flux quantization}   
\label{fluxquantization}  

In this chapter we discuss general global properties of compact Calabi-Yau 
manifolds $M_n$, mainly for complex dimension $n=4$.  
We refer already to some results of section~\ref{globalfiber}, which 
is devoted to global properties that do depend on the fibration structure.          

Let $\chi_q=\sum_{p=0}^{{\rm dim}(M_n)} (-1)^p h_{p,q}$ be the arithmetic genera.  
Then we have from the Hirzebruch-Riemann-Roch (H-R-R) theorem~\cite{Hirzebruch} for 
$B_3$ a rational surface, i.e. with first arithmetic genus $\chi_0=h_{0,0}=1$, that  
\be 
1=\chi_0=\frac{1}{24}\int_{B_3} c_1 c_2 \ .
\label{chi0} 
\ee
For a Calabi-Yau fourfold with first arithmetic genus $\chi_0=h_{0,0}+h_{4,0}=2$ the H-R-R theorem gives 
\be 
2=\chi_0=\frac{1}{720}\int_{M_4}(3 c_2^2-c_4),\quad \chi_1=\frac{1}{180}\int_{M_4}(3 c_2^2-31 c_4),
\quad \chi_2=\frac{1}{120}\int_{M_4}(3 c_2^2+79 c_4),  
\label{hrr} 
\ee
which yields 
\begin{equation} 
\begin{array}{rl}
h^{2,2}=&2(22 + 2 h^{1,1}+ 2 h^{3,1}- h^{2,1})\, ,\\
\chi(M_4)=&6(8 + h^{1,1}+ h^{3,1}- h^{2,1}) \ . 
\end{array} 
\label{index} 
\end{equation}
I.e. to infer the full Hodge diamond of a fourfold $M_4$, we can use its Euler number 
and the Hodge numbers $h^{n-1,1}$, $h^{1,1}$. $H^{4}(M_4)$  splits into a 
selfdual ($*\alpha =+\alpha$) and an anti-selfdual ($*\alpha=-\alpha$) subspace
\begin{equation}
H^{4}(M_4)= H^{4}_+(M_4)\oplus H^{4}_-(M_4), 
\end{equation}
whose signature follows from the Hirzebruch signature theorem~\cite{Hirzebruch} as 
\begin{equation} 
\sigma = h^{4}_+(M_4)-h^{4}_-(M_4)=
\int_{M_4} L_2=\frac{\chi}{3}+32\ ,
\label{signature} 
\end{equation}
where the second $L$ polynomial is $L_2=\frac{1}{45} (7 p_2- p_1^2)$ 
with the Pontryagin classes $p_1=c_1^2- 2 c_2$ and $p_2=c_2^2- 2 c_1 c_3+ 2 c_4$. 
$H^{4}(M_4)$ also splits into an horizontal part generated by 
${\underline \nabla}_{\underline a}^{|\underline a |}\Omega_n$, see section~\ref{KSRST}, 
which is  primitive $J\wedge \alpha_{\rm prim}=0$, and a vertical part generated by Lefschetz $SL(2)$ 
action with $J$ as the raising operator~\cite{MR1288523}. For fixed moduli the dual cycles can 
be calibrated symplectically with ${\rm Re}(e^{i\theta} \Omega_n)$ or holomorphically with 
$\frac{J^n}{n!}$. In addition there can be {\sl Cayley cycles} calibrated with 
$\frac{1}{2} J^2+ {\rm Re} (e^{i\theta} \Omega_n)$   on fourfolds, see~\cite{MR2103716} for a review 
on calibrated geometries.   These  structures are  exchanged by 
mirror symmetry. The only vertical part sits in $H^{2,2}(M_4)=H^{2,2}_{H}(M_4)
\oplus H^{2,2}_{V}(M_4)$. On the middle dimensional cohomology
one has a bilinear form $Q:H^n\otimes H^n\rightarrow \mathbb{C}$, defined by 
\begin{equation} 
Q(\alpha,\beta)=\int_{M_n} \alpha\wedge \beta
\label{bilinear} 
\end{equation}  
with the property
\begin{equation} 
Q(H^{p,q},H^{r,s})=0 \quad {\rm unless} \ \ p=s \ {\rm and} \ q=r\, ,
\label{tranversality} 
\end{equation}
and for primitive forms in the middle cohomology $\alpha\in H^{p,q}$ the positivity of the {\sl real structure}  
\begin{equation} 
R(\alpha,\alpha)= i^{p-q} Q(\alpha , \bar \alpha) >0 \ ,
\label{realstructure}
\end{equation} 
equips $H^n(M_n)$ with a {\sl polarized Hodge structure}.  
For $n=3$ or more generally odd all $\alpha$ are primitive and $Q$ becomes the  
familiar symplectic pairing~\footnote{For $n=3$ one has $(-i,i,-i,i)$ on 
$(H_{3,0},H_{2,1},H_{1,2},H_{0,3})$. This leads the definition of ${\cal N}_{\Sigma\Lambda}$ 
to make the graviphoton kinetic term in $N=2$  supergravity positive, see~\cite{Louis:1996ya} for a review. For $n=2$, i.e. K3 
one has from the HST with $L_1=\frac{1}{3}(c_1^2-2 c_2)$ that $\sigma =\int_{K3} L_1=-16$ 
and $-1,1,-1$ on $(H_{2,0},H^{\rm prim}_{1,1},H_{0,2})$, and $H_2(K3,\mathbb{Z})=H^{\oplus 3}\oplus E_8^{\oplus 2}(-1)$.}, 
while in the $n=4$ case one gets the following signature eigenspaces
\begin{equation} 
\begin{array}{ccccccccccc}
H^{4,0}&\oplus& H^{3,1} &\oplus &(H_{\rm prim}^{2,2}, J^2)&\oplus& H_{\rm notprim}^{2,2}&\oplus& H^{1,3} & \oplus& H^{0,4}\\
+ & & - & & + & & -& & -& & +
\end{array}
\end{equation}
Here the $h^{1,1}-1$ not primitive $(2,2)$-forms $J\wedge \alpha_{\rm notprim}\neq 0$ are 
generated as $J\wedge H_{\rm prim}^{1,1}$~\cite{MR1288523}.

Physically, one has from the equation of motion for the $G_4$-flux 
on $M_4$
\begin{equation}
d*G_4=\frac{1}{2} G_4\wedge G_4- I_8(R)+ \sum_i \delta^{(8)}_i Q^i_{2} +\sum_i T^3\wedge\delta^{(5)}_i Q^i_{5},   
\label{tadpolecondition1}       
\end{equation}
which implies ($24 \int_{M_4} I_8(R)=\chi(M_4)$) the global tadpole condition
\begin{equation} 
\label{tadpolecondition} 
\frac{1}{2} \int_{M_4} G_4\wedge G_4 + N_{M2} =\frac{\chi(M_n)}{24} \ ,          
\end{equation}
 where we exclude the $M_5$ branes   
This can be lifted to F-theory turning the $M2$- to $D3$-branes. Of course consistency 
requires $N_{M2}=N_{D3}\in \mathbb{Z}$. Let us summarize some facts:

\begin{itemize} 
\item i) The second equation in~(\ref{index}) implies $\chi(M_4)= 0\ {\rm mod}\ 6$. 
One finds $\chi(M_4)\neq  0\ {\rm mod}\ 24$ for roughly a fourth of the Calabi-Yau 
fourfolds in weighted projective space~\cite{Klemm:1996ts} independently 
of the fibration structure. These cases {\sl require} half integer fluxes.     
\item ii) One finds that the clean sheet models over almost Fano basis, i.e. 
without non-abelian gauge groups, have always $\chi(M_4)=0\ {\rm mod}\ 24$ and require no fluxes. 
This follows for the $E_6-E_8$ fibre type from (\ref{chi0},\ref{euler4}), while for 
the $D_5$ fibre type one has to invoke in addition (\ref{c1h3}) for $n=3$.           
\item iii) On the other hand Wu's formula $[x]^2=c_2\wedge [x]\ {\rm mod}\ 2$ and the first 
equation in (\ref{hrr}) implies that any flux obeying 
$[x]=[G_4- c_2/2]\in H_4(M_4,\mathbb{Z})$ leads to  $N_{D3}\in \mathbb{Z}$~\cite{Witten:1996md}. 
\item iv) $H_4(M_4,\mathbb{Z})\sim H^4(M_4,\mathbb{Z})$ is {\sl unimodular} by the 
Poincar\'e pairing. Hence if $c_2$ is even, i.e. $c_2= 2 y$ with $y\in H_4(M_4,\mathbb{Z})$, then $H^4(M_4,\mathbb{Z})$ 
is an even unimodular lattice. Even unimodular lattices with trivial signature exist only in 
dimension $d=0\ {\rm mod}\ 8$. Moreover the negative eigenvalues in $H^4(M_4,\mathbb{Z})$ 
pair with positive ones into hyperbolic rank two lattices 
$H=\left(\begin{array}{cc} 0&1 \\ 1&0 \end{array}\right)$~\cite{MR1662447} 
and the others form the even selfdual unimodular $E_8$ lattices so that
 \begin{equation}
 H_4(M_4,\mathbb{Z})=H^{\oplus m}\oplus E_8^{\oplus k}\ ,  
\end{equation}
with $m=(b_4-\sigma)/2$. One corollary for even lattices using (\ref{signature}) 
is that $\sigma=0\ {\rm mod}\ 8$ so that $\chi(M_n)=0\ {\rm mod}\ 24$ and $k=\sigma/8$~\cite{Klemm:1996ts}. 
This is true for the clean sheet models. If $H^4(M_4,\mathbb{Z})$ is not even, it is 
diagonalizable w.r.t. to the bilinear form $Q$ over the integers~\cite{MR1662447}.    
\item v) Using (\ref{relation}), Table 3.2 and explicit calculations of the base one 
can easily determine whether $c_2$ is even for the clean sheet models. 
Example calculations for certain  basis given in (\ref{P1overP2c2},\ref{P1overP1P1c2}) 
show that $c_2$ is even. 
\item vi) It was claimed for the $E_8$ fibre in~\cite{Collinucci:2010gz} that this follows from the expression 
for $ c_2(M_4)$ in terms of base and fibre classes, see Table 3.2  as well as general properties of  
almost Fano basis. The expression for $c_2(M_4)$ for the $E_7$ , $E_6$ and the $D_5$ 
fibres in  Table 3.2 show that the argument extends to these fibres types. This would imply
a remarkable abundance of $E_8$ lattice factors in this class of models.  It would be interesting 
to check it for fibre types with more than four sections.   
\item vii) By the Lefshetz theorem every element in $H^2(M_n,\mathbb{Z})\cap H^{1,1}(M_n)$ 
is represented by an algebraic cycle  specified by the first Chern class of a divisor. However 
for fourfolds is it not know wether $H^4(M_4,\mathbb{Q})\cap H^{2,2}(M_2)$ is 
representable by an algebraic cycle, but suggested by the Hodge Conjecture. In this 
paper we work only with cycles that come from the ambient space and for which the 
HC  holds. The question could be more interesting for the twisted homology elements and the 
Caley cycles.   
\item viii)  Mathematically an integral $G_4$ flux is represented\footnote{On smooth closed 
fourfold we can work with the naive definition given in the introduction. However on singular 
fourfolds the differential cohomology plays a r\^ole.}    by  a 
class in {\sl differential cohomology} given by a {\sl closed} closed differential 
cochain of degree 4 $(\alpha,C,\frac{G_4}{2 \pi})$ where $\alpha\in C^4(M_4,\mathbb{Z})$ is an 
integral 4-cocycle, $C\in C^3(M,\mathbb{R})$ is 3-cochain, $\frac{G_4}{2 \pi}\in 
\Omega^4(M_4)$ is a 4-form and  the boundary operator $d^2=0$ 
is given by $d(\alpha,C,\frac{G_4}{2 \pi} )= (\delta \alpha, \frac{G_4}{2\pi} -
\alpha- \delta C,  \frac{d G_4}{2\pi} )$, while a half integral flux is represented by a 
twisted  closed differential cochain, see~\cite{HS} and references therein.                       
\item ix)  In general not all relevant (half) integer homology classes that support fluxes can lie 
either entirely in $H^{2,2}_{H}(M_4)$ or in $H^{2,2}_{V}(M_4)$.   E.g. for the clean sheet 
models discussed in iv) and (\ref{P1overP2c2},\ref{P1overP1P1c2}) and eventually cases with 
small gauge groups~\cite{Collinucci:2010gz}  $c_2$ is be even, while $H_V^n(M_n,\mathbb{Z})$ 
is small  and has neither the right signature nor dimension to support an even self-dual lattice. 
Therefore one cannot turn on an integer flux on $H_V^n(M_n)$ or $H_H^n(M_n)$ separately.
Similar arguments can be  found  in~\cite{Intriligator:2012ue}.
\item x) For negative Euler number, which exists for fourfolds~\cite{Klemm:1996ts},
supersymmetry is broken due anti $D3$ branes.
\end{itemize}

\subsection{Batyrev's construction with fibration structures}  
 In this section we describe the construction of Calabi-Yau mirror pairs with elliptic 
fibrations in toric ambient spaces.  Starting with~\cite{Klemm:1996ts}  most 
compact  F-theory compactifications to 4d with different fibre types are based 
on  the mirror symmetric construction of Batyrev, as well as Batyrev and Borisov for the 
complete intersection case, with reflexive polyhedra. To extract the information 
encoded in homological mirror symmetry to find the integral basis, the description of the moduli spaces, the induced symmetries 
on it, the Picard-Fuchs ideal and its restrictions the construction is very 
useful and only in section \ref{hemisphere} we come to a more 
general construction, which can yield similar information.       

\subsubsection{Calabi-Yau mirror pairs in toric ambient spaces}   
\label{batyrev}
First we recall the construction of Calabi-Yau hypersurfaces and complete intersections  
$W_n$ and its mirrors $M_n$ in toric ambient spaces due to Batyrev~\cite{Batyrev:1994hm}. $W_n$ ($M_n$) 
is defined as a complete intersection of $r$ suitable Cartier divisors 
in the toric almost Fano variety $\mathbb{P}_{\nabla_{n+r}}$ ($\mathbb{P}_{\Delta^*_{n+r}}$) 
given by reflexive lattice polyhedra $\nabla_{n+r}$ ($\Delta^*_{n+r}$). 
In the simplest case $r=1$~\cite{Batyrev:1994hm} $W_n$ ($M_n$) is defined as 
the hypersurface represented by a generically smooth section of the ample 
anticanonical bundle of $\mathbb{P}_{\Delta_{n+1}}$ ($\mathbb{P}_{\Delta^*_{n+1}}$), 
defined from a reflexive pair of lattice polyhedra 
$(\Delta_{n+1},\Delta^*_{n+1})$ as in (\ref{PDelta}).    

A $d$ dimensional polyhedron $\Delta_d$ is a lattice polyhedron if 
$\Delta_d\subset\Gamma_{\mathbb{R}}$, where $\Gamma_{\mathbb{R}}$  is 
the real extension of a $d$ dimensional 
lattice $\Gamma$ and $\Delta_d$ is spanned by lattice points. 
Reflexivity means that $\Delta_d^*:=\{x \in \Gamma^*_{\mathbb{R}}| 
\langle x, y\rangle\ge -1, \forall y\in \Delta_d\}$ 
is a lattice polyhedron in the dual lattice. Note $(\Delta_d^*)^*=\Delta_d$ 
and the origin $\nu_0$ ($\nu^*_0$) is the only inner\footnote{I.e. lying in codimension $d$ 
inside $\Delta_d$. There are many points inside bounding faces of codimension $1,\ldots,d-1$ 
in $\Delta_d$.} point in $\Delta_d$ ($\Delta_d^*$). In this case $\Delta_d$ ($\Delta_d^*$) 
define fans~\cite{Fulton,CoxKatz,MR2810322} $\Sigma$ ($\Sigma^*$), in particular the one 
dimensional cones $\Sigma(1)\subset \Sigma$ are spans by the points 
$\{\nu_i,i=1,..,|\Delta_d|-1\}= \{\Delta_d\cap \Gamma\}\setminus \{\nu_0\}$.     

To give the ambient space the required fibration (the mathematical 
condition is spelled out in section~\ref{elliptictwistingdata}) such that  
the embedded Calabi-Yau manifold has an elliptic fibration with a 
holomorphic section, we combine a base polyhedron $\Delta^{B*}$ and a reflexive 
fibre polyhedron $\Delta^{F*}$ into the polyhedron $\Delta^*_{n+r}$, $n\ge 1$ as follows 
\begin{equation} 
  \footnotesize 
  \begin{array}{|ccc|ccc|} 
   \multicolumn{3}{c}{ \nu^*_i\in \Delta_{n+r}^*} &\multicolumn{3}{c}{ \nu_j\in \Delta_{n+r}}  \\ 
    &            &\nu_i^{F*} &             &\nu_j^{F} & \\
    & \Delta^{B*}_{n+r-s-2} & \vdots                &s_{ij}\Delta^{B}_{n+r-s-2}&\vdots & \\
    &            &\nu_i^{F*} &                   & \nu_j^{F}& \\
    & 0 \ldots 0      &                       & 0\dots 0              &                          & \\
    & \vdots     &\Delta^{*F}_{2+s}             &   \vdots          & \Delta^{F}_{2+s}            & \\
    & 0 \ldots 0      &                       & 0\ldots 0              &                          & \\
  \end{array} \, .
\label{polyhedrafrombaseandfibre}
\end{equation} 
This describes a reflexive pair $(\Delta_{n+r}\subset \Gamma_{\mathbb{R}}, \Delta^*_{n+r}
\subset \Gamma^*_{\mathbb{R}})$ given by the complex hull of the points specified in 
(\ref{polyhedrafrombaseandfibre}). Here we defined $s_{ij}= \langle \nu_i^F,\nu_j^{F*} \rangle+1\in \mathbb{N}$ 
and scaled $\Delta^B \rightarrow s_{ij}\Delta^{B}$. If  $\Delta^{*F}$ and $\Delta^{*B}$ 
are reflexive then $\Delta^*_{n+r}$ is reflexive. For $\Delta^{*F}$ reflexivity 
is required by the desired elliptic fibration structure. For $\Delta^{*B}_{n-1}$ 
it is not a necessary condition, see~\cite{Huang:2013yta} for more details 
on this construction. In most of our general discussion of the moduli space we focus 
on the hypersurface case $r=1$ and $s=0$ (one has always $s<r$), however in section \ref{fourfoldbasis} we find the integral 
monodromy basis also for seven complete intersections and the formulas in sections \ref{gaussmaninS}, \ref{Gammaclass} and \ref{hemisphere}  apply to hypersurfaces as well as to  complete intersections.     

\begin{itemize} 
\item  In the {\sl hypersurface case} $r=1$, $\Delta^*_{n+1}:=\Delta^*$ and the family of elliptically 
fibred Calabi-Yau\footnote{For $n=1$ the basis is trivial and we just get the elliptic 
mirror of the local del Pezzo discussed in~\cite{Huang:2013yta}.} 
$M_n$ is given as a section of the anticanonical bundle  $K=\sum_{i} D_i$ of $\mathbb{P}_{\Delta^*}$. $K$ is an ample Cartier divisor whose 
upper convex piecewise linear support function $\phi:\Gamma_{\mathbb{R}}\rightarrow \mathbb{R}$  
in $\Sigma(\Delta^*)$ is simply defined by the supporting polyhedron $\Delta$. 
A generic section in ${\cal O}(K)$ can be written in the coordinate ring 
$\{Y_k\}$ of $\mathbb{P}_{\Delta^*}$ as polynomial constraint     
\be 
W_\Delta=\sum_{\nu_i \in \Delta}a_i \prod_{\nu^*_k\in\Delta^*} Y_k^{\langle \nu_i, 
\nu^*_k\rangle+1}=: \sum_{\nu_i\in \Delta}a_i  M_i(Y)=0
\label{WDelta}
\ee  
and the mirror is defined, by exchanging ${\Delta^*}$ against ${\Delta}$ 
in (\ref{WDelta}). 
\item
For the {\sl complete intersections case} one needs $r$ semi-ample Cartier divisors corresponding 
to $r$ upper convex piecewise linear support functions $\phi_l$, which define a 
nef-partition $E=E_1\cup\dots\cup E_r$ of the vertices $\rho^*$  of $\Delta^*:=\Delta^*_{n+r}$ 
into disjoint subsets $E_1,\dots,E_r$ as
\be
\phi_l(\rho^*)=
\begin{cases}
1& \text{if $\rho^*\in E_l$,}\\
0& \text{otherwise.}
\end{cases}
\ee
Each $\phi_l$ defines a semi--ample Cartier divisor $D_{0,l}=\sum_{\rho^*\in E_l} D_{\rho^*}$
on $\mathbb{P}_{\Delta^*}$, where  $D_{\rho^*}$ is the divisor corresponding  
to the vertex $\rho^*\in E_l$. The family of Calabi-Yau manifolds $M_n$ is given as  a
complete intersection $M_n=D_{0,1}\cap\dots\cap D_{0,r}$  of codimension $r$ in $\mathbb{P}_{\Delta^*}$.
Each $\phi_l$ defines the lattice polyhedron $\Delta_l$ as $\Delta_l=\{x\in \Gamma_\mathbb{R}:(x,y)\geq -
\phi_l(y)\ \forall\ y\in \Gamma^*_\mathbb{R}\}$. These $\Delta_l$ support global 
sections of the semi-ample invertible sheaf ${\cal O}(D_{0,l})$, whose explicit 
form is given by (\ref{WDelta}) with $\Delta$ replaced by $\Delta_l$. Note that 
$\sum_{l=1}^r\phi_l=\phi $ yields the support function of $K$ and that  the Minkowski sum is  
$\Delta_1+\dots+\Delta_r=\Delta$. Moreover giving a partition 
$\Pi(\Delta)=\{\Delta_1,\dots,\Delta_r\}$ of supporting polyhedra $\Delta_l$ 
is equivalent to give $E_1,\dots,E_r$ and is therefore also called nef-partition. 
$\nabla_l=\langle\{0\}\cup E_l\rangle \subset \Gamma^*_\mathbb{R}$ defines also 
a nef-partition $\Pi^*(\nabla)=\{\nabla_1,\dots,\nabla_r\}$ and 
$\nabla=\nabla_1+\dots+\nabla_r$ is also reflexive polyhedron with the 
following duality relations 
\begin{eqnarray}
\label{eq:polyDN}
    \Gamma_{\mathbb{R}}\qquad\qquad && \qquad\qquad \Gamma^*_{\mathbb{R}}\nonumber\\
    \Delta=\Delta_1+\ldots+\Delta_r && \Delta^*=\langle\nabla_1,\ldots,\nabla_r\rangle
        \nonumber\\[-5pt]
        &~~~~ (\Delta_l,\nabla_{l'})\ge-\delta_{l\,l'} ~~~~&\\[-5pt]\nonumber
        \nabla^*=\langle\Delta_1,\ldots,\Delta_r\rangle &&
                \nabla=\nabla_1+\ldots+\nabla_r
\end{eqnarray}
where the angle-brackets denote the convex hull of the inscribed polyhedra. By a conjecture due to~\cite{Batyrev:1994pg}
this construction leads mirror pairs $(M_n, W_n)$ of families of complete intersections 
Calabi-Yau varieties where $M_n$ is embedded into $\mathbb{P}_{\Delta^*}$ 
as complete intersection of the sections $W_{\Delta_l}$ of the line bundles 
associated to $D_{0,l}$ specified by $\Delta_l$, while $W_n$ is embedded into 
$\mathbb{P}_{\nabla}$ as complete intersection  of the sections of ${\cal O}(D^*_{0,l})$ 
specified  by $\nabla_l$.           
\end{itemize}

For $r=0$ it is explained in~\cite{Batyrev:1994hm}\footnote{See Cor. 4.5.1, Cor 4.5.2, Thm 4.5.3.}  
how to calculate the Euler number and $h_{n,1}$ and $h_{1,1}$ from the polyhedra.
This determines all Hodge numbers for $n\le 3$ and for $n=4$ the other Hodge numbers 
follow from (\ref{index}). The Euler number and $h_{q,1}$ for $0\leq q \leq d-r$ can also be 
directly calculated by the formulas\footnote{This gives again the full Hodge diamond for $n\leq 4$, 
which is implemented in the software package PALP.} given for $r>0$ in~\cite{Batyrev:1994pg}. The above mentioned  
formulas relate toric divisors and intersections thereof, as well as deformations of (\ref{WDelta}) 
to representatives in the homology groups, while the $E$-polynomial~\cite{Batyrev:1994ju} 
yields for more homology groups only information about the dimensions.

\subsubsection{Fibrations and twistings}
\label{fibretwistingdata}

For Calabi-Yau manifolds defined in toric ambient spaces, as above, 
the fibration structure descends from a toric morphism from 
the ambient space. Denote by $\Sigma$ the fan 
in $\Gamma$ generated from $\Delta$ and by $\Sigma_B$ the fan defined from 
$\Delta_B$ in the lattice $\Gamma_B$ (generated by $\Delta_B$) 
and identify  $\mathbb{P}_\Sigma$ with $\mathbb{P}_\Delta$ etc. 
Here are the two conditions for a fibration map $\tilde \phi$ from 
the ambient space $\mathbb{P}_\Delta$ to $\mathbb{P}_{\Delta_B}$ 
with fibre $\mathbb{P}_{\Delta_F}$~\cite{Fulton}\footnote{See exercise p.49, where the statements 
are made at the level of the fans.}
\begin{itemize}
 \item {\bf F1.)} There exist a lattice morphism $\phi:\Gamma\rightarrow \Gamma_B$. This is the 
 case if $\Delta_F$ is a reflexive lattice sub-polyhedron of the lattice polyhedron 
 $\Delta$ and both share the unique inner point\footnote{This applies in (\ref{polyhedrafrombaseandfibre}) 
 to the quantities with and without a star.}. The lattice $\Gamma_F$ is then in the kernel of $\phi$.
 \item {\bf F2.)} There exists a triangulation  of $\Sigma$ so that every cone 
  $\sigma\in \Sigma$ is mapped under  $\phi$ to a cone $\sigma_B\in \Sigma_B$. In this case 
  there is an $\mathbb{T}_d$-equivariant morphism $\tilde \phi: \mathbb{P}_{\Delta}\rightarrow  
\mathbb{P}_{\Delta_B}$~\footnote{This applies to (\ref{polyhedrafrombaseandfibre}) 
for the quantities with the star. It can also apply to $\Delta$  if $s_{ij}=1$.}.
\end{itemize}
For the manifolds described in section~\ref{batyrev} these two criteria apply to $\Delta^*$. 
It is easy to see that the Calabi-Yau manifolds $\{Y\subset \mathbb{P}_{\Delta^*}| W_{\Delta}(Y)=0\}$~\cite{Kreuzer:1997zg}
and $\{Y\subset \mathbb{P}_{\Delta^*}| W_{\Delta_1}(Y)=0,\ldots, W_{\Delta_r}(Y)=0 \}$~\cite{Klemm:2004km} 
inherit the fibration structure from $\mathbb{P}_{\Delta^*}$. In particular (\ref{WDelta}) is in a 
generalized Weierstrass form.

A slight generalization of (\ref{polyhedrafrombaseandfibre}) is hence to chose instead of 
$\nu_i^{F*}\in \Delta_F^*$ with $i$ fixed more generally $\nu_i=(m^i_1,\ldots,m^i_{2+s})$ 
for $i=1,\ldots, |\Delta^{B*}_{n-1}|-n$. These twisting data of the fibration of    
$\mathbb{P}_{\Delta^{*F}}$ over $\mathbb{P}_{\Delta^{B*}}$ have a simple bound in terms 
of the canonical class of $\mathbb{P}_{\Delta^{B*}}$ for $\mathbb{P}_{\Delta^*}$ to be 
almost Fano~\cite{MoriMukai}  or equivalently $\Delta^*$ to be reflexive, which we discuss 
in section \ref{heteroticbasis}. In the heterotic/F-theory 
dictionary worked out in~\cite{Friedman:1997yq}  only the trivial 
twisting are considered and it would be interesting to 
complete this dictionary. This leads in general to rational instead of holomorphic sections. 
I.e. the coefficients of the fibre coordinates might be all non-trivial section over the base and one 
can encounter the situation discussed above, namely that these sections all vanish at 
special points in the base. In this case one gets generically non-flat fibres~\cite{Cvetic:2013uta}.  

It is mathematically and physically useful to distinguish between flat and and 
non-flat fibrations. The fibre in a flat fibration has  {\sl fixed dimension}, which it keeps in particular at higher  
codimension loci in the base where the fibre degenerates into several 
irreducible components. E.g. the generic elliptic fibre ${\cal E}$   degenerates in a flat 
fibration to the singular one dimensional configuration ${\cal E}_{\hat g}$ described in section \ref{elementsofF}.  
In a non-flat fibration  some of these components of the singular fibres will have higher dimensions. This change of dimension can even happen for the 
$\mathbb{P}_{\Delta_F}$ fibration of the ambient space. The corresponding flatness 
criterium is spelled out in \cite{Hu:2000pr}. Let us call it {\bf Fl1.)}. The question 
of flatness or non-flatness of the elliptic fibre is analyzed using $W_\Delta$  or $W_{\Delta_l}$, $l=1,\ldots, r$ 
in the case of the complete intersections. These equations involve the coordinates of the fibres $x,y,z,w,\ldots$, as in
(\ref{fibertypes},\ref{tateform},\ref{e7form},\ref{e6form}) 
corresponding to points of $\Delta^{*F}$. The monomials $M( x,y,z,w,\ldots)$ in these coordinates are multiplied by 
coefficients that take value in line bundles ${\cal L}_{M}$ over the base and will in addition 
depend on `blow up coordinates' corresponding to the blow up divisors, see e.g. 
(\ref{e7form},\ref{e6form})~\footnote{In fact for the non-trivial twistings described below  
all monomials in the fibre coordinates might be multiplied by coefficients transforming in non-trivial 
line bundles over the base.}. Let us assume that {\bf Fl1.)} holds. The condition for the flatness 
of the elliptic fibre {\bf Fl2.)}  is as follows: If these coefficients do not vanish in a 
sublocus of the base so that the constraint on the fibre coordinates becomes trivial, 
the fibration is flat. 

The toric blow up description involves adding points to $\Delta^*$ to make it into  
$\ci \Delta^*$,  which adds the corresponding `blow up coordinates', see section \ref{transitions}. 
In the general terminology of resolutions the  $W_{\ci \Delta}$ or $W_{\ci \Delta_l}$, $l=1,\ldots, r$ in the blow up 
coordinate ring  are called the {\sl proper transforms} of the $W_\Delta$ or $W_{\Delta_l}$ defined 
in the original coordinate ring. 

The fibrations described by (\ref{polyhedrafrombaseandfibre})  fulfill {\bf Fl1.)} 
and the elliptic fibrations over the almost Fano basis described in section \ref{heteroticbasis} 
and \ref{toricfanobases} are generically flat. However if we degenerate the complex 
structure to obtain gauge groups and higher codimension structure and resolve, the 
elliptic fibration might not stay flat.  Examples for the flatness condition in non-toric 
resolution of the $E_8$ codimension 3 singularity  are discussed in section \ref{case2}.  
Non-flat fibres are rather ubiquitous in elliptic fourfolds with gauge groups~\cite{Candelas:2000nc}.            

The physical significance of the non-flat fibration is that  in the limit of vanishing 
fibre volume a holomorphic surface $S$ collapses and the BPS states coming from p-branes 
of wrapping $S$ as well the holomorphic curves $C$ inside $S$ lead in general to a non-local 
quantum theory including tensionless extended objects~\cite{Witten:1995zh} as some light states can 
have electric while others have magnetic charges as in the QFT~\cite{Argyres:1995jj}. Part of the spectrum 
of the tensionless string can be analyzed using their WS CFT~\cite{Maldacena:1997de} or upon further 
dimensional reduction to a quantum field theory~\cite{Klemm:1996hh}.  Since the fibre is involved 
in the present case this is to be analyzed first in the M-theory picture, where one gets as 
in~\cite{Maldacena:1997de} a (0,4)- in 5d or (0,2)-supersymmetric tensionless string in 3d from 
the M5 brane wrapping $S$ as well as an infinite tower of massless particles from M2 branes wrapping 
all $C$'s. The infinite tower of massless states survives the M-theory/F-theory lift to 4d 
and can pose a thread to F-theory phenomenology. It is not very well understood whether 
background fluxes can be used to avoid the vanishing of the sections in ${\cal L}_{M}$ 
or to lift the masses of these states.

\subsection{The complex moduli space of toric hypersurfaces}
\label{complexmoduli}
Having immediately the mirror construction  and therefore 
the description of the K\"ahler- and complex moduli on the same footing is very useful to 
study the corresponding moduli spaces 
\be 
{\cal M}_{cs}(M_n)={\cal M}_{K\ddot ahler\ structure}(W_n)\, , 
\ee
which ultimately in the physics context one wants to lift by a superpotential. 
Let us focus on the hypersurface case $r=1$ and $s=0$ and recall the features of $W_{\Delta}$, 
the {\sl Newton polynomial} of $\Delta_{d=n+1}$, and its complex deformations parametrized by the $a_i$.   
Its independent deformations ${\cal M}_{W_{\Delta}}\subset {\cal M}_{cs}$, whose infinitesimal 
directions correspond to $H^1(M_n,TM_n)$, are given by a projectivization of 
the $a_i$ modulo automorphisms of $\mathbb{P}_{\Delta^*}$\footnote{These are called monomial
deformations in the complex moduli space or toric deformations in K\"ahler moduli space. There can 
be non-monomial (non-toric) deformations which  indicated in brackets whenever we specify 
$h^1(M_n,TM_n) \sim h_{n-1,1}(M_n)=\# monomial(\# non-monomial)$ or  $h_{1,1}=\# toric(\# non-toric)$. 
Note that the monomial deformations define a good subspace in ${\cal M}_{cs}(M_n)$ and in the 
following we assume for simplicity that there are no non-monomial 
deformations.}. 

We want to determine possible group actions on ${\cal M}_{cs}(M_n)$ in 
order to argue that gauge symmetry enhancements can be 
induced by turning on fluxes on the invariant- or sometimes the 
non-invariant periods of $M_n$ respectively, as discussed in 
section~\ref{minimizingW}. 
In the construction of mirror pairs of Calabi-Yau manifolds by 
reflexive polyhedra and by orbifolds such group actions  
appear  naturally and one can define a variation 
of the mixed Hodge structure in terms of invariant periods on  
the invariant locus ${\cal S}\subset {\cal M}_{cs}(M_n)$. 
The monodromy group acting on the periods of a family of Calabi-Yau manifolds $M_n$ 
forms a subgroup $\Gamma_{\cal M}$ of linear integer transformations 
respecting a quadratic intersection form $Q$ on $H_n(M_n,\mathbb{Z})$. For $n$ odd the 
$Q$ is symplectic and $\Gamma_{\cal M} \subset {\rm Sp}(b_n(M_n),\mathbb{Z})$, 
which is not necessarily of finite index for $n\ge 3$, see e.g. \cite{HvS} for 
recent progress on this question for one parameter CY 3-folds. For $n$ 
even $Q$ is symmetric with signature $\sigma=\int_{M_{2 m}} L_m$ 
(cf. (\ref{signature})). The realization of a group action on 
${\cal M}_{cs}(M_n)$ leads to a sub-monodromy problem on the 
invariant sub-locus ${\cal S}\in {\cal M}_{cs}$ for which $\Gamma_{\cal S} 
\subset \Gamma_{\cal M}$ can e.g. be a finite index subgroup 
in products of $SL(2,\mathbb{Z})$ as in the example in section 
(\ref{stabledegenerationlimit}). Studying the various $\Gamma_{{\cal S}_i}$ 
is obviously an important tool to get a handle on $\Gamma_{\cal M}$, 
with many implications, e.g. that the functions that determine the 
effective action in section \ref{effectiveaction} are organized in 
terms of automorphic forms w.r.t. $\Gamma_{{\cal S}_i}$. 
In all known cases the variation of the mixed Hodge structure related 
to $\Gamma_{\cal S}$ is the variation of mixed Hodge 
structure of an actual geometry. Either of a Calabi-Yau manifold  
$M_n$  of the same dimension possibly after a transition or, if 
the flux violates the tadpole condition~(\ref{tadpolecondition})  and 
decompactifies $M_n$, of a lower dimensional geometry, e.g. a 
Seiberg-Witten curve or a lower dimensional Calabi-Yau manifold. 
It is clear that the tadpole condition~(\ref{tadpolecondition}) 
cannot be realized in general on both sides of a transition 
without tuning the fluxes. The simplest examples are transitions 
between Calabi-Yau manifolds where $\chi \ mod \ 24$  differs on 
both sides. From the pure geometrical point of view 
such transitions are possible~\cite{Klemm:1996ts}.  
A detailed analysis of this question of fluxes in local 
transitions has been made in~\cite{Intriligator:2012ue}.  In the applications in 
F-theory moduli stabilization it however is not necessary 
to actually go through the transition. We just want the moduli to settle 
close to the transition point.

To describe ${\cal M}_{cs}$ it is useful to extend the spaces in which the polyhedra live by one dimension  
and to define $(\bar \Delta, \bar \Delta^*)$ as the embedding of 
$(\Delta,\Delta^*)$ in hyperplanes $\mathbb{H}$ at distance one 
from the origin, i.e. as the complex hull of $\{\bar \nu_i\}:=\{(\nu_i,1)\}$ 
and $\{\bar \nu_i^*\}:=\{(\nu_i^*,1)\}$. Each of 
the $k=|\Delta^*|-(d+1)$ linear relations 
\be
\sum_{i} \bar \nu_i^* l^{*(p)}_i=0, \qquad p=1,\ldots, k\, ,
\label{linerarelations}
\ee  
with $l^{*(p)}_i\in \mathbb{Z}$ and $\sum_i l^{*(p)}_i=0$ between the points 
of the polyhedron $\bar \Delta^*$ yields an action of $\mathbb{C}^*$ 
on the coordinates of $Y_j$ defined by 
\be 
Y_j\mapsto Y_j \mu_p^{l^{*(p)}_j}, \quad j=1,\ldots,|\Delta^*|-1\, ,  
\label{scalings} 
\ee 
with $\mu_p\in \mathbb{C}^*$ under which $W_{\Delta}=0$ has to be invariant\footnote{If one keeps $Y_0$ where $\nu^*_0={\underline 0}$ is the origin 
then $W_\Delta$ itself is invariant, not just its vanishing locus.}. The latter property defines it and is 
necessary make it well defined in the ambient space      
\be
\mathbb{P}_{\Delta^*}=(C[Y_1,\dots,Y_{k=|\Delta^*_{d}|-1}] \setminus {\cal SR}^*)/ G,
\label{PDelta}
\ee
where $G= {\rm Hom}_{\mathbb{Z}} (A_{d-1},\mathbb{C}^*)$. The Chow group $A_k$ is 
generated by the orbit closures of $d-k$ dimensional cones. I.e. $A_{d-1}$ is generated by one dimensional 
cones $\Sigma(1)$, which are the Weyl divisor modulo linear equivalence.    
Hence for reflexive polyhedra $G=(\mathbb{C}^*)^{|\Delta^*|-(d+1)} \times G_{tor}$, where the 
finite group $G_{tor}={\rm Hom}_{\mathbb{Z}}( A_{n-1}(\mathbb{P}_{\Delta_*})_{tor},
\mathbb{Q}/\mathbb{Z})$ and $(\mathbb{C}^*)^{|\Delta^*|-(d+1)}$ is 
generated by (\ref{scalings}). The Stanley-Reisner ideal  ${\cal SR}^*$ depends on a 
subdivision $S^*$ of $\Sigma^*$ into $d$-dimensional simplicial cones $\Sigma(d)$ of volume 1 
in $\Gamma^*$, the toric description of resolving the singularities of 
$\mathbb{P}_{\Delta^*}$ completely\footnote{A coarse subdivision $S^*_c$ is given by a complete 
subdivision of $\Delta^*$  into $d$-dimensional simplices, where each 
simplex has the origin as a vertex. Not all simplicial cones 
obtained in this fashion have volume 1, so that rays through 
points outside $\Delta^*$ have to be added to make 
$\mathbb{P}_{\Delta^*}$ smooth, however $W_{\Delta}=0$ 
misses the corresponding singularities and is already 
smooth with the coarse subdivision.}. From this 
subdivision the Stanley Reisner ideal is given simply 
combinatorial: Let $I^*$ be any index set of points in $\Delta^*$ with 
$|I^*|\le d$ then the $B_{d-|I^*|}=D_{i_1} \cap \ldots \cap D_{i^*_{|I|}}=
(\{Y_{i^*_1}=0\}\cap \ldots \cap  \{Y_{i^*_{|I^*|}}=0\}$ is in  
${\cal SR}^*$, iff $\{\nu^*_{i_1}, \ldots, \nu^*_{i_{|I|}}\}$ 
are not in a cone $\Sigma(|I^*|)$ of $S^*$.

We have to understand in a first step the action of the 
automorphism group of $\mathbb{P}_\Delta^*$, because it 
reduces the naive moduli space parametrized by the $a_i$, 
see~\cite{CoxKatz}. According to this description generators of 
automorphisms come in three types:
\begin{itemize}
\item A1) The action of the algebraic torus $\mathbb{T}_d=(\mathbb{C}^*)^d$. 
Note that $\mathbb{P}_{\Delta_*}$ contains the algebraic torus $\mathbb{T}_d$ 
as an open dense subset and the natural $(\mathbb{C}^*)^{|\Delta^*|-1}$ action 
on $\mathbb{C}[Y_i]$ is reduced by the identifications (\ref{scalings}), as expressed by the exact sequence  
\be 
1 \rightarrow G\rightarrow (\mathbb{C}^*)^{\Sigma(1)}\rightarrow \mathbb{T}_d\rightarrow 1\ ,
\ee
to an action of $\mathbb{T}_d$ on $\mathbb{P}_{\Delta^*}$ that extends 
the natural action of $\mathbb{T}_d$ on $\mathbb{T}_d\subset \mathbb{P}_{\Delta^*}$.
\item A2) The second type are weighted homogeneous 
coordinate transformations 
\be
Y_i \mapsto b^{(i)}_0 Y_i + \sum_k b^{(i)}_k m^{(i)}_k(\underline Y)\, ,   
\label{weightedhomogeneoustransformations} 
\ee
of $\mathbb{P}_{\Delta^*}$, where $b^{(i)}_l\in \mathbb{C}$ and the 
monomials $m^{(i)}_k(\underline Y)$ do not contain $Y_i$ and are of the 
same degree as $Y_i$ so that both sides of (\ref{weightedhomogeneoustransformations})
transform equal under (\ref{scalings}) and $W_\Delta$ stays   
well defined under (\ref{weightedhomogeneoustransformations}). 
Pairs $(Y^i,b^{(i)}_k)$ are called roots. 
\item A3 ) There can be symmetries of 
the toric polyhedron, which according to~\cite{CoxKatz} have 
to be identified.    
\end{itemize}

The actions  A1-A3 do not leave the general $W_{\Delta}$ invariant. 
They have to be compensated by actions on the $a_i$. Dividing the space 
of $a_i$ by the latter action leads to a model for the complex moduli space. 
To construct this quotient~\cite{CoxKatz} work with a {\sl Laurent 
polynomial} in $d$ variables instead of the Newton polynomial $W_\Delta$. 
First notice that any lattice polyhedron $\Delta_d\in \Gamma$, which contains the origin, 
comes with a very ample divisor $k D_\Delta$ in $\mathbb{P}_{\Delta}$. 
It is  given by its support function $\phi:\Gamma^*_\mathbb{R}\rightarrow \mathbb{R}$  
to be $\phi_{k \Delta}(v)={\rm min}_{\nu\in {k \Delta}} \langle m,v\rangle$.  
Here $k=1$ for reflexive polyhedra. By definition $\mathbb{P}_{\Delta}$ can then 
be embedded in a projective space. Concretely this is done
by taking the points $\{\nu_1,\ldots, \nu _s\} = \Delta \cap \Gamma$ 
and map $\mathbb{T}_d\rightarrow  \mathbb{P}^{s-1}$  by sending 
$t\in \mathbb{T}_d $ to $(t^{\nu_1},\ldots,t^{\nu_s})$, where $t^{\nu_i}=
\prod_{j=1}^d t_j^{\nu_{i,j}}$ and $t_j$ are a coordinate 
basis for $\mathbb{T}_d$. $\mathbb{P}_\Delta$ is the 
completion  of the image of this map in $\mathbb{P}^{s-1}$. One can define a vector space of Laurent polynomials
\be
L(\Delta\cap \Gamma)=\{w_{\Delta}:w_{\Delta} =\sum_{{\nu_i} \in \Delta\cap \Gamma} a_i t^{\nu_i}, a_i \in \mathbb{C}\}\ .
\label{laurant} 
\ee 
The coordinate ring $Y_i$ captures all the blow up coordinates,
but as far  as the complex structure deformations of $M_n$  go, the 
$d$ variables $t_i$ are sufficient and we need not to distinguish 
between $w_{\Delta}$ and $W_\Delta$\footnote{Physically this independence 
of complex parameters from the blow ups moduli reflects, e.g. the decoupling of vector- and 
hypermultiplets in type IIb compactifications on $M_3$ to 4d at generic 
loci in the moduli space.}. The statement about the moduli space can now 
be phrased as 
\begin{equation}  
{\cal M}_{W_\Delta}=\mathbb{P}(L(\Delta\cap \Gamma))/Aut(\mathbb{P}_{\Delta^*})\ .
\label{Msc}
\end{equation}

It has been further shown by Batyrev that any complex structure 
deformation has a representative under the (gauge) orbits 
(\ref{weightedhomogeneoustransformations}), which corresponds to 
the restricted Newton polynomial of $\Delta$ in which only such 
monomials in (\ref{WDelta}) are considered that correspond to 
points in $\nu^{(i)}$ not inside co-dimension one faces 
of $\Delta$, we call $\Delta$ without those points $\Delta_0$. This can be viewed 
as a gauge fixing.  This leads to the definition 
\begin{equation}  
{\cal M}^{simp}_{W_\Delta}=\mathbb{P}(L(\Delta_0\cap \Gamma))/\mathbb{T}_d,
\label{batyrevgauge}  
\end{equation}
and one can show that the map $\phi: {\cal M}^{simp}_{W_\Delta} 
\rightarrow {\cal M}_{W_\Delta}$ is at most a {\sl finite cover}. 
Note that not all symmetries of $M_n$ might be manifest in a chosen gauge.

As it turns out the most interesting  points in the analysis below  
are precisely related to the nature of the {\sl finite covers}. In order 
to get the right description for ${\cal M}_{cs}$ we propose to divide  
$Aut(\mathbb{P}_{\Delta^*})$ by the discrete group $G$ described in section \ref{groupactions}.     

\subsubsection{Large complex structure coordinates and the point of  maximal unipotent 
monodromy}   
\label{mumpoints}
One can introduce coordinates, which eliminate the $\mathbb{T}_{d=n+1}\times \mathbb{C}^*$ action 
(the $\mathbb{C}^*$ action is the one that scales $W_\Delta$)    
\begin{equation} 
z_k=(-1)^{l_0^{(k)}}\prod_{i=0}^{|\Delta_{0}|} a_i^{l_i^{(k)}}, 
\forall k=1,\ldots ,|\Delta_{0}|-(d+1)=h_{n-1,1}\ .      
\label{largevolumecoords}  
\end{equation}
A mirror conjecture of  Batyrev states that if the $l^{(k)}$ 
are the vectors spanning the Mori cone of $\mathbb{P}_\Delta$, 
which descends to $W_n$, then $z_k=0$ $\forall k$ is a point of 
{\sl maximal unipotent monodromy} in the complex moduli space of $M_n$, 
which maps under the mirror map to a large volume point of $W_n$ 
inside the K\"ahler cone of $\mathbb{P}_{\Delta}$. The 
topological data of $W_n$ determine the degenerations of 
the period vector at this point, which is not the the only but the most 
important datum to fix an integral basis for the periods.  The K\"ahler cone is dual to the Mori 
cone. In the case of toric varieties $\mathbb{P}_{\Delta}$, the Mori cone 
can be calculated from a maximal star (K\"ahler) triangulation 
of $\Delta$~\footnote{As encoded in the secondary fan, 
there can be many K\"ahler triangulations 
of $\Delta$ and correspondingly many points of maximal unipotent 
monodromy in the family $M_n$.} and of course due to mirror 
symmetry the formalism described in great detail in~\cite{MR2810322} 
for the K\"ahler cone applies on both sides.

\subsection{Group action on ${\cal M}_{cs}(M_n)$} 
\label{groupactions}
Now that we have a model for ${\cal M}_{cs}(M_n)$ we can 
come to the main point namely the induced group actions 
on it. 
\subsubsection{Discrete group actions and Orbifolds}
\label{discretegroups}

Group actions on ${\cal M}_{cs}$  can be induced  from symmetry acting
on $M_n$. Generic Calabi-Yau  manifolds with full $SU(n)$-holonomy have 
no continuous symmetries, but one easily finds discrete symmetry groups $G$ acting on them. 
To define a Calabi-Yau orbifold, whose moduli space is the invariant  subspace ${\cal S}\subset 
{\cal M}_{cs}$, they have to leave the holomorphic $(n,0)$-form $\Omega_n$ invariant. The latter condition 
implies that $G$ acts like  a discrete subgroup 
of $SU(n)$ in each coordinate patch of $M_n$. For $n<4$ this guarantees  
that the fixed sets of the $G$ action on $M_n$ can be geometrically 
resolved without changing the triviality
of the canonical class, i.e. maintaining the CY condition on the resolved space 
$\widehat{M_n/G}$ . For $n=4$ there can be terminal singularities 
remaining on the CY orbifold, which do not render string and F-theory 
compactifications inconsistent.     
Dividing discrete symmetries and resolving the singular space if the symmetries have fix sets 
to $\widehat{M_n/G}$ was described in~\cite{Fuchs:1989yv} in comparison with symmetries that 
lead orbifolds of Gepner models with $(2,2)$ world-sheet supersymmetry. 
For the latter the condition $G\subset SU(n)$ ensures that the $(2,2)$ 
super charges are not projected out. The symmetries considered in~\cite{Fuchs:1989yv} 
act on the ambient space coordinates with two properties 
\begin{itemize}
\item G1.) They leave $W_\Delta$ invariant in the sense 
that they can be compensated by an action on the $a_i$~\footnotemark[1]
\item G2.) They leave $\mu$ defined in (\ref{mu}) invariant\footnotemark[1].
\end{itemize}
\footnotetext[1]{Up to scaling 
$\lambda \in \mathbb{C}^*$.}
The geometrical condition G2.) is controlled  by the explicit
form (\ref{mu}) for toric varieties with a dominant  
weight vector~\footnote{In the general case one can control 
this condition by establishing an \'Etale map between two 
reflexive toric polyhedra.}. For example, $G$ can be 
generated by phases $\alpha_{k}^{n}=\exp(2 \pi i n/k)$, $k,n\in \mathbb{Z}$   
acting on the coordinates as  $x_i\mapsto \alpha^{n_i}_{k} x_i$, $i=1,\ldots,n+2$ where 
$\prod_i \alpha^{n_i}_{k}=1$ or by {\sl even} permutations on coordinates with equal weights. 
Such permutation orbifolds including non-abelian orbifolds have been considered 
in~\cite{Klemm:1990df}. These permutations  are special 
cases of discrete root operations (\ref{weightedhomogeneoustransformations}), 
which can contain more general generators.  It was observed in~\cite{Greene:1990ud} 
that dividing the maximal phase symmetry group $G^{max}_{ph}$ from 
Fermat hypersurfaces yields a mirror geometry\footnote{This a special case of Batyrev's construction 
and an example is discussed in section \ref{orbifoldpoints}.} $\widehat{M_3/G^{max}_{ph}}$, a 
statement that can be generalized to higher dimensional $M_n$.      

The $G$ action $G:M_n\rightarrow M_n$ induces an action 
$G_*:H_*(M_n)\rightarrow H_*(M_n)$ and properties of this map are
captured by the Lefschetz fixed-point formulas. Here we need a 
much more trivial fact. The above $G_*$ action will decompose  $H_n(M,\mathbb{Z})$ 
in  cyclic orbits of finite length. Let us focus on one divisor of $|G|$ 
and denote the orbit of length $k$ by  
\begin{equation}
\Gamma_0\mapsto \Gamma_{1}\mapsto \ldots \mapsto \Gamma_{k-1} 
\mapsto \Gamma_0 \ .
\label{cycle}
\end{equation}
Note that there can be relations in homology among the cycles $\Gamma_i$. 
The eigenvectors with eigenvalue $(\alpha_k)^p$  in this orbit are 
$\Gamma^{diag}_p=\frac{1}{k}\sum_{l=0}^{k-1} (\alpha_k)^{l(k-p)} 
\Gamma_l$ with $\alpha_k$ the elementary $k$'th root of unity. I.e. 
$\Gamma^{diag}_0$, the invariant cycle can be defined after 
rescaling in $H_n(M,\mathbb{Z})$, but to diagonalize the 
non-invariant  cycles $H_n(M)$ w.r.t. to $G$ if $k$ is prime and there are 
no homological relations, one has to extend the number field $\mathbb{Z}$ to 
$\mathbb{Z}\oplus \{\alpha_{k}\}$\footnote{ In fact extending the number field  is 
an interesting logical possibility. It would still represent  a discrete choice 
of fluxes and if the flux disappears at the attractor point there 
seems no physical contradiction.}.           

We assume now that the automorphisms $\mathbb{T}_{n+1}\times \mathbb{C}^*$ and A2.) have been 
fixed to leave $a_i$, $i=1,\ldots, h_{n-1,1}$ as coordinates on $M^{simp}_{W_\Delta}$. 
 G1.) defines an action  $\tilde G:{\cal M}^{simp}_{W_\Delta}\rightarrow 
{\cal M}^{simp}_{W_\Delta}$. Let  us consider a generator of order 
$k$,  i.e. a phase action $g:x_i\mapsto \alpha^{n_i}_{k} x_i$  that rotates  
monomials $M_i \rightarrow  \alpha^{m_i}_k M_i$ in $W_\Delta$ by a nontrivial $k$-th root 
of unity. For simplicity we first assume $k$ to be prime. Latter on we discuss 
implications if that is not the case.  The phase rotations induce an action on 
the non-invariant  moduli $\tilde a_i$    
\begin{equation} 
\tilde g:\tilde a_i \rightarrow  \alpha^{-m_i}_k\tilde  a_i \ .
\label{moduliaction} 
\end{equation}
Let us call $W^{inv}_{\Delta}=\sum_{k} \ci a_i M^{inv}_k$  
the  polynomial containing only invariant monomials. We assume that $W^{inv}_{\Delta}=0$ 
is transversal, i.e. $d W^{inv}_{\Delta}=0$ and $W^{inv}_{\Delta}=0$ have generically no solution,   
to have a good sub-monodromy problem. By Bertinis-theorem this can be checked on the 
base locus of $\mathbb{P}_{\Delta^*}$. A combinatorial criterium is given in~\cite{MR1798982} and 
is equivalent to the condition that $W^{inv}_{\Delta}$ is the Newton polynom of a reflexive 
polyhedron.    

Now note that the action of $\tilde g$ (\ref{moduliaction}) 
is not visible in the coordinates (\ref{largevolumecoords}). This is due to the fact 
that the actual moduli space is parametrized by the $a_i$ variables that 
multicover the $z_i$. Let $\Pi_p=\int_{\Gamma_p} \Omega$, $p=0,\ldots, k-1$ be 
periods over the cycles $\Gamma_p$, $p=0,\ldots, k-1$ in (\ref{cycle}). We like 
to study them as solutions of the Picard-Fuchs differential system and their 
dependence on one non-invariant modulus $\tilde a$ on which $\tilde g$ acts by 
(\ref{moduliaction})   near the orbifold point in the moduli space  $\tilde a=0$. Here their power series expansion  
has to have the form $\Pi_p=\sum_{n=0}^\infty c_n(\check a) (\tilde a \alpha^p_k)^n$. The 
occurrence of $\alpha_k$ is necessary to realize the transformation (\ref{cycle}).
The $c_n(\check a)$ depends on moduli that do not transform under the generator  $\tilde g$ and  
contains in particular the totally invariant moduli  under $G$ which we call $\ci a$.  
Since by assumption  $W^{inv}_{\Delta}=0$ is transversal $c_0(\check a)\neq 0$ for generic $\check a$.
Otherwise the invariant period would vanish generically on ${\cal S}$, in contradiction to the assumption. 
One concludes that the series expansion of the periods  $\Pi^{diag}_p$  near $\tilde a=0$ is 
\begin{equation}
 \Pi^{diag}_p= \sum_{n=0}^\infty c_{p+ k n} (\check a) \tilde a^{p+k n}, \qquad p=0,\ldots,k-1    \ .               
\label{orbifoldbehaviour}  
 \end{equation}
Let $\tilde z_i$ be a variable in the base, which is by (\ref{largevolumecoords}) $k$-times 
covered by the $\tilde a_i$ on which (\ref{moduliaction}) act. Then the periods $\tilde \Pi_p$ over the non-invariant 
cycles $\Gamma_p$, $p=1,\ldots,k-1$, will have the typical orbifold behaviour at 
${\cal S}$, i.e. they go in leading order with $\Pi^{diag}_p=\tilde z_i^{p/k}+ {\cal O}(\tilde z_i^{p/k+1})$,
$p=0,\ldots, k-1$. We can conclude some more facts. The monodromy $M_{\mathbb{Z}_k}\in \Gamma_{\cal M}$ around 
${\cal S} $ is of order $k$ and acts on $k-1$ integer periods, which  span a subspace of all  periods.  
The invariant submoduli space ${\cal S}\subset {\cal M}_{W_\Delta}$ is the vector space $L(\Delta^{inv}\cap \Gamma)$ 
divided by $G$ invariants roots $(x_i,b^{(i)}_k)$ and $\mathbb{T}_{n+1}\times \mathbb{C}^*/G$. 

We have further seen that the relevant automorphism group is  
\begin{equation}
Aut(W_\Delta)=Aut(\mathbb{P}_{\Delta^*})/G\  , 
\end{equation}
where $G$ is the maximal group obeying G1) and G2).  Accordingly we define the moduli space 
of the polynomial deformations as  
\begin{equation}  
{\cal M}_{W_\Delta}=\mathbb{P}(L(\Delta\cap \Gamma))/Aut(W_\Delta) \ .
\label{Msc}
\end{equation}
This has the correct multicovering structure and we get an induced group action of 
$G$ on ${\cal M}_{W_\Delta}$  as well as on the periods 
\begin{equation}
\Pi(\tilde g a)=  M_g \Pi( a) \ .
\label{actiononperiods}  
\end{equation}
In particular $M_g$  acts on $\Gamma$ in $H_n(M_n)$ as a 
monodromy matrix respecting the dual intersecting pairing 
\begin{equation}
Q^*(M_g\Gamma,M_g\Gamma)= Q^*(\Gamma,\Gamma) \ .
\label{Kaehlerinvarianz}
\end{equation} 
E.g. if $n$ is odd  $M_g$ is a symplectic matrix in ${\rm SP}(b_n,\mathbb{Z})$.
We assumed in (\ref{orbifoldbehaviour}) that there are no homological 
relations between the $\Gamma_p$. In compact cases there will be generically 
such relations and one has to calculate the transition matrix between the  
local basis $\Gamma_p$, $p=0,\ldots, k-1$ and the global integral basis, 
see e.g. (\ref{continuationsextic}). Since $G$ is globally defined one 
expects this matrix to be  rational rather than transcendental. This is one 
key ingredient to have an integral  superpotential that drives the moduli 
to the orbifold point and  will be established rigorously for important 
general cases in section  \ref{hemisphere}. For the sextic  and other 
one parameter models we establish this fact as well  as  
(\ref{Kaehlerinvarianz}) explicitly for  fourfold orbifolds e.g. in 
(\ref{sexticmonodromy2}). Many threefold cases can be found 
in~\cite{Font:1992uk,Klemm:1992tx}.  E.q. (\ref{Kaehlerinvarianz})
means that the K\"ahler potential (\ref{Kaehlerpotential}) is  
monodromy invariant.   

The ideas discussed in this section are related to a general principle to characterize the monodromy group $\Gamma_{{\cal M}(M_n)}$ of a 
family of Calabi-Yau manifolds $M_n$ by the monodromy group $\Gamma_{{\cal M}(U_n)}$ of a more universal 
family of Calabi-Yau manifolds $U_n$, whose generic member has a smaller symmetry group $G(U_n)\subset G(M_n)$, 
as 
\begin{equation} 
\Gamma_{{\cal M}(M_n)} \sim \Gamma_{{\cal M}(U_n)}\frac{ G(U_n)}{G(M_n)}\ .
\end{equation}  
This is very natural for elliptic curves, where the universal family is (\ref{Weierstrass}) with 
$f\sim E_4(\tau)$ and $g\sim E_6(\tau)$ and the monodromy of  algebraically realized families $M_1(a)$ 
can  be characterized as $PSL(2,\mathbb{Z})/G(M_1)$. In this case $G(M_1)$ is the {\sl Galois group} 
of the map $a\mapsto j$. For an early discussion in a physical context see~\cite{Giveon:1990ay}.

\subsubsection{A dual operation: adding points to $\Delta^*$ and omitting points of $\Delta$.}
\label{transitions}
One can define a dual operation on the category of reflexive pairs $(\Delta,\Delta^*)$, which simply 
consists of omitting sets of points $\{\nu^{(1)},\ldots,\nu^{(N)}\}$ from $\Delta$, which do no contain 
$\nu_0$, so that $\Delta^{inv}=\ci \Delta=\Delta\setminus \{\nu^{(1)},\ldots,\nu^{(N)}\}$ 
in the minimal lattice $\ci \Gamma\subset \Gamma$ spanned by the points of $\ci \Delta$ 
from $\nu_0$ is a reflexive polyhedron. It follows from the definition of the dual 
polyhedron $\ci \Delta^*$ that $\ci \Delta^*\supset \Delta^*$ in the 
dual lattice $\ci \Gamma^*\supset\Gamma^*$. I.e. we consider two reflexive pairs  
$(\Delta,\Delta^*)$  and $(\ci \Delta,\ci \Delta^*)$    
with the following inclusion
\begin{equation}
\begin{array}{ccc} 
 (\ci \Delta,\ci \Gamma)& \leftarrow\ {\rm reflexive}\ \rightarrow& (\ci \Delta^*,\ci \Gamma^*)\\ 
             \cap             &                                         &    \cup\\ 
   (\Delta,\Gamma)            & \leftarrow\ {\rm reflexive}\ \rightarrow & (\Delta^*,\Gamma^*) \ .
   \end{array} 
\label{embedding}     
\end{equation}
The first observation made by Batyrev is that the orbifoldization by phases 
is a special case of this construction. For example for the Fermat polynomials 
the Newton polytope of the invariant constraint $W^{inv}_{\Delta}$ under such 
a phase symmetry $G$ as discussed in $\ref{discretegroups}$  defines 
$\ci \Delta=\Delta^{inv}$ and one has $G\sim \Gamma/\ci \Gamma\sim 
\ci \Gamma^*/\Gamma^*$~\cite{Batyrev:1994hm}. The more general construction 
has been used in many places in the literature. For example~\cite{greenekantor} argue
that the moduli space of all Calabi-Yau threefolds defined as the anti 
canonical bundle in $\mathbb{P}_{\Delta_4}$  are connected by transitions 
induced by the inclusion (\ref{embedding}) by showing that all pairs  
$(\Delta_4,\Delta_4^*)$ in the class of 4-dim reflexive polyhedra 
can be related by chains of such inclusions, which gives some support for a 
conjecture of Miles Reid~\cite{MR909231} in a special class of 3-folds. With  
suitable choice of the gauge it was shown in~\cite{Berglund:1996uy}  that 
the subspace ${\cal S}$ of the invariant model is characterized by    
\begin{equation} 
{\cal S}={\cal M}_{W^{inv}_\Delta}= 
\mathbb{P}(L(\Delta^{inv} \cap \ci \Gamma))/Aut(\mathbb{P}_{\ci \Delta^*})
\subset {\cal M}_{W_\Delta}\ .  
\label{S} 
\end{equation}
In general this cannot be directly seen in the gauge (\ref{batyrevgauge}). An 
application to non-abelian Higgs-chains on threefolds was given 
in~\cite{Klemm:1996kv}\cite{Berglund:1996uy}.

Given the polyhedron $\ci \Delta$, one can write computer programs, which test 
reflexivity by constructing $\ci \Delta^*$ and check whether 
it contains the unique inner point\footnote{The authors have developed such programs, 
but is also a functionality of the SAGE packages for Polyhedra~\cite{BraunSage}.}. 
These were heavily used to construct the explicit examples in section \ref{compactCY}.

Let us give another interpretation of ${\cal S}$ involving a stratification 
of ${\cal M}_{cs}$ by $\mathbb{C}^*$ actions. We add a point 
${\bar \nu}^{*(K)}\in \Gamma_\mathbb{Q}$ in $\mathbb{H}$ to 
$\Delta^*$ and include $Y_K$ as  parameter in $W_{\Delta}$, 
i.e. we consider 
\be 
W_\Delta=\sum_{\nu^{(i)}\in \Delta} ( a_i Y_K^{\langle \nu_i,\nu^{*(K)}\rangle+1})  \prod_{\nu^{*(k)}\in\Delta^*} Y_k^{\langle \nu^{(i)}, 
\nu^{*(k)}\rangle+1}=0\ . 
\label{WDeltas}
\ee  
The point ${\bar \nu}^{*(K)}\in \mathbb{H}$ defines 
a new linear relation $\ci l^{(r+1)}$ in $\langle \Delta^{*},\nu^{*(K)}\rangle$, 
which we let act in the usual fashion on all $Y_i$ by (\ref{scalings}). 
Under this scaling $\mu_k$ the constrained $W_{\Delta}$ is not invariant. 
It can be kept invariant by rescaling the $a_i$. This defines uniquely an $\mathbb{C}^*$-action 
on the moduli space $\ci \mu_K:{\cal M}_{W_\Delta}\rightarrow {\cal M}_{W_\Delta}$. 
Moreover the fixed locus  of the induced action on  ${\cal M}_{W_\Delta}$  
is simply defined by setting the coefficients $a_i$ of those monomials 
to zero, which are not invariant under $\mu_k$. In general we add 
several points ${\bar \nu}^{*(K)}$, $K=1,\ldots, N^*$ and get  
$N^*$-actions $\ci \mu_K$, whose fixed locus is is obtained by setting more 
coefficients $a_i$ to zero. The invariant polynomial is 
$W_{\Delta^{inv}}$ and $\ci \Delta^{*}=\langle \Delta^{*}, 
\nu^{*(1)},\ldots \nu^{*(N)}\rangle$ is a new lattice polyhedron in the new 
lattice $\ci \Gamma^*$. The characterization of the fix point 
locus ${\cal S}$ is again given by (\ref{S}).

We claim that ${\cal S}$ is a good subspace of ${\cal S}\subset 
{\cal M}_{W_\Delta}$ in the following sense. The Gauss Manin 
connection closes on ${\cal S}$ for a subset of invariant
periods. The monodromy group acting on the invariant periods 
$\Gamma_{\cal}$ is a proper subgroup of $\Gamma_{{\cal M}_{cs}}$.  
This will be further discussed in section (\ref{gaussmaninS}). 
Of course by mirror symmetry there is a dual inclusion of the  
quantum corrected K\"ahler moduli space ${\cal S}^*={\cal M}_{W_{\Delta^*}}=
\mathbb{P}(L(\Delta^* \cap \ci \Gamma^*))/Aut(\mathbb{P}_{\Delta})
\subset {\cal M}_{W_{\ci \Delta}^*}$. For future reference we will 
call the manifold defined  as anticanonical bundle in  
$\mathbb{P}_{\ci \Delta}$  as $\ci M_n$.

\subsection{The variation of Hodge structures: degenerations and sub-monodromy systems}
\label{gaussmaninS} 
The primitive horizontal subspace in the cohomology of a Calabi-Yau $n$-fold $M_n$ comes with a polarized 
Hodge structure, see \cite{CoxKatz}\cite{MR2931227}\cite{MR0419438} for  reviews.  For families of Calabi-Yau $n$-folds  
$\pi: {\cal  M}_n \rightarrow {\cal M}_{cs}$   over their complex moduli space ${\cal M}_{cs}$ with fibre $\pi^{-1}(t)=M_t$ 
one captures  the Hodge decomposition  $H^n(M,\mathbb{C})=\oplus_{k=0}^n H^{n-k,k}$ with 
$\overline{ H^{p,q}(M)}=H^{q,p}(M)$, with respect to the real structure $H^n(M_n,\mathbb{R})$, 
in terms of Hodge filtrations\footnote{Sometimes, especially in the physics literature,  $F_p=F^{n-p}$ is used.} $F^\bullet={\{F^p} \}_{p=0}^n$  with 
$F^p=\oplus_{l\ge p}^p H^{l,n-l}$ so that  $F^p\oplus \overline{F^{n-p+1}}=H^n(M,\mathbb{C})$ , 
$H^{p,q}(M)=F^p(M)\cap \overline{{ F^q}(M)}$ and $H^{p,q}(M)=F^p/F^{p+1}$. 
Together  with the lattice $H^n(M_n,\mathbb{Z})$ this defines the {\sl Hodge structure}.
Unlike the $H^{p,q}(M_n)$ the  $F_p(M_n)$ vary holomorphically with the complex structure and 
fit into locally free sheaves with inclusion  ${\cal F}^p\subset {\cal F}^{p-1}$, which defines 
a {\sl decreasing or Hodge filtration}.  In particular
${\cal F}^0=R^n\pi_* \mathbb{C}\otimes {\cal O}_{M_{cs}}$ is the Hodge bundle ${\cal H}$,
which has a locally constant subsheaf $ R^n\pi_*\mathbb{C}$. Taking this as the flat section 
of ${\cal F}^0$  defines a flat connection, called the {\sl Gauss Manin connection} 
\begin{equation} 
 \nabla (s\otimes f) =s \otimes {\rm d} f \ . 
\label{gaussmanin} 
\end{equation}
The Gauss Manin connection fulfills the Griffiths transversality  
\begin{equation} 
\nabla{\cal  F}^p \subset {\cal F}^{p-1}\  
\label{griffithstransversality} 
\end{equation}  
and is equivalent to the Picard-Fuchs differential ideal ${\cal I}_{PF}$, which 
is {\sl the kernel} of the map $\phi(X_1,\ldots, X_r)=\nabla_{X_1}\ldots \nabla_{X_r} \Omega(t)$ 
from the  sheaves of linear differential operators ${\cal D}_i$ on $M_{cs}$ 
to the Hodgebundle ${\cal F}^0$, which makes the latter into a 
{\sl  ${\cal D}$-module}.
Here  $X_i=\partial_{a_i}$, $i=1,\ldots,r={\rm dim}({\cal M}_{cs})$ is a basis of 
vector fields on ${\cal M}_{cs}$.  As 
\begin{equation}
\partial_{a_i} \int_{\Gamma_t} \Omega(t)=\int_{\Gamma_t} \nabla_{\partial_{a_i}} \Omega(t)
\label{GMdiff} 
\end{equation}
the ${\cal D}_i$ annihilate the periods and determine them. 
 
${\cal D}$-modules are a notion to study monodromy systems. In particular  they 
were used in~\cite{MR579742,MR743382} to prove the Riemann-Hilbert problem for holonomic 
system in more variables, see~\cite{Kashiwarabook} for a review.  A criterium when the ${\cal D}_i$ 
generate ${\cal I}_{PF}$ was given in~\cite{Hosono:1993qy}: This is the case  if 
the symbols of the ideal ${\cal I}_{PF}$ at a point of maximal unipotent monodromy of $M_n$ determine the 
classical intersection ring of $W_n$ up to one normalization, see (\ref{ringrelation}). Iff this is the case, then one 
can obtain  exactly all  $3$-point couplings from  ${\cal I}_{PF}$ 
using Griffiths transversality~\cite{Hosono:1993qy}. These exact couplings  are in the A-model language 
the fully instanton corrected ones.  A {\sl sub-monodromy system}  is in this language a 
restriction of the dependent variables onto a subspace  ${\cal S}\subset {\cal M}_{cs}$, so that  
${\cal I}_{PF}|_{\cal S}$ is Picard-Fuchs ideal of another Calabi-Yau geometry. This restricts the 
${\cal D}$-modul to sub-modul  and the restricted Calabi-Yau  geometry stays  compact, if  the period of the 
Hodge bundle that maps to the structure sheaf of $W_n$  restricts to a period contained in 
the sub-modul, otherwise the restricted Calabi-Yau geometry becomes non-compact.        

The $h_n(M_n)$ solutions of the Picard-Fuchs 
equations  are independent  sections of the Hodge bundle 
that extend globally over all ${\cal M}_{cs}$.
We  group the solutions into the period vector $\Pi(a)=(\int_{\Gamma_1} \Omega_n(a),\ldots, \int_{\Gamma_{b_n}} \Omega_n(a))^T$ .                                        
 Because of the flatness of $\nabla$ the relevant data determining $\Pi(a)$ 
and (also the  dual $\tilde \Pi(a)$, see (\ref{picardfuchsfirstorderform})) are the monodromies of $\Pi(a)$  around the 
components of the discriminant  locus $\Delta=0$  in ${\cal M}_{cs}$.  The 
latter has to  be resolved  to $\overline{ {\cal M}}_{cs}$  so that the components $\Delta_i=0$ of $\Delta=0$ have all 
normal  crossings and the monodromies can be defined. The theorem of 
Landman describes the behavior of the periods at $\Delta=0$. Essentially 
they can have at most $\log(\Delta_i)^n$ singularities and/or finite order $k$ branch 
cuts. This is encoded in (Lefschetz)  monodromy  matrix $M_i$ that describes the 
transport of $\Pi(a)$  around $\Delta_i=0$. Landman's monodromy theorem 
states that $M_i$ has the property~\cite{MR0344248}
\begin{equation}
(M_i^k-{\bf{1}})^{p+1}=0, \ \ {\rm with} \ p\le n \ . 
\end{equation}          
Here $p$ is the smallest integer so that the r.h.s. is zero.  For $p=0$, $k>1$ one has 
an $\mathbb{Z}_k$ orbifold singularity. The case is discussed in sections \ref{discretegroups} 
and \ref{orbifoldpoints}. The cases  $k=1$ are the unipotent cases. The conifold in $n$ odd has $p=1$ 
and the  most relevant case  for mirror symmetry is the maximal unipotent case $p=n$. Mixed
cases are also relevant in physics, e.g. the $SU(N)$ Seiberg-Witten embedding has in the 
asymptotic free region $k=N$ and $n=1$. The degeneration of the Hodge structure is 
studied extensively in mathematics. However this is still a local  analysis and as such 
a linear analysis. In order to enjoy the full modularity properties of the amplitudes like 
the superpotential one has to do the global construction outlined in section 
\ref{threefoldbasis} and  \ref{fourfoldbasis}. An intermediate concept is local mirror 
symmetry,  which  deals with a sub-monodromy problem of the finite periods in a 
semi local limit, where part of $M_n$ decompactifies, see (\ref{locallimit}) and  section 
(\ref{stabledegenerationlimit}).  

As a local analysis  we recall as a simple example the limiting mixed Hodge structure 
for the unipotent cases, see~\cite{MR0417174}\cite{MR2931227} for review. Let $z_i\sim\Delta_i$ be the coordinates 
on the ith  punctured disk $D_i^\circ$  so that $(D^\circ)^r$ is a local neighborhood 
at a normal crossing point. The mixed case is reduced to unipotent case by 
changing  to multicover variabels $z^\frac{1}{k}\mapsto z$. Since $N_i=M_i-{\bf 1}$ 
is nilpotent one can define the Lie algebra generator $\EuFrak{N}_i=\log(M_i)$ as a 
finite expansion in $N_i$. $N_i$ and $\EuFrak{N}_i$ have the same kernel 
and cokernel. Schmid's nilpotent orbit theorem~\cite{MR0382272}  provides an extension $\overline {\cal F}^p$  
of ${\cal F}^p$  from $(D^\circ)^r$ to the product of full disks  $D^r$. In particular he extends the Gauss Manin 
connection to a map $\overline{\nabla}:\overline{{\cal F}}^0\rightarrow \overline {{\cal F}}^0\otimes \Omega^1_{\overline{\cal M}_{cs}} (\log(\Delta_i))$.
Here  the sheaf of  rational one forms $\Omega^1_{\overline{\cal M}_{cs}} (\log(\Delta_i))$ is in the case of $l$ unipotent divisors locally generated 
by $\frac{{\rm d }z_1}{z_1},\ldots,\frac{{\rm d }z_l}{z_l}$, $ {\rm d} z_{l+1},\ldots,{\rm d}z_r$.   On  $D^r$ he introduces  
single valued periods 
\begin{equation}  
\Pi_S(z)=\exp\left(-\frac{1}{2 \pi i} \log( z_i) \EuFrak{N}_i\right) \Pi(0)\ ,
\end{equation}   
extends that to a section of $\overline {\cal F}^p$ and shows that this is a leading order 
approximation to the periods on the disk. From the degenerations of the periods found in 
mirror symmetry calculations that is no surprise and contrary to the claim in~\cite{Donagi:2012ts} it  
gives not a very useful approximation to the superpotential as it misses  the terms  from 
the $\hat \Gamma$ class in the periods, i.e. not only terms that are  exponentially 
suppressed, but for the fourfolds terms that actually diverge logarithmically at $z=0$. 

An important implication of the work of ~\cite{MR0382272}\cite{MR0498552} is 
the ability to define the {\sl limiting mixed Hodge structure}, which describes how 
the integral Hodge structure of the singular model sits inside the integral Hodge 
structure of the smooth case. In particular at $0\in D^r$ there is a limiting Hodge 
filtration ${\overline {\cal F}}^\bullet={\cal F}_{\rm lim}^\bullet$ with 
$ \EuFrak{N}_j ({\cal F}_{\rm lim}^p) \subset {\cal F}_{\rm lim}^{p-1}$. Both $ \EuFrak{N}_i$ and the extension 
$\overline{ \nabla}_{\theta_{z_j}}  ( {\cal F}_{\rm lim}^p)\subset {\cal F}_{\rm lim}^{p-1}$ 
induce a linear map ${\cal F}_{\rm lim}^{p}/ {\cal F}_{\rm lim}^{p-1}
\mapsto {\cal F}_{\rm lim}^{p-1}/ {\cal F}_{\rm lim}^{p}$ and are identified as  
$\overline{ \nabla}_{\theta_{z_j}}=-\frac{1}{2 \pi i}  \EuFrak{N}_j$. Moreover if $\Pi_S(z)$  is a 
multivalued flat integer section of ${\cal H}_\mathbb{Z}$ then $\Pi_S(0)$ is an 
integral element over $0$. The mixed Hodge structure comes from the {\sl monodromy  weight filtration} 
$W^\bullet$ with $W_0\subset W_1\subset \ldots \subset W_{2n}=H^n(M_t,\mathbb{C})$. For 
any linear combination $ \EuFrak{N}$ of the $ \EuFrak{N}_i$ with strictly positive coefficients one defines 
$W_0={\rm Im}( \EuFrak{N}), W_1={\rm Im}( \EuFrak{N}^{n-1})\cap  {\rm Ker}( \EuFrak{N}),W_2=
{\rm Im} ( \EuFrak{N}^{n-2} ) \cap {\rm Ker} (\EuFrak{N})+{\rm Im} ( \EuFrak{N}^{n-1} ) \cap {\rm Ker} (\EuFrak{N}^2), 
 \ldots, W_{2n-1}={\rm Ker} (\EuFrak{N}^n)$. Let ${\rm Gr}_k=W_k/W_{k-1}$. It is easy to see that $ \EuFrak{N}(W_k) \in W_{k-2}$ and 
it follows from the Jacobson-Morozov Lemma that $ \EuFrak{N}^l:{\rm Gr}_{k+l}\sim {\rm Gr}_{k-l}$. Much more non-trivially ${\cal F}^\bullet_{\rm lim}$ is a 
Hodge structure of weight $k$ on ${\rm Gr}_k$, which means that $({\cal F}^\bullet_{\rm lim},W_\bullet)$ 
is a mixed Hodge structure. It can be shown that $ \EuFrak{N}$ is the lowering operator of a $SL(2,\mathbb{C})$ 
action on the LMHS~\cite{MR0382272}. 

The easiest example is the nodal degeneration of  a genus $g$  
Riemann surface $\Sigma_g$. One can chose a symplectic basis $A^i,B_i$ $i=0,\ldots, g-1$ 
such that $A^0$ degenerates. By the Lefschetz formula (\ref{conifoldmonodromy}  the only 
cycle which is not monodromy invariant is $M:B_0\mapsto A^0+B_0$. $N= \EuFrak{N}$ is nilpotent $\EuFrak{N}^2=0$. One has 
$ \EuFrak{N}:B_0\mapsto A^0$ while all others cycles are annihilated. $W_{-1}:=0$, 
so ${\rm Gr}_0=\mathbb{Q}\cdot A^0$ of grade $0$ must be of type $(0,0)$, 
$ {\rm Gr}_1={\rm span}_\mathbb{Q}\{A^1,\ldots,A^{g-1},B_1,\ldots,B_{g-1}\}=
H_n(\Sigma_{g-1},\mathbb{Q})$  and  ${\rm Gr}_2=\mathbb{Q}\cdot B_0$ of grade $2$ must be of type $(1,1)$. 
I.e. over $\mathbb{Q}$ the cohomology of $\Sigma_g$ splits  and on  $H_n(\Sigma_{g-1},\mathbb{Q})$ and  one 
has a closed  sub-monodromy problem on the latter.  The situation is very similar for the conifold transition 
in Calabi-Yau  3-folds, up to the fact that in a real transition $m$ $S^3$ with $r$ relations 
shrink~\cite{Anderson:2013rka}. 

Another typical situation in which a sub-monodromy problem arises is for the central fibre 
in a stable degeneration limit, see section \ref{stabledegenerationlimit} for the explicit solution 
of the latter for the $\lim_{het} K_3={\rm dP}_9\cup_{T^2} {\rm dP}_9$ example. In addition to 
the limiting mixed Hodge structure described above, the Clemens-Schmid exact sequence 
relates the Hodge structure on the central fibre to the one of the smooth space, see D. Morrison's 
article in~\cite{MR0419438} for a review and~\cite{Anderson:2013rka} for examples.  
The  approach to glue in the complex degeneration limit  two or more components of $M_n$  
with positive first  Chern class  along the central fibre with vanishing  first chern class  is a very old idea  from perspective of 
Gromov-Witten invariants on $M_n$, as the latter are deformation invariant and the calculation in 
the degeneration limit is at least possible. There has been a lot of recent progress in this 
program by Gross and Siebert, see~\cite{MR3011419} for review. Maulik and Pandharipnade used it 
to get information about higher genus invariants  on the degenerate quintic, where the central 
fibre is a K3. However working purely in  the A- model with relative Gromov-Witten invariants
is combinatorial extremely  complicated and the authors obtain explicitly  only the degree 1 genus two 
Gromov-Witten invariant on the quintic~\cite{MR2248516}. It is reasonable to assume that the B-model perspective, 
i.e. the  modularity in the sub-mondromy problem and integrality of the BPS invariants is of great help 
to tackle the combinatorics.  In fact  Maulik, Pandharipande  and Thomas~\cite{MR2746343} use 
this approach in the stable degeneration limit of F-theory to the heterotic string, 
discussed more in section \ref{stabledegenerationlimit}, where the modular properties 
are best understood, to prove integrality of the BPS invariants.

In order to use the localization argument w.r.t. to the group action $G$ 
its fix points ${\cal S}$  must be a sub locus in ${\cal M}_{cs}$ that leads to a well defined 
variation of the Hodge structures of the invariant periods. If we include 
all orbifold divisors in $\Delta$  then ${\cal S}\in \Delta$ then ${\cal S}\subset \Delta$. 
This structure in particular defines a sub-monodromy problem in the invariant subspace ${\cal S}$.     
We therefore discuss in the rest of the section two and a half methods for deriving the differential 
ideal ${\cal I}_{PF}$ that makes the occurrence of the sub-monodromy problem obvious in the 
hypergeometric case.

\begin{itemize}
\item (i) {\sl Gelfand-Graev-Kapranov-Zelevinsky method:} The ``periods''
\be 
\hat \Pi_i(a)=\int_{\Gamma_i} \Omega_n=\int_{\Gamma_i}  \oint_\gamma \frac{\mu}{W_{\Delta}(x,a)},
\label{period} 
\ee
where $\gamma$ is a path in $\mathbb{P}_{\Delta^*}$ around  
$W_{\Delta}=0$, $\Gamma_i\in H_n(M_n,\mathbb{Z})$ and~\footnote{The $x_i$ are obtained from the 
$Y_i$ by setting all but $n+1$ suitable ones to 1. The choice 
is canonical if $\mathbb{P}_{\Delta^*}$ is the resolution 
of a weighted projective space $\mathbb{P}^{n+1}(w_1,\ldots, w_{n+2})$. 
Then the $x_i$ are its coordinates. The issue can be avoided altogether by working in the Laurent  
polynomial (\ref{laurant}) defining 
$\Pi_i(a)=\oint_\gamma \int_{\Gamma_i}\frac{1}{w_{\Delta}(a,t)} \bigwedge_{i=1}^{n-1} \frac{d t_i}{t_i}$.}       
\be 
\mu_{n+1}=\sum_{j=1}^{n+1} (-1)^j w_i x_j d x_1\wedge \ldots \wedge {\widehat {dx_j}} \wedge \ldots   d x_{n+1}\ ,
\label{mu} 
\ee
are annihilated by a restricted set of GKZ differential 
operators of~\cite{MR902936,MR948812} appropriate for the resonant case   
\be
{\cal D}_{l^{(k)}} =\prod_{l^{(k)}_i>0} \left(\frac{\partial}{\partial a_i}\right)^{l^{(k)}_i} - \prod_{l^{(k)}_i<0} 
\left(\frac{\partial}{\partial a_i}\right)^{-l^{(k)}_i}\ , \quad  
{\cal Z}_i=\sum_{j=0}^r {\bar \nu}_{j,i} \theta_j-\beta_i\ ,
\label{GKZ} 
\ee
where $1\le k\le |\Delta_0|-d-1$ , $1\le i\le n+1$,   $r=|\Delta_0|$ and $\beta=(-1,0,\ldots,0)$ and 
$\theta_i=a_i\frac{\partial}{\partial a_i}$.
The first set of operators $k=1,\ldots,h_{11}(W)$ follows from the linear relation between the points in 
$\bar \Delta$, which are encoded in $\{l^{(i)}\}$ obtained from an analogous 
construction to the one that lead to (\ref{linerarelations}).  The second set of 
operators $j=0,\ldots,d$  eliminates the $\mathbb{C}^*$ scaling symmetry of $W_{\Delta}$ and  
the $\mathbb{T}^d$ redundancy A1) in the parametrization of ${\cal M}_{cs}$  in terms of the $a_i$. 
It was observed in~\cite{Hosono:1993qy} that~(\ref{GKZ}) are sufficient to solve model of type I and that for type II models  the 
infinitesimal version of the invariance of (\ref{period}) under each transformation 
corresponding to the roots~(\ref{weightedhomogeneoustransformations}) leads to 
first order differential operators ${\cal Z}'_k$, which supplement (\ref{GKZ}) and that 
for general type III models one needs further differential equations   ${\cal Z}''_k$ coming from 
relations between monomials corresponding to points at height modulo  $\partial_i W_{\Delta}$. 
Such relation can be found algorithmically using the Groebner basis for the Jacobian 
ideal $J$, see below. The main advantage of using the differential symmetry~(\ref{GKZ}) 
is that solutions can be very explicitly written down near the large complex structure 
point ${\underline z}=0$, where $z(a)$ is defined in (\ref{largevolumecoords}). One finds~\cite{Hosono:1994ax} with the definition 
\begin{equation} 
\omega_0({\underline{z}},{\underline{\rho}}) = \sum_{{\underline{n}} } c({\underline{n}},{\underline{\rho}}) z^{{\underline{n}}+{\underline{\rho}}}\ \ 
{\rm where} \ \  c({\underline{n}},{\underline{\rho}})=\frac{\prod_j \Gamma(\sum_\alpha l_{0j}^{(\alpha)} ( n_\alpha+\rho_\alpha)+1)}{
\prod_i \Gamma(\sum_{\alpha} l_i^{(\alpha)}( n_\alpha+\rho_\alpha)+1)}\  
\label{solution} 
\end{equation}
that $\omega_0({\underline{z}},{\underline{\rho}})|_{{\underline{\rho}}=0}$ is a solution 
and all other solutions are given by derivatives w.r.t. 
$\rho_i$ at ${\underline{\rho}}=0$ by the Frobenius method,  
see section~\ref{threefoldbasis}. Note that in  formula (\ref{solution})  
we have included as a slight generalization of the discussion in 
section \ref{complexmoduli} (in particular (\ref{linerarelations})) the possibility of 
complete intersections in toric ambients spaces. In this case the $l_{0j}$ equal $-d_i$, where $d_i$ 
are  the degrees of the $W_{\Delta_i}$,  $i=1,\ldots,r$ in section~\ref{batyrev}, see~\cite{Hosono:1994ax}.         

The important point regarding the sub-monodromy  system is as follows. 
If we restrict to $W_{\ci \Delta}$ we get a subset ${\cal D}_{\ci l^{(k)}}$  
of differential operators  with $\{\ci l\}\subset \{l\}$, which define a 
closed differential  ideal $\ci {\cal I}_{PF}$, if $\ci \Delta$ 
is reflexive. In particular these operators depend only on the 
relation of the invariant points and are independent 
of the blow ups by the $Y_{K_i}$. Therefore  they define a closed  
sub-monodromy system on the invariants locus ${\cal S}\subset 
{\cal M}_{cs}$, namely the one of periods of the Calabi-Yau $\ci M_n$.

Let us finally note that $a_0\Omega$, with $a_0$ the 
coefficient of the inner point in (\ref{WDelta}), is the right measure and 
\begin{equation} 
\Pi_i(a)=a_0 {\hat \Pi}_i(a)\, ,
\label{innerpoint}
\end{equation}
the right period~\footnote{$\Omega$ is not 
invariant under $T_d$, because the $x_i$ $(Y_i)$ is not invariant. $W_{\Delta}$ is invariant 
because the $a_i$ compensate the action. In particular $a_0 \Omega$ is invariant. 
The point of working with $\hat \Pi_i(a)$ is that the differential operators 
(\ref{GKZ}) are easier to state.}.  The operators  readily 
convert to operators annihilating $\Pi_i(a)$ by commuting $a_0$ 
in from the left. The GKZ system has more solutions than the periods 
of $M_n$ and hence does not determine them. However one can factorize 
the operators to create a differential ideal, that singles out the right 
solutions~\cite{Hosono:1993qy}. 
\item (ii) {\sl Dwork-Griffiths reduction method:}\cite{MR0188215,MR0242841}
Consider the graded ring defined by the Jacobian ideal  $J$
generated by the partial derivatives of the weighted homogeneous 
polynomial $W_{\Delta}(x)$ of degree $d=\sum_{i} w_i$ in the weighted 
projective space  $\mathbb{P}^{n+1}(w_1,\ldots, w_{n+2})$
\begin{equation} 
{\cal R}=\frac{\mathbb{C}[x_1,\ldots, x_{n+2}]}{\{\partial_{x_i} W_{\Delta}(x)\}}\ ,    
\label{monomialring}  
\end{equation}
with elements $\phi^{i_{k}}_{d k}(x)$, of weighted degree $d k$ with $k=0,1,\ldots, d$ 
and $i_{k}=1,\ldots,h^{hor}_{n-k,k}$. Indeed this ring elements span the 
horizontal cohomology~\footnote{ Throughout the section we assume 
that all deformations of $W_{\Delta}(x)$ are polynomial, i.e. 
no ``twisted fields'' in other places indicated by $(h^{twisted}_{pq})$.}             
$H^{hor}_{n-k,k}(M_n,\mathbb{C})$  by the Griffiths residuum formula 
\begin{equation}
\chi_{i_{k},n}=\int_\gamma \frac{\phi^{i_{k}}_{dk} a_0 \mu }{ W_{\Delta}(x)^{k+1}}\  
\label{ringelement}  
\end{equation}
that yields a basis of rational cohomology. It is easy to calculate the number of  
$\phi^{i_{k}}_{d k}(x)$ for a given degree by the  Poincar\'e polynomials, which 
were used to count the chiral (and anti-chiral) fields  in Calabi-Yau/ Landau 
Ginzburg models~\cite{Vafa:1989pa}. In fact the E-polynomials~\cite{Batyrev:1994ju} 
are based on the same idea.  
 
The point is that any other numerator that lies in $J$ can be 
reduced modulo exact forms by the formula ($deg(P_j(x))=(l-1)d+w_j$ to make the following well defined)  
\begin{equation}
d \phi =\frac{l \sum P_i \partial_{x_i} W_{\Delta}}{W_{\Delta}(x)^{l+1}}\mu_{n+1}
-\frac{ \sum \partial_{x_j}  P_j }{W_{\Delta}(x)^{l}} \mu_{n+1} \ .  
\label{partialintegration} 
\end{equation}
Using Buchbergers algorithm one can chose a Groebner basis for 
the Jacobian ideal $J$.   The properties of the Groebner basis  allow to decompose 
any monomial $m_{kd}^{(i)}(x)$  of degree ${k d}$ uniquely as follows 
\begin{equation}
m_{kd}^{(i)}=q^{(i)}_j(a) \tilde M_j (x)+\sum_{j} P^{(i)}_j (a,x)\partial_{x_j} W_{\Delta}\ .
\label{reduce}
\end{equation} 
Here $\tilde M_j(x)$ are degree $k d$ monomials  in the multiplicative 
ring ${\cal MR}(M_j(x))$  generated by the $M_j(x)$ in (\ref{WDelta}), 
$q^{(i)}_j(a)$ is an unique rational function and the $ P^{(i)}_j (a,x)$ are 
likewise uniquely determined. In practice one uses (\ref{reduce}, \ref{partialintegration})  as follows: 
One takes $n+1$ derivatives of (\ref{period}) w.r.t. to any of $a_i$ in (\ref{WDelta}). 
This produces an integrand $\tilde M_i(x)/W_{\Delta}^{n+2}$, whose numerator is 
completely  reducible by  (\ref{reduce}) to the last term on the r.h.s. The first term on the r.h.s 
of (\ref{reduce}) is zero as the ring ${\cal R}$ is empty at this degree. Using (\ref{partialintegration})  
one can reduce the integrand, up to exact terms which do not  affect the period integral,   
to sums of $m^{(i)}_{nd }/W_{\Delta}^{(n+1)}$.  Repeating the procedure reduces the  $n+1$th 
derivative to lower derivatives of (\ref{period})  with rational coefficients. This produces a differential
operator for (\ref{period}) as in the GKZ method.   Again the point is that this procedure 
closes into a closed sub-monodromy problem, if the Newton polytope $\ci \Delta$ 
is reflexive.   
\item (iii) {\sl The Mellin-Barnes integral  method:}
At  least for the toric cases this method is very similar to the GGKZ method and relies on taking  
derivatives of the explicite Mellin-Barnes integral representations~\cite{Candelas:1990rm},\cite{Hori:2013ika}, 
which depend only on the $l^{(\alpha)}$-vectors  in~\cite{Candelas:1990rm} and manipulating 
the Mellin-Barnes integral.  We discuss the formalism  once the necessary background has been 
developed in  item (i) of  section \ref{hemisphere}. 
\end{itemize}

The formulas (\ref{period},\ref{monomialring},\ref{ringelement},\ref{partialintegration}) 
have generalizations to the complete intersection case, see e.g.~\cite{Hosono:1994ax} also 
(\ref{GKZ}) can be generalized~\cite{Batyrev:1994pg,Hosono:1994ax,Klemm:2004km}, see also 
section~\ref{hemisphere}. 

The  period vector $\tilde \Pi(a)$ defined as $\tilde \Pi_i(a)=\int_{\Gamma} \chi_{i,n}$  
is annihilated by the connection
\begin{equation} 
\nabla_i\tilde \Pi(a)=[\partial_{a_i}+ C_{a_i} (a)] \tilde \Pi(a)=0\,.       
\label{picardfuchsfirstorderform}
\end{equation}
 Here 
$\chi_{i,n}$ spans a rational basis of the horizontal cohomology of 
$M_n$, see (\ref{ringelement}). The connection (\ref{picardfuchsfirstorderform}) 
is obtained from ${\cal I}_{Pf}$  derived  with methods (i) - (iii) by rewriting it as a system of 
first order equations in the obvious way. 

As we have seen in Calabi-Yau manifolds embedded in  toric spaces with embedded polyhedra as 
in (\ref{transitions})  it is  quite simple to conclude that there is a sub-monodromy problem
on the invariant subspace ${\cal S}$. This is  due to the special properties of generalized hypergeometric 
differential systems with coefficients given in (\ref{solution}), which govern the periods of 
torically embedded Calabi-Yau spaces. This methods does not exhaust all consistent sub 
monodromies problems of a given Calabi-Yau in a toric ambient space as many sub-monodromy  
problems are of the Apery type~\cite{Zagier}\cite{MR2454322} and at least for the one moduli 
problems the vast majority of cases is of the Apery type.  This is discussed for the $n=1$ case 
in~\cite{Zagier} and for the $n=3$, $h_{n-1,1}=1$ case in~\cite{MR2454322} 
and references therein. They come sometimes from much more exotic 
construction like from determinantal or Pfaffian CY-embeddings 
in Grassmannians or Flag-manifolds or don't have a known geometrical
interpretation at all.

After introducing the concept of the flux super potential  we
give in section~\ref{submonodromy},   based on the integral structure of the 
monodromy, arguments that the idea  of landscaping by fluxes applies to these cases as well.


\subsection{The calculable terms in the effective action}     
\label{effectiveaction} 

We review the simplest analytic quantities that depend on 
the complex moduli and can be calculated from the topological 
string point of view and their interpretation in F-theory. 
All these can be calculated in the B-model from the period integrals 
w.r.t. to an integral basis in homology or in the A-model by 
localization w.r.t. to the torus action of the ambient 
space. 
The methods discussed in section~\ref{gaussmaninS} 
yield linear differential equations for the periods, i.e.
it remains to find those linear combination in the 
solution space, which correspond to the integer basis.    

\subsubsection{K\"ahler potential, superpotential and Ray-Singer torsion} 
\label{KSRST}

Our main interest is in the quantities which are 
involved in minimizing the flux superpotential.  
Since all gauge symmetry enhancements are at high 
codimension in the generic moduli space of the 
``clean sheet''  Calabi-Yau manifolds one might 
conclude that F-theory predicts that there is 
no unbroken gauge group at low energies. 
However as it was demonstrated in~\cite{Curio:2000sc} 
for the case of Seiberg-Witten gauge symmetry 
embedding in Calabi-Yau geometries a flux 
superpotential can at least reduce this 
choice of o set of measure zero in the continuous 
moduli space to the discrete choice of a flux, 
which in view of (\ref{tadpolecondition}) might 
even be a discrete choice in a {\sl finite set} of 
consistent theories. 

We  are interested in F-theory super potentials on 
fourfolds and Type II super potentials on threefolds and 
start with elements of the theory, which apply to the 
complex moduli spaces of Calabi-Yau spaces of any 
dimensions namely the Weil-Petersson metric 
on the complex moduli space ${\cal M}_{cs}$, which 
determines the metric in front of the kinetic terms  
of the moduli fields. The latter exist geometrically, 
because by the Tian-Todorov theorem the moduli space of 
Calabi-Yau manifolds  is unobstructed~\cite{MR915841,MR1027500}. Its 
tangent space are described by  $H^1(M_n,T M_n)$, 
which can be identified with $H^{n-1,1}(M_n)$ by contracting with $\Omega_n$.  
The metric on the complex moduli space called Weil-Petersson metric 
is K\"ahler and the real K\"ahler potential is written in terms of $\Omega_n$ as~\cite{MR915841,MR1027500}
\be
\exp(-K(a,\bar a))=(-1)^{(n(n-1)/2)} (2 \pi i )^n \int_{M_n} \Omega_n(a)\wedge \bar  \Omega_n(\bar a)\  .
\label{Kaehlerpotential} 
\ee 
As the no-where vanishing holomorphic $\Omega_n$ form lives in a holomorphic line 
bundle ${\cal L}$ over ${\cal M}_{cs}$ transforming as $\Omega(a) \rightarrow e^{f(a)} \Omega(a)$,
the K\"ahler form transforms in the K\"ahler line bundle with 
K\"ahler transformations $K(a,\bar a)\rightarrow K(a,\bar a)  
- f(a) -\bar f(\bar a)$ so that $e^{-K}$ is a section of ${\cal L}\otimes \bar {\cal L}$.
One has natural covariant derivatives w.r.t. to the W-P metric connection and the K\"ahler 
connection that  act on sections $V_{j\bar \jmath}\in T_{1,0}^*{\cal M}_{cs} 
\otimes  T_{0,1}^*{\cal M}_{cs}\otimes  {\cal L}^{\otimes n}  \otimes \bar {\cal L}^{\otimes m}$ 
as 
\begin{equation}
D_i V_{j\bar \jmath} = \partial_i V_{j\bar \jmath}-\Gamma^{l}_{ij} V_{l \bar \jmath} + n K_i V_{j\bar \jmath},
\qquad   
D_{\bar \imath} V_{j\bar \jmath} = \partial_{\bar\imath} V_{j\bar \jmath}-\Gamma^{\bar l}_{\bar i\bar j} V_{i \bar l} + m K_{\bar \imath}  V_{j\bar \jmath},
\end{equation}
with $K_i=\partial_i K$  and $K_{\bar \imath}= \partial_{\bar\imath} K$.  Griffiths transversality states 
that any combination  $\nabla_{i_1}\ldots\nabla_{i_{r}}$    of the application of the Gauss-Manin connection in the direction  of complex moduli $a_{i}$, 
$i=1,\ldots,{\rm dim}({\cal M}_{cs})$  to  $\Omega_n$  has the property 
\begin{equation}
{\underline {\nabla}}_{\underline {i}}\Omega_n=\nabla_{i_1}\ldots\nabla_{i_{r}} \Omega \in {\cal F}^{n-r}  \  ,
\end{equation}  
cff section \ref{gaussmaninS}.
As a corollary to the T-T theorem $\chi_i=(\nabla_i+K_i) \Omega$ spans $H^{n,n-1}(M_n)$. All derivatives 
of $\Omega$ generate the primary  horizontal subspace $H^n_H(M_n)$. By consideration of Hodge type and by 
(\ref{tranversality})  one gets 
\begin{equation} 
Q({\underline {\nabla}}_{\underline {i}}\Omega_n,\Omega)= 
\int_{M_n} (\nabla_{i_1} \ldots \nabla_{i_r} \Omega_n)\wedge \Omega_n
= \left\{\begin{array}{ll} 0 & \ \ {\rm for}\  0 \le r <n \\ 
         C_{{i_1} \ldots {i_n}} & \ \ {\rm for}\  r=n
        \end{array}\right.
\label{griffith} 
\end{equation}
        
The $n$-point coupling  $C_{{i_1} \ldots{i_n}} $ transforms as a section 
in ${\cal L}^2 \times {\rm Sym}^n(T^* {\cal M}_{cs})$.  The underlying field theory
of the non-linear $\sigma$ model has $(2,2)$ supersymmetry and can be 
twisted to the $A$ and the $B$ model so that the chiral-chiral $(c,c)$-ring
and the chiral-antichiral $(c,a)$-ring are identified with 
$H^p(M_n,\wedge^q T M_n)$ and $H_{dR}(M_n)$ respectively~\cite{Witten:1991zz}, 
where we focus on $p=q=1,\ldots, n$, but only allow deformations for the $p=q=1$  
part of the cohomology~\footnote{It was claimed in \cite{Cecotti:2009zf} that the more general 
deformations introduced by Kontsevich lead to residua calculation that yields exact 
expressions for the Yukawa couplings.}   
The holomorphic flux super potential is given by   
\be 
W=\int_{M_n} \Omega_n(a)\wedge G_n, \
\label{superpotential} 
\ee 
where $G_n$ has for fourfolds the properties discussed 
section~\ref{fluxquantization}, while for threefolds it 
is just integer quantized in $H^3(M_3,\mathbb{Z})$.  The superpotential 
$W$ obviously transforms in ${\cal L}$. Generically 
one can show using world-sheet arguments and mirror symmetry 
that $\Omega$ has to be viewed as the dilaton field 
in topological string theory so that an amplitude, which can 
get genus $\chi=2 -2 g -h$ instantons corrections, transforms 
in ${\cal L}^\chi$~\cite{Bershadsky:1993cx}, e.g. $W$ can in fact be 
interpreted as a disk instanton potential.  As a consequence there is a natural 
${\cal L}$ covariant derivative acting on $W(z)$ 
\be 
D_i=\partial_i + \partial_i K\ , 
\ee
with $\partial_i=\partial_{t_i}$, where $t_i$ is flat w.r.t. the Weil-Petersson connection.  

To organize  possible holomorphic terms related to the Griffiths Frobenius 
structure  one can use mirror symmetry and  look at the index theorem 
that constrains the  topological string $A$-model amplitudes related to 
maps $\Phi:\Sigma_{g} \rightarrow C_\beta$  from the worldsheet $\Sigma_g$  
to curves $C_\beta\subset W_n$ in the class $\beta\in H_2(W_n,\mathbb{Z})$ 
whose virtual dimension is   
\begin{equation}          
{\rm dim}_{\mathbb{C}} {\cal M}_{g,\beta}(W_n)=\int_{C_\beta} c_1(TW_n)-(g-1) ( n-3)\ .
\end{equation} 
For Calabi-Yau $n=3$-folds  ${\rm dim}_{\mathbb{C}} {\cal M}_{g,\beta}(W_3)=0$ and 
there is a rich Gromov-Witten theory on the generic CY 3-fold in each genus. 
For $g=0$ and $n=4$   one has  ${\rm dim}_{\mathbb{C}} {\cal M}_{0,\beta}(W_4)=1$. 
To get an enumerative problem one has to put the constraint $C_{\beta}\cap H_4\neq 0$, where 
$H_4\in H^{prim}_4(W_4,\mathbb{Z})$. The corresponding  generating functions of restricted 
genus zero instantons is encoded  the period in $H^p$ in (\ref{exapndomega40}) 
whose integer cohomology cycle is determined in section  (\ref{fourfoldbasis}). It 
enjoys an integral genus zero instanton expansion ($q=e^t$)
\begin{equation}
H^p(q)={\rm quad}^p(t)+ \sum_{\beta\in H_2(W,\mathbb{Z})} n_\beta^p {\rm Li}_2(q^\beta)\ .
\label{integerfourfold}
\end{equation} 
This is the same integrality property exhibited by the superpotential counting 
holomorphic disks ending on special Lagrangians in Calabi-Yau 3-fold. In fact for Harvey 
Lawson type branes in non-compact toric Calabi-Yau fourfold spaces can be found such that the periods of the non-compact fourfold 
do give the disk superpotential~\cite{Mayr:2001xk}. In this identification 
some of the closed fourfold  moduli become brane moduli other map  to 
closed moduli of the 3-fold.           
                     
Besides the superpotential there is another holomorphic quantity in $N=1$ theories, which is the 
gauge kinetic function, which in the brane setting is generated
by the annulus amplitude. On the topological string side there is
the genus one amplitude the generating function for elliptic curves  
that is non-trivial on all Calabi-Yau manifolds, because  
${\rm dim}_{\mathbb{C}} {\cal M}_{1,\beta}(W_n)=0$. In the $B$-model it corresponds to  
the Ray-Torsion analytic torsion, given concretely as~\cite{Bershadsky:1993cx,Klemm:2007in}
\begin{equation}
F_1=\frac{1}{2}\sum_{p,q} (-1)^{p+q} \left(p+q\over 2\right) {\rm Tr}_{p,q} [\log (g)]-\frac{\chi(M_4)}{24} K +\log |f|^2 \ ,
\label{integratedF1}
\end{equation}
which evaluates on a fourfold with $h_{21}=0$ to~\cite{Klemm:2007in} to 
\begin{equation}
F_1= \left(2+h_{11}(M_4)-\frac{\chi(M_4)}{24} \right) K -\log \det G+\log |f|^2 \ ,
\label{integratedFFF1}
\end{equation}
where $K$ is the K\"ahler potential on $\cM_{cs}$, $G$ the K\"ahler metric
and $f$ an holomorphic ambiguity in the definition of $F_1$.
$F_1$ transforms as ${\cal L}^0$. This suggests that derivatives of it 
w.r.t. to the those complex  moduli that corresponds to brane position  
moduli on a degenerate elliptic Calabi-Yau manifold should be identified with 
the annulus contributions between the corresponding branes and give 
the exact gauge kinetic terms in the corresponding  limit. For generic  
fourfolds there  are no higher genus ($g>1$) Gromov-Witten invariants
because  ${\rm dim}_{\mathbb{C}} {\cal M}_{g>1,\beta}(M_{n>3})<0$. Moreover 
as there are open and closed higher genus Gromov-Witten in the 3-fold geometry 
we expect that the degenerate limit has symmetries that create 
additional zero modes which   
make ${\rm dim}_{\mathbb{C}} {\cal M}_{g>1,\beta}(W_4^{deg})\ge 0$ .

\subsection{Fixing an integral monodromy basis using mirror symmetry  and Griffiths-Frobenius geometry}
\label{integralmonodromie}
The real structure  (\ref{realstructure},\ref{Kaehlerpotential}) and the  pairing $Q(\cdot,\cdot)$ with the Griffiths transversality 
(\ref{griffith})  condition put  strong restrictions on the period geometry  of Calabi-Yau n-folds. For $n=3$  the geometry is called {\sl special geometry} and has been 
found  in~\cite{BG}. In physics the same structure was found to be an intrinsic feature of 
$N=2$ supergravity theories~\cite{Strominger:1990pd}\cite{Castellani:1990tp}.  
For fourfolds  (\ref{griffith}) the even bilinear form $Q(\cdot,\cdot)$ 
gives algebraic as well  differential relations among the periods. The Frobenius structure  
is  more interesting as its  maximal grade is $n$, which implies that for $n\ge 4$ it allows 
for various types of three point functions and non-trivial associativity constraints. In addition 
the real structure (\ref{Kaehlerpotential}) gives  non-holomorphic relations 
between the two types of couplings and the metric structure. 
Even though (F-theory)  compactifications on $4$-folds lead to $N=1$  supergravity and 
the structure is in part similarly realized in open string compactifications with  $N=1$  
supergravity, it is not intrinsic  in $N=1$ supergravity. We  will refer to the 
structure therefore as Griffiths-Frobenius geometry with the understanding that 
it is $N=2$ special geometry for compactifications on $M_3$. Finding the 
Griffiths-Frobenius geometry in  4d $N=1$ physics on the other hand would be 
a hint for string theory.   

Combined with mirror symmetry, which maps the Griffiths-Frobenius geometry  on the $(c,c)$ 
B-model ring to the one on the  $(a,c)$ A-model ring, the constraints turn out to be strong 
enough to find an integral monodromy basis. For CY 3-folds with one parameter this was  
done in~\cite{Candelas:1990rm} \cite{Font:1992uk}\cite{Klemm:1992tx} and for arbitrary 
number of parameters in~\cite{Hosono:1994ax}.  We give a result oriented review of 
this in section \ref{threefoldbasis}, where we stress the elements of the theory  
that apply to any dimension.

Even though genus zero instantons for one parameter families of Calabi-Yau spaces in 
various dimensions have been calculated in~\cite{Greene:1993vm} and for multi parameter 
fourfolds in~\cite{Mayr:1996sh,Klemm:1996ts} and even genus one 
instantons  and their  multi covering formula of the genus zero curves has been 
found and confirmed in several multi parameter fourfolds~\cite{Klemm:2007in} 
the integral monodromy basis on the primary horizontal subspace of $H_4(M_4)$ has 
not been determined.  The structures discussed partly in~\cite{Greene:1993vm} and in 
\cite{Klemm:1996ts,Mayr:1996sh} are a prerequisite to determine this basis and will be 
discussed and extended in the  section \ref{frobeniusstructure}.   Expressions for the 
K\"ahlerpotential the large complex structure points can be obtained from the sphere
partition function~\cite{Doroud:2012xw,Benini:2012ui}\cite{Jockers:2012dk}\cite{Honma:2013hma}.  

In section \ref{fourfoldbasis} we use the structures described above 
to  find this basis. We exemplify this first with the sixtic in 
$\mathbb{P}^5$ and eight other examples and generalize this to  the most general
 constructions of fourfolds  using the hemi-sphere partition function in 
section~\ref{hemisphere}.  Once this integral basis has been 
determined one can also set up the $tt^*$ structure and study it at  those 
degenerations of fourfolds that are interesting for F-theory.       

\subsubsection{The Frobenius algebra  and the $tt^*$ structure} 
\label{frobeniusstructure} 
A Frobenius algebra has the following elements. It is a graded vector space 
$\cA=\oplus \cA^{(i)}$, $i\ge 0$  
with a symmetric non-degenerate bilinear form $\eta$ and a cubic form 
\begin{equation} 
C^{(i,j,k)}: \cA^{(i)}\otimes \cA^{(j)} \otimes  \cA^{(j)} \rightarrow  \mathbb{C}.      
\end{equation}
Here is  the list of defining properties:       
\begin{itemize} 
 \item FAs) Symmetry: $C^{(i,j,k)}_{abc}=C^{(\sigma(i,j,k))}_{\sigma(abc)}$ under any permutation of indices. 
 \item FAd) Degree:\footnote{Because of this property the last index indicating the degree dropped in the following.} $C^{(i,j,k)}=0$ unless $i+j+k=n$. 
 \item FAu) Unit:  $C^{(0,i)}_{1ab} =\eta^{(i)}_{ab}$.
 \item FAnd) Non-degeneracy:  $C^{(1,j)}$ is non-degenerate in the second slot.  
 \item FAa) Associativity: $C_{abp}^{(i,j)} \eta^{pq}_{(n-i-j)} C^{(i+j,k)}_{qcd}=C^{(i,k)}_{acq} \eta^{qp}_{(n-i-k)} C^{(i+k,j)}_{pbd}$.            
\end{itemize}
Here the latin indices $a,b,c,\ldots$ refer to a choice of basis $\cA^{(i)}_a$, $a=1,\ldots, {\rm dim}(\cA^{(i)})$ of $\cA^{(i)}$.
The above defines a commutative algebra as follows  
\begin{equation} 
 \cA^{(i)}_a \cdot \cA^{(j)}_b=C_{abq}^{(i,j)} \eta_{(i+j)}^{qp}\cA^{(i+j)}_p= C_{ab}^{(i,j)p}  \cA^{(i+j)}_p\ .    
\end{equation}
These structures are intrinsic to the $(c,c)$- and $(a,c)$-rings of the worldsheet $(2,2)$ superconformal 
theory. The charges of the $U(1)_A$ and $U(1)_V$ currents in the $(2,2)$ supersymmetry algebra define the grading in the $(c,c)$- and the $(a,c)$ rings respectively and the rest
 of the  Frobenius structure follows  from the axioms of conformal field theory and the closing of the rings 
up to $Q_B$, $Q_A$ exact terms, see~\cite{MR2003030}.  Note that in families of Frobenius 
algebras  $\eta^{(p)}_{ab}$ is a constant topological pairing, while the general 
$C_{abc}^{(i,j)}({\underline{t}})$ varies with the deformation parameter.   

In the $B$-model the $(c,c)$ ring is identified for fixed complex structure with the
elements $B^{(p,q)}_a$  in $H^p(M,\wedge^q TM)$. We consider only those  elements  
which are mapped by contraction with $\Omega$ to elements ${\cal B}^{(p)}=\Omega(B)$ in 
$H^n_{\rm prim}(M_n)$, i.e. in particular $p=q$. The Hodge type for this complex structure, 
which can be taken at the point of maximal unipotent monodromy, defines the grading, so that one gets 
\begin{equation} 
C^{(p,q)}_{abc}=Q(\Omega(B^{(p)}_a\wedge B^{(q)}_b\wedge B^{(n-p-q)}_c),\Omega)\ ,
\label{coeffdef1}  
\end{equation}
which depends also on the K\"ahlergauge of $\Omega$.  For complex families, i.e. deformations w.r.t. 
elements with $(p=q=1)$, the grade of  ${\cal B}^{(p)}$  is encoded in the 
filtration parameter of  ${\cal F}^{n-p}$ and we give a covariant definition of  (\ref{coeffdef1}) below.

The $(a,c)$ ring  is mapped to the quantum  cohomology 
extension of   $H^*_{deRham}$. On the vertical cohomology in $H^{p,p}(M_n)$  the grading of the $A$-model is simply 
identified with the form degree. We allow again only the complexified K\"ahler deformation family w.r.t. 
to elements with $p=1$. These deformation families of rings are pairwise identified by mirror symmetry 
on $M$ and $W$ and  at the point of maximal unipotent monodromy the gradings can be 
matched using the monodromy weight filtration. This gives important information about the  
basis of the cohomology and homology groups in $H^n_{\rm prim}(M_n)$. 

One can chose a basis $A^I_{(p)}$ of the homology  $H_n(M_n,\mathbb{Z})$  
and a dual one $\alpha_I^{(p)}$  for cohomology of the primary horizontal 
subspace so that  
\be 
\int_{A^I_{(p)}} \alpha^{(q)}_J=\delta^I_J\delta^q_p,\qquad 
\int_{M_n} \alpha^{(q)}_I \wedge \alpha^{(p)}_J= 
\left\{\begin{array}{cc} 
0&\quad {\rm if}\ \  p+q> n,\\ 
\eta^{(q)}_{IJ} &\quad  {\rm if}\  p+q=n  \ .                                                         
\end{array}\right.                                                     
\label{basisinH}  
\ee 
Here $p=0,\ldots, n$ denotes a grading which can be related to 
the Hodge type given a point in the moduli space. As mentioned above 
the  most useful one is the large complex structure point. 

The information in the Gauss-Manin connection  (\ref{gaussmanin}) and in the 
Picard Fuchs ideal ${\cal I}_{PF}$ is equivalent. Combined with Griffiths 
transversality this information  determines the Frobenius structure. 
From  ${\cal I}_{PF}$ and the differential and algebraic relations that follow from  
Griffiths transversality one can calculate the Frobenius structure constants 
explicitly, see~\cite{Candelas:1990rm} for the quintic and~\cite{Hosono:1993qy} 
for any Calabi-Yau manifold.   More abstractly the Frobenius structure can 
be identified on the $B$-model cohomology ring by choosing appropriate 
basis vectors ${\cal B}^{(p)}_a$ in ${\cal F}^{n-p}$ with the properties
\begin{equation} 
\eta_{ab}^{(p)}=Q({\cal B}^{(p)}_a,{\cal B}^{(n-p)}_b), \qquad C^{(1,p)}_{abc}=Q(\nabla_a {\cal B}^{(p)}_b,{\cal B}^{(n-p-1)}_c) \ . 
\label{ringindentification1}
\end{equation} 
The last equation of the two equations defining the ring homomorphism may be stated shortly 
\begin{equation} 
{\cal A}^{(1)}_\alpha \cdot \ \  \leftrightarrow \ \ \nabla_\alpha \cdot \ \ .
\end{equation}
FAs) is fulfilled because  the Gauss-Manin connection is flat $[\nabla_a,\nabla_b]=0$, FAnd) 
because of the T-T Lemma and the rest  of the axioms  follow from Griffiths transversality. 
Note  associativity determines all possible couplings and that  ${\cal B}^{(p)}_a$ can 
be readily  expanded in the $\alpha^{(p)}_I$ basis  with the following upper triangular property  
\begin{equation}
\int_{{A^I}_{(p)}}  {\cal B}^{(q)}_J=  \left\{\begin{array}{cc} 0&\quad {\rm if}\ \  p<q\\ 
\delta^I_J &\quad  {\rm if}\  p=q   \end{array}\right. \ .
\end{equation}

We can  restate the Gauss-Manin connection in an  easy form. Using the operator state correspondence in 2d field theory we write 
$({\cal B}^{(0)}=\Omega_n, {\cal  B}_{\alpha_1}^{(1)}, {\cal  B}_{\alpha_2}^{(2)}, \ldots , {\cal  B}_{\alpha_{n-2}}^{(n-1)}, {\cal B}^{(n)})$ 
as $(|0\rangle$, $|\alpha_1 \rangle$,    $|\alpha_2 \rangle,\ldots$, $|\alpha_{n-2}\rangle$, $|n\rangle)$. Since  the
the Gauss-Manin connection becomes the ordinary derivative in flat coordinates, which  are given by a ratio of  $t^\kappa=X^\kappa/X^0$ 
of  the projective complex coordinates  $X^I$ (cff.  (\ref{expandomega03},\ref{exapndomega40}), as e.g. the mirror map. 
Using the Griffiths-Frobenius structure on the B-model  one can write the GM connection as
\begin{equation} 
\partial_{t_{\kappa}}
\left(\begin{array}{l}
|0\rangle\\  
|\alpha_1 \rangle \\  
|\alpha_2 \rangle\\ 
\vdots\\
 |\alpha_{n-3}\rangle\\
 |\alpha_{n-2}\rangle\\ 
|n\rangle
\end{array}\right)
=
\left(\begin{array}{cccccccc} 
0&\delta_{\kappa,\alpha_1}&0&0&\ldots &0&0\\
0&0&C^{(1,1)\ \alpha_2}_{\kappa,\alpha_1}&0&\ldots &0&0\\
0&0&0&C^{(1,2)\ \alpha_3}_{\kappa,\alpha_2}&\ldots &0&0\\
&&&&&&\\
0&0&0&0&\ldots &  C^{(1,n-2)\ \alpha_{n-2} }_{\kappa,\alpha_{n-3}}&0\\
0&0&0&0&\ldots &0&\delta_{\kappa,\alpha_{n-2}}\\
0&0&0&0&\ldots &0&0\\
\end{array} \right) 
\left(
\begin{array}{l}
|0\rangle\\  
|\alpha_1 \rangle \\  
|\alpha_2 \rangle\\ 
\vdots\\
 |\alpha_{n-2}\rangle\\
|\alpha_{n-2}\rangle\\ 
|n\rangle
\end{array}\right)\ ,
\end{equation} 
for $\kappa=1,..., h_{n-1,1}(M_n)$. This specializes for the 3-fold case 
given  in~\cite{MR1416353}, see also~\cite{CoxKatz} section 5.6  
and~\cite{Ceresole:1992su} for a physics derivation from ${\cal N}=2$ special geometry. 

The {\sl  real structure}   
\begin{equation} 
g^{(q)}_{\alpha\bar \beta}=\langle \bar  \beta,q|n-q,\alpha  \rangle = R({\cal B}_\alpha^{(n-q)} , {\cal B}_\beta^{(q)})\ ,
\label{realstructure} 
\end{equation}      
and the worldsheet parity operation 
\begin{equation} 
 \langle \bar \alpha, q | =\langle \beta | M^{{(q)}\,  \beta  }_{\bar \alpha}    \ , 
\label{worldsheetpartity} 
\end{equation}   
which fulfills the worldsheet  $CPT$  constraints  $M M^*=1$, extend  the Griffiths-Frobenius 
package on the mixed Hodge structure  to the $tt^*$- structure~\cite{Bershadsky:1993cx}.

In particular one can chose  the basis ${\cal B}_{\alpha}^{(p)}$ compatible with the real structure, i.e. 
\begin{equation} 
{\cal B}^{(p)}_\alpha={\overline{{\cal B}^{(n-p)}}}_{\bar \alpha}\ . 
\end{equation}   
The degree one  elements are the well known tangents vectors to the complex deformation space of 
Tian and Todorov 
\begin{equation} 
 {\cal  B}^{(1)}_\alpha=D_\alpha \Omega, \qquad   {\overline {\cal  B}^{(1)}}_{\bar \alpha}={\bar D}_{\bar \alpha} \bar\Omega  \ .                 
\end{equation}  
Note that $g^{(q)}_{\alpha\bar \beta}$   is the Zamolodichkov metric, $g_{i \bar \jmath}$  is related to the 
Weil-Petersson metric $G_{i\bar\jmath}$   by  $g^{(1)}_{i\bar \jmath}= e^{-K} G_{i\bar \jmath}=\langle \bar 0|0  \rangle G_{i\bar \jmath}$ and has 
blockform w.r.t. to the grading.
The higher degree operators are given for $p=2,\ldots,\lfloor\frac{n}{2}\rfloor$  by 
\begin{equation}
{\cal B}^{p}_{\alpha}={\cal D}^{(p)}_{\alpha} \Omega, \qquad  {\overline {\cal  B}^{(p)}}_{\bar \alpha}={\overline {\cal D}}^{(p)}_{\bar \alpha}  \bar\Omega,  
\end{equation} 
with ${\cal D}^{(p)}_{\alpha}= \frac{1}{p!} \kappa_{\alpha;i_1,\ldots,i_p} D_{i_1}\ldots D_{ i_p}$. 
These operators are closely related to the {\sl Frobenius operator} ${\cal D}^{(p)}(\underline{\rho})$ 
and as the latter determined at the point of maximal unipotent 
monodromy from the {\sl symbol} of the Picard-Fuchs differential ideal ${\cal I}_{PF}$ on 
$M_n$ as in  (\ref{ringrelation}) or  equivalently from the information in the Chow ring of $W_n$. 
In order to fix  them  completely in the $n>3$ cases one has to construct the integer basis, 
which we do in sections \ref{fourfoldbasis}, with the generalizations in sections 
\ref{HMS},\ref{hemisphere}.   

Note that the  metric connection to $g_{i\bar \jmath}$   is K\"ahler, i.e. pure  $\gamma_{ij}^k=g^{\bar l k} \partial_j g_{j \bar l }$, 
$\gamma_{\bar \imath \bar \jmath}^{\bar k}=g^{\bar k l} \partial_{\bar \imath}  g_{l \bar \jmath }$ and defines  metric connections 
$d_i=\partial_i-\gamma_i$ and $d_{\bar \imath}=\partial_{\bar \imath}- \gamma_{\bar \imath}$. 
The $tt^*$  connection is defined as 
\begin{equation} 
\nabla_i=d_i+C_i\ ,   \qquad \nabla_{\bar i}=d_{\bar \jmath} + C_{\bar \imath}\, ,
\end{equation}
$C_{\bar \jmath}$ denotes the complex conjugated Frobenius structure constants, they have pure holomorphic or 
anti-holomorphic  indices, which are raised and lowered with the constant  topological pairing $\eta_{ij}=\eta_{\bar \imath\bar \jmath}$.      
All expressions inherit the grading from the Frobenius algebra, which we suppress in this paragraph.   
One has  the $tt^*$ relations  of Cecotti and Vafa, see \cite{Hori:2003ic} for a review for CY-3 folds,
\begin{equation}
\begin{array}{rl}  
[d_i,d_j]&=[d_{\bar \imath} , d_{\bar \jmath}]=[d_i,C_{\bar \imath}]= [d_{\bar \imath},C_{i}]=0,\\[2 mm]
[d_i,C_j]&=[d_j,C_i],   \qquad \quad  [d_{\bar \imath} ,C_{\bar \jmath}]=[d_{\bar \jmath},C_{\bar \imath}],\\[ 2 mm] 
[d_i,d_{\bar \jmath}]&=-[C_j,C_{\bar \imath}],  \quad {\rm hence} \quad  [\nabla_i,\nabla_j]= [\nabla_{\bar \imath},\nabla_{\bar \jmath}]
= [\nabla_i,\nabla_{\bar \jmath}]=0 \ .   
\end{array}
\end{equation} 
The connection $\nabla_i$ is the Gauss-Manin connection and the  last equation promotes that structure  to the  flat $tt^*$ 
connection.

Mathematically the $tt^*$-structure is related to the TERP-structure, see review of Hertling and 
Sabbah~\cite{MR2806464}, and the non-commutative Hodge structure~\cite{MR2483750}. It 
was in the latter context that the $\hat \Gamma$ classes were found in~\cite{MR2483750}, 
that we use in section \ref{HMS} to confirm the integral basis that we determine in  
section~\ref{fourfoldbasis}.

\subsubsection{Integral basis for the middle homology of Calabi-Yau threefolds} 
\label{threefoldbasis}

For $n=3$ and more generally $n$ odd $H^n(M_n,\mathbb{Z})$ is primitive, $\eta$ is 
antisymmetric and one can chose a  symplectic  basis $\alpha_I$, $\beta^I$, 
in $H^n(M_n,\mathbb{Z})$  and a dual one $A^I$, $B_I$, $I=0,\ldots, h=b_n/2-1$ 
in $H^n(M_n,\mathbb{Z})$ with a symplectic pairing 
\begin{equation}
\Sigma= \left(\begin{array}{cc} 
        0 & \mathbbm{1}_{k \times k}\\ 
        -\mathbbm{1}_{k \times k}  & 0 
        \end{array}\right)\ . 
\end{equation}
Then $\Omega$ can be expanded in the symplectic basis of $H^n(M_n,\mathbb{Z})$  
in terms of the period vector $\Pi(a)=(X^I=\int_{A^I} \Omega(a) ,F_I=\int_{B_I}\Omega(a) )^T$ over the 
symplectic basis of $H_n(M_n,\mathbb{Z})$ as 
\begin{equation}
\Omega = X^I \alpha_I - F_I \beta^I \, ,
\label{expandomega03}  
\end{equation}
and the K\"ahler potential becomes
\begin{equation}
e^{-K}=i \Pi^\dagger(\bar a) \Sigma \Pi(a),
\label{Kaehlerpotentialperiods}  
\end{equation}
while the superpotential is 
\begin{equation}
W= q_I X^I - p^I F_I, 
\label{superpotentialperiods}        
\end{equation}
where $q_I$  and $p^I$ in $\mathbb{Z}$ are electric and magnetic charges.
Using (\ref{griffith}) one derives for $n=3$ the existence of special geometry~\cite{Candelas:1990pi}
and in particular an holomorphic prepotential $F$,  homogeneous degree two in the 
$X^I$ periods with $F_I=\frac{\partial F}{\partial X^I}$, which determines the periods, 
the couplings $C_{ijk}$, as well as the K\"ahler potential, hence the name.  Based on the seminal work of~\cite{Candelas:1990rm} 
and the mirror symmetry conjecture it was pointed out in~\cite{Hosono:1994ax} how 
to fix in general an integer symplectic basis for the periods at the large 
complex structure point from the topological data of the mirror manifold $W_n$~\footnote{This large volume point might not be 
unique. There are as many such points as there different Calabi-Yau triangulations of $\Delta^*$, 
and hence different Mori vectors. Each defines by (\ref{largevolumecoords}) a large volume point, 
which correspond to different phases of $W_n$.} 
where $i,j,k=1,\ldots,h$     
\begin{eqnarray}
{ F}
&=&-{C^0_{ijk} X^i X^j X^k \over 3! X^0}+ n_{ij} {X^i X^j \over 2}+ c_i X^i X^0-i{\chi 
\zeta(3)\over 2 \pid}(X^0)^2+ (X^0)^2 f(q)\nonumber\\
&=& (X^0)^2{\cal F}= (X^0)^2\left[-{C_{ijk} t^i t^j t^k\over 3!}
+n_{ij} {t^i t^j \over 2}+
c_A t^A-i{\chi \zeta(3)\over 2 \pid} +f(q)\right]\ . \label{geomprep}
\end{eqnarray}
It defines an integral basis for the periods at the large complex structure points 
given by
\beq
\Pi_{lcs}=\left(\begin{array}{c} 
X^0  \\ 
X^i \\
{F_0 }\\
{F_i}
\end{array}\right)=X_0\left(\begin{array}{c} 
1   \\
t^i \\
2 {\cal F}- t^i  \pa_i {\cal F }\\
{\pa {\cal F}\over \pa t^i}\end{array}\right)=X^0\left(\begin{array}{c}
1\\
t^i\\
{C^0_{ijk}\over 3!} t^i t^j t^k+c_i t^i-i{\chi \zeta(3)\over \pid}
+f(q)\\
-{C^0_{ijk}\over 2} t^i t^j+{n_{ij}} t^j+c_i+\pa_i f(q)
\end{array}\right)\ .
\label{periodbasis3fold} 
\eeq

Here we defined the mirror map  $t^i(a)$ as  
\begin{equation} 
t^i(a)=\frac{X^i(a)}{X^0(a)}=\frac{1}{2 \pi i} \left(\log(z_i)+ \Sigma^i({\underline z})\right), \quad i=1,\ldots, h_{n-1,1} \ .    
\label{mirrormap} 
\end{equation}
Note that $t^i$ has in 
the $A$-model, the  interpretation of a complexified area $t^i=\int_{{\cal C}^i}( J + i B_2)$, where 
$J$ is the K\"ahler form and $B_2$ is Neveu-Schwarz two form on $W_n$.
The map $t^i(a)$ can be made explicit at the large complex structure point using 
\begin{equation} 
X^0(a)=\omega_0({\underline{z(a)}},{\underline{\rho}})|_{{\underline{\rho}}=0},
\label{analyticperiod} 
\end{equation}
with $\omega_0({\underline{z(a)}},{\underline{\rho}})$ defined in (\ref{solution}) and $z(a)$ 
defined in (\ref{largevolumecoords}) as well as 
\begin{equation} 
 X^i=\frac{1}{2 \pi i}\partial_{\rho_i}\omega_0({\underline{z(a)}},{\underline{\rho}})|_{\underline{\rho}=0}\ .
\label{linlog} 
 \end{equation} 
This is the first step in the construction of solutions by the Frobenius 
method, which relies on the fact that certain differential operators, the {\sl Frobenius operators}  
${\cal D}^{(k)}({\underline \rho})$ in $\rho$ with constant coefficients~\footnote{Which 
are given by the classical intersection of the $A$ model, see below.}    
of degree $k=1,\ldots,n$ commute    
\begin{equation}
[D_i({\underline z}),{\cal D}^{(k)}({\underline \rho})]|_{\underline{\rho}=0}=0, 
\label{commute}  
\end{equation}
on the solution space with the operators in ${\cal I}_{PF}$. ${\cal I}_{PF}$ 
determines the structure constants of the Frobenius algebra that appears 
in the periods. This fact  determines the  ${\cal D}^{(k)}({\underline \rho})$~\cite{Hosono:1993qy}.   
In particular due to FAnd)  (\ref{commute}) holds for all operators of 
degree $k=1$, which yields (\ref{linlog}) in any dimension.  Another universal solution 
is~\footnote{Here we sum over equal indices. We denote by $C^0_{i_1,\ldots, i_n}$ the classical intersections, 
which are a valid approximation only at the large radius point, while $C_{i_1,\ldots, i_n}$ 
are the intersections in quantum cohomology, which are exact expressions throughout the 
complexified K\"ahler- or complex moduli space.}                   
\be 
{\cal D}^{(n)} ({\underline \rho})\omega_0({\underline{z(a)}},{\underline{\rho}})|_{{\underline{\rho}}=0}=\frac{C^0_{i_1,\ldots, i_n}}{n! (2 \pi i)^n} \partial_{\rho_{i_1}}\cdots \partial_{\rho_{i_n}} 
\omega_0({\underline{z(a)}},{\underline{\rho}})|_{{\underline{\rho}}=0}\  , 
\label{highestsolution}  
\ee 
where the $C^0_{i_1,\ldots, i_n}=\int_{W_n} J_{i_1} \wedge \ldots  \wedge J_{i_n}\ge 0$ are 
the classical intersection numbers and $J_i$ are $(1,1)$-forms in $H^2(W_3,{\bf Z})$, which span 
the K\"ahlercone. The solution (\ref{highestsolution})  has  in the $A$-model  the interpretation 
of a $2n$-volume. $F_0=\int_{A^{(n)}} \Omega$ contains (\ref{highestsolution}) and 
lower logarithmic solutions. In the homological mirror symmetry interpretation $F_0$ 
is the central charge of the $D_{2n}$ brane, which correspond to the structure sheaf on 
$W$. The lower logarithmic solutions are the $K$-theoretic corrections due to induced lower 
brane charges.  For $n=3$ one has $c_i={1\over 24}\int_{W_3} c_2 \wedge J_i$ and 
$n_{ij}=\int_{W_3}i_* c_1(D_i)\wedge J_j$.  Special geometry implies that  
${\cal D}_k^{(2)}({\underline \rho})=\frac{C^0_{ijk}}{3! (2 \pi i)^2} \partial{\rho_i}\partial{\rho_j} \omega_0({\underline{z(a)}},
{\underline{\rho}})|_{\underline{\rho}=0}$, $h=1,\ldots, h_{21}$  are also solutions.  
The $q$ expansion of $F_I$ around the large volume is the worldsheet instanton expansion 
on $W_3$. Note that the  explicit form of $f(q)$ is completely determined by (\ref{solution}) and 
its derivatives with respect to $\rho$, whose precise structure is fixed by the $t^i$ 
polynomials. The methods  fixes the period vector in an integer basis near the large complex 
structure point with a finite radius of convergence and hence  everywhere else in 
${\cal  M}_{cs}$ by analytic continuation.

Beside $F_0=\int_{A^{(n)}} \Omega$, the other universal solution $
X^0=\int_{A^{(0)}} \Omega$ is the $D_0$-brane central charge, which 
corresponds in the A-model to the sky-scraper sheaf on $W$. 
The cycle $A^{(n)}$  has the topology of an $S^{n}$ sphere,  which is the one that  
vanishes at the generic {\sl conifold divisor} compare (\ref{leading}) and is in the 
class of the base of  Strominger-Yau-Zaslow fibration, while $A^{(0)}$ has the 
topology of a $T^n$ and  is in the class of the SYZ torus fibre. 
According to SYZ mirror symmetry is $T$-duality on $T^n$  and maps 
Type II A/B on $M_n$  to Type II B/A on $W_n$ for $n$ odd and  Type II A/B on 
$W_n$ for $n$ even. 

An explanation of the $\zeta(k)$ terms from the Frobenius method was given
in~\cite{Hosono:1994ax}.  In particular the term  $i{\chi \zeta(3)\over 2 \pid}$ 
comes from  the third derivative of the $\Gamma$-function in (\ref{solution}) 
due to ${\cal D}^{(3)}$. This logic applies also to the B-model in higher dimensional  
Calabi-Yau space. Recently it has been realized in the context of homological mirror 
symmetry~\cite{IRITANI1,MR2483750,Kontsevichgamma} that  these subleading terms 
can be obtained  by modifying  the Chern-Character map in the $K$-theory group of the 
A-model by the $\hat \Gamma$ classes.  In section \ref{Gammaclass} we summarize 
this formalism as a check of the integral monodromy basis determined that we determine 
on the $B$-model side in the next section.          

\subsubsection{Integral basis for the primitive horizontal homology of fourfolds} 
\label{fourfoldbasis}
If $n$ is even then the primitive part of $H_H^n(M_n)$  that is 
accessible to the analysis by period integrals on $M_n$ is a subspace of $H^n(M_n)$. 
By mirror symmetry the  vertical subspace $H_V(M_n)$ is  accessible  to the analysis by 
period integrals on $W_n$. Often in the toric context  the intersection of these spaces gives 
 all of $H^n(M_n)$. The techniques that we develop below allow to find the integral monodromy basis for both cohomologies.
These integer structures are necessary to make these cohomology elements well defined  
on the corresponding deformation family.  By the dimensional argument  in viii) of section 
\ref{fluxquantization}  it is in general not possible to find an integral element in $H^*(M_n,\mathbb{Z})$ 
that lies entirely in  $H_H^n(M_n)$ or $H_V^n(M_n)$~\footnote{The argument  does  not exclude 
the possibility to find such a splitting over $\mathbb{Z}/2$. Even  if the dimensional 
argument does not forbid a separation, it is still unlikely to occur, 
because it is physically expected for ${\cal N}=1$ theories that the 
K\"ahler and complex moduli sector do couple and the available 
concrete discussion of conditions on fluxes is still very incomplete.}. 
In any case the elements might be rational in  $H^*(M_n,\mathbb{Z})$, but 
they can be added to an integer basis of  $H^*(M_n,\mathbb{Z})$ with 
coefficients one that can be reconstructed from the lattice structure and the 
monodromy information.     

The two subspaces  govern the vacuum structure of the complex moduli  and the K\"ahler moduli
by the $F$-term superpotentials respectively. To actually calculate those $F$-terms  that restrict 
the complex moduli from the fluxes on the primitive horizontal space on $M_4$ we determine now 
the integer monodromy basis that allows to evaluate the formulas (\ref{supo},
\ref{Kaehlerpotential},\ref{superpotential}) explicitly.

This requires to delve deeper into the structure of periods on fourfolds discussed in section \ref{frobeniusstructure}. 
We expand the $(4,0)$ form $\Omega$ in terms of the basis $\alpha^{(p)}_I$ of the primary horizontal subspace as 
\begin{equation} 
\Omega=X^0 \alpha^{(0)}+ X^i  \alpha_i^{(1)}+ H_p  \alpha^{(2)p}  + F_i  \alpha^{(3)i}+ F_0  \alpha^{(4)}\ ,
\label{exapndomega40} 
\end{equation}
where $i=1,\ldots,h=h_{3,1},p=1,\ldots,k=h_{2,2}^{\rm prim}$. Due to the grading the bilinear form $Q$ on $H^{\rm prim}_4(M_4)$, 
has the form   
\begin{equation} 
Q^*(\underline{\Gamma},\underline{\Gamma})=\Gamma_\alpha \eta^{\alpha \beta}  \Gamma_\beta    = {\underline \Gamma}^T\left(
\begin{array}{ccccc} 
*&*&* &* & 1\\
*&*&* &  \eta^{(1)}_{h \times h}&  \\
*&*& \eta^{(2)}_{k \times k}&&  \\
*&(\eta^{(1)})^T_{h \times h}&&&  \\
1&&&& \\
\end{array}\right){\underline \Gamma} \ .       
\label{Qpairing}
\end{equation}
We claim that with a ${\rm GL}(h_4,\mathbb{Z})$ transformation we can eliminate the $*$ entries and 
bring  $Q$ in a block anti-diagonal form.  Instead of special geometry Griffiths transversality implies  
\begin{equation} 
\begin{array}{rl} 
 &2 X F + H^2 =0, \ \  (\eta^{(1)} F)_i + X \partial_i F + H \partial_i H=0,\ \    X \partial_i\partial_j F + H \partial_i\partial_j H  =0, \\ 
 &X \partial_i\partial_j\partial_k F + H \partial_i\partial_j\partial_k  H=0, \ \ X \partial_i\partial_j\partial_k \partial_l F + H \partial_i\partial_j\partial_k \partial_l H=C_{ijkl}    ', \\ 
\label{GT4fold}
\end{array}
\end{equation}
where we used a vector notation for the periods $\{X,H,F\}=\{\int_{\Gamma_i} \Omega\}$  and the pairing on the periods 
induced from (\ref{Qpairing}) in the block anti-diagonal form.

Analogously to section 4.2 in~\cite{Hosono:1993qy} (\ref{GT4fold}) and the 
differential equations derived from (\ref{GKZ}) can be used to obtain the 
$C_{i_1,\ldots,i_r}$ and match the Frobenius structure as~\cite{Mayr:1996sh,Klemm:1996ts} 
\begin{equation} 
 C_{ijkl}=C^{(1,1,2)}_{ijp} \eta_{2}^{pq} C^{(2,1,1)}_{pkl}=\partial_i \partial_j H \partial_k \partial_l H \ .
\label{matching}  
\end{equation}
We will now show how this structure and the observations discussed 
in section~\ref{nodalpoints} can be used to fix an integral basis 
for the primitive  homology of a fourfold. Let us treat the 
sextic in $\mathbb{P}^5$ and develop the general picture 
along the way. After dividing by $\mathbb{Z}_6^4$ 
acting as $x_i\rightarrow \exp( 2 \pi i r^{(l)}_i)  x_i$ with 
$r^{(1)}=\frac{1}{6} (1,5,0,0,0,0)$, $r^{(2)}=\frac{1}{6} (1,0,5,0,0,0)$, 
$r^{(3)}=\frac{1}{6} (1,0,0,5,0,0)$ and $r^{(4)}=\frac{1}{6} (1,0,0,0,5,0)$ 
one gets the invariant sextic constraints defined in ${\widehat{\mathbb{P}^5/\mathbb{Z}_6^4}}$ as  
\begin{equation}
W_{\Delta^{\rm inv}}=P=\sum_{i=1}^6 a_i x_i^6- a_0 \prod_{i=1}^6 x_i . 
\label{sextic} 
\end{equation}
The same combinatorics is encoded in the vertices of $\Delta^*$ spanned 
by $\nu_i=e_i$, $i=1,\ldots,5$ and $\nu_6=-\sum_{i=1}^5 e_i$ so that $P=W_{\Delta^*}$ and 
$l^{(1)}=(-6;1,1,1,1,1,1)$ and the Picard-Fuchs follows from (\ref{GKZ}) to be  in the large complex structure variable (\ref{largevolumecoords}) 
given as  (\ref{diffHP}) with $k=6$. The $a_i$, $i=1,\ldots,6$ are gauged to $1$ by the $\mathbb{T}_{n+1}\times \mathbb{C}$ action and we call  $a_0$ 
simply $a$ in the following.  The complex moduli space  is parametrized by $z=1/a^6$ and can be compactified to ${\cal M}_{cs}=\mathbb{P}^1\setminus \{ z=0,z=\frac{1}{6^6},1/z=0\}$, 
where the critical points are the large complex structure point $z=0$, the conifold point $z=\frac{1}{6^6}$ and an $\mathbb{Z}_6$ 
orbifold point. The mirror\footnote{According to~\cite{Batyrev:1994hm} $\chi=2160$ moreover from Thm. 4.3.7 of~\cite{Batyrev:1994hm}  ($n$ in~\cite{Batyrev:1994hm} 
is the dimension of $\mathbb{P}_{\Delta}$) one has $h_{31}(M_4)=426$, $h_{11}(M_4)=1$. 
Hence $h_{21}(M_4)=0$ from (\ref{index}), moreover from section \ref{fluxquantization} one gets $\sigma(H_{22})=1752$, 
where the primitive part contributes $1751$ and the vertical part $1$ to $H_+(M_4)$. Hence $H^{\rm prim}_{22}(W_4)=1$.} has $h^{\rm prim}_4(W_4)=5$  
and the five solutions of (\ref{diffHP}) corresponds to the integrals over the five dual 4-cycles. Using mirror symmetry and the intersections of the 
Chow ring of $M_4$ one can anti-diagonalize the pairing over $\mathbb{Z}$ to get 
\begin{equation} 
\eta^{-1}=\eta^{\alpha \beta} =\left(
\begin{array}{ccccc} 
&& &  & 1\\
&& &  1& \\
&& C^{0}&&  \\
&1&&&  \\
1&&&& \\
\end{array}\right)\ ,
\label{LSintersection}
\end{equation} 
where $C^{0}=\int_{M_4} J^4$ is the classical intersection. 
 
It has been pointed out in~\cite{Grimm:2009ef} that the period $X_\nu$ that vanishes 
at the conifold $\delta_c=(1-6^6 z)$ as~\footnote{Different and sign conventions and normalizations have 
been used in~\cite{Grimm:2009ef}.}    
\begin{equation} 
X_\nu= \frac{2\delta_c^\frac{3}{2}}{\sqrt{3} \pi^2}(1+ \frac{17}{18}\delta_c+ \frac{551}{648} \delta_c^2+ {\cal O}(\delta_c^3)),   
\label{Xnu}
\end{equation}
i.e. when $W_n$ acquires a nodal singularity, has the following simple relation to the $D0$ period 
$X^0=\Pi_{reg}+X_\nu/2$ and the $D8$ period $F_0=-\Pi_{reg}+X_\nu/2$ at the large complex structure point. 
Here $\Pi_{reg}$ is a period that stays finite at $\delta_c=0$. We will use this 
observation, (\ref{GT4fold}) and the requirement of an integer monodromy around $z=0$ 
now to construct explicitly the period vector that corresponds to an integer basis at $z=0$. 
In particular one has from (\ref{LSintersection},\ref{Snintersection})
\begin{equation}
F_0=X_\nu-X^0\ .
\label{shift} 
\end{equation}
The analytic continuation of $X_\nu$ from $\delta_c=0$ to $z=0$ yields~\cite{Grimm:2009ef} 
\begin{equation} 
F_0=X^0\left(\frac{1}{4!} C^0_{ijkl} t^i t^j t^l t^k+\frac{1}{3!} c_{ijk} t^i t^j t^k +\frac{1}{2} c_{ij} t^i t^j + c_i t^i + c_0 + f_0(q)\right),    
\label{predictionhighest}     
\end{equation}
where\footnote{In~\cite{Grimm:2009ef} we used the first eq. (\ref{hrr}) $1440=\int_{W_4}(3 c_2^2 - c_4)$ to eliminate $\int_{W_4} c_4$ from the equation for $c_0$ yielding $c_0=\frac{\zeta(4)}{2^2 (2 \pi i)^4 } \int_{M_4} 5 c_2^2$. Here we write it in the 
form (\ref{prediction})  to make the comparison with section \ref{Gammaclass} easier.}  
\begin{equation}
c_0=\frac{\zeta(4)}{2^2 (2 \pi i)^4 } \int_{M_4} (7 c_2^2 - 4 c_4)-1,\quad c_1=-\frac{\zeta(3)}{(2 \pi i)^3} \int_{M_4} c_3\wedge J_i,\quad c_{ij}=-\frac{\zeta(2)}{2 ( 2 \pi i)^2} \int_{M_4} c_2 \wedge J_i\wedge J_j\, ,          
\label{prediction}
\end{equation}
and $c_{ijk}=0$. We claim that the rest of the 
periods is determined by (\ref{GT4fold},\ref{matching}) and the requirement that all monodromies are 
integer and leave the bilinear form $Q$ invariant up to an ${\rm GL}(h_4,\mathbb{Z})$ transformation, 
that leave  $Q$ invariant. We check this now explicitly for the sextic\footnote{Other examples 
are worked out in~\cite{KlemmYeh}.}, which has $\int_{M_4} c_2^2=1350$, $\int_{M_4} c_3 J=-420$ and 
$\int_{M_4} c_2 J^2=90$. This is the first time such a basis has been found for fourfolds. The period 
vector at $z=0$ is 
\begin{equation}
\Pi_{z=0}=\left(\begin{array}{c}  
X^0\\
X^1\\
H\\
F_1\\
F_0\end{array}\right)=  
X^0\left(\begin{array}{c} 
1\\
t\\ 
-\frac{1}{2} t^2 + \frac{1}{2} t + \frac{5}{8}+h(q)\\ 
-t^3 +\frac{3}{2} t^2 - \frac{3}{4} t - \frac{15}{8} +\frac{105 i \zeta(3)}{2 \pi^3}+f_1(q) \\
\frac{1}{4} t^4 + \frac{15}{8} t^2-\frac{105 i \zeta(3)}{2 \pi^3} t-\frac{75}{64}+ f_2(q) \end{array}\right) \ . 
\label{piinf} 
\end{equation}
Note that these are exact expressions, since the leading logarithm terms fix 
the Frobenius operator ${\cal  D}^{(k)}_i$ that acts on  $\omega({\underline a},{\underline \rho})$ 
to yield the full solution. The parameter $q=e^{2 \pi i t}$ 
is monodromy invariant under $t\sim \frac{1}{2 \pi i} \log(z)\rightarrow  t+1$. Hence 
monodromies encircling in the mathematically positive direction from the reference point $p_0$ the critical points  
$z=0$ or $\delta_c=0$ are 
\begin{equation} 
M_{z=0}= \left(\begin{array}{rrrrr} 
1& 0&0&0&0\\
1& 1&0&0&0\\
0&-1&1&0&0\\
-4&-3&6&1&0\\
3&7&-6&-1&1\\
\end{array}\right),\qquad M_{\delta_c=0}= 
\left(\begin{array}{rrrrr} 
0& 0&0&0&-1\\
0& 1&0&0&0\\
0&0&1&0&0\\
0&0&0&1&0\\
-1&0&0&0&0\\
\end{array}\right)\ ,
\label{sexticmonodromy1}   
\end{equation}
where the latter monodromy behaviour is predicted by (\ref{conifoldmonodromy}) the  compatibility of  (\ref{LSintersection},\ref{Snintersection}) 
which lead to (\ref{shift}). In the A-model language of the derived categories of coherent sheaves it is the four dimensional version of 
the Seidel-Thomas  auto equivalence of the derived  category involving only the skyscraper sheaf ${\cal Q}_{pt}$ and the 
structure sheaf ${\cal O}_W$ of the mirror manifold~\cite{MR1831820}.                 
The first very non trivial check is that 
\begin{equation}
 M^{-1}_{\delta_c=0} M^{-1}_{z=0}=M_{1/z=0}=\left(\begin{array}{rrrrr} 
-4& 4&0&-1&-1\\
-1& 1&0&0&0\\
-1&1&1&0&0\\
7&-3&-6&1&0\\
-1&0&0&0&0\\
\end{array}\right)\, ,
\label{sexticmonodromy2}   
\end{equation}
is an orbifold element of order six. Now we have determined the basis in which the 
K\"ahler potential for the complex structure moduli of the metric is given up to an 
irrelevant constant by
\begin{equation} 
K=-\log( \Pi^\dagger \eta^{-1} \Pi) .    
\label{kahler}
\end{equation}
Note that $M^t \eta^{-1} M =\eta^{-1}$ for all monodromy matrices $M$, which is a necessary condition 
for the  K\"ahler potential to be well defined in ${\cal M}_c$.   At the large complex structure point we obtain
\begin{equation} 
K=-\log\left((\bar t - t)[ (\bar t - t)^3+ 420 i \frac{\zeta(3)}{\pi^3}]\right)+{\cal O}(e^{2 \pi i t})\  .
\end{equation}    
Note the factorization of the threefold contribution, with its $\alpha'$ corrected volume 
$(\bar t-t)^3-\frac{i \zeta(3)}{\pi^3}  \int_{W_4} J\wedge c_3)$ in the large volume.     
The subleading terms are instanton corrections, which are determined exactly using the Frobenius 
methods, which gives  $h(q),f_1(q),f_2(q)$. These series converges for $|z|< 1/6^6$ and  gives  
the exact expression for the periods and hence for $K$ in this region. To obtain it in all ${\cal M}_{cs}$   we evaluate next 
the transition matrix between given local basis of solutions to the Picard-Fuchs 
equations. At the orbifold we define $w=1/z$. The transition matrix can 
be calculated analytically using the Mellin-Barnes integral representation of 
the periods, which for $0\le  {\rm a}rg(a)< \frac{\pi}{3}$,  is given by   
\begin{equation} 
\varpi(a)=\frac{1}{2 \pi i}\int_{\cal C} \dd s \frac{\Gamma(-s) \Gamma(6 s +1)}{\Gamma^5(s+1)} e^{i \pi s} a^{-6 s} \ .  
\label{barnes}
\end{equation}
It is easy to see that if one closes the contour ${\cal C}$  
\begin{itemize} 
\item a.) along $a=-\epsilon$ to the right to enclose all the poles at $s\in \mathbb{N}$  one gets $\varpi(a)=X_0(z)$, convergent for $|z|<\frac{1}{6^6}$ , 
\item b.) while if  one closes it to left to enclose all the poles at $s=-l/6$ with $l\in \mathbb{N}_+$ one gets a solution at the orbifold convergent for $|a|<6$
\begin{equation}
\varpi_0(a)=-\frac{1}{6} \sum_{m=1} e^{\pi i m 5/6} \frac{\Gamma\left(\frac{m}{6}\right)}{\Gamma(m)\Gamma^5\left(1-\frac{m}{6}\right)} a^m\  .
\end{equation} 
\label{varpi}
\end{itemize}
The other four solutions at the orbifold are given by $\varpi_i(a)=\varpi_0( \beta^i a)$, $i=1,\ldots,4$  with 
$\beta=e^\frac{2 \pi i}{6}$. In the $\varpi_i(a)$, $i=0,\ldots,4$ basis of solutions the  monodromy upon 
$a\rightarrow  \beta a$ is obviously given by
\begin{equation} 
\tilde M_{1/z=0}=    
\left(\begin{array}{rrrrr} 
0& 1&0&0&0\\
0& 0&1&0&0\\
0&0&0&1&0\\
0&0&0&0&1\\
-1&-1&-1&-1&-1\\
\end{array}\right)
\label{sexticmonodromy3}  \ .  
\end{equation}        
To analytically continue the other four  solutions to $z=0$ one notes that
\begin{equation}
\varpi_k(a)=-\frac{1}{6 (2 \pi i)^5 }\sum_{p=1}^5 \beta^{k p}(\beta^p-1) \tilde \varpi(a), 
\end{equation}  
where 
\begin{equation}
\tilde \varpi_k(a)=-
\int_{\cal C} \frac{\dd s}{e^{2 \pi si}-1}\frac{\Gamma^6\left(s+\frac{s}{6}\right)}{\Gamma(6s+k)}  a^{6s+k}\, ,
\end{equation} 
with the contour of type b.). Choosing the contour along the path described as in a.) a slightly tedious calculation allows to match 
the  logarithm of $z$  and get  the transition matrix $\Pi_{z=0}=m (\varpi_0,\varpi_1,\varpi_2,\varpi_3,\varpi_4)^T$ by comparing with (\ref{piinf}) as 
\begin{equation}
m=\left(\begin{array}{rrrrr} 
1&0 &0&0&0\\[1 mm]
-\frac{1}{6}& \frac{2}{3}& \frac{1}{2}& \frac{1}{3}& \frac{1}{6}\\ [1 mm]
\frac{1}{4}& \frac{2}{3}& \frac{1}{4}& 0& -\frac{1}{12}\\ [1 mm]
\frac{1}{3}& -\frac{4}{3}& 1& \frac{4}{3}& \frac{2}{3}\\ [1 mm]
0& -1& 0& 0& 0\\[1 mm]
\end{array}\right).
\label{continuationsextic}
\end{equation}
It is a nontrivial check that $\tilde M_{1/z=0}= m^{-1} M_{1/z=0} m$. It is harder to determine completely the transition 
matrix from the large complex structure point to the conifold exactly $\Pi_{z=0}=n(1-\frac{1}{3888}\delta_c^4 +{\cal O}(\delta_c^5) ,$ $\delta_c+\frac{1981}{19440}\delta_c^4 +{\cal O}(\delta_c^5),$ 
$\delta_c^2-\frac{41}{48}\delta_c^4 +{\cal O}(\delta_c^5),$   $\delta_c^3+\frac{125}{72}\delta_c^4 +{\cal O}(\delta_c^5), X_\nu)$ . We reduced  the $50$ real constants of $n$ 
to $14$ which we determined numerically.  
\begin{equation}
n=\left(\begin{array}{rrrrr} 
a&b &c&d&1\\[1 mm]
i a_1& ib_1& i c_1&id_1 &0\\ [1 mm]
a_1+ i \frac{a_2}{2}&b_1+ i \frac{b_2}{2}    &c_1+ i \frac{c_2}{2} &d_1+ i \frac{d_2}{2} & 0\\ [1 mm]
- 3 a_2 +i a_3&-3 b_2+ i b_3 &-3 c_2+ i c_3 &-3 d_2+ i d_3 &0\\ [1 mm]
-a& -b& -c& -d& 1\\[1 mm]
\end{array}\right)\ .
\end{equation}
$a_3=\frac{3 a_2}{a_1}- \frac{3 a_1}{4}-\frac{ a^2}{a_1}$ and   $b_3=\frac{3 b_2}{b_1}- \frac{3 b_1}{4}-\frac{ b^2}{b_1}$, with\footnote{We determined them with over 30 significant 
digits, but give only $6$ below.}   
$a=1.02252$, $b= -0.0515973$, $c= -0.0577426$, $d= -0.0524450$,  $a_1=1.71697$, $b_1= 0.155077$, $c_1= 0.0787695$, $d_1= 0.0527232$
$a_2=2.09001$, $b_2= 0.271880$, $c_2 =0.148585$, $d_2= 0.103310$m $c_3= 0.772689$ and $d_3= 0.561544$.

We can now study the superpotential throughout the complex moduli space in the integer moodromy 
basis  basis of $H^{\rm prime}_4(M_4)$, which can readily  be extended to an (half)integral basis of  
$H_4(M_4,\mathbb{Z})$ by multiplying by a finite integer.  In particular all vacua of  the flux superpotentials 
in ${\cal M }_{cs}$ can be found.            

We checked (\ref{piinf})  explicitly by  analytic  continuation of  the period that corresponds  
to the  vanishing $S^4$ at the conifold from the conifold  divisor to the large complex structure 
point for the following one parameter models.   
\begin{table}[hbt]
\begin{center}
\caption{Topological data of one moduli fourfolds given by complete intersections $X_{d_1,\ldots,d_r}(w_1,\ldots,w_{r+5})$  of degree $d_1,\ldots,d_r$ in weighted projective space $\mathbb{P}(w_1,\ldots,w_{r+5})$. 
The conifold behaviour is $X_{\nu}=C_c\delta^{3/2}+{\cal O}(\delta^{5/2})$ with $\delta=(1-C_d z)$. }
\label{onemoduli4f}\footnotesize
\begin{tabular}{|c|c|c|c|c|c|c|c|c|c|} \hline
                & $ \! \!  X_6(1^6) \! \!$\ &  $\! \! X_{10}(1^55) \! \!$ & $ \! \! X_{3,4}(1^7) \! \!$ &$ \! \!X_{2,5}(1^7) \! \!$ & $ \! \!X_{4,4}(1^6,2) \! \!$  &  $ \! \!X_{2,2,4}(1^8) \! \!$&$ \! \!X_{2,3,3}(1^8) \! \!$& $ \! \!X_{2,2,2,3}(1^9) \! \!$& $ \! \!X_{2^5}(1^{10})\! \!$\\ \hline
 $C^0$ & $6 $ &  $2 $ & $12 $ & $ 10$ & $8 $  &  $16 $&  $18 $& $24 $& $32  $\\  
$\int c_4 $ & $2610 $ &  $5910 $ & $1476 $ & $2190 $ & $1464^* $  &  $1632 $&  $ 1206$& $ 1152$& $960^*$\\ 
$\int c_2^2$ & $1350 $ &  $2450 $ & $972 $ & $1210 $ & $968 $  &  $1024 $&  $882 $& $864 $& $800  $\\ 
$\int c_3 J $     & $ -420$ &  $-580 $ & $-336 $ & $-420 $ & $-304 $  &  $-384 $&  $-324 $& $-336 $& $-320  $\\  
 $\int c_2 J^2$ & $90 $ & $ 70$ &  $108 $ & $110 $ & $88 $ & $128 $  &  $126 $&  $144 $& $160 $\\ 
 $C_c$ & $\frac{2}{\sqrt{3} \pi^2} $ & $ \frac{2}{3 \pi^2} $ &  $\frac{4}{\sqrt{6} \pi^2}  $ & $\frac{10}{3 \sqrt{5} \pi^2} $& $ \frac{4}{3  \pi^2} $ & $ \frac{8}{3 \sqrt{2} \pi^2} $ & $\frac{2}{ \pi^2}  $  &  $\frac{4}{\sqrt{3} \pi^2}  $& $ \frac{2}{3 \pi^2} $\\
 $C_d$ & $6^6 $ & $ 20^5$ &  $ -2^83^3 $ & $-5^5 4 $& $2^{14}   $ & $ 2^{12} $ & $2^43^6  $  &  $-2^6 3^3 $& $2^{10}  $\\
\hline  
\end{tabular}
\end{center}
\end{table} 
The Picard-Fuchs operators are derived using the vector $l^{(1)}=(-d_1,\ldots,- d_r, w_1;\ldots, w_{r+5})$ from 
the reduction of the GGKZ system as described in section \ref{gaussmaninS}:     
\begin{equation} 
\begin{array}{rl}
{\cal D}(  X_6) = &\theta^5- 6 z\prod_{k=1}^5 (6 \theta+k), \quad    {\cal D}(  X_{10})=\theta^5-32 z\prod_{k=1}^5 (10 \theta+ {2 k}-1), \\ [3 mm]
{\cal D}(  X_{3,4})=&\theta^5+12 z\prod_{k=1}^3 (4 \theta+ k)\prod_{k=1}^2 (3 \theta+ k),\quad {\cal D}(  X_{2,5})=\theta^5+10 z (2 \theta+1)\prod_{k=1}^4 (5 \theta+ k),\\ [3 mm]
 {\cal D}(  X_{4,4})=&\theta^5+16 z (4 \theta+2) (4 \theta+ 1)^2 (4 \theta+3)^2,\quad {\cal D}(  X_{2^24})=\theta^5+32 z (2 \theta+1)^3(4 \theta+ 1) (4 \theta+3),\\ [3 mm] 
 {\cal D}(  X_{3^22})=&\theta^5-18 z (2 \theta+1) (3 \theta+ 1)^2 (3 \theta+2)^2,\quad {\cal D}(  X_{2^33})=\theta^5+24 z (2 \theta+1)^3(3 \theta+ 1) (3 \theta+1),\\ [3 mm]
  {\cal D}(  X_{2^5})=&\theta^5-32 z (2 \theta+1)^5 .
\end{array}
\end{equation} 
We can easily calculate the monodromies for these models using  (\ref{GT4fold}) and (\ref{prediction}). E.g. for 
the $X_{10}$ case we get      
\begin{equation} 
M_{z=0}= \left(\begin{array}{rrrrr} 
1& 0&0&0&0\\
1& 1&0&0&0\\
-1&-1&1&0&0\\
-4&-1&2&1&0\\
3&3&0 &-1&1\\
\end{array}\right),\qquad M_{1/z=0}=\left(\begin{array}{rrrrr} 
0& 0&0&0&-1\\
0& 1&0&0&1\\
0&1&1&0&0\\
0&-1&-2&1&-3\\
-1&-4&-2&1&-3\\
\end{array}\right), 
\label{decticmonodromy2}   
\end{equation}
where it is again an excellent check that $M^{10}_{1/z=0}=1$. For the $X_{2^5}$ case 
\begin{equation} 
M_{z=0}= \left(\begin{array}{rrrrr} 
1& 0&0&0&0\\
1& 1&0&0&0\\
1&1&1&0&0\\
-24&-16&-32&1&0\\
8&8&0&-1&1\\
\end{array}\right),\qquad M_{1/z=0}=\left(\begin{array}{rrrrr} 
0& 0&0&0&-1\\
0& 1&0&0&1\\
0&-1&1&0&0\\
0&-16&32&1&-8\\
-1&-24&32&1&-8\\
\end{array}\right)\ .
\label{x22222monodromy2}   
\end{equation}
Now the situation at $1/z=0$ is completely different as here the indices 
are completely degenerate and we get a second point with maximal unipotent
monodromy which is in fact conjugate to the one at $z=0$.    

\begin{table}[hbt]
\begin{center}
\caption{The low degree genus zero BPS invariants $n_d^{H^2}$ for the the one moduli models  according to (\ref{integerfourfold}), 
where $H^2$ is the restriction of the Chow ring element $H^2$ generated by the hyperplane class $H$ of the ambient space to $M_4$.   
By the Griffiths-Frobenius structure they  define all instanton corrections to the classical period expressions at large radius. 
The triple  coupling $C^{2,1,1}$ follows by  (\ref{matching}). The  $n_d^{H^2}$ for $X_{10}(1^55)$ are divided by $800$ to fit the table.  }
\label{onemoduli4f}\footnotesize
\begin{tabular}{|c|c|c|c|c|c|} \hline
   $M_4$      & $d=1$& $d=2$ & $d=3$& $d=4$& $d=5$ \\ 
 \hline
$X_6(1^6) $    &60480& 440884080& 6255156277440& 117715791990353760& \
2591176156368821985600       \\  
$ X_{10}(1^55)$    & 1978&  989931233&  979521830499958& 1289166560052720145740& \
1986508760963967874295408390     \\  
$ X_{3,4}(1^7) $    &  16128& 17510976& 36449586432& 100346754888576& 322836001522723584     \\  
$X_{2,5}(1^7) $    &24500& 48263250&181688069500& 905026660335000&5268718476406938000    \\  
$ X_{4,4}(1^62) $    &27904& 71161472& 354153540352& 2336902632563200& 18034529714742555392    \\  
$X_{2,2,4}(1^8)  $    &11776& 7677952& 9408504320& 15215566524416& 28735332663693824      \\  
$X_{2,3,3}(1^8) $    & 9396& 4347594& 3794687028& 4368985908840& 5873711971817268    \\  
$X_{2,2,2,3}(1^9) $    & 6912& 1919808& 988602624& 669909315456& 529707745490688    \\  
$X_{2^5}(1^{10}) $    & 5120& 852480& 259476480& 103646279680& 48276836019200 \\  
\hline  
\end{tabular}
\end{center}
\end{table}  

In table~3.2 we list the BPS invariants for genus zero instantons  for the nine 
one parameter families  discussed in this section.    We checked that multi covering at genus one  fulfills the 
prediction of~\cite{Klemm:2007in}.  Since the normalization of the topological metric $\eta$ and $H$ is very 
important to get the minimal integral  monodromy basis,  we confirmed  the number of lines $L$  by an elementary intersection 
calculation on the moduli  space of  lines in the ambient space. We make the argument only explicit 
for the embeddings in the projective spaces $\mathbb{P}^k$, where this moduli space is (in our 
notation)  the Grassmannian ${\rm  Gr}(2,k+1)$ of dimension $2(k-1)$. Lines $L$ that lie in $M_4$  and meet  the restriction 
of the hyperplane class $H^2$ of $\mathbb{P}^k$ to $M_4$, correspond to the sublocus of the moduli  space that 
is specified by the following two classes: The top Chern class of a global  section in  the universal quotient
 bundle $c_{top}(\prod_{i=1}^r c_{d_i+1} (S^{d_i} Q))$. It is given in terms of symmetric functions 
in two variables say $x,y$ as the homogeneous top degree in $\prod_{i=1}^r \prod_{j=0}^{d_i} (1+ j x+ (d_i-j) y)$ 
expressed  in terms of the Schubert cycles classes $\sigma_1=(x+y)$ and $\sigma_{1,1}=\sigma_1^2-\sigma_2=xy$. 
The  second class,  which represents the condition that $L$ meets $H^2$, is simply $\sigma_1$. 
For example for the penta-quadric in $\mathbb{P}^9$ we have hence to evaluate the number of lines $L$  by
\begin{equation} 
n_1^{H^2}(X_{2^5}(1^{10 }))=\int_{{\rm Gr}(2,10)} \sigma_1 \cdot  (c_3(S^2 Q))^5=2^{10} 
\sum_{k=0}^5 (-1)^k \left(5\atop k\right)\int_{{\rm Gr}(2,10)} \sigma_1^{6+2k}  \sigma_2^k=5120\  ,
\end{equation}                  
which perfectly matches our normalizations.  The evaluation of the actual intersection of the Schubert cycles  
$\sigma_1^{16}=1430$, $\sigma_1^{14} \sigma_2=1001$, $\sigma_1^{12} \sigma_2^2=704$, 
$\sigma_1^{10} \sigma_2^3=492$, 
$\sigma_1^{8} \sigma_2^4=352$ and  
$\sigma_1^{6} \sigma_2^5=250$ is done by Pieri's formula\footnote{Which is a special case of the 
Littlewood-Richardson  rule for Young tablauxs.}  and the fact that $\sigma_{8-b,8-a}$ is Poincar\'e dual to 
$\sigma_{a,b}$ in the 16 dimensional Grassmannian ${\rm Gr}(2,10)$. Similarly we checked all  predictions for the 
embeddings in $\mathbb{P}^k$ of the lines in the table.  The two weighted cases require additional techniques.

Some large radius expansion for multi moduli case with elliptic fibration 
can be also found in~\cite{Klemm:2007in} .  Genus zero BPS expansion in multi moduli fourfolds  
appear in~\cite{Mayr:1996sh,Klemm:1996ts,Klemm:2007in}.

\subsection{The basis of the primitive horizontal  subspace and homological mirror symmetry}   
\label{HMS}
It is likely that the theory of  lattices for given intersection forms  and homological mirror symmetry 
will lead to a complete understanding of the integer basis of the cohomology of fourfolds. Here we 
restrict ourself  to  primitive horizontal  subspace and confirm and extend  the calculations  
of the previous section using considerations of homological mirror symmetry and supersymmetric  
localization on the hemi-sphere in 2d $(2,2)$ gauged linear $\sigma$ models.            

\subsubsection{The $\hat \Gamma$ classes and homological mirror symmetry} 
\label{Gammaclass}
It has been noticed  in~\cite{Hosono:2000eb,IRITANI1,MR2483750,Kontsevichgamma}, 
that all coefficients of the logarithmic terms and the constant term of 
the solutions at the large complex structure point can be obtained by homological mirror 
symmetry in the A-model. In particular the even  $D$-brane charges are described by the $\hat \Gamma$-classes 
and involve $\zeta(2n+1)$ values. Similar interesting arithmetic values appear in the open 
string periods as has  been studied  in~\cite{Jefferson:2013vfa}.  Let us recall the 
following multiplicative characteristic classes. The well known Chern class the, A-roof genus, the Todd class and 
the $\Gamma$-roof class introduced in~\cite{MR2483750,Kontsevichgamma}
\begin{equation} 
{\rm ch}(x)=e^x, \qquad \hat A (x)=\frac{x/2}{\sinh(x/2)}, \qquad {\rm td}(x)=e^{x/2} \hat A(x) , 
\qquad   \hat \Gamma(x)=\Gamma\left(1-\frac{x}{2 \pi i} \right)\ . 
\label{multclasses}
\end{equation} 
 For a vector  bundle  $V$ we may expand the $\hat  \Gamma$  class as\footnote{Here $\gamma$ 
is the Euler-Mascheroni constant. The term drops out in the application below since 
${\rm ch}_1(TW)=c_1(TW)=0$.}      
\begin{equation}
\begin{array}{rl} 
\hat \Gamma(V)&=\displaystyle{\exp\left( \gamma \frac{{\rm ch}_1(V)}{2 \pi i}+\sum_{k\ge 2} \zeta(k) (k-1)! \frac{{\rm ck}_k(V)}{(2 \pi i)^k }\right)}\\ [3 mm]
                         &=\displaystyle{1\!-\! \frac{i c_1 \gamma }{2 \pi }\! -\! \frac{1}{48} \left(c_1^2 \left(\frac{6 \gamma^2}{\pi ^2}+1\right)\!-\!2 c_2\right)\!+\!\frac{i (2 c_1^3 \gamma^3\!+\!\pi^2 (c_1^2\!-\!2
   c_2) c_1  \gamma+4 (c_1^3\!-\!3 c_2 c_1\!+\!3 c_3) \zeta(3))}{96 \pi ^3}}\\
   &+O(4) \, .\\ [ 3 mm]                        
\end{array}
\label{hatG}
\end{equation}   
For the tangent bundle of a Calabi-Yau manifold one has hence 
\begin{equation}
 \hat \Gamma(TW)=1+\frac{1}{24} c_2+\frac{i c_3 \zeta(3)}{8 \pi ^3}+\frac{\left(7 c_2^2-
4 c_4\right)}{5760}+\frac{i\left(\pi^2 c_2 c_3 \zeta (3)+6 (c_2 c_3-c_5) \zeta (5)\right)}{192 \pi ^5}+ {\cal O}(6)\ .
\end{equation}

As has been explained in~\cite{IRITANI1,MR2483750,Kontsevichgamma}  due to the property 
\begin{equation} 
\hat A(V)=e^{c_1(V)/2} {\rm td}(V)=\hat \Gamma(V) \hat \Gamma^*(V)\ ,
\end{equation}       
the $\hat \Gamma$ class can be  viewed as an alternative definition of the square root of the A-roof  genus of the tangent bundle $V=TW$ of 
the mirror in Mukai's modified Chern-Character map, which turns out to be the correct one to get integral 
auto equivalences (twists) on the $K$-theory classes of the objects in the $A$ brane category and 
the central charges that determine Bridgeland stability~\cite{MR2373143}.   This fact makes the 
notion very useful for the comparison with the integral basis obtained in the last subsection. 

The classical Chern character map from topological $K$-theory to even 
cohomology  ${\rm ch}:K(W)\rightarrow H^{2*} (W,\mathbb{Q})$ and the Hirzebruch Riemann Roch 
index theorem
\begin{equation} 
\begin{array}{rl} 
\eta_{\alpha\beta}=& \displaystyle{Q(\Gamma_\alpha,\Gamma_\beta)=
\chi( {\cal E}_\alpha, {\cal E}_\beta)=
\int_{W_n} {\rm ch}({\cal E}^*_\alpha)\wedge   {\rm ch}({\cal E}_\beta) \wedge {\rm td}(TW_n)} \\[ 3 mm]
=&\displaystyle{\sum_{p=0}^n {\rm dim   }{\rm Ext}^p_{{\cal O}_{W_n}}( {\cal E}_\alpha, {\cal E}_\beta)} (-1)^p   \  ,
\end{array} 
\label{pairing}  
\end{equation} 
suggested  Mukai to define a  modified Chern-Character map as $\mu({\cal E})= {\rm ch}({\cal E}) \sqrt{\rm td(TW)}$. 
However as observed by~\cite{IRITANI1,MR2483750,Kontsevichgamma}  in order to get the right integer 
auto equivalences on the $K$-theory classes one must replace  $\sqrt{\rm Td(TW)}$ by 
$\hat \Gamma(TW)$. The modification of~\cite{IRITANI1,Kontsevichgamma}  leads to the central charge formulas of the $A$-branes given by 
\begin{equation} 
Z({\cal O}_{s_i})=\int_{W_n} e^{tJ} \wedge \hat \Gamma(TW_n)\wedge  {\rm ch}({\cal O}_{s_i}), \qquad i=0,\ldots, n\ ,      
\label{Abasis}  
\end{equation}     
where $s_i$ are holomorphic sub-varieties of complex dimension $i$. The point is that the twist 
${\cal E}\rightarrow  {\cal E}\times {\cal O}(1)$ comes out correctly integer  with both definitions
\begin{equation} 
 \chi({\cal E} \otimes {\cal O} (1) , {\cal F}\otimes {\cal O}(1))=\chi({\cal E} , {\cal F})\ ,
\label{twist1}  
\end{equation} 
 but the Seidel-Thomas twist~\cite{MR1831820}
\begin{equation} 
\chi(\Phi_{\cal O}( {\cal E}) , \Phi_{\cal O}( {\cal F}))=\chi({\cal E} , {\cal F})\ ,
\label{seidelthomastwist}
\end{equation} 
 which corresponds to the conifold monodromy,   requires the use of the $\hat \Gamma$  class in a further 
modified Chern character map  $\mu_{\hat \Gamma} ({\cal E})= {\rm ch}({\cal E}) \hat\Gamma(TM_n)$ to
 be integer in the same basis. In particular for $i= n$ the $K$-theory charge 
of the top dimensional brane is obtained as 
\begin{equation} 
\begin{array}{rl}
Z({\cal O}_W)=&\displaystyle{\int_{W_n} e^{tJ} \wedge \hat \Gamma(TW_n)= \int_{W_n}\biggl(1 +  J t +\left(\frac{J^2 t^2}{2} + \frac{c_2}{24}\right) + 
\left (\frac{J^3 t^3}{6}  + \frac{1}{24} t J  c_2 + \frac{i  c_3 \zeta(3)}{8 \pi^3}\right)}\\ [ 3 mm]   
 &+ \displaystyle{ \left(\frac{J^4 t^4}{4!} + \frac{1}{48} J^2 t^2 c_2 + \frac{7 c_2^2- 4 c_4}{5760}+ \frac{i J t c_3 \zeta(3)}{8 \pi^3}\right) }+ {\cal O}(5)\biggr) +{\cal O}(e^{-t}) \ ,
\label{gammalargevolume} 
\end{array}        
\end{equation}   
where we understand that we restrict to the term of order $n$ for Calabi-Yau $n$-folds and integrate the class $J$, which 
is dual to the K\"ahler class $t$ , and its wedge products with the Chern  classes over $W_n$. 
All terms are corrected by instantons effects of order ${\cal O}(e^{-t})$.  Similar one gets for the $D0$  and $ D2$ 
brane charges universal formulas     
\begin{equation}
\begin{array}{rl} 
 Z({\cal O}_{pt})=&1+{\cal O}(e^{-t})\, ,\\
Z({\cal O}_s [-1])=&t+{\cal O}(e^{-t}) \ . 
\end{array}
\end{equation}   
$Z({\cal O}_{s_2})$ and  $Z({\cal O}_{s_3})$ are compatible with the monodromy calculation on the 
complex structure side.

Because of the requirement that (\ref{seidelthomastwist}) and (\ref{twist1}) are integer in the 
same basis, the  appearance  of the coefficients of the $\hat \Gamma$  class, like $\zeta(k)$,  in the central 
charge formula, follows also from the analytic continuation of the period $X_\nu$ over vanishing cycle $\nu$ at the conifold point 
to the large complex  structure point and the fact that this period maps under homological 
mirror symmetry to the structure sheaf on $W_n$. 
This was used in~\cite{Candelas:1990rm} and \cite{Grimm:2009ef} to determine these 
coefficients for threefolds and fourfolds respectively.
It is possible to prove (\ref{Abasis}) using the hemi-sphere partition function obtained by localization~\cite{Hori:2013ika}.  
As was pointed out in~\cite{Halverson:2013qc} the $\hat \Gamma$ class is also compatible with  
the interpretation of the sphere partition function  of~\cite{Benini:2012ui,Doroud:2012xw} as $e^{-K}$~\cite{Jockers:2012dk}.  

The basis (\ref{Abasis}) is not anti-diagonal because the structure sheaf on a fourfold 
has self-intersection  
\begin{equation}
\chi({\cal O}_{W_4}, {\cal O}_{W_4})= \sum_{p=0}^4 H^p({\cal O}_{W_4} \otimes {\cal O}^*_{W_4} )(-1)^p=2 \ ,         
\end{equation}  
 which is just the self-intersection of the $S^4$-sphere given in (\ref{Snintersection}). Similarly     
\begin{equation}
\chi({\cal O}_{pt}, {\cal O}_{W_4}) =\sum_{p=0}^n H^p({\cal O}_{pt} \otimes {\cal O}^*_{W_4} )(-1)^p=1 \ .         
\end{equation} 
So in oder to make the metric anti-diagonal  we have chosen the class ${\cal O}_{W_4}-{\cal O}_{pt}$ as basis element. 
That corresponds to (\ref{shift}) and  lead to the  $-1$ shift in $c_0$ of (\ref{prediction}).

\subsubsection{The hemi-sphere partition function from supersymmetric localisation } 
\label{hemisphere}  
One new outcome of the  supersymmetric localization~\cite{Benini:2012ui,Doroud:2012xw} 
are the formulas  for partition function of the $(2,2)$ gauged linear $\sigma$ model~\cite{Witten:1993yc} 
on the hemi-sphere with $A$- and $B$-type  boundary 
conditions~\cite{Hori:2013ika}. Even though the final formulas for 
the  hemisphere partition function for the $\sigma$-model with abelian gauge 
groups are familiar period expressions, the formalism yields new conceptual 
insights in the problem of finding a rational basis $H^n(M_n,\mathbb{Q})$ 
due to its relation to homological mirror symmetry. It is in addition very general and 
for non abelian gauge groups it describes Calabi-Yau manifolds  embedded in Grassmannians, flag-varities and   
those determinantal and more general non-complete intersection embeddings of Calabi-Yau manifolds  
that are  e.g. needed to get F-theory with more than $4$ global $U(1)$ factors.  	   

Let\footnote{Our notation follows~\cite{Hori:2013ika} and lectures of K. Hori at the University of Michigan in Ann Arbor.}  
$G$ be the rank $l_G$ gauge group of the 2d  $(2,2)$ linear $\sigma$ 
model and  $T\subset G $ its maximal torus. The complex matter fields $\phi_i$ transform
in a $G$ representation space $V$ and carry a charge with respect to $T$ called  $Q_i$.  
$V$ is also acted on by an  $R$ symmetry representation ${\rm End}(V)$  with charge $R_i$, i.e. 
$V|_{\mathbb{C}_R\times T}=\oplus_i \mathbb{C}(R_i,Q_i)$. One has an embedding $e^{\pi i R}=:J\in G$.     
The superpotential  $W$ is a gauge invariant function of the matter fields  and  homogeneous of degree 
two w.r.t.  the $R$ charge, i.e. ${\cal W}(\lambda^R \phi)=\lambda^2 {\cal W}(\phi)$. We need to consider only 
the situations where the twisted superpotential is linear and contains  just the
term $\tilde {\cal W}=-\frac{1}{2 \pi} t(s)$, where $t=\zeta- i\theta$  is a complex combination 
of the Fayet-Iliopoulus parameter, identified with the K\"ahler parameter in the CY phase, and 
the theta parameter in $\mathbb{R}/\mathbb{Z}^{l_G}$, identified with the periodic 
$B$ field in the CY phase.  

The boundary data for the hemi-sphere are specified by  ${\cal R}=(M,{\cal Q},\rho , r^*)$,  
where $M$ is the $\mathbb{Z}_2$ graded vector space of Chan-Paton factors 
$M=M^{ev} \oplus M^{od}$  over $\mathbb{C}$ and $\rho$ and $r^*$  are 
representations  of $G$ and $R$ on $M$, i.e. $M|_{\mathbb{C}^R\times T}=
\oplus{C}(r_i,q_i)$.  It has the charge integrality property 
\begin{equation}
e^{\pi i r^*} \rho(J)=\begin{cases} 1 & \mbox{on}\ \ M^{ev}  \\  -1 & \mbox{on} \ \ M^{od} 
\end{cases} \ .
\label{chargeintegrality}
\end{equation}
${\cal Q}$ is a ${\rm End}^{od}(M)$ valued holomorphic 
function in $V$, which is gauge invariance $\rho^{-1}(g) {\cal Q}(g \phi) \rho(g)={\cal Q}(\phi) $, 
homogeneous $\lambda^{r^*} {\cal Q}(\lambda^R \phi) \lambda^{-r^*}=\lambda {\cal Q} $ 
and has the matrix factorization property of $W$ 
\begin{equation}
{\cal Q}(\phi)^2={\cal W}(\phi) {\id}_M\ .
\label{matrixfactorisation} 
\end{equation} 

The general form for the semi sphere partition is calculated by  
supersymmetric localization and reads for the boundary data ${\cal R}$~\cite{Hori:2013ika} 
\begin{equation} 
Z_{D^2}({\cal R})=c (\Lambda,r)\int_{\gamma} {\rm d}^{l_G} s \prod_{\alpha>0} \alpha(s) {\rm sinh} (\pi \alpha(s)) 
\prod_{i} \Gamma\left(i Q_i (s)+\frac{R_i}{2}\right)  e^{t(s)} {\rm Tr}_M\left( e^{\pi i r^*} e^{2 \pi \rho(s)})\right)\ ,
\label{hemispherepartitionfunction} 
\end{equation}  
 where $\alpha$ are the roots of  $G$. The contour $\gamma$ is chosen in a multidimensional 
generalization of the contour used section (\ref{fourfoldbasis}) (rotated by $\frac{\pi}{2}$ to 
the left), so that it is a deformation if the real locus $i t \subset t_{\mathbb{C}}\sim \mathbb{C}^{l_G}$  
so that 
\begin{itemize}
\item C1.)  the integral is convergent and 
\item C2.) the deformations does not cross poles of 
the integrand.   
\end{itemize}
We specialize to the Calabi-Yau n-fold  case where the axial $U(1)_A$ 
anomaly is cancelled, $t$ is not renormalized,  $\hat c={\rm tr}_V(1-R)-{\rm dim}{G}=n$, the 
dependence  on the scale $\Lambda$ and radius of the hemi-sphere $r$ disappears and $c(\Lambda,r)$ 
becomes  a normalization constant.  We focus  the attention to $A$-branes on $W_n$, which 
correspond to twisted  line bundles 
${\cal O}(q^1,\ldots,q^{l_G})$, with $l_G=h_{1,1}(W_n)=h$ with Chern character 
\begin{equation} 
{\rm ch}({\cal O} (q^1,\ldots,q^h))=\exp\left({\sum_{\alpha=1}^h q^\alpha J_\alpha }\right) \ . 
\label{cherntwistedbundles} 
\end{equation} 
The range of independent $\{ q_i\}$ is restricted by  the relations in 
the Chow ring of $W_n$ and the corresponding branes for a $\mathbb{Q}$-basis in the $K$-theory  
classes of the derived category of coherent sheaves. In particular with standard intersection 
calculations we can determine  (\ref{pairing}) in this basis. In order not to clutter the notations, we consider 
abelian gauge groups $G=U(1)^h$ and only matter fields representations  and superpotentials that 
lead to complete intersections. In this case the first product  in  (\ref{hemispherepartitionfunction}) 
is trivial and yields one and we get 
\begin{equation} 
Z_{D^2}({\cal O}(q^1,\ldots, q^h))=\frac{2^r i}{ (2\pi i)^{n+r}}  \int_{\gamma} {\rm d}^h s \prod_{j=1}^r \Gamma\left( i l^{(\alpha)}_{0j} s_\alpha +1\right) 
\prod_{j=1}^k\Gamma(i l^{(\alpha)}_j s_\alpha)   e^{2\pi (t^\alpha+q^\alpha) s_\alpha}  \prod_{j=1}^r  \sinh( \pi  l^{(\alpha)}_{0j} s_\alpha)\ ,
\label{hemispherepartitionfunctionu1} 
\end{equation}             
where the sum over $\alpha$ is implicit and we restrict the arguments of $e^{ \pi t_{\alpha}}$ analogously  to (\ref{barnes}) 
to be in the ranges  $0\le {\rm arg}(e^{t_{\alpha}})< {\rm min} \left(\frac{2 \pi}{-l^\alpha_{0,j}}\right)$.          

It is not  hard to show that the  formulas (\ref{hemispherepartitionfunctionu1})  
are indeed a Mellin-Barnes integral representation of (\ref{solution}) and certain combinations  
of its $\partial^{\underline r}_{\underline \rho}|_{\underline \rho}$ derivatives.         
For  example we can  rewrite (\ref{barnes}) and its  transforms $\varpi_0(\beta^k a)$  
using the identity $\Gamma(1-s)\Gamma(s) =\frac{\pi}{\sin(\pi s)}$  in the form
(\ref{hemispherepartitionfunctionu1})  
\begin{equation} 
Z_{D^2}({\cal O}(q))=\frac{ i}{ (2 \pi i)^5}  \int_\gamma {\rm d}s  \Gamma( -6 i s +1) \Gamma^6( i s) e^{ 2 \pi  t s}( e^{2 \pi (q-3) s}-e^{2 \pi  ( q+3) s})\ .
\label{sixtichoribranes}
\end{equation}
The the contour $\gamma$ can be closed according to the conditions C1, C2     
\begin{itemize} 
\item a.) along $s=-i \epsilon$ in the upper halfplane to enclose all the poles at $s= i l$ with $l\in  \mathbb{N}$. This  gives the solutions that converge at the large radius 
 or large complex structure point, 
\item b.) in the lower half plane to enclose all the poles at $s=-i l/6$ with $l\in \mathbb{N}_+$, to  get the  solutions that converge 
at the orbifold/Landau-Ginzburg or Gepner point.
\end{itemize}
Of particular interest is the identification of the  structure sheaf ${\cal  O}_M$, which, as we already  have
seen in the last two sections, corresponds to  the maximal logarithmic solutions in the large radius limit and the 
vanishing cycle corresponding to the Seidel-Thomas  auto equivalence. For the sextic Calabi-Yau fourfold $G=U(1)$, 
$V=\mathbb{C}(-6,2)\oplus \mathbb{C}(2,0)^{\oplus 6}\supset (P,x_1,\ldots, x_6)$ and ${\cal W}=PW(x)$, 
where $W(x)$ is defined in (\ref{WDelta}) is the constraint of homogeneous degree six. The fermionic 
partners  of the $\phi_i$ are $\eta_i$ and $\bar \eta_{\bar \imath}$. They have gauge charge $(-1,1)$ and $U(1)_v$ charge $(1,-1)$ and fulfill 
the Clifford algebra $\{\eta_i, \bar \eta_{\bar \imath}\}=\delta_{i\bar \imath}$ and $\{\eta_i,  \eta_j\}=\{\bar \eta_{\bar \jmath}, \bar \eta_{\bar \imath}\}=0$.         
Following the work of~\cite{Herbst:2008jq} one can identify the matrix factorization for the structure sheaf as  
\begin{equation}
\mathbb{C}(0,0) {\xrightarrow{W(x)} \atop \xleftarrow[\ \   P \ \  ]{}}   \mathbb{C}(1,d) \ .
\end{equation}   
These maps represent the matrix as  ${\cal Q}=\left(\begin{array}{cc} 0 & W(x)\\ P&0\end{array}\right).$ 
One has  $\{{\cal Q}, {\cal Q}^\dagger\}=(|P|^2+|W(x)|^2){\bf 1}$ and since for $\zeta \gg 0$ this vanishes 
at the CY $W_n$ and the brane is identified at large radius with the structure sheaf. Algebraically this can be seen form 
the exactness of the following infinite sequence 
\begin{equation} 
{\cal O}(0)\xrightarrow{W}{\cal O}(d)  \xrightarrow{P}{\cal O}(d) \xrightarrow{W}{\cal O}(2d)  \xrightarrow{P}{\cal O}(2d)  \xrightarrow{W}\ldots
\label{sequencestructure} 
\end{equation}  
At the Landau Ginzburg point $\zeta \ll 0$,   $\{{\cal Q}, {\cal Q}^\dagger\}$ vanishes nowhere, hence there 
is no brane with the ${\cal Q}$ matrix factorization data.  It is an easy exercise to see  that  closing the contour a.) for the integral 
(\ref{sixtichoribranes}) or more generally   (\ref{hemispherepartitionfunctionu1}) for the 
structure sheaf reproduces the last entry of  (\ref{piinf})  and more generally  (\ref{predictionhighest}) 
once the shift (\ref{shift})  has been taking into account. The matrix $\tilde m$  as in 
$\Pi_{z=0}=\tilde m (Z({\cal O} (0)), \ldots,  (Z({\cal O} (4))^T$ is readily calculated. We report it together
with (\ref{pairing}), which is likewise  readily calculated  using (\ref{cherntwistedbundles}), and 
${\rm td}(TW_n)$ from (\ref{multclasses}) in terms of the Chern classes  for the sextic obtained from the adjunction formula 
\begin{equation} 
\tilde m= \left(
\begin{array}{rrrrr}
 \frac{1}{6} & -\frac{2}{3} & 1 & -\frac{2}{3} & \frac{1}{6} \\
 -\frac{5}{12} & \frac{3}{2} & -2 & \frac{7}{6} & -\frac{1}{4} \\
 -\frac{35}{72} & \frac{49}{36} & -\frac{4}{3} & \frac{19}{36} & -\frac{5}{72} \\
 \frac{5}{3} & -\frac{4}{3} & -2 & \frac{7}{3} & -\frac{2}{3} \\
 \frac{5}{6} & \frac{2}{3} & -1 & \frac{2}{3} & -\frac{1}{6} \\
\end{array}
\right),\quad 
 \eta^{\alpha\beta}=\chi({\cal O}(\alpha),{\cal O}(\beta) )  = \left(\begin{array}{ccccc}
 2 & 6 & 21 & 56 & 126 \\
 6 & 2 & 6 & 21 & 56 \\
 21 & 6 & 2 & 6 & 21 \\
 56 & 21 & 6 & 2 & 6 \\
 126 & 56 & 21 & 6 & 2 \\
\end{array}\right) \ .   
\label{tildemsixtic} 
\end{equation} 
Now we calculate as a check that $({\tilde m}^{-1})^{\alpha \mu}  \eta_{\mu \nu}  ({\tilde m}^{-1})^{\nu\ beta}$ yields the 
form of the pairing (\ref{LSintersection}) of course with $C^0=6$, which perfectly closes the chain of arguments. Note that 
$X_\nu=\tilde F_0+X^0$ is the structure sheave. We could have chosen $F_0=X_\nu$ in (\ref{piinf}).

 This is reflected in $\tilde m$ as the first and the last line add up to $(1,0,\ldots,0)$. So 
the information in (\ref{predictionhighest}) reflecting the $\hat\Gamma$ class contributions can be also established from the 
analytic continuation of $Z({\cal O}_W)$ instead of $X_\nu$ to the large complex structure point as in~\cite{Grimm:2009ef}.  
One  can easily calculate $\tilde m$ for the other one parameter models  and show that it is rational.  For example 
for the  dectic in $\mathbb{P}(1^5,5)$ one has the data
\begin{equation} 
\tilde m=\left(
\begin{array}{rrrrr}
 \frac{1}{2} & -2 & 3 & -2 & \frac{1}{2} \\
 -\frac{5}{4} & \frac{9}{2} & -6 & \frac{7}{2} & -\frac{3}{4} \\
 \frac{5}{8} & -\frac{15}{4} & 7 & -\frac{21}{4} & \frac{11}{8} \\
 -\frac{5}{4} & \frac{13}{2} & -9 & \frac{9}{2} & -\frac{3}{4} \\
 \frac{1}{2} & 2 & -3 & 2 & -\frac{1}{2} \\
\end{array}
\right), \quad
  \eta^{\alpha\beta}=\left(
\begin{array}{ccccc}
 2 & 5 & 15 & 35 & 70 \\
 5 & 2 & 5 & 15 & 35 \\
 15 & 5 & 2 & 5 & 15 \\
 35 & 15 & 5 & 2 & 5 \\
 70 & 35 & 15 & 5 & 2 \\
\end{array}
\right)\ .
\label{tildemdectic} 
\end{equation}
The formalism is universal and applies to complete intersections with logarithmic singularities at the origin as well.  
For the five  quadrics in $\mathbb{P}^{10}$ the superpotential is ${\cal W}=\sum_{i=1}^4 P_i W_i$ and the matrix 
factorisation is ${\cal Q}=\sum_i W_i \eta_i+\sum_i P_i \bar \eta _i$   and   one can establish correct 
weights  of the structure sheaf in (\ref{hemispherepartitionfunctionu1}) from the Koszul sequence
\begin{equation}
\mathbb{C}(0,0) {\xrightarrow{W} \atop \xleftarrow[\ \   P_1 \ \  ]{}}   \mathbb{C}(1,2)^{\oplus 4}  {\xrightarrow{W} \atop \xleftarrow[\ \   P_2 \ \  ]{}}  \mathbb{C}(2,4)^{\oplus \left(4\atop 2\right)}  {\xrightarrow{W} \atop \xleftarrow[\ \   P_3 \ \  ]{}}  \mathbb{C}(3,6)^{\oplus \left(4\atop 3\right)}  {\xrightarrow{W} \atop \xleftarrow[\ \   P_4 \ \  ]{}}    \mathbb{C}(4,8),
\end{equation} 
where the terms in the sequence from left to right represent the Clifford vacua  $|0\rangle$,  $\eta_i|0\rangle$,    
$\eta_i\eta_j|0\rangle$,    $\eta_i\eta_j\eta_k|0\rangle$ and $\eta_1\eta_2\eta_3\eta_4 | 0\rangle$. From this 
series one can find a  sequence  similar to (\ref{sequencestructure}) see~\cite{Herbst:2008jq} for 
a similar example, which established  the weights for the structure sheaf in (\ref{hemispherepartitionfunctionu1}). Using  
this  formula we get 
\begin{equation} 
\tilde m=
\left(
\begin{array}{ccccc}
 \frac{1}{32} & -\frac{1}{8} & \frac{3}{16} & -\frac{1}{8} & \frac{1}{32} \\
 -\frac{5}{64} & \frac{9}{32} & -\frac{3}{8} & \frac{7}{32} & -\frac{3}{64} \\
 \frac{5}{128} & -\frac{5}{64} & \frac{1}{32} & \frac{1}{64} & -\frac{1}{128} \\
 \frac{5}{8} & \frac{1}{4} & -\frac{3}{2} & \frac{3}{4} & -\frac{1}{8} \\
 \frac{31}{32} & \frac{1}{8} & -\frac{3}{16} & \frac{1}{8} & -\frac{1}{32} \\
\end{array}
\right), \qquad 
 \eta^{\alpha\beta}=
\left(
\begin{array}{ccccc}
 2 & 10 & 50 & 170 & 450 \\
 10 & 2 & 10 & 50 & 170 \\
 50 & 10 & 2 & 10 & 50 \\
 170 & 50 & 10 & 2 & 10 \\
 450 & 170 & 50 & 10 & 2 \\
\end{array}
\right)\ .
\label{tildem}  
\end{equation}
In fact it is an important general feature of all cases including the multimoduli cases. The reason is  that this matrix can be 
calculated  from the {\sl grade restriction rule}~\cite{Herbst:2008jq}. For orbifold singularities it can also be
seen in the closely related approach~\cite{Mayr:2000as}.  One advantage of the  basis 
 ${\cal O}({\underline q})$  is that  allows  most easily to calculate (\ref{pairing})  and hence the 
K\"ahler potential  can be readily given albeit only w.r.t.  rational basis, which nevertheless 
allows to evaluate  everywhere  in the complex  moduli space.  The analytic continuation between 
the large complex structure point and its inverse basically defined by changing from contour a) to contour 
b)  is relatively simple. Two moduli  examples where treated in~\cite{Candelas:1993dm,Candelas:1994hw}. 
Elements of a general theory are outlined in~\cite{ MR1472476,MR1619440}. 

Let us finish the section with three remarks. 
\begin{itemize} 
\item (i) It might sound anti climatic,  but  it is useful and  at least in the $U(1)^h$ cases  
straightforward,  to derive from (\ref{hemispherepartitionfunctionu1})  the  full system of Picard-Fuchs 
equations ${\cal D}_i$, $i=1,\ldots, s$ . The reason for the usefulness of the remark 
is that it is hard to read from  (\ref{hemispherepartitionfunctionu1}) all components of the critical locus of the 
periods, which follow on other hand immediately from the resultant of the  
symbols of the  ${\cal D}_i$, $i=1,\ldots, s$. Moreover the expression (\ref{hemispherepartitionfunctionu1}) 
is  not useful to analytically continue to most of these components, e.g. to the conifold
loci. To derive the  ${\cal D}_i$, $i=1,\ldots, s$ that generate the ideal ${\cal I}_{PF}$, see section \ref{gaussmaninS},  we follow 
the  observation made in~\cite{Hosono:1993qy} that the classical intersection ring  
${\cal R}(\underline \xi)=c_{i_1,\ldots, i_n} \xi^{i_1}\cdots\xi^{i_n}$   at a large radius point in 
coordinates\footnote{Note that in this section $t_\alpha$ denotes not the quantum corrected 
K\"ahlerparameter, which is defined in \ref{mirrormap}.}   at $z_\alpha=e^{-t_\alpha}$ is given by 
\begin{equation}
{\cal R}({\underline \xi})=\mathbb{C}[\xi_1,\ldots, \xi_h]/{\rm Id}({\cal S}_i({\underline \xi}) ,i=1,\ldots,s),
\label{ringrelation} 
\end{equation} 
where ${\cal S}_i$ is obtained  as ${\cal S}_i=\lim_{t_\alpha\rightarrow \infty} 
{\cal D}_i(\partial_{t_\alpha}=\xi_{\alpha})$ and Id denotes the multiplicative 
ideal and the $\xi$ are the {\sl symbols} of the differential ideal ${\cal I}_{PF}$. Consider now  the monomials $D_{i_1}\cdot D_{i_{|I^*|}}$ representing the 
Stanley Reisner ideal for a given triangulation. Pick a basis $K_i$ of the Chow ring and express the 
$D_{i_1}\cdot D_{i_{|I^*|}}$  in terms of the  $K_j$, $j=1,\ldots, h$. This yields polynomials 
${\cal S}_i(K_i=\xi_i)$, $i=1,\ldots, s-\delta$  which generate  part of the ideal Id. The full  ideal can be 
obtained by completing the  ${\cal S}_i(K_i=\xi_i)$, $i=1,\ldots, s-\delta$ minimally  so that   
(\ref{ringrelation}) holds. Now we can act with the ${\cal S}_j(\xi_i=\partial_{t_i})$, $j=1,\ldots, s$ 
on any period say  the one corresponding to the structure sheaf  $Z({\cal O}_W)$. This brings down 
$s_i$ monomials in the integrand, whose  exponents  can be lowered by the relations 
$x\Gamma(x)=\Gamma(x+1)$, $\Gamma(x) \Gamma(1-x)=\frac{\pi}{\sin(\pi x)}$  
and a redefinition of the  integration variables. This yields a relation between maximal order 
derivatives and lower order derivatives of  
$Z({\cal O})_W$ with polynomial coefficients in the $e^{2 \pi i t_\alpha}$  and 
constitutes a linear differential operator annihilating  all $Z({\cal O})_W(q^1,\ldots,q^h)$, 
i.e. a Picard-Fuchs operator ${\cal D}_i$. 
The differential ideal of the latter completely determines these periods,  
if (\ref{ringrelation}) holds. The latter point should also hold in the case of 
non-abelian gauged  linear $\sigma$-models, which leads not to differential 
systems  of generalized  hypergeometric type, but rather to the Apery 
type~\footnote{Only the  one moduli cases of Apery type, like the Grassmannians 
for which higher genus invariants have been calculated in~\cite{Haghighat:2008ut},  
have conifold loci in different distance from the large complex structure point 
$\lim_{t \rightarrow \infty}$. This is a necessary condition for fast enough 
convergence  of the analytic continuation from the conifold to the large 
complex structure point, that would be needed  to prove the irrationality 
of $\zeta(2m+1)$  occurring in  the periods of CY $2m+1$-folds for $m>1$ at 
infinity due to (\ref{hatG}) in (\ref{Abasis}). We thank Sergei Galkin to point this fact out.}. 
The $\alpha(s)$ factor in (\ref{hemispherepartitionfunction}) makes it slightly 
more non-trivial to lower the powers and rewrite the integral 
 in the standard form described above.        
   
\item (ii) We can study the period system on local Calabi-Yau spaces, by  
replacing the sections $W_{\Delta_l}=0$ of $D_{0,l}$ by the 
total space of the bundles $\oplus_{l=1}^r {\cal O}(-D_{0,l})$ over 
$\mathbb{P}_{\Delta^*}$. This is the obvious generalization of 
taking instead of the elliptic curve (quintic hypersurface) defined as a section of the 
canonical bundle $K$ in $\mathbb{P}^2$ ($\mathbb{P}^4$) the
noncompact Calabi-Yau 3(5)-manifold defined as the total space of the 
anti-canonical line bundle ${\cal O}(-3)\rightarrow \mathbb{P}^2$  
(${\cal O}(-5)\rightarrow \mathbb{P}^4$).   This is implemented
by the following change in the integrand of (\ref{hemispherepartitionfunctionu1})
\begin{equation} 
 \sinh( \pi  l^{(\alpha)}_{0j} s_\alpha)\rightarrow   \frac{\sinh( \pi  l^{(\alpha)}_{0j} s_\alpha)}{s_\alpha}\ . 
\label{locallimit}
\end{equation}                
 This process can be done successively  leading to an increasing number of 
non-compact directions.      
\item (iii) Finally we like to point out that by applying  a Legendre 
transform  to the $e^{t_\alpha}$ variables in the  integrals 
(\ref{hemispherepartitionfunctionu1}) one  gets periods governing 
the $I$-functions of the $A-$model  on $\mathbb{P}_{\Delta^*}$. 
This can be done  partially on some of the $e^{t_{\alpha}}$. In this
case one gets more involved del Pezzo manifolds $W$ 
given by constraints that render $c_1(W)>0$. In this case one gets 
instead of hypergeometric or Apery-like functions Bessel like 
functions that have different asymptotic expansion. The transitions 
matrices $m$, $\tilde m$ and  $n$  can be related to the Stokes matrices 
of that system.          
\end{itemize}

\subsection{Minimizing the flux superpotential}  
\label{minimizingW}    

The superpotential in type IIB and F-theory can be expanded in 
terms of a basis of periods as 
\be 
W=\int_{M_n} \Omega_n(z)\wedge G_n=\sum_{\alpha} n^\alpha \Pi_\alpha(z) \ ,
\ee
where 
\be 
\Pi_\alpha(z)=\int_{\Gamma_\alpha} \Omega_n(z), \quad \ \Gamma_\alpha \in H^{\rm prim} _n(M_n)\ . 
\ee
In the presence of space filling  branes internally wrapped on  special 
Lagrangian branes $L$, $\Gamma_\alpha$ can be more generally viewed  
as a chain integral in relative homology $H^{rel}_n(M,L)$ 
ending on $L$ and contributing likewise to the space-time 
superpotential. The superpotential $W$  transforms in ${\cal L}$ and 
the covariant derivative  $D_i=\partial_i + \partial_i K$ enters the 
minimization conditions for the superpotential spelled 
out for type IIB vacua in~\cite{Curio:2000sc}. If one denotes 
by $a_i$ the chiral ${\cal N}=1$ superfields the corresponding 
parts of the effective action can be expressed in terms of a 
generalized K\"ahler function~\footnote{That transforms     
with $G(a,\bar a)\rightarrow  G(a,\bar a)+ f(a) + \bar f(\bar a)$ 
under gauge transformations  $\Omega\rightarrow \Omega e^{f(a)}$.}       
\begin{equation} 
G(a,\bar a)=K(a,\bar a)+ \log |W(a)|^2,    
\end{equation}
which determines the scalar potential as 
\begin{equation} 
V=e^G \left( G_i G_{\bar \jmath} G^{i {\bar \jmath}} -3\right) \ .     
\end{equation}
Note that in general the $a_i$ are not only the complex moduli, but all 
moduli of the effective  theory, i.e. also the K\"ahler moduli and the dilaton. 
These can be treated separately as long as they contribute at least approximately 
additively to $G$. I.e. if $K=K_{cs}+K_{ks}+K_d$ and  $W=W_{cs}W_{ks}W_d$, which assumes 
decoupling properties. In particular as far as the dilaton is concerned 
this is a weak coupling approximation\footnote{It is only in this section that we 
distinguish between $K$ $(W)$  and $K_{cs}$ $(W_{cs})$ in the other 
sections $K$ $(W)$ simply mean $K_{cs}$ $(W_{cs})$.}. In this case the
contribution of the complex moduli $a$ is 
\begin{equation} 
V_{cs}=e^{K_{cs}} G^{a\bar a}|D_a W|^2 \ .    
\label{Vcs}   
\end{equation}
The mass of the ${\cal N}=1$ gravitino is given by 
\begin{equation} 
 m_{3/2}=\exp(G)= \exp(K/2) |W|\ .     
\end{equation}
Generically minimisation of $V$, i.e. $\frac{{\rm d}}{{\rm d} a_i} V=0$ 
leads to anti de Sitter space at $V(a_{min})$. To find a 
${\cal N}=1$ supersymmetric vacuum one needs all auxiliary fields to vanish  
\be 
H^{\bar \imath}=G^{\bar \imath j} |W| e^{K\over 2} (\partial_j K + \frac{1}{W} \partial_j W)=0\ . 
\label{susycondition}  
\ee
Since these terms are all positive they have to vanish {\sl individually}. In particular if 
$G^{\bar \imath j} e^{K\over 2}$ as well as $\partial_j K$ are generic  at the minima this means 
\begin{equation}
W|_{min}=0, \qquad \partial_i W|_{min} =0 \     
\label{minimisation1} 
\end{equation}
is  a supersymmetric vacuum.  More generally one searches for  
\be 
D_i W=0 \ .  
\label{minimisation2} 
\ee
If the vacuum manifolds represent smooth Calabi-Yau fourfolds one has therefore  
the following conditions at the minima of the superpotential~\cite{Gukov:1999ya}, 
see also~\cite{Denef:2008wq}. 
\begin{itemize}
\item M1)  For the superpotential (\ref{supo})  $W_{cs}({\underline{a}})=\frac{1}{2 \pi } \int_{M_4} 
\Omega({\underline{a}}) \wedge G_4$, the first condition in (\ref{minimisation1})  
means by  (\ref{bilinear}) that $G_4$ cannot be of type $(0,4)$. The second equation 
and (\ref{GMdiff}) and Griffiths transversality (\ref{griffithstransversality}) tells that 
it cannot be of type\footnote{If one imposes just $D_i W_{cs}=0$ one can, as $D_i\Omega $ spans 
$H^{3,1}$ by the T.-T. Lemma,  only conclude that $G_4$ is not of type $(3,1)$ and $(1,3)$. 
Still $G_4$ will be selfdual as $(4,0)$ and $(0,4)$  are selfdual.}  $(3,1)$ and since $W$ is holomorphic $G_4$ has to be of type $(2,2)$.  
\item M2) 
For the superpotential (\ref{supo2}) $W_{ks} =\frac{1}{4 \pi} \int_{M_4}  J^2 ({\underline t})\wedge G_4$ 
the second condition in~\ref{minimisation1} implies that $J\wedge G_4=0$, i.e. that $G_4$ is primitive. 
It follows then by M1.) and (\ref{signature}) that $G_4$ is selfdual. The first condition in (\ref{minimisation1})  imposes  
no  further constraint on $W_{ks}$.     
\end{itemize} 

It is hence possible to restrict many complex moduli by turning  
on a primitive flux $G_4$ on an algebraic $(2,2)$ cycle $[G_4]$  as long 
as the minimum configuration is smooth. Like for K3 the existence of an 
algebraic cycle restricts the complex family of Calabi-Yau fourfolds to the 
analog of the Noether-Lefshetz locus~\footnote{We would 
like  to thank Sheldon Katz for pointing out many examples of 
this kind, including a preliminary analysis that by this method one 
can restrict the sixtic to the parameter family that we analyze 
in section~\ref{fourfoldbasis}. We also were not aware of the following  
deformation argument.}.  It can be argued from deformation theory that 
a single $(2,2)$ class could in principle restrict all complex moduli.       
In practice one try to find restricted but smooth configuration of the algebraic constraints, 
which allow the algebraic cycle to exist. E.g. one can insist that there is an 
algebraic  $\mathbb{P}^2$ in the sixtic in $\mathbb{P}^5$ by restricting the 
general constraint to the form 
\begin{equation} 
W_{\ci \Delta}=x_1 f^{(1)}_5({\underline{x}}) + x_2 f_5^{(2)} ({\underline{x}})+  x_3 f^{(3)}_5({\underline{x}})=0\ ,
\end{equation}  
where the $f_5^{(i)}$ are arbitrary homogeneous polynomials of 
degree $5$. For this constraint the  $\mathbb{P}^2$  specified by  
$x_1=x_2=x_3=0$ is always on the sixtic and some complex  structure 
deformations are absent. Setting up the  same calculation we used 
at the end of section~\ref{fourfoldbasis} to find the number of 
lines on the sixtic one finds that the rank of the quotient bundle 
$S^6 Q_{\mathbb{P}^2}$ describing $\mathbb{P}^2$  that lie 
on the sixtic is 28, while the dimension of $Gr(3,6)$ is 9. Hence 
the algebraic  cycle obstructs $19$ complex structure moduli. 
This can be also counted by counting the polynomial defomations of 
$W_{\ci \Delta}$ modulo  automorphisms.         

This method is closely related to the one discussed 
in (\ref{groupactions}) and \ref{submonodromy}). In particular these 
configurations are also at  fixpoint of the induced automorphisms 
discussed in (\ref{groupactions}).  In F-theory the approach gives a 
simple characterization of the allowed fluxes. These are those primitive  
algebraic cycles, which preserve the Weierstrass forms.   In order to 
study  the actual scalar potential at the fixed loci in ${\cal M}_{cs}$, to 
find the flux quanta relative to an integral basis or to find the K\"ahler 
potential one has to determine the integral basis at the degeneration 
locus.

Note that in ${\cal N}=2$ backgrounds the conditions (\ref{minimisation1}) 
are equivalent to the conditions for an attractor point for a 
supersymmetric black hole of mass 
\begin{equation} 
 M^2_{BPS}=e^{K_{cs}}|W(a)|^2 ,
 \label{MBPS} 
\end{equation}
where the gravitino mass 
is identified with the Bekenstein-Hawking entropy of the black hole at 
the attractor point~\cite{Curio:2000sc} 
\begin{equation} 
S= \pi e^{-K_H} m^2_{3/2}\ . 
\label{BHEntropy}  
\end{equation} 
Here $H$ are the hypermultiplets, i.e. the dilaton plus K\"ahler moduli in 
type IIB compactifications. This correspondence allows to use 
some results on attractors in the next sections.   

Now suppose that there is a symmetry $\Gamma:{\cal M}_{cs}\rightarrow {\cal M}_{cs} $ 
acting on the complex moduli space  and $H^{\bar \imath}$ are generic enough equivariant 
Morse functions w.r.t. $\Gamma$, then by the standard argument $H^{\bar \imath}$ 
vanish only at the fix-point locus $f_\Gamma$  of $\Gamma$. However the local Torelli 
theorem holds for Calabi-Yau manifolds and assures that one can pick certain 
$\Pi_i(z)$, $i=1,\ldots, h_{n,1}$ to be faithful local coordinates near $f_\Gamma$ 
so that one can always pick $n^\alpha$ in $W$ to $H^{\bar \imath}$ generic 
enough Morse functions.

\subsubsection{Arithmetic of orbifolds and supersymmetric vacua  }
\label{orbifoldpoints} 
We have seen in section (\ref{discretegroups}) that finite order discrete symmetries 
that act  subject to the conditions  G1) and G2) on $M_n$ have an induced action 
on the single cover description of  ${\cal M}_{cs}$. 

The first observation is that by turning on fluxes on the 
invariant periods $\ci W=|G|\sum_i n_i \Pi^{{\rm diag}\ i}_0 $  under $G$ 
the theory is driven to the sublocus ${\cal S}$ in the moduli space. 
Here we multiplied by the order $|G|$ of the group $G$ 
and pick  $n_i\in \mathbb{Z}$, which guarantees that the flux is integral.  
Let us denote as before the  invariant moduli $\ci a_i$ and the  non-invariant moduli $\tilde a_i$. 
Then (\ref{orbifoldbehaviour}) is the model for the local behavior of the 
periods near an orbifold point $\tilde a_i=0$ written in terms  
the local vanishing non invariant modulus $\tilde a$. Hence by (\ref{orbifoldbehaviour}) 
we can conclude that
\begin{equation}
\partial_{\tilde a_i}\ci W(\tilde a_i, \ci a_i)  =0 \ . 
\label{restrictioninv}  
\end{equation}
Because of the positivity of the individual terms in (\ref{susycondition}) e.q.  
(\ref{restrictioninv}) is a necessary condition to have 
a supersymmetric vacuum and drives the theory to the invariant
subspace  ${\cal S}$~\cite{Curio:2000sc}. Moreover at an orbifold point the K\"ahler 
metric is regular as $W^{inv}_{\Delta}=0$ is transversal. 
Now if the flux conditions $\partial_{\ci a_i} \ci W(\tilde a_i, \ci a_i)=
\ci W(\tilde a_i, \ci a_i)=0$ drive the flux to the supersymmetric 
locus $W(\tilde a_i, \ci a_i)=0$, then all conditions for the 
supersymmetric vacuum are met. 

If $k>2$ then we can also put the flux on $\tilde W=\sum_i c_i \Pi^{{\rm diag}\ i}_{k>1} $
and get 
\begin{equation}
\partial_{\tilde a_i}\tilde W(\tilde a_i, \ci a_i)_{\cal S}=\tilde W(\tilde a_i, \ci a_i)_{\cal S}= 0 \ . 
\label{restrictioninvb}  
\end{equation}
Again since at the  orbifold point the K\"ahler metric is regular this 
solves the supersymmetry condition {\sl already} on the entire 
subspace ${\cal S}$. Of course the question here is whether one can 
pick the $c_i$ so that $\tilde W=\sum_i c_i \Pi^{{\rm diag}\ i}_{k>1}$  
is an integral flux potential. If there is just one prime orbit in $G$ of order $p$  
without homological relations among the cycles $\Gamma_i$ this is impossible, because of the roots of $p$  
in the coefficients  $c_i$ as  explained in section~\ref{discretegroups}. In this case one can approximate 
the roots of $p$  to arbitrary precision by a rational number. The corresponding large denominators 
can be cancelled by choosing a large common factor in the integer  flux quanta.  This can lead to 
hierarchically small  breaking of supersymmetry.              

Let us describe simple situations in which the theory can be 
driven to an orbifold point 
through integral fluxes. Based on the results of~\cite{Font:1992uk,Klemm:1992tx} 
the orbifold point $a_0=0$ of the degree six hypersurface $\sum_{i=1}^4 x_i^6+ 2 x_5^3- 6 a_0 
\prod_{i=1}^5 x_i$=0 in $\widehat {\mathbb{P}(1,1,1,1,2)/(G^{\rm max}_{ph}
=\mathbb{Z}_6^2\times \mathbb{Z}_2)}$ was found to be 
an attractor point in~\cite{Moore:1998pn}\footnote{There are four families of 
hypersurfaces of degree $k=5,6,8,10$ in weighted projective space $\mathbb{P}^4(\underline{w})$~\cite{Font:1992uk,Klemm:1992tx} 
and many formulas can be written in terms of $k$ and $w_i$, $i=1,\ldots, 5$.}. The latter can be viewed as 
the mirror of the degree six hypersurface in $\mathbb{P}(1,1,1,1,2)$. 
The model has an additional $\mathbb{Z}_6$ symmetry $x_4\rightarrow x_4 \alpha$, 
$a_0\rightarrow a_0 \alpha^{-1}$ with $\alpha=\exp(2 \pi i/6)$. This fulfills G1) in 
section~\ref{discretegroups}, but not G2). Hence there will not be 
a resolvable Calabi-Yau orbifold $M_3/\mathbb{Z}_6$. The rotation of 
$a_0$, the coefficient of the inner point $\nu_0$ cff. (\ref{WDelta}), 
is a special case for two reasons~\footnote{For the cases discussed 
in~\cite{Fuchs:1989yv} it is also a point with an exactly solvable rational super-conformal 
field theory called the Gepner model.}.
Firstly even though  G2) is violated $\Omega$ is invariant because of (\ref{innerpoint}). Secondly the invariant 
subspace in the moduli space cannot be the moduli space of another Calabi-Yau 
manifold, as eliminating the inner point leaves the class of reflexive polyhedra. 
It follows in particular that $\Gamma_0$ must be trivial in homology so 
that there will be no invariant period. From these facts and the ones stated in section 
(\ref{discretegroups}) it follows that solutions near  $a_0=0$ take the 
form  $\Pi_p=\sum_{n=1}^\infty c_n (a_0 \alpha^p)^n$~\footnote{The $c_n=-i\frac{2^3 \pi^4}{k|G^{max}_{ph}|}
\frac{\gamma^n e^{i\frac{\pi}{k} (k-1)n}}{\sin\left(\frac{\pi n}{k}\right)\Gamma(n) 
\prod_{i=1}^5\Gamma\left(1-\frac{n w_i}{k}\right)}$ with 
$\gamma=k\prod_{i=1}^5 w_i^{- w_i/k}$ are calculated  
as the solutions to the PF-equation at this point. $|G^{max}_{ph}|=3^3 2^2$ for the $k=6$ case.}. One can pick 
a basis $\Pi_{orb}=(\varpi_2,\varpi_1,\varpi_0,\varpi_5)^T$ and using the  
Mellin-Barnes integral continuation one finds~\cite{Klemm:1992tx} 
\begin{equation} 
\Pi_{lcs}=\left(\begin{array}{rrrr} 
                  0&0&-1&0\\
               -\frac{1}{3}& -\frac{1}{3}& \frac{1}{3}&\frac{1}{3} \\ 
                 0&-1&1&0\\ 
                1&0&-3&-2\\  
                \end{array}\right) \left(\begin{array}{c} 
                 \varpi_2\\
                 \varpi_1\\ 
                 \varpi_0\\ 
                 \varpi_5\\  
                \end{array}\right)\ .  
\end{equation}
So we learn that $\tilde W=k(F_1-3 X^0)+n F_0+ m X^1\sim a_0 (3 k- m-n)+{\cal O}(a_0^2)$ with $(3 k- m-n)=0$ 
and $k,m,n\in \mathbb{Z}$ subject to this condition generates  a two parameter lattice of flux superpotentials which 
drives the theory to the orbifold point $\sigma=\frac{\varpi_1}{\varpi_0}\sim a_0  =0$ where one has $\tilde W=0$ 
and $\partial_{\sigma} W=0$  at $\sigma =0$ hence a minimum that is equivalent to an attractor point of  
an ${\cal N}=2$ black hole with electric and magnetic charges. Here we introduced the flat orbifold 
coordinate $\sigma$\footnote{See \cite{Bouchard:2008gu}  how to chose flat closed and open orbifold coordinates.} .   
 The reason behind that solution is of course just the arithmetic relation $\alpha^5=-\alpha^2=\alpha^0-\alpha^1$ of sixth 
order roots. For $k=5$ there is no arithmetic relation  between the fifth roots and for the $k=8$ and 
$k=10$ cases the arithmetic relations between the roots that would lead to integral 
fluxes correspond to cohomologically trivial fluxes. Note that in these 
special cases with no invariant period $e^{K_{cs}}\sim 1/|\sigma|^2$ has a 
singularity at $\sigma=0$, so that the contribution of the complex structure moduli to $V$ 
(\ref{Vcs}) is positive despite the fact that $\partial_{\sigma} \tilde W\sim \sigma$ 
and $K_{\sigma} \tilde W \sim \sigma$ vanish at $\sigma=0$~\cite{Curio:2000sc}. For the general orbifold  
one has a non-vanishing invariant period. This implies that $e^{K_{cs}}$ is 
regular at ${\cal S}=0$ and one gets  a supersymmetric vacuum manifold.                           

For the degree six  Calabi-Yau fourfold $X_6(1^6)$  in $\mathbb{P}^5$ we can use (\ref{continuationsextic}) to  
establish a flux potential $\tilde W(k,m,n)$ that behaves like    
\begin{equation}
\tilde W= 12 (m +  n) X^0 +2( 5 k + 3 m +  n) X^1 + (7 k + 5 m + 3 n) H +4 (6 m+5 n -2k) F_1 -12 k F_0\sim a_0^2+{\cal O}(a_0^3)\  ,
\end{equation}   
for $k,n,m\in \mathbb{Z}$ arbitrary.  Using the solutions(\ref{varpi})  one concludes 
that within this three parameter family of flux potentials there is a one parameter family
\begin{equation}
\tilde W(0,m,m)\sim a_0^3+{\cal O}(a_0^4)\  .
\end{equation}    
The explicit expression for the integer basis of periods and the K\"ahler potential 
shows  with $\sigma=\frac{\varpi_1}{\varpi_0}\sim a_0$ that 
$D_{\sigma} W\sim \sigma^2$, while the asymptotic behavior of $G_{a\bar a} $ is regular 
$G_{a_0,\bar a_0}=\frac{\Gamma^6\left(\frac{5}{6}\right) \Gamma^6\left(\frac{1}{3}\right)}
{ \Gamma^6\left(\frac{1}{6}\right) \Gamma^6\left(\frac{2}{3}\right)}$  and 
 $e^{K_{cs}}\sim 1/|\sigma|^2$. Therefore the scalar potential
\begin{equation} 
V=e^{K_{cs} } G^{\sigma \bar \sigma}|D_{\sigma} \tilde W|^2\sim|\sigma|^2\, ,     
\end{equation} 
vanishes at the orbifold point $\sigma=0$. It is quite significant that we can first reduce 
the dimension  of the moduli space of the sextic to a one dimensional  monodromy invariant 
subspace ${\cal S}$ by the $\mathbb{Z}_6^4$ maximal phase symmetry or eventually by finding
appropriate algebraic $(2,2)$  and fix the last modulus at the $\mathbb{Z}_6$ orbifold or Gepner point  
with {\sl vanishing scalar potential}.   With a different choice of $m,n,k$ one can add 
a positive constant to $V$.    Because of the general rationality of the transition matrix $\tilde m$ due to the grade restriction rule 
cff. (\ref{tildem}), this result  generalizes immediately to other Fermat hypersurface  in weighted projective spaces  
$X_d(w_1,\ldots,w_6)$,  i.e.  with $w_i|d$, because the latter condition together with 
the Calabi-Yau condition $\sum_{i} w_i=d$ implies that $d\in 2 \mathbb{Z}$,
which makes  the cancellation of non-trivial roots possible. Since  among the Fermat  
hypersurfaces one has hundreds of  elliptic fibrations the mechanism is relevant for F-theory, 
with the additional bonus that the discrete  orbifold symmetries have the potential to 
create selection rules that might  lead  to approximate hierarchies among  
the Yukawa couplings.             

Let us conclude the section with the general criterium to have an 
integral superpotential on the non-invariant cycles: 
{\sl If the finite group $G_{\rm fix}$ that fixes ${\cal S}$ has 
arithmetic relations among its roots that do not correspond to 
cohomological relations between the cycles in its orbit it is 
possible to find an integral flux superpotential $\tilde W$ 
that has ${\cal S}$ as its supersymmetric moduli manifold.}

\subsubsection{The nodal points and supersymmetric vacua}
\label{nodalpoints}  
Let us discuss now the behavior of the superpotential at the 
most generic local singularities 
\begin{equation}
\sum_{i=1}^{n+1} \zeta_i^2=\epsilon^2=\delta_c\, ,
\label{conifoldsingularity}  
\end{equation}
of a $n$ dimensional complex manifold, i.e. when a cycle $\nu$ with the 
topology of a sphere $S^n$, which is the real section of (\ref{conifoldsingularity}), 
shrinks to zero size. The singularity is called a node and the 
codim one sublocus $\delta_c=0$ in the complex moduli space is 
called the {\sl conifold divisor}. The integral over the cycle $\nu$ can be 
calculated iteratively in $\epsilon$ as was done for $n=3$ 
in~\cite{Candelas:1990rm,Font:1992uk,Klemm:1992tx}. To 
give a dimensional argument for its leading behavior in $\epsilon$ 
for various $n$, it is convenient to integrate over $\gamma$ in 
(\ref{period}). One can do this in a patch of the toric 
variety where the local form of $\Omega$ becomes 
$\Omega=a_0 \frac{{\rm d} x_1\wedge\ldots \wedge  {\rm d} x_n}{\partial_{x_{n+1}} W_{\Delta}(x,a)}$.
Moreover one always finds a coordinate transformation 
$x_i\rightarrow c_i+\sum_{k=1}^{n+1} c^k_i \zeta_k$ so 
that locally $W_{\Delta}(x,a)= -\epsilon^2 + \sum_{i=1}^{n+1} \zeta_i^2+{\cal O}(3)$ 
and  $\eta=\partial_{x_{n+1}} W_{\Delta}(x,a)=\sum_{i=1}^{n+1} d_i \zeta_i +{\cal O}(2)$, 
where $\zeta$, $\eta$ and $\epsilon$ have the dimension of a length which sets the 
order parameter. It follows from the dimension of 
$\Omega=a_0 {\rm Jac}\left(\frac{\partial x}{\partial \eta_i}\right) \frac{{\rm d} \eta_1\wedge\ldots \wedge  {\rm d} \eta_n}{\eta}$ 
that to leading order~\footnote{The proportionality constant is easily calculated by making 
a further orthogonal transformation on $\zeta_i\rightarrow \eta_i$ so that 
$\eta$ becomes one of the coordinate axes. Then one can introduce 
spherical coordinates and perform the integral. This can be turned 
into an iterative scheme, but it is easier to get the subleading 
terms from the PF equation.}                   
\begin{equation} 
\int_{S^{n}} \Omega\sim  \epsilon^{n-1}= \left(\delta_c\right)^\frac{{n-1}}{2}\ . 
\label{leading} 
\end{equation}
This behavior at the conifold divisor can be readily checked for the 
mirrors of degree $k$ hypersurfaces in $\mathbb{P}^{k-1}$, where the PF-differential 
operator is derived from the GGKZ method to be 
\begin{equation} 
 D^{k}=\theta^{k-1}+ (-1)^{k-1} k z 
\prod_{i=1}^{k-1}( k \theta+i),
\label{diffHP} 
\end{equation}
where $\theta=z \frac{\rm d}{{\rm d}z}$ and 
$z=\frac{1}{k^k a_0^k}$ and $\delta_c=(1+(-1)^{k-1} k^kz)$, 
by calculating  the  indicial equations for the solutions near $\Delta_{c}(a_0)=0$. 
Examples of shrinking $S^4$'s in more general fourfolds are discussed in section \ref{fourfoldbasis}.

One of the most famous predictions of ${\cal N}=2$ Seiberg-Witten theory
is the existence of massless BPS monopoles~\cite{Seiberg:1994rs}. The corresponding one 
loop $\beta$-function for the gauge coupling, as encoded in the derivative of $F_\Gamma$ w.r.t. 
$X_\nu$, generates a shift monodromy like in (\ref{shift}). Eq. (\ref{leading}) does 
not allow vanishing periods on elliptic curves $n=1$ for the canonical $(1,0)$ form. The 
conclusion is avoided in Seiberg-Witten by using a meromorphic one form 
$\lambda$, which descends from the $(3,0)$ form of a Calabi-Yau 3-fold 
in the local limit. In this case (\ref{leading},\ref{MBPS},\ref{shift}) 
allows massless BPS states as well as logarithmic shifts. One may see 
this as a dimensional argument that these gauge theory effects find 
their natural explanation in string theory.

According to the Lefschetz formula with some sign corrections by Lamotke~\cite{MR592569}, an n-cycle $\Gamma$ 
transforms along a path in the moduli space encircling the 
conifold divisor $\epsilon=0$, where the $S^n=:\nu$ vanishes, 
with positive orientation with the  monodromy action on $\Gamma$ 
that is either a {\sl symplectic} - for $n$- 
odd or an {\sl euclidean} reflection $w$ for $n$-even, i.e. 
\begin{equation}
 w(\Gamma)=\Gamma+ (-1)^{(n+2)(n+1)/2} \langle \Gamma, \nu\rangle \nu \ .
 \label{conifoldmonodromy}  
\end{equation}
The self intersection of the sphere itself is given by 
\begin{equation} 
 \langle \nu, \nu\rangle = 
 \left\{\begin{array}{ll} 0   &, \  n\ {\rm odd} \\ 
         (-1)^{n/2}\cdot 2   &, \  n\ {\rm even} \ .
        \end{array}
   \right. 
\label{Snintersection}
\end{equation}
Let us now discuss the two cases in turn: 
\begin{itemize} 
\item $n$ odd: Due to the non-degenerate symplectic pairing the vanishing cycle $\nu $ 
intersects a dual cycle $\Gamma$ and in order to realize (\ref{conifoldmonodromy}) the periods 
over these cycles degenerate in the local parametrization $\delta_c=0$ for $n$ odd like  
\begin{equation}
\begin{array}{rl}
X_\nu=&\ds{\int_\nu \Omega=\delta_c^\frac{n-1}{2}\sum_{k=1}^\infty c_k(\check a) \delta_c^n}\, ,\\ [2 mm]
F_\Gamma=&\ds{\int_{\Gamma} \Omega= \frac{(-1)^{(n+2)(n+1)/2} }{2 \pi i } X_\nu \log(X_\nu)+ \sum_{k=0}^\infty b_k(\check a)\delta_c^k}      .
\end{array}
\label{shift2}
\end{equation}
Usually one cannot determine the integral and cycles directly and rather 
solves the Picard-Fuchs equations near $\epsilon=0$, but the Lefschetz formula fixes 
the relative normalization of the solutions so that 
$T=\left(\begin{array}{cc} 1& 0 \\ 1 & 1\end{array}\right)$ is the unipotent  
monodromy when for a path that encircles $\delta_c=0$ counter clockwise.

It has been argued \cite{Polchinski:1995sm},\cite{Curio:2000sc} that 
putting a flux on the $S^3$ in $M_3$, i.e.  the non-invariant 
periods drives the theory towards the conifold\footnote{Note that in  
general at the conifold locus of compact examples one has 
several $S^n$ with topological relations vanishing, which in the transition 
are blown to $\mathbb{P}^1$ with relations. This fits perfectly  
the ${\cal N}=2$ Higgs mechanism with abelian gauge groups~\cite{Greene:1995hu}. 
More complicated situations with non-abelian gauge symmetry 
enhancements are discussed in~\cite{Klemm:1996kv},\cite{Berglund:1996uy}.}.

\item $n$ even: The $S^n$ intersects with itself and so the typical local behaviour 
for the $K3$, 4-folds etc is $w(\nu)=-\nu$, 
for a single $S^n$ and hence 
\begin{equation}
X_{\nu}= \delta_c^\frac{n-1}{2}\sum_{k=1}^\infty d_k(\check a) \delta_c^k \ ,       
\end{equation}
e.g. for 4-folds one finds $X_{\nu}= \delta_c^\frac{3}{2}$ etc. This leads to a $\mathbb{Z}_2$ monodromy. 

If we put a  $G_4$ flux on the $S^4$, which is at least a rational cohomology cycle, 
then  
\begin{equation}
W|_{\delta_c=0}=0, \qquad \partial_{\delta_c} W|_{\delta_c=0} =0 \ .    
\label{minimisation1b} 
\end{equation}
More accurately in the right local coordinate $X_\nu$ one has from the local 
limit of  the K\"ahler potential
\begin{equation} 
D_{X_\nu}W=0 \ . 
\end{equation}    
{\sl 
This means that we can always find a rational flux, which drives  the 
theory towards the most generic singularity of the  4-fold, where we 
find a supersymmetric vacuum.
}     
\end{itemize}

\subsubsection{Flux superpotential and  sub-monodromy problem}
\label{submonodromy}  

In the section \ref{gaussmaninS} we explained how to get the Picard-Fuchs or equivalently 
the Gauss-Manin connection for $W^{inv}_\Delta(a,Y)$ on the  subspace  ${\cal S}$ 
of the moduli space and argued that the latter defines  a sub-monodromy  problem 
on the invariant periods.  Here we want to investigate the properties of the flux 
superpotential that can drive the complex moduli to the invariant subspace ${\cal S}$.

\noindent
{\sl Local considerations:} 
Let us start with the local properties of the Gauss-Manin connection locally near ${\cal S}$.   
We  introduce the following notations: $a_i$, $i=1,\ldots, h_{n-1,1}(M_n)$  are gauged fixed complex  
structure variables. We split them into variables along the invariant subspace $\ci a_i$,$i=1,\ldots, {\rm dim}({\cal S})$ and 
the ones transversally to ${\cal S}$  $\tilde a_k$, $k=1,\ldots, {\rm codim}({\cal S})$  so that $\tilde a_k=0$ characterizes ${\cal S}$.
The Frobenius algebra property FAnd.) 
is equivalent to the local Torelli and so we can 
introduce locally  periods  $\overset{\circ}{\Pi}_i$ along and  $\tilde \Pi_k$ transversally to ${\cal S}$ 
and write $\Pi(a)=(\overset{\circ}{\Pi}(a) ,\tilde \Pi(a))$. Using the pairing between $\chi_{i_k,n}$ and $\Gamma_i$ 
we write the Gauss-Manin connection (\ref{picardfuchsfirstorderform})\footnote{In fact this connection can be easily numerically 
integrated along paths around $\Delta_i$ to obtain all monodromy matrices $M_i$ as the precision even for multimoduli 
problems is good enough to approximate the integers in $M_i$. We thank Duco von Straten for pointing this out to us.}   for 
the non-invariant variables $\tilde a_i$
\begin{equation} 
\left[\partial_{\tilde a_i}  +\left(\begin{array}{rl} 
\Gamma_{\tilde a_i, \ci a}^{\ci a} &   \Gamma_{\tilde a_i, \tilde  a}^{\ci a}   \\
\Gamma_{\tilde a_i, \ci a}^{\tilde a} &   \Gamma_{\tilde a_i, \tilde  a}^{\tilde a} \end{array}\right)\right]  \left( \begin{array}{c} \ci \Pi\\ \tilde \Pi\end {array}\right)=0 \, ,
\label{connection1} 
\end{equation}  
and the one for the invariant variables
\begin{equation} 
\left[\partial_{\ci a_i}  +\left(\begin{array}{rl} 
\Gamma_{\ci a_i, \ci a}^{\ci a} &   \Gamma_{\ci a_i, \tilde  a}^{\ci a}   \\
\Gamma_{\ci a_i, \ci a}^{\tilde a} &   \Gamma_{\ci a_i, \tilde  a}^{\tilde a} \end{array}\right)\right]  \left( \begin{array}{c} \ci \Pi\\ \tilde \Pi\end {array}\right)=0 \, ,
\label{connection2} 
\end{equation}
For $\Gamma_{\cal S}$ to be a closed sub-monodromy problem the terms 
$\Gamma_{\tilde a_i, \ci a}^{\ci a}=\Gamma_{\ci a_i, \tilde  a}^{\ci a}=\Gamma_{\ci a_i, \ci a}^{\tilde a}=0$ 
have to vanish at ${\cal S}$. 
\begin{itemize} 
\item Let us now consider a flux on the non-invariant periods $\tilde \Pi$. From equation (\ref{connection1})  
and the fact  that $\Gamma_{\tilde a_i, \ci a}^{\tilde a}|_{{\cal S}}=0$ one sees that 
$\partial_{\tilde a_i} \tilde \Pi$ has to vanish, when $\Gamma_{\tilde a_i, \tilde  a}^{\tilde a}$ 
does not have a pole in $\tilde a_i$ of order one or higher. This is not the case at orbifold 
divisors, irrespectively of the fact whether the latter is abelian or non-abelian.
Similar $\partial_{\ci a_i} \tilde \Pi=0$, because $\Gamma_{\ci a_i, \ci a}^{\tilde a}|_{{\cal S}}=0$ 
and  $\ci \Pi$ stays generically regular as $\ci M_n$ is transversal. Differently as in the somewhat 
particular  orbifolds discussed in (\ref{orbifoldpoints})  $e^{K_{cs}}$  and $\partial_{a_i} K_{cs}$ have no pole 
generically at ${\cal S}$, again because $\ci M_n$ is transversal for generic 
values of $\ci a_i$. Therefore if the superpotential is entirely put on the $\tilde \Pi$ periods  
$D_i W|_{\cal S}=0$ and also $V|_{\cal S}=0$. In the case of the conifold singularities 
we refer to~\cite{Polchinski:1995sm},\cite{Curio:2000sc} and the result in section  
\ref{nodalpoints} to reach the same conclusion, namely that putting the fluxes 
on the non-invariant periods drives the theory to the supersymmetric locus 
${\cal S}$. One can also argue that the  auxiliary fields (\ref{susycondition}) 
are  equivariant Morse functions under the actions defined  
in section (\ref{discretegroups}) and (\ref{transitions}) and 
therefore have to vanish at the fix point set ${\cal S}$  
of these actions.             
\item Next we consider a flux on the invariant periods $\ci \Pi$. Again we conclude 
that  $\partial_{\tilde a_i}\ci \Pi|_{\cal S}=0$ and since $e^K$ is bilinear in 
the periods also $\partial_{{\tilde a}_i} K|_{\cal S}=0$ hence $D_{\tilde a_i} W|_{\cal S}=0$. 
Of course this means only that we found a necessary condition to have a vacuum.  
In addition one  has to have $D_{\ci a_i} W=0$.  Again one makes a simple 
localization argument~\cite{Giryavets:2003vd}, provided that one has established 
a group action on the periods like in (\ref{actiononperiods}) and the invariance of $K$ under it. 
Then one can differentiate $\ci \Pi(\tilde g\cdot a) = \ci \Pi(a)$ and $K(\tilde g\cdot a)=K(a)$ 
with respect to $\tilde a_i$ to conclude from $\tilde g  \partial_{\tilde a_i} \Pi(\tilde g \cdot a) 
= \partial_{\tilde a_i} \Pi(a)$ and  $\tilde g  \partial_{\tilde a_i} K(\tilde g \cdot a) 
= \partial_{\tilde a_i} K(a)$  that $\partial_{\tilde a_i} \Pi(a)$ and $\partial_{\tilde a_i} K(a)$ 
has to vanish at the fix set $\tilde g\cdot  a=a$ and so again that $D_{\tilde a_i} W|_{\cal S}=0$. 
Such arguments are used also in~\cite{Denef:2004dm}\cite{Louis:2012nb}.    
\end{itemize} 

\noindent
{\sl Global considerations:} 
While $\Gamma_{\cal S}$ is a subgroup of $\Gamma_{\cal M}$ it should be made 
clear that $\ci \Pi$ is not an invariant subspace under the action of
$\Gamma_{\cal M}$.  An illustrative example for this was given in~\cite{ Kachru:1995fv}, where 
the periods of the pure $N=2$ SU(2) Seiberg-Witten appear in a decompactification limit of 
a global Calabi-Yau 3-fold  defined as  hypersurface $X_{12}(1,1,2,2,6)$ in the weighted 
projective space $\mathbb{P}(1,1,2,2,6)$ . In this case the invariant periods transform under the 
$\Gamma_{\cal S}=\Gamma_0(4)\in {\rm SL}(2,\mathbb{Z})$ sub-monodromy group in the Seiberg-Witten
 limit within the compact Calabi-Yau 3-fold. Eq. (5.4) shows that invariant and  
non-invariant  periods mix under the full original monodromy group $\Gamma_{{\cal M}_{cs}}$  
with rational coefficients. Such a rational mixture should be generic, when a sub-monodromy 
problem on the invariant periods can be defined  in the  sublocus ${\cal S}$.  It is therefore 
natural to conjecture that the flux quanta on the non-invariant periods, that drive the 
theory towards ${\cal S}$ by a flux superpotential  $\tilde {\cal W}$ can always be 
chosen to be integral.                       

Based  on the examples and the symmetry considerations we close the section with  
a conjecture concerning the sublocus ${\cal S}$ and the sub-monodromy system 
described in section \ref{gaussmaninS}. Whenever there is a sublocus ${\cal S }$ 
so that ${\cal I}_{PF}|_{\cal S}$ describes a closed  sub-monodromy system and 
${\cal S}$ is not at the boundary of ${\cal M}_{cs}$ where $M_n$ degenerates, 
then ${\cal S}$ is at the fix point of a group action on ${\cal M}_{cs}$ induced 
from discrete group action on  $M_n$. Depending on the arithmetic there might be an 
integral flux that drives the moduli to ${\cal S}$. All other integral 
fluxes drive the moduli to a subloci  ${\cal S}$ at the boundary of 
${\cal M}_{cs}$.

\section{Compact Calabi-Yau manifolds with no gauge group}  
\label{elliptictdata}

We will construct families of toric hypersurface with 
elliptic fibrations of type $D_5$, $E_6$, $E_7$ and $E_8$, which have 
no non-abelian gauge symmetry enhancements. 
The significance of these families is that for a given base and 
fibre choice they have the most general complex structure. 
In this sense they can be seen as clean sheets on which by suitable 
specializations of the complex structure induced by $G_4$ fluxes, 
singularities along divisors yield non-abelian gauge symmetry enhancement 
and by further enhancement of the singularity in higher codimension the 
matter sectors and Yukawa couplings might be imprinted. It is a general 
strategy to analyze these singularities by blowing them up to a smooth variety, 
which is a valid approach in our case as well that we pursue in section~\ref{SU5andE8yukawa}. 
However in F-theory one has to keep in mind that these blow ups 
which involve finite volumes exceptional divisors in the fibre 
are unphysical. The actual breaking of the gauge symmetry 
is by deforming the singularities i.e. by Higgsing, by flux 
breaking or by Wilson lines.

As explained in section~\ref{compactification} the anticanonical 
divisor has to be semipositive on any effective curve. The 
simplest possible choice for the basis are 84 Fano basis $B_3$~\cite{MoriMukai}. 
A much more wider class are the almost toric basis $\mathbb{P}_{\Delta}$ 
corresponding to the 4319  3d reflexive polyhedra that are classified 
to construct K3 as hypersurfaces~\cite{Kreuzerliste}\footnote{The 18 toric almost Fano $B_2$ were classified, e.g. by Batyrev.}. 

The advantage of the construction is that there exists always 
triangulations of $\Delta_5$ compatible with the fibration 
structure and that all classes of the base are torically 
realized as classes of the fourfold $M_4$.   

We first will classify the  Fano or almost Fano 
smooth projective  toric varieties, which allow for a  
$\mathbb{P}^1$ fibration over $B_2$ and have therefore a 
dual heterotic description. Ultimately they will be all contained  in the list 
of $4319$  $\mathbb{P}_\Delta$  mentioned above. 
It has nevertheless advantages to start with $B_2$. 
The reason is that, while  F1.) is trivial to establish, 
to check for the compatible triangulations required in F2.) 
for a given $B_3$ takes time.\footnote{This applies similarly to the 
list of~\cite{Morrison:2012js}, which lists all 
possible toric basis which allow for a compact 
Weierstrass model. It should be implicit in the 
complete list of $4d$ reflexive polyhedra but 
checking for F2.)  given F1.) is difficult.}                      

There is no complete list of reflexive polyhedra in 
5d and we are not trying to classify all possible toric 
3d basis as in~\cite{Morrison:2012js}, which lead to 
a Weierstrass with smoothable Kodaira fibers, only the 
ones which are already smooth. The reason why this is 
almost as useful is that giving blow ups in the base 
or any number of divisors on which to impose a gauge 
group leads to a sub-point configuration of points in
$\Delta$ or $\Delta^*$ and there are programs like Palp, 
which can check whether these points can be completed 
to a reflexive polyhedron.

\subsection{Rational fibred basis over almost Fano varieties}  
\label{heteroticbasis} 

For comparison with the heterotic string we construct 
bases $B_{n-1}$ that exhibit a $\mathbb{P}^1$ fibration 
$\pi_r:=\mathbb{P}^1\rightarrow \bar B_{n-2}$, which 
by the adiabatic extension of the heterotic type II duality is 
identified with the base of the compactification manifold 
$Z_{n-1}$ $\pi_h:{\cal E}^{het}\rightarrow \bar B_{n-2}$. 

The gauge and gravitational anomaly condition is  the  
Donaldson-Uhlenbeck-Yau  condition for stable heterotic bundles 
\be 
\lambda(V)= \lambda(V_1)+ \lambda(V_2)+[W]= c_2(T Z_{n-1})\ , 
\ee 
where the bundle $V_1$ is the one of the first $E_8$ 
and $V_2$ of the second $E_8$ and $\lambda(V)$ is in the 
fourth real cohomology, e.g. the second Chern class $\lambda(V)=c_2(V)$ 
for $SU(n)$, but for $Spin(n)$ the first Pontryagin class 
$\lambda(V)=p_1(V)$. Further to admit spinors one needs
\be 
c_1(V_{1,2})= 0\ {\rm mod} \ 2\ . 
\ee 
For the case $n=3$ this yields the 6d F-theory 
compactification dual to the heterotic string compactified 
on an elliptically fibred K3 and the only option for the  
$\mathbb{P}^1$ fibration over $B_{1}$ are the Hirzebruch 
surfaces $\mathbb{F}_m$. Since $\int_{K3} c_2(TK3)=24$, 
one denotes 
\be 
t=12-\pi_{h*}(\lambda(V_1))\, ,
\label{defn} 
\ee 
and considers for symmetry reasons only $0 \le t  \le 12$. 

\begin{figure}[htdp] 
\begin{center} 
\includegraphics[width=.76\textwidth]{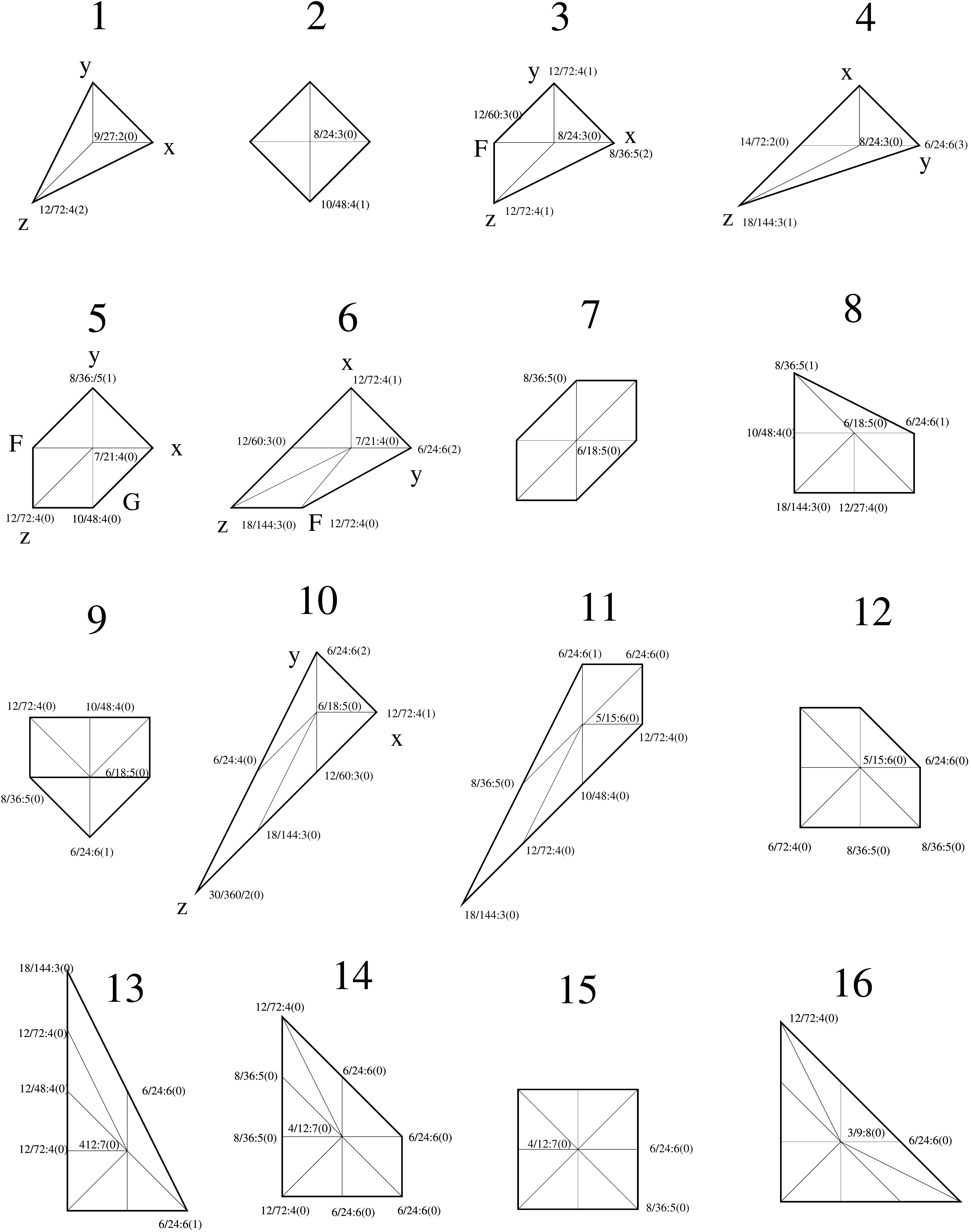} 
\begin{quote} 
\caption{\label{fig1}The 16 reflexive polyhedra $\Delta_2$ in two dimensions. 
When we refer to the points we denote the origin as $\nu_0$ the  
one to the right (for 13 right down) of the origin as $\nu_1$ and  
numerate the rest counter-clockwise. The polyhedra  are relevant for two 
considerations in this paper: Firstly they are the toric basis for the 
F-theory basis $\Delta_3^{B*}:\mathbb{P}^1 \rightarrow ({\overline B_2}=\Delta_2)$ 
for those clean sheet models that  have heterotic duals in the classification 
in table \ref{constantlist}.  Secondly they classify toric elliptic fibres $\Delta^{*F}_2=\Delta_2$ in 
(\ref{polyhedrafrombaseandfibre}). In this  application one can read  from the  labels on 
the inequivalent points  the Euler number and the number of K\"ahler classes 
according to formula (\ref{explanationhodge}) for 3-folds and  4-folds. 
\vspace{-1 cm}} 
\label{poly} 
\end{quote} 
\end{center} 
\end{figure}

It has been observed in~\cite{Morrison:1996pp} and proven 
in~\cite{Friedman:1997yq} that  one can identify $t$ with the 
twisting of the fibre $\mathbb{P}^1$ in $\mathbb{F}_t$. 
Moreover only for $t=0,1,2$ the bundles $V_1$ and $V_2$ can be generic 
enough in both $E_8's$ to break the gauge group completely. 
The goal of the subsection is to find the analogous geometric  
condition for hypersurfaces in toric ambient space of F-theories 
with heterotic duals in 4d. The data of $\pi_r$ are a projectivized 
bundle ${\cal O}(1)\otimes {\cal T}$, where ${\cal T}$ is a line bundle over $\bar B_{n-2}$.  
Let us denote $\alpha= c_1({\cal O}(1))$ on $B_{n-2}$ and $t=c_1( {\cal T})$ and specialize to 
$n=4$, i.e. compactifications to 4d.  
Then by adjunction $c(B_3)=(1+c_1(\bar B_2)+c_2(\bar B_2))(1+\alpha)(1+ \alpha + t)$ 
and using $\alpha(\alpha +t)=0$ one can calculate by integrating over 
the $\mathbb{P}^1$ fibre the integral
\be 
\int_{B_3} c_1^3 = 6 \int_{\bar B_2} c_1^2 + \int_{\bar B_2} t^2\, ,
\label{intb3} 
\ee 
in terms of integrals over the base $\bar B_2$.  In the most general toric 
construction $\bar B_2$ is defined as one of the sixteen toric almost Fano 
varieties $B_2=\mathbb{P}_{\Delta^{(i)}}$, $i=1,\ldots, 16$ classified 
by Batyrev and displayed in figure \ref{fig1}. We like to know which line bundle 
${\cal T}$ leads to F-theory without gauge groups.\footnote{By 
Hirzebruch-Riemann Roch index theorem all rational 3 folds have $\int_{B_3} c_1 c_2=24$, 
hence (\ref{intb3}) is the only variable contribution to the Euler number (\ref{euler4}).}

The simplest example $\bar B_2=\mathbb{P}^2$~\cite{Klemm:1996ts} is given torically 
as $\mathbb{P}_{\Delta^{*B}_{\mathbb{F}_m}}=\mathbb{F}_m$ defined by 
the complex hull of the points $\{
u_1=(1,0,0),
u_2=(0,1,0)$,
$u_3=(-1,-1,-m),$
$w_1=(0,0,1),
w_2=(0,0,-1)
\}$. 
It has $\chi(\mathbb{F}_m)=6$. The first Chern class  of the bundle ${\cal T}$ 
specified by $\Delta^{*B}$ integrates to $\int_{B_2} t^2 = m^2$.       
Here $m$  is an example of  a twisting datum and without restriction we can chose 
$m\ge 0$. Let us denote the divisors associated to the points $\nu_i$ by $D_{\nu_i}$. Between 
toric divisors one has the relations $\sum_i \nu_{i,k} D_{\nu_i}=0$ 
for $k=1,2,3$, e.g. it follows $D_{u_1}=D_{u_2}=D_{u_3}=F$. The canonical 
class is $K=\sum_{i}  D_{\nu_i}=(3+m)F+2S$, where $S=D_{w_2}$, a section of the fiber $\mathbb{P}^1$, 
and $F$ are a basis of the Picard group. The Mori vectors representing dual curves 
can be found at the circuits of the unique star triangulation of $\Delta^{*B}$~\cite{CoxKatz} and are in this case $(1,1,1,0,-m;m-3)$ and $(0,0,0,1,1;-2)$. 
In particular there  is a curve $h=S F$, the hyperplane in the base $\mathbb{P}^2$,  
and the evaluation of $K$ on $h$ is $K h =3-m$. Hence ampleness of $K$ yields the 
bound $m<3$ and semi positivity  $m\le 3$ and precisely for 
these values $\Delta^{*B}$ is reflexive. Evaluating the contribution 
to $c_2(M_4)$ from the base in table \ref{tab:elliptic}, we find e.g. for the $E_8$ 
fibre 
\begin{equation}
11 c_1^2 + c_2= (11 m^2+69 m+102) F^2 + 138 F S \ . 
\label{P1overP2c2} 
\end{equation}
We can conclude that $c_2(M_4)$ is even in terms of an integral basis of 
algebraic four-cycles  for all $m$. This is true as long as the coefficient of $c_1^2$ 
in~$c_2(M_4)$ is odd, otherwise $m$ has to be odd.

For the general case let $\nu^{(i)}_j$, $j=1,\ldots |\Delta_2^{(i)}\setminus \{\nu_0\}|=r$ 
be the points of ${\Delta_2^{(i)}}$ defining the one dimensional cones, then a $\mathbb{P}^1$ 
fibration over $\mathbb{P}_{\Delta_2^{(i)}}$ is defined by the polyhedron 
$\Delta^{(i,m)}_3 $, which is the complex hull of the points
\be 
  \footnotesize 
  \begin{array}{|lr|}  
     u_1=&(\nu^{(i)}_1,-m_1) \\
     \vdots &  \vdots  \\
      u_r=&(\nu^{(i)}_{r}, -m_{r}) \\
      w_1=&(0,0 ,\ \ \ \ \ 1)\\
      w_2=&(0,0 ,\ \  -1)
  \end{array} \ , 
\label{polyhedrabase}
\ee
where $m_i\in\mathbb{Z}$ are the twisting  data, which have to be
small enough for $\mathbb{P}_{\Delta^{(i,m)}_3}$ to be (almost) Fano. 
Of course in view of the relations between the divisors in the 
base only $r-2$ twisting data are relevant. 

Let us discus the cases with two divisors in the base as examples. For 
$\bar B_2=\mathbb{F}_0=\mathbb{P}^1\times \mathbb{P}^1$ the situation is a
straightforward generalization of the above. Since $D_{\nu_1}=D_{\nu_3}=F_1$ 
and $D_{\nu_2}=D_{\nu_4}=F_2$ one needs only say $m_3$ and $m_4$ and 
the Picard group is generated by $F_1,F_2$ and $S=D_{w_2}$. The canonical 
class is $K=(2+m_3)F_1+(2+m_4)F_2+ 2S$ and the  Mori vectors are 
$(1,0,1,0,0,-m_3;m_3-2)$, $(0,1,0,1,0,-m_4;m_4-2)$ and $(0,0,0,0,1,1;-2)$. 
The strongest restriction on positivity of $K$ comes for the curves  
$h_1=S F_1$ and  $h_2=S F_2$ namely $m_3\le 2$ and $m_4\le 2$, which 
corresponds to entry 2 of table \ref{constantlist}. Again we calculate the contribution to  
$c_2(M_4)$ from the base in table \ref{tab:elliptic}  for the $E_8$ fibre type 
\begin{equation}
11 c_1^2 + c_2=2 (11 m_3 m_4+23(m_3+m_4)+46) F_1 F_2 + 92 S (F_1+F_2)\, ,
\label{P1overP1P1c2} 
\end{equation}
and get an even expression. 

The  Fano basis $\mathbb{F}_1$  is somewhat more interesting. We can pick the fibre of 
$\mathbb{F}_1$, $D_{\nu_2}=D_{\nu_4}=F$ and a section $T=D_{\nu_3}$ of the fibre
of $\mathbb{F}_1$  and the section $S=D_{w_2}$ and get from 
$h_1=S T$ and $h_2=S F$ constraints $m_3\le 2$ and   $m_4\le 3$. 
However the case $m_3=2$ and $m_4=0$ is not admissible as this 
polyhedron does not admit a K\"ahler star triangulation fulfilling F2.) 
of section (\ref{fibretwistingdata}). The only one that 
exist  is $\{\{3, 4, 6\}, \{1, 2, 6\}, \{2, 3, 6\},$ $\{1, 4, 6\}, \{1, 4, 5\},\{1, 2, 5\}, 
\{2, 3, 4\}, \{2, 4, 5\}\}$ and the  offensive
tetrahedron is\footnote{In a star triangulation 
all tetrahedra contain the origin which is therefore not indicated.}  $\{2, 3, 4\}$ as 
it involves three points in the base at different heights. In fact it has volume 
two and the associated toric variety is therefore singular. It is interesting 
to note that the anti canonical hypersurface $M_4$ in 
$\mathbb{P}_{\Delta_5^*}$ with $\Delta_5^*$ build in  (\ref{polyhedrafrombaseandfibre})  
avoids that singularity and has the Hodge data~\footnote{We give the complete Hodge 
data in the form $^{h_{31}(t_{31})}_{\ \ \ \ \ h_{21}}M^{\chi}_{h_{11}(t_{11})}$, where $t_{11}$ 
means twisted contributions from points in dual one and three faces, which indicate 
non rational divisors Poincar\'e dual to the 
corresponding $(1,1)$ forms and vice versa for $h_{11}(W_4)=h_{31}(M_4)$.}    
$^{2464(2)}_{\ \ \ \ 0} M^{14856}_{4(0)}$. I.e. from $h_{11}(M_4)=4=h_{11}(B_3)+1$ we learn that the 
fibration has no further singularities. However from $(14856-12 \times 24)/(12\times 60)= \frac{607}{15}$  
we see by comparison  with (\ref{euler4}), which holds if $M_4$  has only $I_1$ fibres  
that $M_4$ must have other non-singular fibre types. For the almost Fano basis 
$\mathbb{F}_2$ we get form the semi positivity of $K$, $m_3\le 2$,\ $m_4 \le  2 m_3$.

We find the following necessary conditions for the smoothness of 
$M_4$ 
\begin{equation} 
\int_{\bar B_2} t^2 \le \int_{\bar B_2} c_1^2 \ .   
\label{twistingconditionbase} 
\end{equation}
However non-singularity and the existence of a triangulation 
respecting the fibre structure further restrict the values to the ones 
listed in table~\ref{constantlist}.                 
\begin{table}[hbt]
\begin{center}
\begin{tabular} {|c |c |c| }
\hline
  $B_2$ &  range of twist data  & $\int_{B_2} \tau^2$ \\ \hline
 1  & $m_1\le 3$ &  $m_1^2 $\\ \hline
 2  & $m_3\le 2$,\ $m_4\le 2$  &  $2 m_3 m_4$ \\ \hline
 3  & $m_3\le 2$,\ $m_3 \le  m_4+1$,\ $m_4\le m_3+1$  &  $2 m_3 m_4-m_3^2$ \\ \hline
 4  & $m_3\le 2$,\ $m_4 = 2 m_3 $ &  $2 m_3 m_4-2 m_3^2$ \\ \hline
 5  & \footnotesize{$\{m_3,m_4,m_5\}={\{\{0, 0, 1\}, \{0, 1, 2\}, \{1, 1, 1\},\atop  \{1, 2, 1\}, \{1, 2, 2\}, \{2,3, 2\}\}}$} &  $2 m_4 m_3-m_3^2-(m_4-m_5)^2$ \\ \hline
 6  & \footnotesize{$\{m_2,m_3,m_4\}={\{\{1, 0, 0\}, \{1, 2, 1\}, \{2, 2, 1\}, \atop  \{3, 4, 2\} \}}$} &  $2 m_3 m_2-m_2^2-m_3^2-2 m_4^2+2 m_3 m_4$ \\ \hline
 7  & \footnotesize{$\{m_1,m_2,m_3,m_4\}={\{ \{0, 0, 0, 1\}, \{0, 1, 1, 1\}, \{0, 1, 2, 1\}\atop \{1, 0, 0, 1\}, \{1, 2, 2, 1\} \}}$} &  $2 m_2 m_1-m_1^2-m_2^2-m_3^2-m_4^2+2 m_2 m_3+2 m_3 m_4 $ \\ \hline
 8  & \footnotesize{$\{m_3,m_4,m_5,m_6\}={\{\, \{0, 0, 1, 2\}, \{1, 2, 2, 2\}, \{2, 4, 3, 2\}\}}$} &  $2 m_4 m_3-2 m_3^2-m_4^2-2 m_5^2-m_6^2+2 m_4 m_5+2 m_5 m_6$ \\ \hline
 9  & \footnotesize{${\{m_1,m_3,\tilde m_4=\atop m_4+\frac{m_5}{2},\tilde m_6=m_6+\frac{m_5}{2}\}}={\{0,1,0,0\},
 \{0,1,1,1\},\{0,2,2,0\} \atop \{0,2,3,1\},\{1,1,0,0\} \}}$} & $2 \tilde m_6 m_1-m_1^2-m_3^2-(\tilde m_4^2+{\tilde m_6}^2)/2+2 m_3 \tilde m_4+\tilde m_4 \tilde m_6$ \\ \hline
 10  & \footnotesize{$\{m_4,m_5,m_6\}=\{\,\{1,2,3\}\}$} &  $ 2(m_5 m_4-m_4^2-m_5^2+m_5 m_6) $ \\ \hline
 11  & \footnotesize{$\{m_4,m_5,m_6,m_7\}=\{\{0, 0, 0, 1\}, \{1, 2, 3, 2\}\}$} &  $ 2 m_5 m_4-2 m_4^2-2 m_5^2-m_6^2-m_7^2+2 m_5 m_6+2 m_6 m_7 $ \\ \hline
 12  & \footnotesize{${\{m_2,\tilde m_3=\sum_{i=3}^7 m_i,\atop \tilde m_4=m_3+m_7+2(m_4+m_6)+3 m_5\}}={\{\{1,0,0\},\{0,1,0\},\atop \{0,0,1\}\}}$} &  $ \tilde m_3- m_2+5 \tilde m_4    $ \\ \hline
 13  & \footnotesize{ only\ trivial\ twist } &  $ 0$ \\ \hline
 14  & \footnotesize{$\{m_4,m_5,m_6,m_7,m_8\}=\{1,2, 3, 2, 1\}$} & $0,4$ \\ \hline
 15  & \footnotesize{$\{m_3,m_4,m_5,m_6,m_7\}=\{ 1, 2, 2, 2, 1, 0\}$}&  $0,4$ \\ \hline
 16  & \footnotesize{$\{m_4,m_5,m_6,m_7,m_8\}= \{1, 2, 3, 2, 1, 0\}$} &  $0,3$ \\ \hline
\end{tabular}
\caption{ The $B_3=\mathbb{P}\rightarrow {\overline B}_2$ almost Fano toric basis with all possible 63 twisting data.}
\label{constantlist}
\end{center}
\end{table}

Notice that in this construction the polyhedra of the base $\bar{B}_2$ 
represent all possible del Pezzo surfaces albeit the members of the 
series $d_n \mathbb{P}^2$,  i.e.  $0\le n\le 8$ blow ups of $\mathbb{P}^2$, 
are for $n>3$ realized at special values of their moduli. 
In particular the reflexive polyhedra $15,16,13,10$ in figure \ref{poly} 
are specializations of the $D_5,E_6,E_7,E_8$ del Pezzo's.  
The smooth elliptic fibred CY 4-fold  will have $h_{11}(M_4)=h_{11}(B_2)+2+s$, 
where $s$ is the number of sections of the elliptic fibre.

One might ask whether one can also realize the half K3 or $d_9 \mathbb{P}^2$  as 
heterotic basis. Then $c_1^3=0$ and one expects  from (\ref{euler4}) that 
$\chi=288$  independent of the fibre type. This is not describable as a 
toric base, e.g. because of formula (\ref{c1h3}), but it can be realized as an 
$r=2$ and $s=0$  complete intersection in (\ref{polyhedrafrombaseandfibre}). 
An interesting case was constructed  in~\cite{Donagi:1996yf} as a complete intersection 
of two polynomials in $\mathbb{P}^2\times \mathbb{P}^1  \times \mathbb{P}^1 \times 
\mathbb{P}^2$ of degree $(3,1,0,0)$  and $(0,1,2,3)$ respectively.  
This corresponds to the $E_6$ fibre type. In the 
language  of section  (\ref{batyrev}) we get $\Delta^*$ as the convex 
hull of the points which describe the $\mathbb{P}^k$ factors, each 
given by a simplex with vertices, $e^{(l)}_1,\ldots, e^{(l)}_{k},
-e^{(l)}_1-\ldots- e^{(l)}_{k}$. We consecutively  label the vertices  
with the indices  $1,\ldots, 10$. The nef-partition is given by 
$D_{0,1}= D_1+D_2+D_3+D_4$ and $D_{0,2}= D_5+D_6+ D_7+D_8 + D_9+ D_{10}$ 
and the Euler number is indeed $\chi=288$, while $h_{11}=12$, $h_{3,1}=56$ 
and $h_{21}=28$. A remarkable property in these models is that 
they  allow an infinite number of rigid divisors, which all  contribute 
to a non-perturbative  superpotential~\cite{Donagi:1996yf}. The existence of the divisors can be seen 
by lifting rational curves in the del  Pezzo to divisors. Since the latter  are counted 
by  $E_8$ theta functions, see section \ref{E8BPSstates}, one gets a superpotential with 
modular properties  and non-trivial minima.  Similarly we can consider other fibre types e.g. the 
$E_8$ fibre as complete intersection $(6,1,0,0)$ and  $(0,1,2,3)$ in the ambient space  
$\mathbb{P}^2(1,2,3)\times \mathbb{P}^1  \times \mathbb{P}^1 \times \mathbb{P}^2$  
and find also $\chi=288$ with  $h_{11}=15$, $h_{3,1}=44$ and $h_{21}=19$.

The embedding map $i:B_{n-1} \hookrightarrow M_n$ has an 
induced  map on the second cohomology 
\be 
i^*:H_2(M_n) \rightarrow H_2(B_{n-1})\ , 
\label{induced}
\ee
which can have a non-trivial co-kernel. As a consequence not all K\"ahler moduli 
of a del Pezzo surfaces embedded in $M_n$ are also K\"ahler moduli of $M_n$. 
In the construction we presented in (\ref{constantlist}) there are as many 
classes in the del Pezzo torically realized  as possible and these classes are 
also classes in $M_n$.

\subsection{Toric Fano and almost Fano basis $B_3$}
\label{toricfanobases}
Almost toric Fano basis are  classified by the classification  
3d reflexive polyhedra $\Delta_3$, which were mainly constructed as 
ambient spaces to study K3, which can be realized as toric hypersurfaces. 
There are $4319$ such polyhedra, which can be constructed by the software 
PALP. From them there are $100$ polyhedra $\Delta_3^*$, with no points in 
codimension $1$ and $2$ faces, which implies that $\mathbb{P}_{\Delta_3*}$ are 
strictly Fano. In view of the classification of~\cite{MoriMukai}, which contains 
only $84$ Fano varieties, this list must be redundant and in fact there are only 
$36$ cases~\footnote{With respect to the numbering in~\cite{Kreuzerliste} these are 
1, 127, 129, 235, 380, 418, 484, 490, 492, 779, 975, 1141, 1152, 1339, 1343, 
1349, 1878, 2290, 2292, 2304, 2555, 2573, 3315, 3321, 3327, 3346, 3394, 3998, 
4024, 4032, 4036, 4102, 4270, 4309, 4310, 4318.}, which lead to 
topological different elliptic fourfolds $M_4$.

The latter are obtained  by the fibre construction as explained above for each 
fibre type $E_8,E_7,$ $E_6,D_5$ with $s=1,2,3,4$ sections. If we dot not consider 
the twisting data of section~\ref{fibretwistingdata}, we get hence for each 
fibre type $4319$ Calabi-Yau fourfolds with only $I_1$ fibres and $804$ different 
Hodge diamonds. The Euler numbers $\chi(M_4)$ and $h_{11}(M_4)=
h_{11}(\mathbb{P}^3_{\Delta_3^*})+s= (|\Delta_3^*|-4)+s$ are 
correlated by a neat mirror symmetry formula for projective varieties defined by reflexive polyhedra
\be
\int_{\mathbb{P}^n_{\Delta^*_n}} c_1^n = (n-1) (|\Delta_n|-n+1)\ .
\label{c1h3} 
\ee
This formula can be proved by calculating the global section of the 
canonical bundle of $\mathbb{P}^n_{\Delta^*_n}$ using the 
Hirzebruch-Riemann-Roch theorem to be $|\Delta_n|$ which implies (\ref{c1h3})
\footnote{We are grateful for a correspondence of 
Victor Batyrev regarding this matter.}.  Using this with $n=3$ and (\ref{euler4}) fixes $\chi(M_4)$. 
Since the list of  $4319$ 3-d  reflexive polyhedra 
is mirror symmetric (with 2120 pairs of 
reflexive polyhedra and 79 self-dual polyhedra)  we only need to list the  
occurring $h_{11}(\mathbb{P}^3_{\Delta_3^*})$ with their multiplicity  $\{(h_{11})^{mult}\}= \{1^1,2^7,3^{23}, 
4^{54}, 5^{135}, 6^{207}, 7^{314}, 8^{373}, 9^{416}, 10^{413}, 11^{413},$ $12^{348}, 
13^{334}, 14^{274}, 15^{234}, 16^{179}, 17^{151}, $ $ 18^{117}, 19^{87}, 
20^{66}, 21^{40}, 22^{42}, $ $23^{27}, 24^{18}, 25^{8}, $ $26^{13}, 
27^9, 28^4, 29^2,$ $ 30^2, 31^5, 32^1,35^2\}$ as it follows from the 
above that there is a list of $\int_{\mathbb{P}^3_{\Delta_3}} c_1^3$ determining 
$\chi(M_4)$ that runs over $\{2 (h_{11}( \mathbb{P}^3_{\Delta_3^*})+1)\}$, $h_{11}(B_3)=1,\ldots,35$  with the same multiplicities. 
This does not completely fixes the multiplicities of the Hodge diamond since 
$h_{21}$ is non trivial in this class and occurs with the multiplicities  
$\{h_{21}^{mult}\}=\{0^{899}, 1^{1051}, 2^{823}, 3^{599}, $ $4^{362}, 5^{231}, 
6^{133}, 7^{84}, 8^{56}, 9^{33}, 10^{22},  $ $ 11^{10}, 12^6, 
13^5, 14^2, 15^1, 16^1, 17^1\}$.

\subsection{Global properties related to the fibre types} 
\label{globalfiber}

Let us start the section with a short review of the fibre types. 
While all elliptic curves can be transformed into Weierstrass form, 
if the coefficients are in a chosen number field or in polynomial ring 
over the base in an algebraic description of an elliptic fibration as 
in~\ref{WDelta}, it is not possible to transform it to Weierstrass form and  
keeping the coefficients in the same number field or 
ring~\footnote{E.g. in Nagells algorithm for the general cubic 
one has to take roots of the coefficients.}.

Since global properties, like the Euler number of $M_n$ in terms of the base and 
fiber data, the Mordell-Weyl group $MW(M_n/B_{n-1})$ of the sections and 
the monodromy group $\Gamma={\rm PSL}(2,\mathbb{Z})/\Gamma_M$ depend on 
the specific algebraic realization, many such realizations appear 
in the explicit study of elliptic fibrations. 

It is natural to label the fibre types by $\mathfrak{g}$ indicating a Lie Weyl Group  
action on the cohomology of closely related del Pezzo surfaces,
in fact a relation that enables one to construct fibrations with 
up to $s=8$ sections. For $s=1,\ldots,3$, i.e. $\mathfrak{g}=E_8,E_7,E_6$, 
they can be realized as hypersurfaces for $s=4$, i.e. $\mathfrak{g}=D_5$ 
as complete intersection of two polynomials, for $s=5$ as determinantal 
varieties and for $s>6$ by more general prime ideals.            

Let us give the geometric and topological properties for generic elliptic fibrations   
with no non-abelian gauge symmetry enhancement. They  depend only on topological 
data of the almost Fano base, the fibre type and eventually the twisting data. 
In~\cite{Klemm:1996ts} four fibre types have been discussed
\be
\begin{array}{rl} 
 D_5\ :& \quad X_{2,2}(1,1,1,1)=\left\{ \begin{array}{l} x^2+y^2-szw=0\\   z^2+w^2-sxy=0 \end{array}\ \bigg|\ (x,y,z,w)\subset \IP^3(1,1,1,1)\right\},\\ [2 mm]
 E_6\ :& \quad X_{3}(1,1,1)=\{
z^3+x^3+y^3-s w x y =0\ |\ (z,x,y)\subset \IP^2(1,1,1)\},\\ [2 mm]
 E_7\ :& \quad X_{4}(1,1,2)=\{
z^4+x^4+y^2-s x y z=0\ |\ (z,x,y) \subset \IP^2(1,1,2)\},\\ [2 mm]
 E_8\ :& \quad X_{6}(1,2,3)=\{
z^6+x^3+y^2-s x y z=0\ |\ (z,x,y) \subset \IP^2(1,2,3)\},
\end{array}
\label{fibertypes}
\ee 
for which we give here a standard transversal complex family. We denote the degree 
of the polynomials by $d_1,d_2$ or $d$ respectively.

The simplest compactification of (\ref{Weierstrass}) is with the $E_8$ fiber. 
Topological properties of this type have been discussed first in~\cite{Sethi:1996es}, 
however with a homogenization, which makes the generalization to the other 
fibre types more inconvenient to state. The toric description is 
given by polyhedron number 10 in figure \ref{fig1}.  As indicate there  we 
associated the coordinates to the points as follows $\{z=(-2,-3),x=(1,0),y=(0,1)\}$ 
and choose in  (\ref{polyhedrafrombaseandfibre}) $\nu_i^{F*}=(-2,-3)$. 
The hypersurface (\ref{WDelta})  written  in the coordinate 
ring $\{Y_k\}=\{z,x,y,\underline {u}\}$ is then automatically in the \textit{Tate Form}    
\begin{equation}
 y^2 + x^3 + a_6(\underline {u}) z^6 + a_4(\underline {u})  x z^4 + a_3(\underline {u})  y z^3 + a_2 (\underline {u}) z^2 x^2 + 
 a_1(\underline {u})  z x y= 0 \ .
\label{tateform} 
 \end{equation}  

By the construction of $\Delta^*$ the relations (\ref{scalings}) ensure 
that $z\in {\cal O}(1)$, $x\in {\cal O}^2(1)\otimes K_B^{-2}$  and 
$y\in{\cal O}^3(1)\otimes K_B^{-3}$. Further the $a_i({\underline {u}},{\underline \bf {a}})$, weighted 
polynomials in the base coordinates ${\underline {u}}$, transform consistently 
as sections of line bundles over the base, whose properties are obvious by 
noting that since $d=6$ (\ref{tateform}) is a section in ${\cal O}^6(1)\otimes K_B^{-6}$.   
Our choice  $\nu_i^{F*}=(-2,-3)$ corresponds to the canonical 
twisting data mentioned in section~\ref{fibretwistingdata}, 
i.e. the projective space $\mathbb{P}^2(1,2,3)$ is fibred over 
$B_n$ to yield the ambient space of the elliptic fibrations 
as a projectivization of powers of the canonical line 
bundle of $B_n$. It is clear  from (\ref{tateform}) that the fibre has an unique holomorphic 
section at $z=0$, that restricts the elliptic curve to the 
point\footnote{Note that other solutions to (\ref{tateform}) 
for $z=0$ lie in the same equivalence class as 
$(-1,1)$ in $\bbP^2(1,2,3)$.} $(x,y)=(-1,1)$. The section 
can be thought as the internal space of type IIB string theory.

The $E_7$ fibration type  is defined by the polytope number 4 with $\nu_i^{F*}=(-1,-2)$ 
so that (\ref{WDelta}) becomes a section in ${\cal O}^4(1)\otimes K_B^{-4}$
\begin{equation}
y^2+y x^2 + x^4 + a_1 x^3 z+ b_1 x y z +  a_2 x^2 z^2 + b_2 y z^2 + a_3 x z^3  + a_4 z^4 = 0 \ ,
\label{we7} 
\end{equation}  
where $z\in {\cal O}(1)$, $x\in {\cal O}(1)\otimes K_B^{-1}$  and $y\in{\cal O}^2(1)\otimes K_B^{-2}$, 
while the $E_6$ fibration type corresponds to polytope number 1 with $\nu_i^{F*}=(-1,-1)$ 
and the  section in ${\cal O}^3(1)\otimes K_B^{-3}$ is given by 
\begin{equation}
y^3+ x y^2 +y^3 +a_1 x^2 z+b_1 x y z+c_2 y^2 z + a_2 x z^2+b_2 y z^2+ a_3 z^3\ .
\label{we6} 
\end{equation}  
As in the (\ref{tateform}) the indices $k$ on the $a_k,b_k,c_k$ indicate 
that they transform with $K_B^{-k}$. In both cases $z=0$ are sections 
and it is easy to see that they have multiplicity  two and three 
respectively.               

Let us denote the class in which $z$ is a section by $\alpha={\cal O}(1)$ and by  
$c_1=-K_B$. Since divisors $z=0,x=0,y=0$ have no intersection, i.e. 
$\alpha( \alpha + c_1)( \alpha + c_1)=0$ in the cohomology ring of 
total space  of the projective bundle $\mathbb{P}$ so that on $M_n$ one has 
the relation in cohomology   
\begin{equation} 
\alpha^2 = -\alpha c_1\ .
\label{relation} 
\end{equation}
Using this in the adjunction formula  one can write 
down an expression for the total Chern class of the Calabi-Yau 
space $M_n$~\footnote{In the $D_5$ complete intersection case $d_1=d_2=2$.  One has to add a factor 
$(1+ \alpha+ c_1)$ in the numerator and a factor $(1+2 \alpha+ 2 c_1)$ in the denominator.} as 
\begin{equation}  
C(TM_n) =\left(1+\sum_{i=1}^{n-1}c_i\right) \frac{(1+\alpha)(1+w_2 \alpha+ w_2 c_1) (1+w_3 \alpha+w_3 c_1)}{1 + d \alpha + d c_1}\ .      
\label{chern} 
\end{equation} 
The Chern forms $c_k(TM_n)$  are the coefficients  
in the formal expansion of (\ref{chern}) of the degree $k$ in terms of  
$a$ and the monomials of the Chern forms  $c_i$ of base $B$. For the convenience 
of the reader the result is given in table \ref{tab:elliptic}.  

\begin{center} 
\begin{table}[h!]           
$$
  \begin{array}{|c|ccc|} 
  \hline 
 \text{Fibre}& c_2(TM_n)  & c_3(TM_n) & C_4(TM_n) \\[1mm] 
  \hline  
  D_5  &  3 \alpha c_1+(2 c_1^2+c_2)&-4 \alpha c_1^2-(4 c_1^3+c_2 c_1-c_3)&3 \alpha c_1 (3 c_1^2+c_2)\\[1mm] 
  E_6  &  4 \alpha c_1+(3 c_1^2+c_2)&-8 \alpha c_1^2-(8 c_1^3+c_2 c_1-c_3)&4 \alpha c_1 (6 c_1^2+c_2)\\[1mm] 
  E_7  &  6 \alpha c_1+(5 c_1^2+c_2)& -18 \alpha c_1^2-(18 c_1^3+c_2 c_1-c_3)&   6 \alpha c_1(12 c_1^2+c_2)\\[1mm] 
  E_8  &  12 \alpha c_1+(11 c_1^2+c_2)& -60 \alpha c_1^2-(60 c_1^3+c_2c_1-c_3)&  12 \alpha c_1(30 c_1^2+c_2)\\[1mm] 
 \hline 
  \end{array} 
 $$ 
  \caption{\label{tab:elliptic} Chern classes $c_k(TM_n)$ of regular  
  elliptic Calabi-Yau manifolds in terms of the Chern classes of the base and the fibre class. 
  Integrating $\alpha$  over the fibre yields a factor $a=\frac{\prod_i {d_i}}{\prod_i{w_i}}$, 
  i.e. the number of sections $1,2,3,4$ for the four fibrations in turn.}                
\end{table} 
\vskip -10mm
\end{center} 

In particular the Euler number for $n=3$ is given by 
\begin{equation}  
\chi(M_3)=-2{\rm c}(\mathfrak{g})\int_{B_2} c_1^2\, ,
\label{euler3}
\end{equation}
where  ${\rm c}(\mathfrak{g})$ is the Coxeter number of the Lie algebra labeling the fibre type. 
For $M_4$ one has  
\be 
\chi(M_4)=12 \int_{B_3} \left(c_1 c_2 + 3 b(\mathfrak{g}) c_1^3\right)\ ,
\label{euler4} 
\ee
with $b(D_5)=1$,  $b(E_6)=2$, $b(E_7)=6$ and $b(E_8)=10$. 
We note here that this result of the discussion is 
already summarized in figure \ref{fig1} where we indicate on each point 
of the fiber polynomial the data
\begin{equation}
{\rm c}(\mathfrak{g})/36 {\rm b}(\mathfrak{g}):h_{11}(\mathbb{P}^{n-1})(h^{twist}_{11}(\mathbb{P}^{n-1}))\ . 
\label{explanationhodge} 
\end{equation}
Independent of the Lie algebra meaning ${\rm c}(\mathfrak{g})$ and $36 {\rm b}(\mathfrak{g})$ 
denotes just the coefficient in (\ref{euler3},\ref{euler4}) and $h_{11}(\mathbb{P}^{n-1})$ 
denotes the number of $h_{11}$ for the basis $\mathbb{P}^{n-1}$, 
while $h^{twist}_{11}+1$ denotes the number of irreducible components 
of $z=0$ hyperplane. In the special class of Calabi-Yau n-folds 
the C.T.C Wall data that specifies the topological type of $M_n$ and 
even the quantum cohomology can be fixed from base and fibre data. 
On the other hand starting from the generic model it is possible to specify 
the corrections to~(\ref{euler4}) and the other mentioned data due to 
singular fibres.


\section{Compact Calabi-Yau manifolds with gauge groups}  
\label{compactCY}
Here we discuss first  the non-abelian gauge symmetry enhancements 
that follows from the Kodaira classification~\cite{kodaira1}\cite{kodaira2}   
and  monodromy data that can be obtained for the $E_8$ fibre from the Tate 
algorithm~\cite{Tate}\cite{ Bershadsky:1996nh}\cite{Katz:2011qp}.  
 
Further global gauge $U(1)$ gauge symmetry enhancement come from additional 
global sections of the elliptic fibre as was predicted in~\cite{Morrison:1996pp}, explicitly 
constructed in~\cite{Klemm:1996hh} and interpreted in the heterotic limit as 
split spectral cover in~\cite{Grimm2010c}.     

We argue that the loci of enhanced gauge symmetry 
corresponds to sub-monodromy problems in the complex moduli space  
that can hence be reached by turning on the corresponding fluxes 
on the vanishing periods.

\subsection{Non-abelian gauge symmetries}   
Gauge symmetry enhancements are obtained  by specializing the complex 
structure in the generic clean sheet  model so that the local behaviour of 
the base and moduli dependent coefficients in the elliptic fibre, which are given for 
the $(E_8,E_7,E_6)$ fiber types  in (\ref{tateform},\ref{e7form},\ref{e6form}), near a 
given divisor $D_g$ in the base is of the desired form. The non-abelian gauge group $G$ follows from 
\begin{itemize} 
\item g1) The Kodaira singular fibre type, which is specified by the Weierstrass data $[f,g,\Delta=f^3-27 g^2]$ 
according to Kodaira's classification reproduced in table \ref{table:kodaira}. The Weierstrass data are obtained using Nagells algorithm 
and are given for the $(E_8,E_7,E_6)$ fibres in (\ref{E8fg},\ref{E7fg},\ref{E6fg})\footnote{One can take any 
of the 2d reflexive polyhedron in figure \ref{fig1} 
 as fibre polytope and obtain their Weierstrass data from trivial  
specializations of the Newton polynomial for the polyhedra $10$ (E8), $13$ (E7), $15$ (D5) and $16$ (E6). 
The Weierstrass data of  polyhedra $10$, $13$, $15$ and $16$  are given in appendix  A of~\cite{Huang:2013yta}. } 
\item g2) The possible  monodromy in $D_g$ acting  by an automorphism on the homology 
of  the fibre. This  reduces the Lie algebra $g$ of the Kodaira type to a non simply 
laced one with lower rank: $\mathbb{Z}_3:D_4\rightarrow G_2$, $\mathbb{Z}_2:E_6\rightarrow 
F_4$,  $\mathbb{Z}_2:D_n\rightarrow B_{n-1}$ , $\mathbb{Z}_2:A_{2n+1}\rightarrow C_k$.    
\end{itemize}    
For the $E_8$ fibre Tate's algorithm relates  the monodromies  to the  leading 
coefficients in the Tate form (\ref{tateform}) that vanish generically in 
codimension one in $B_{k>1}$~\cite{Tate}\cite{ Bershadsky:1996nh}\cite{Katz:2011qp}, 
summarized in table \ref{table:tatenonsplit} and \ref{table:tatesplit}. Higher codimension singularities occur if the divisors $D_g$ 
are singular or cross each other and  have not been treated systematically.     

If we consider toric divisors $D_g$, the specialization of the complex structure
amounts to set $a_i$ in (\ref{WDelta}) to zero, i.e. to exclude points $\nu_{i_1},\ldots, \nu_{i_s}$   
of $\Delta$. This reduces ${\cal M}_{cs}$ to a monodromy invariant subspace ${\cal S}$  
if  $\tilde \Delta=\langle \Delta \setminus \{\nu_{i_1},\ldots, \nu_{i_s}\}\rangle\subset 
\Gamma_{\mathbb{R}}$ is a reflexive polyhedron in the reduced lattice 
$\tilde \Gamma$. This implies that $\tilde \Delta^*\subset \Gamma^*_{\mathbb{R}}$ 
exists and resolves the singularities of $W_{\tilde\Delta}=0$ torically. As we argued in 
the previous sections that implies the gauge symmetry enhancement occurs at the minimum of 
a flux superpotential.    

For the clean sheet models the coefficients of the fibrations are usually generic enough so 
that if one enforces vanishing orders of $[f,g,\Delta]$ at a single divisor for a low 
rank gauge group, the change of the Hodge number $\Delta(h_{11})$ in the torically resolved 
model is entirely due to homological non-trivial\footnote{I.e. the eventual outer automorphism 
is already divided out in the global toric construction.}  divisors of the form of a $\mathbb{P}^1$ 
fibration over $D_g$, where the  $\mathbb{P}^1$ are the blow ups of the Kodaira 
singularity, whose intersections yield then the negative Cartan matrix of $\hat g$. In particular $\Delta(h_{11})={ \rm rank}(g)$ 
and this information combined  with the original  Kodaira singularity type fixes the 
possible outer automorphism. 

Of course if the rank of the gauge group exceeds a bound 
the polyhedron ceases to be reflexive and if one approaches 
this bound from below the coefficients of the corresponding fibre types are not 
generic enough so that the toric resolution produces in addition to the  
gauge divisors, exceptional divisors, whose shrinking lead e.g. to 
tensionless strings. For example at a single toric divisor in a clean sheet 
model with fibre type $E_n$ leading to a 6d or 4d theory it is not 
possible to enhance the singularity just to $E_n$. Similarly we have  
studied intersecting toric gauge divisors in the base. It is possible to 
enforce low rank gauge groups on three  intersecting divisor and 
just get the exceptional divisors that correspond to the gauge symmetry 
enhancements including the standard model gauge group, but  as expected 
higher rank groups on intersecting divisors produce additional singularities. 

For the $E_8$ fiber we choose a  construction with heterotic duals and 
investigate the effects of non-generic coefficients for the resolution of  
higher rank gauge groups in~\ref{tate} as compared to the generic Tate algorithm. 
For the 4-fold $\overline{B}_2$ is determined by the polytopes of Figure~\ref{poly} 
such that $B_3=\mathbb{P}^1\times \overline{B}_2$ and for the 3-fold 
case the base is given by $\tilde{B}_2=\mathbb{P}^1\times\overline{\mathbb{P}}^1$. 
One finds that even in the most generic global  models the coefficients are 
usually less generic than assumed in local model building.   

As mentioned  the $E_7$  (\ref{e7form}), $E_6$ fibre (\ref{e6form})  and $D_5$  fibre~\cite{Huang:2013yta} 
can be locally brought into Weierstrass form using Nagells algorithm, see (\ref{E7fg},\ref{E6fg}) and~\cite{Huang:2013yta}.
This information, i.e. the vanishing orders of $[f,g,\Delta]$ at $D_g$  gives 
the Kodaira fibre type over the generic point in $D_g$ that  yields the
upper bound on the rank of the gauge group. Often one can understand 
the additional effects due to the non-generic coefficients and using $\Delta(h_{11})$ 
from the toric resolution one can fix the outer automorphism. One can then 
turn Nagells formalism around and specify  similarly like in table \ref{table:tatenonsplit} and
\ref{table:tatesplit} the leading behaviour of the coefficients in  (\ref{e7form},\ref{e6form}) that 
leads to the desired gauge group, see tables  \ref{E7U1vanish} and \ref{E6U1vanish}. This reproduces 
the results obtained from the ``tops'' introduced in~\cite{Candelas:1996su} and 
used in~\cite{Braun:2013nqa} in the same context, but is slightly more 
general, because it is not restricted to the toric embeddings and also applies to crossing toric divisors.

\subsection{The $E_8$ fibre and Tate's algorithm} 
\label{tatem} 

Since for the $E_8$ fibre (\ref{WDelta}) is in the Tate form (\ref{tateform}), which we reproduce for convenience
$$
 y^2 + x^3 + a_6(\underline {u}) z^6 + a_4(\underline {u})  x z^4 + a_3(\underline {u})  y z^3 + a_2 (\underline {u}) z^2 x^2 + 
 a_1(\underline {u})  z x y= 0 \ ,
$$
one  can restrict the complex structure moduli $\underline{z}$~\footnote{From now on we use $z_i$  instead of $a_i$  
to denote  the complex structure variables, to avoid confusion with the $a_i({\underline {u}},v,w,{\underline z})$, which are moduli 
dependent sections over the base. Coordinates on the base  $B_{n-1}$ are denoted generically by ${\underline{u}}$,  but  if $B_{n-1}$ has a 
fibration structure $\mathbb{P}^1\rightarrow \overline{B}_{n-2}$ we call the coordinates pf $\mathbb{P}^1$ $v,w$ and the coordinates of $B_{n-2}$ $\underline u$ .} 
to achieve singular fibers at a given divisor in the base. 
For example consider cases with a 
$\mathbb{P}^1$ fibration over a heterotic base $B_{n-2}$ and call the 
coordinates $\{Y_k\}=\{z,x,y,u_1,\ldots u_m,w,v\}$, where 
$u_i$ are now coordinates of the heterotic base  $B_{n-2}$ and $w,v$ are homogeneous 
coordinates of the $\mathbb{P}^1$. E.g. for $w=0$ the whole $B_{n-2}$ becomes a gauge divisor $D_g$ 
and by setting the  coefficients of the monomials in $a_i({\underline {u}},v,w,{\underline z})$
to zero. i.e. by specializing the complex structure moduli, 
one can achieve that  $a_i({\underline {u}},v,w,{\underline z})$  are of the form
\begin{equation}
a_1 = \alpha_1 w^{[a_0]} \, , \quad 
a_2 = \alpha_2 w^{[a_2]} \, , \quad 
a_3 = \alpha_3 w^{[a_3]} \, , \quad 
a_4 = \alpha_4 w^{[a_4]} \, , \quad 
a_6 = \alpha_6 w^{[a_6]} \, ,
\label{ai}  
\end{equation}
where $\alpha_i({\underline {u}},v,w,{\underline z})$ are of order 
zero in $w$. Choosing this leading behaviour at $w=0$ or at any other divisor $D_g$ of 
the base $B_{n-1}$  leads by Tate's algorithm to singular fibres and hence a gauge group along $D_g$. 
The association of the leading powers of $[a_i]$ with the singularity is given by Tate's  algorithm~\cite{Tate},
which is  reproduced in~\cite{Bershadsky:1996nh}. The Kodaira type follows already from the Weierstrass 
data (\ref{Weierstrass})  $[f,g,\Delta=f^3-27 g^2]$, which we give for completeness
\begin{equation} 
f=\left(a_1^2+4 a_2\right)^2-24 (a_1 a_3+2 a_4),\quad g=\frac{1}{3^\frac{3}{2}}(36 (a_1 a_3+2 a_4) \left(a_1^2+4 a_2\right)-216 \left(a_3^2+4 a_6\right)-\left(a_1^2+4 a_2\right)^3)\ .
\label{E8fg}
\end{equation}

\subsubsection{The stable degeneration limit and the moduli space of the central fibre}
\label{stabledegenerationlimit}   
We demonstrate the formalism of landscaping by fluxes and the occurrence of the sub-monodromy problem  
with the enforcement of the stable degeneration limit in the generic clean sheet elliptic K3, given 
by the member $M^A_1$ of the A-series $X_{12}(1,1,4,6)$ considered in sect~\ref{extremalbundle}. 
In the eight dimensional case the F-theory base is  $B_1=\mathbb{P}^1$ parametrized by  
$v,w$  while the heterotic base $\bar B_{n-2}=B_0=pt$ is trivial. However the essential
features carry over to the situation when the $\mathbb{P}^1$ is a fibration over $\bar B_1$ or $\bar B_2$.
The polyhedron $\Delta^*$ is defined by the points   $\{\{1,-2,-3\}$,$\{-1,-2,-3\}$,$\{0,-2,-3\}'$, $\{0,1,0\}$, $\{0,0,1\}\}$ 
corresponding to coordinates  $v,w,z,x,y$. Of course $\Delta^*$ 
contains  the inner point $\nu_0^*$ and the  points above  are corners except 
for the primed one, which it lies on an edge. The section of the anti canonical bundle 
characterizing the K3  defined by  (\ref{WDelta})  reads  
\begin{equation} 
\begin{array}{rl}
W_{\Delta}=& y^2 + x^3 + 
(a_1 v^{12}+ 
a_2 v^{11} w+
a_3 v^{10} w^2+
a_4 v^9 w^3+
a_5 v^8 w^4+
a_6 v^7 w^5+
a_7 v^6 w^6+ \\ & 
a_8 v^5 w^7+  
a_{9} v^4 w^8+  
a_{10} v^3 w^9+
a_{11}v^2 w^{10}+ 
a_{12}v w^{11}+ 
a_{13}w^{12})z^6+ 
(a_{14}v^8+ 
a_{15}w^8) x z^4 +\\&   
(a_{16}v^6+
a_{17} w^6) y z^3+ 
(a_{18} v^4+a_{19} w^4) z^2 x^2 + a_{20}(v w) x y z \ . 
\end{array} 
\end{equation} 
From the 20 $a_i$, two are redundant, but we keep the full flexibility to set 
them to zero and impose  Kodaira fibres at $v=0$ and (or) at $w=0$ 
using (\ref{ai}).  Passing from the restricted polynomial to its Newton polytope and 
checking whether the restricted polytope is reflexive yields the sub-monodromy problems related 
to the gauge groups at  $v=0$, $w=0$  of the elliptic K3 hypersurface in this toric ambient 
space. In particular  we can enforce two $E_8$ fibres at $w=0$ and $v=0$. This  adds 
$16$ points to $\Delta^*$ to complete the polytope 
$\ci \Delta^*= \{\{6,-2,-3\}$,$\{-6,-2,-3\}$,$\{0,-2,-3\}'$, $\{0,1,0\}$, $\{0,0,1\}\}$, 
see~\cite{Candelas:1996su}\cite{Kachru:1997bz}\cite{Berglund:1998ej} for explicit lists 
of these points. The $16$ new points correspond to the toric resolution of the two $E_8$'s. The restricted 
polynomial is 
\begin{equation} 
W_{\ci \Delta}= a_1 y^2 + a_2 x^3 + a_3 v^{7} w^5 z^6+ a_4 v^5 w^{7} z^6
+ a_5 v^6 w^6 z^6 + a_0  v w x y z \ .
\label{WtD} 
\end{equation}  
It is combinatorial equivalent to the invariant polynomial under the maximal phase 
symmetry group $\mathbb{Z}_2\times\mathbb{Z}_3$ generated by 
$(y\rightarrow -y,v\rightarrow -v)$ and $(x\rightarrow \exp(2\pi i/3) x,v\rightarrow 
\exp(-2\pi i/3)v)$
\begin{equation} 
W_{\ci \Delta_{ph}}= a_1 y^2 + a_2 x^3 + a_3 v^{12} z^6+ a_4  w^{12} z^6
+ a_5 v^6 w^6 z^6 + a_0  v w x y z \  , 
\label{WtD2} 
\end{equation}         
in the coarse invariant lattice $\ci \Gamma_{ph}$. In either formulation of the restriction 
one uses the arguments given before, to conclude that there is the heterotic  $G$-flux in F-theory, 
which drives the complex moduli in the fibred 4d version to the stable degeneration limit.    
The Gauss-Manin connection for the sub-monodromy problem is readily  
obtained using GGKZ method described in section  (\ref{gaussmaninS}).  It depends 
on the linear relations among the points of the Newton polyhedron, which are    
\begin{equation} 
l^{(e)}=(-6,3,2,0,0,1),\qquad  l^{(b)}=(0,0,0,1,1,-2)\ ,
\label{baseandfibre} 
\end{equation}
where $l^{(e)}$ is the Mori vector associated to the elliptic fibre class, while $l^{(b)}$ is 
associated to the base class.  
This leads straightforwardly~\cite{Hosono:1993qy} to the Picard-Fuchs differential 
ideal ${\cal I}_{PF}$ of the sub-monodromy problem of the central fibre generated  by
\begin{equation} 
{\cal D}_1=\theta_e(\theta_e- 2 \theta_2)- 12 z_e (6 \theta_e+5)(6 \theta_e+1), \quad  
{\cal D}_2=\theta_b^2- z_b (2 \theta_b-\theta_e)( 2 \theta_b-\theta_e +1)\ .
\label{k3pf} 
\end{equation}
According to~\cite{Morrison:1996pp} (see also the discussion~\cite{Friedman:1997yq}) 
the stable degeneration is where the K\"ahler  class $\rho$ in the dual heterotic string is 
large $\rho \rightarrow i \infty$, while the complex structure of the F-theory 
torus $\tau$  is identified with the complex structure of the heterotic torus $\tau$. 
The sub-monodromy problem was solved in~\cite{Klemm:1995tj}. In fact in 
an elliptically fibred 3-fold with $\bar B_1=\mathbb{P}^1$, which has a heterotic
dual, in the eight dimensional limit ${\rm vol}(\bar  B_1)\rightarrow \infty$. The 
equivalence  is immediate by noticing that (\ref{k3pf}) is equivalent to 
the first two differential operators in eq. (9) in~\cite{Klemm:1995tj}, which 
determine the finite periods in the eight dimensional limit. Hence the map from the F-theory moduli  
to the heterotic theory is completely precise and one can identify the stringy gauge theory  
enhancements of the heterotic $T_2$ compactifications in  the  F-theory moduli space. It is 
fixed by  the modular properties of the mirror map and given in~\cite{Klemm:1995tj} as
\begin{equation}
\begin{array}{rl}  
z_e&=\displaystyle{ \frac{ j(\tau)+ j(\rho) -\mu} { j(\tau)j(\rho)+\sqrt{j(\tau)(j(\tau)-\mu)} \sqrt{j(\rho)(j(\rho)-\mu)}}}\ ,\\[3 mm]
z_b&=\displaystyle{ 2\frac{ \bigl[j(\tau)j(\rho)+\sqrt{j(\tau)(j(\tau)-\mu)} \sqrt{j(\rho)(j(\rho)-\mu)}\bigr]^2 } 
{ j(\tau)j(\rho)[j(\tau)+ j(\rho) -\mu]^2}}\ ,\\
\label{mapFtheoryheterotic} 
\end{array}
\end{equation}
where $\mu=1728$.  Calculating the discriminant of (\ref{WtD}) yields after some scaling $\delta_1=(1-4 z_b)$ and 
$\delta_2=(1-432 z_e)^2-1728 z_e^2 z_b$. Comparing with heterotic string at $\delta_1=0$  
one has at $\rho=\tau$  an $SU(2)$ enhancement while for $\delta_2=0$ we have solutions 
$\rho=\tau=i$ corresponding to the $SU(2)\times SU(2)$- and at  $\rho=\tau=\exp(2 \pi i/3)$ to an 
$SU(3)$  stringy heterotic gauge theory enhancement~\cite{Klemm:1995tj}. 
Specializing the complex structure to get the $E_8$ Kodaira singularities at antipodal points of the base $\mathbb{P}^1$
is the first step  to reach the stable degeneration limit. In addition one has to chose the volume of the base 
$\mathbb{P}^1$ to be infinite, which according to (\ref{baseandfibre}) means $z_b=0$. As explained 
in some detail in~\cite{Friedman:1997yq} in complex geometry this means that the base $\mathbb{P}^1$ splits   
into two $H_i\sim\mathbb{P}^1_i$, $i=1,2$,  which share one common point $Q$. One has $\pi^{-1}(Q)=T^2_{het}$ 
and $\pi: {\cal E}\rightarrow H_i$ are two rational elliptic surfaces. Since they have $4$ $I_1$ fibres and one 
$\hat E_8$ Kodaira fibre or at more generic points in their complex structure moduli space $12$ $I_1$ 
fibers they are called $\frac{1}{2}K3$. $H_i$ is a section of the corresponding elliptic fibration with 
projection map $\pi$. Blowing down  this section yields an $dP_8$ del Pezzo surface obtained by eight 
successive blow ups of $\mathbb{P}^2$. Instead of an elliptic fibration the $dP_8$ is an elliptic 
pencil with a base point. Blowing up the latter gives back  the $\frac{1}{2} K3$,  which maybe also
called $dP_9$.                        

The map (\ref{mapFtheoryheterotic})  describes  precisely how the limit decompactifies   
the elliptic torus, since $\rho  \rightarrow i \infty$ ($j(\rho)\rightarrow \infty)$ corresponds 
to $z_b\rightarrow 0$, while $z_e=\frac{1} { j(\tau)+\sqrt{j(\tau)(j(\tau)-\mu)}}$ is the  
map to the complex structure modulus of heterotic torus in the stable degeneration limit.
The map  distinguishes the 2 moduli of the central fibre, the heterotic $T^2_{het}$ whose sub-monodromy 
system we solved above, from the 16 moduli of the heterotic 
$E_8\times E_8$ bundle, whose sub-monodromy system is discussed in the next section.


\subsubsection{The moduli space of gauge bundles and the $[p,q]$ string states of  $E_8$ gauge group}  
\label{E8BPSstates} 
For the $E_8$ fibre type compactified in a toric ambient space the following data describe 
equivalently the gauge theory enhancement  with gauge group $G$ over a codimension 
one toric divisor $D_g$ in the base $B_{n-1}$, $n>2$: 
\begin{itemize} 
\item G1) The classification of the Kodaira fibre types and the action of monodromies in $D_g$  
on its homology  as encoded in the vanishing order of the coefficients 
$[a_0,a_2,a_3,a_4,a_6]$  of the Tate form at the divisor $D_g$. 
\item G2)  The ``tops''  classified in~\cite{Candelas:1996su} constructed torically 
over $D_g$. 
\end{itemize}
Simply restricting the coefficients of the Tate form $W_\Delta\rightarrow W_{\ci \Delta}$ and 
constructing  ${\ci \Delta}^* $  yields the additional points in ``tops" as embedded~\footnote{Constructing the 
additional points is part of the functionality of Sage~\cite{BraunSage}. We wrote a  code, 
which is based upon a mathematica program that P. Candelas released some 12 years ago.} in  ${\ci \Delta}^* $, which are listed  
in~\cite{Candelas:1996su}\cite{Kachru:1997bz}\cite{Berglund:1998ej}.  In the special case of a $\mathbb{P}^1$ fibration
over $\overline{B_k}$ with $(v:w)$ homogeneous coordinates of the fibre $\mathbb{P}^1$   
one gets the ``tops'' by restricting the coefficients of the Tate form near $v=0$ and the 
``bottoms'' by restricting them near $w=0$.  This was used  e.g.  in~\cite{Kachru:1997bz} 
to construct  the same  heterotic  gauge group in each of the heterotic $E_8$ to get a $\mathbb{Z}_2$ 
symmetric model  that allows for a CHL involution. 

As mentioned for  high rank gauge groups close to the bound 
of non-reflexivity the additional effects due to non-generic coefficients in the compactification is 
not systematically explored and the  information G1) is more universal as it can be 
immediately  used to construct  gauge groups on crossing  toric and non-toric divisors.

Taking the stable degeneration limit described in the last section in the cases  with heterotic  duals 
$Z_{n-1}=({\cal E}_{het}\mapsto \overline{B}_{n-2})$, i.e.  for F-theory bases of the 
form $B_{n-1}=(\mathbb{P}^1 \rightarrow \overline{B}_{n-2})$, one gets the  
maps from the heterotic bundle moduli to the F-theory  moduli~\cite{Friedman:1997yq}.
For each heterotic elliptic fibre ${\cal E}={\cal E}_{het}$ the moduli  space of the 
holomorphic $G$-bundle is described by the moduli space of the map from the 
weight lattice $\Gamma_g$ of $G$ to the Jacobian of ${\cal E}$ modulo the  
Weyl group $\phi:\Gamma_g \rightarrow {\rm Jac}({\cal E})/{\rm Weyl}$, see e.g. ~\cite{MR0466134}. 
This moduli  space is  a weighted projective space $\mathbb{P}^r(s_0, \ldots,s_r)$, where $r$ is the rank and $s_i$ are 
the Coxeter numbers of $G$~\cite{MR515632}.  This description can be fibred over the base and 
except for the gauge group $G=E_8$ the  moduli of $G$-bundles on the elliptically fibred Calabi-Yau 
manifold $Z_{n-1}$ are described  by  a projectivization $\cal W$ of the bundle
 ${\cal O}\oplus (\oplus_{i=1}^r {\cal L}^{-d_i})$  
over $ \overline{B}_{n-2}$. Here ${\cal L}^{-1}$ is the normal bundle of the base 
and  $d_i$ are the degrees of the independent Casimir operators of $G$. For $n\le 3$ the 
$G$-bundle can be reconstructed  from a section of $s: \overline{B}_{n-2} \rightarrow 
{\cal W} $, which is called  {\sl spectral cover}  $C$  of $B$  and a line bundle on $C$, 
called ${\cal S}$. E.g. for $G=SU(N)$ the $G$-bundle on ${\cal E}$  is given 
explicitly as  $V=\oplus_{i=1}^n  {\cal N}_i$, where $ \otimes_{i=1}^n {\cal N}_i
={\cal O}$, which implies that the ${\cal N}_i$ vanish of first order at $Q_i$  and have 
a pole at the origin $P$ , i.e.  ${\cal  N}_i={\cal O}(Q_i)\otimes {\cal O}(P)^{-1}$, 
with  $\sum_{i=1}^n Q_i=0$ in the group law  in ${\cal E}$. On ${\cal E}=X_6(2,3,1)$ with coordinates 
$x,y,z$ and $z=0$ is $P$ such  points are specified by the spectral equation
\begin{equation}              
w=a_0+ a_2 x+ a_3 y+ a_4 x^2 +\ldots +\left\{\begin{array}{lr} 
a_{n-1} x^{N/2}& \quad n \ {\rm even} \\ 
a_{n-1} y x^{(N-3)/2}&\quad n \ {\rm odd}\ .
\end{array}\right.  
\label{spectralcoversuN}
\end{equation} 
Let us  take the coordinates for $\Delta^*$ as in the previous section. 
The identification of the moduli is discussed in~\cite{Berglund:1998ej} and amounts to 
change the identification of the $\mathbb{C}^*$ actions in order to 
define $\mathbb{P}_{\Delta}$  in the limit. 
Define the coordinate ring of the dual space by 
$(1,0,0)\sim x_1=:v$,   
$(0,-1,-1)\sim x_2=:\tilde z $,  
$(0,2,-1)\sim x_3=:x $ and 
$(0,-1,1)\sim x_4=:y$. 
This allows to write $W_{\Delta^*}$ in the 
limit as follows~\cite{Berglund:1998ej}
\begin{equation} 
\begin{array}{rl} 
W^{BM}_{\Delta^*}&= w_0(x,y,\tilde z,{\underline{u}}) + w_-+ w_+  \quad  {\rm with} \\  [  2mm]
w_-&=\displaystyle{\sum_{j=1}^J v^j{ w_-^j}(x,y,\tilde z, \underline{u}),\quad  w_+=\sum_{j=1}^J v^{-j} {w^j_+} (x,y,\tilde z, \underline{u})},
\end{array}
\end{equation}    
where the monomials $M_i$  of $W^{BM}_{\Delta^*}$  are 
defined by $M_i=\prod_j x_j^{\langle \nu^*_i, \nu_j\rangle+1-\delta_{j,1}}$.
Because  of the shift $\delta_{1,m}$,  $W^{BM}_{\Delta^*}=0$  is not quite the correct 
global description of $W_n$ as it tempers with scalings on the $\mathbb{P}^1$, which is 
the base of the elliptic fibration, but as long as one focusses on the north- or on the southpole 
of the (fibre) $\mathbb{P}^1$ , e.g. by scaling  $w_-^j\rightarrow \epsilon^j w^j_- $ and sending  
$\epsilon\rightarrow 0$,   one can extract the local limit for patch $v\neq 0$. 
The logic is~\cite{Katz:1996fh} that  as long as  the exponents of the (Laurant) 
monomials of the Newton (Laurant) polynomial obey the linear relations described 
by the charge vectors $l^{(i)}$ 
\begin{equation}
\prod_{l_j^{(i)}>0} M_j^{l^{(i)}_j}=  \prod_{l_j^{(i)}<0} M_j^{l^{(i)}_j}\ ,  \forall \ j\ ,
\label{scalingsdiff} 
\end{equation} 
which  implies that the periods of that geometry are annihilated by the~(\ref{GKZ}) 
operator, the B-model geometry of the local sub-monodromy problem is captured 
faithfully by that Newton (Laurant) polynomial. Moreover by the fibration structure of the base $\mathbb{P}^1\rightarrow 
\overline{B}_{n-2}$, $w_0(x,y,\tilde z,{\underline{u}})=0$ is an elliptic fibration  in the 
Tate form over $\overline{B}_{n-2}$, which is the heterotic Calabi-Yau $Z_{n-1}$.  
In the limit for $n=2$ one gets e.g. for $SU(N)$, where we neglected in $w_0$ the points on codim 1 faces,  
 \begin{equation}
 W_L=(y^2+ x^3+\tilde z^6 + \alpha x y \tilde z)  + v\left( a^{(1)}_0 \tilde z^N + a_2^{(1)} x\tilde z^{N-1} + a_3^{(1)} y\tilde z^{N-2}+\ldots +
\left\{\begin{array}{lr} 
a^{(1)}_{n-1} x^\frac{N}{2}& \ \ n {\rm even} \\ 
a^{(1)}_{n-1} y x^\frac{N-3}{2}& \ \ n \ {\rm odd}
\end{array}\right. \right). 
\end{equation}     
For $SU(N)$ $v$ appears only linearly  and one can argue that the linear $v$ can be integrated 
out to one is left with $w_0(x,y,\tilde z)=0$ and $w_+^1=0$. This is the same 
information as in (\ref{spectralcoversuN}) and is easy to see that the weights $s_i$
of the $a^{(s_i)}_i$ over the base are given by the power $s_i$ of $v^{s_i}$ multiplying 
them. In the polyhedra language this is the height of the additional point.  
The case with the most different heights is the $E_8$ case which are the same than in an $E_8$ del 
Pezzo with fixed canonical section with weights fitting into  $\mathbb{P}^8(1,2,2,3,3,4,4,5,6)$  
\begin{equation} 
\begin{array}{lll}
w^1_+=a_1^{(1)} \tilde z^5, & w^2_+=a_1^{(2)} \tilde z^4+  a_2^{(2)} x^2 & w_+^3= a_1^{(3)} \tilde z^3 + a_2^{(3)}y\\ 
 w^4_+=a_1^{(4)} \tilde z^2 + a_2^{(4)} x , & w^5_+=a_1^{(5)} \tilde z & w_+^6= a_1^{(6)} \ .
\end{array}  
\label{locale8}   
\end{equation}  
This  geometry i.e. an elliptic curve ${\cal E}$ given by the  fixed anti-canonical bundle in an 
$E_8$ del Pezzo surface $S$ was used  by Looijenga~\cite{MR0466134} to describe 
the $E_8$ gauge bundle  on ${\cal E}$. Let $-K=c_1(TS)$ and $L=H_2(S,\mathbb{Z})$, then 
the orthogonal lattice $L'\in L$ with $L'\cdot K=0$ is the weight lattice of $E_8$ 
(which is isomorphic to the root lattice as the $E_8$ lattice is selfdual).  As explained
 in~\cite{MR0466134}\cite{Friedman:1997yq} every $y\in L'$ gives rise to a  line 
bundle on $S$ whose restriction to ${\cal E}$ yields the desired map from the  
weight lattice to the Jacobian of ${\cal E}$  modulo the Weyl  transformation. By 
the Torelli theorem the moduli of that map are the moduli of the pair $({\cal E}, S)$, 
i.e. a projectivization of the parameters in  (\ref{locale8}).\footnote{To see that the form of the
del Pezzo surface and ${\cal E}$ given in~\cite{Friedman:1997yq} is equivalent to 
(\ref{locale8}), identify the $u$ used \cite{Friedman:1997yq} with the $v$ used here and  
identifications of the deformations modulo the ideal  $\{\partial_x W_L, \partial_y W_L \partial_{\tilde z} W_L\}$, 
which do respect the scaling in  $\mathbb{P}^8(1,2,2,3,3,4,4,5,6)$.} Note that the 
dual polynom $W_{\Delta^*}$ enters that construction. This follows from mirror  symmetry 
on the K3, which exchanges the horizontal- and the vertical subspaces of the cohomology. 
In the polarization that is  fixed by the toric embedding of $M_n$  in $\mathbb{P}_{\Delta^*}$ 
and $W_n$  in  $\mathbb{P}_{\Delta}$ these spaces determine, which gauge groups $G$  
and $H$ can be enhanced  by specializing the K\"ahler moduli  and the complex moduli 
respectively. The groups $G$ and $H$ are exchanged by exchanging  $M_n$ with 
$W_n$~\cite{MR0429876}\cite{MR1265318}\cite{MR1420220}. A list of the groups 
$H$ and $G$ for those  K3 polyhedra that  realize the Tate degenerations is 
given in~\cite{Berglund:1998ej}.

Instead of looking at the del Pezzo surfaces it is better to start with the half K3, which contains
all del Pezzo surfaces as blow down limits. This geometry has an elliptic fibration, and   geometrical 
invariants  associated  to curves in it have a very nice relation to the affine $E_8$ group. 
The latter appears also in the work of~\cite{MR0466134} and has been extended from Loop 
groups to torus groups in~\cite{MR791645}.  For smooth elliptic fibrations the  genus g instanton 
generating  function can be organized as follows~\cite{Klemm:2012sx} 
\begin{equation}
F_g(q,Q)=\sum_{\beta \in H_2( B_{n-1},\mathbb{Z})} \left(\frac{q^{\frac{1}{24}}}{\eta(q)}\right)^{p(\beta)} P_{p'(g,\beta)}(E_2,G_{w_1},\ldots,G_{w_d})  Q^\beta , 
\end{equation} 
where $\beta$ is the base degree  and $Q^\beta=\exp(2 \pi i \sum_k \beta_k t_k)$ with $t_k$,  $k=1,\ldots, h_{11}(B_{n-1})$  
are  the K\"ahler classes of the basis,   while the fibre degree is encoded in $q=\exp(2 \pi i \tau)$, where $\tau$ is the dilaton,  and 
given by the Fourier expansion of the modular functions and form. In particular $\eta(q)$ is the Dedekind $\eta$-function, $E_2$ is the second Eisenstein series and $G_{w_1},\ldots , G_{w_d}$ are the generators of the modular forms associated 
to the  elliptic fibre type. They depend on the number of torsion points and hence the number of sections of the elliptic 
fibration. $p,p'$  are linear in their arguments with coefficients spelled out for general basis in~\cite{Klemm:2012sx}. 
The structure was first observed for rational elliptic surfaces in ~\cite{Klemm:1996hh}
\cite{Minahan:1998vr}\cite{Hosono:1999qc}. If the latter have the $E_8$ fibre type, currently under  discussion, there are only no additional torsion poinst 
and  the modular group is $\mathbb{P}{\rm SL}(2,\mathbb{Z})$ with 
the two generators  $G_4=E_4$ and $G_6=E_6$. For rational surfaces $B_{n-1}=\mathbb{P}^1$ with one 
class $\beta$ and $p=12\beta$ and $p'=2 g -2 + 6 \beta$. In this case       
\begin{eqnarray}
F_0(q,Q)=  \frac{q^{\frac{1}{2}}E_4}{\eta^{12}(q)} Q + \frac{q (E_2 E_4^2+2 E_4 E_6) }{24\eta^{24}(q)} Q^2+ {\cal O}(Q^3)\, ,
\end{eqnarray}  
is the genus zero prepotential. One can get this result from a sub-monodromy system in an elliptic 
Calabi-Yau 3-fold. As in the previous section it is trivial to embed it in a 4-fold geometry e.g. by 
fibering the 3-fold over $\mathbb{P}^1$. The reason that one starts with a 3-fold  rather 
than describe the two dimensional geometry directly as for the K3  in previous section  is that 
the half K3 has positive Chern class. Hence we have to consider the non-compact  total 3d  space  
${\cal O}(-K_{\frac{1}{2}K3})\rightarrow K_{\frac{1}{2}K3}$, i.e.  the 
canonical line bundle over the half K3. 3-folds with such a local limit are elliptic 
fibrations with $E_8$ fibre type, i.e. $\Delta_2^{*F}=\Delta^{(10)}_2$  with $\nu^{*F}=(-2,-3)$ over the 
Hirzebruch surface $\mathbb{F}_1=\mathbb{P}_{\Delta^{(3)}_2}$ with the 2d polyhedra 
in  figure \ref{fig1}. Hence the relevant points for the polyhedron $\Delta^*$ are      
{
\begin{equation} 
 \label{dataratfib} 
 \begin{array}{c|crrrr|rrrl|} 
    \multicolumn{6}{c}{\nu^*_i }  &l^{(e)}& l^{(b)}&  l^{(f)} &\\ 
    x_0   &&     0&     0&   0&   0&     - 6&     0& 0&     \\ 
    u_1   &&     1&     0&   -2&   -3&      0 &    1&  0&      \\ 
    u_2   &&    -1&     0&   -2&   -3&       0&    1&  -1&      \\ 
    w      &&     0&     1&   -2&   -3&       0&    0&  1&       \\ 
    v       &&    -1&    -1&   -2&   -3&       0&    0&  1 &     \\
    z       &&    0&    0&   -2&   -3&       1&    -2&  -1 &     \\
    x       &&      0&     0&   1&   0&       2&    0&  0&         \\ 
    y       &&     0&     0&   0&   1&        3&   0&   0&       \\ 
  \end{array} \  
\end{equation} }
and the $l^{(i)}$ vectors generate the Picard-Fuchs ideal ${\cal I}_{PF}$
\begin{eqnarray}
{\cal D}_1&=&\theta_e (\theta_e-2 \theta_b - \theta_f) - 12 z_e (6 \theta_e+5) (6\theta_e+1)\ ,\\   
{\cal D}_2&=&\theta_b (\theta_b- \theta_f) - z_b (2 \theta_b+\theta_f- \theta_e) (2 \theta_b+\theta_f-\theta_e+1)\ ,\\ 
{\cal D}_3&=&\theta_f^2 - z_f ( \theta_f-\theta_b) (2\theta_b+\theta_f -\theta_e) \ . 
\end{eqnarray} 
From the last section we know that  $l^{(e)}$ and $l^{(b)}$ correspond to the moduli 
of the central fibre and that the stable degeneration limit is $z_b\rightarrow 0$ 
or equivalently that the volume of the base of the elliptic K3 is taken to infinity. In 
this limit it is more non-trivial to obtain the sub-mondromy system, as the invariant 
point configuration is not reflexive, but corresponds rather to the non-compact CY 3-fold. 
From the Picard-Fuchs ideal of the local system it can be proven that the period $F_0$ 
is expressible in terms of the modular forms, see \cite{Hosono:1999qc} for the $E_6$-
and~\cite{Klemm:2012sx} for the $E_8$ fibre type. Because of the map (\ref{induced}) the 
specialization  of the global Calabi-Yau Picard-Fuchs system only captures two of the 10 
parameters of  the half K3,  the base with parameter $Q$ and the fibre with 
parameter $q$. The moduli parametrizing  the $E_8$ vector bundle can be 
added using the Weyl symmetry~\cite{Minahan:1998vr}. 
There are nine Weyl invariant Jacobi forms for the $E_8$ lattice $A_i$, $i=1,\ldots, 5$  and $B_2,B_3,B_4$ and $B_6$, 
the simplest  being  $A_1=\Theta(\vec{m},q)_{\vec m=0}=E_4$ 
\begin{eqnarray} 
\label{E8weyl}
\Theta_{E_8}(\vec{m},\tau) =\sum_{\vec{w}\in \Gamma_8}e^{\pi i \tau \vec{w}^2+2\pi i \vec{m}\cdot \vec{w}} =
\frac{1}{2} \sum_{k=1}^4 \prod_{j=1}^8 \theta_k(m_j,\tau)=\sum_{\mathcal{O}_{p,k}}  q^p  \sum_{\vec{w}\in\mathcal{O}_{p,k}}  e^{2\pi i \vec{m}\cdot\vec{w}}, 
\end{eqnarray} 
and transforms with weight four and index one~\footnote{For the other Weyl invariant Jacobi 
forms see e.g. \cite{Klemm:2012sx}.}. Here ${\mathcal{O}_{p,k}}$ are the Weyl orbits. In terms 
of the nine forms  the genus zero $F_0$ prepotential can be written as  
\begin{eqnarray}
F_0(q,m,Q)=  \frac{q^{\frac{1}{2}}A_1}{\eta^{12}(q)} Q + \frac{q (4 E_2 A_1^2+3 A_2 E_6+5 E_4 B_2) }{96\eta^{24}(q)} Q^2+ {\cal O}(Q^3) \ .
\label{supokw} 
\end{eqnarray}  
The first observation is that the leading term is the superpotential of~\cite{Donagi:1996yf}~\footnote{In fact after a 
correction  in the power of $\eta$ that was found in~\cite{Curio:1997rn}, which does not invalidate the minimization 
conditions  stated in~\cite{Donagi:1996yf} .}. The minima of the superpotential  at $d \Theta_{E8}=\Theta_{E_8}=0$ 
were determined in~\cite{Donagi:1996yf}.  The conclusion is that the moduli $q$ and $Q$ are not restricted, but that the minima are at common zeros of the  $\theta_k(m_j,\tau)$. Because of the K3 mirror duality  exchanging $G$ and $H$
one can view this superpotential also as  prepotential  of the complex moduli problem of the generic 
K3 discussed in the last section near one tip of the $\mathbb{P}^1$. The fact that $F_0$ does not obstruct $q$ 
and $Q$  is in perfect agreement with the analysis of the last section, where $q$ and $Q$ become the unobstructed  
moduli of the central fibre.   

The analysis of this section recalls how to identify  the bundle  moduli in 
F-theory so that the flux  superpotential  and the metric can be evaluated for 
these moduli, with the formalism developed  in the previous section.  

A maybe more  exciting and mysterious aspect that shows up in this calculations is 
that the superpotential or more generally the periods ``count something'', when 
expressed in the correct $N=1$ variables.  For example the leading term of 
(\ref{supokw}) counts the rigid  $-2$ curves (and thereby the rigid divisors in 
$M_4$) in the $E_8$ del Pezzo, in other limit it counts disk instantons\cite{Grimm:2009ef}\cite{Grimm:2008dq}\cite{Alim:2009bx}\cite{Alim:2011rp} 
or D-instantons\cite{Jockers:2009ti}.  The structure becomes even more interesting, 
when one includes higher genus invariants. The most  information one gets out, when one uses the refined counting that 
splits the genus in two counting parameters introduced by Nekrasov. The refined 
invariants  are labeled by two spins $j_L/j_R$  and for the half K3 they  where 
obtained in~\cite{Huang:2013yta} and interpreted as the states associated to the $[p,q]$ strings. 
For example for $Q^{d=1}$ one gets:   
\begin{table}[h!]
\begin{center}
\begin{tabular} 
{|c|c|c|} \hline $2j_L \backslash 2j_R$  & 0 & 1 \\  \hline  0 & 248 &  \\  \hline1 &  & 1 \\  
\hline \end{tabular} \vskip 3pt  $d=1$ \vskip 15pt  
\end{center}
\vskip -4 mm
\end{table} 

and in particular the  $248$ in this expression have been interpreted as the ground state of the 
$[p,q]$ of the $7$-branes with mutually non local charges that appear in the $II^*$ 
Kodaira fibre. The splitting of the $E_8$ representations of the genus one 
invariants into $E_{8-k}\times U(1)^k$ will be discussed in the next section.

\subsection{Additional sections and additional $U(1)$}    
\label{addU1}
Let us point out the relation between the existence of 
$k+1$ global sections of the elliptic fibre and the globalization of the 
abelian $U(1)^k$  that emerge in the Wilson line breaking of the $E_8$ 
gauge group  to $E_{8-k}\times U(1)^k$~\cite{Klemm:1996hh}.

\begin{figure}[htdp] 
\begin{center} 
\includegraphics[width=.5\textwidth]{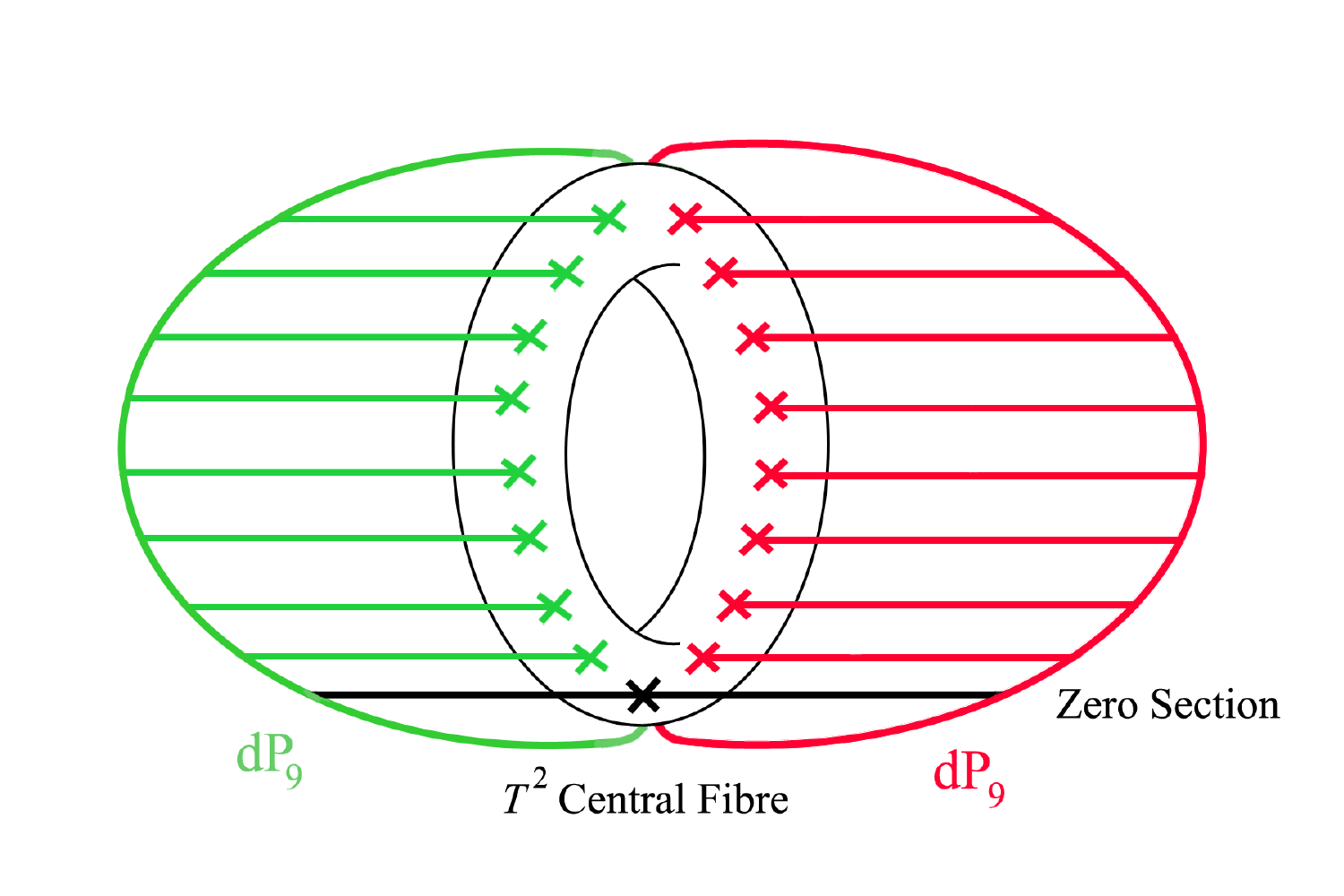} 
\begin{quote} 
\caption{Stable Degeneration limit of a heterotic fibration \label{fig2}} 
\vspace{-1 cm} 
\label{STD} 
\end{quote} 
\end{center} 
\end{figure}

Let us look at one half K3 that arrises in the stable 
degeneration limit. All exceptional divisors $E_i$, $i=1,\ldots, 9$  with 
intersection $E_i E_j=-\delta_{ij}$ in $d_9\mathbb{P}^2$ become 
sections of the elliptic fibration of the rational elliptic surface. This can be seen 
as  follows. The elliptic fibre class is  $e=-K=3H -\sum_{i=1}^9 E_i$, 
where $H$ is the hyperplane class of $\mathbb{P}^2$. 
Using $H^2=3$ and $HE_i=0$  one gets $e^2=0$, i.e. the 
section of the anti-canonical bundle, which is always an elliptic 
curve, becomes now a fibre,  and  $E_i$ intersect the fibre with 
$e E_i=1$  are now sections of the fibration. If we consider the 
stable degeneration that arises in elliptic fibrations over 
$\mathbb{F}_1$ one gets the situation depicted in figure \ref{fig2}.

If we smooth the geometry to the generic fibration with $E_8$ fibre type over 
$\mathbb{F}_1$ the co-kernel of the map $i^*:H_2(M_n)\mapsto H_2(S)$ is eight 
dimensional and one has only the fibre $e$ and the zero section $f$ 
represented in the Calabi-Yau $M_n$. As was explained 
in~\cite{Morrison:1996pp} one can flop the zero section by a transition 
to a second phase, which is described at the level of the Mori cone by 
$l^{(1)}=l^{(e)}+l^{(f)}$,  $l^{(2)}=l^{(b)}+l^{(f)}$ and $l^{(3)}=-l^{(f)}$. 
This flops out the $\mathbb{P}^1$  or a ruled divisor in the  higher dimensional 
case, which corresponds to the zero section. This  makes the half K3 into 
the $d8\mathbb{P}^2$ del Pezzo. The latter can  be shrunken in an 
extremal transition to an elliptic fibration over $\mathbb{P}^2$~\cite{Morrison:1996pp}\footnote{ For $n=3$ the transition is very well understood from the anomaly 
cancellation in the $6d$ effective theory. It 
enforces a relation between number of hyper- vectors 
and tensor multiplets, which is $H-V-T=3 + {\rm c}(\mathfrak{g})(9-T)$. 
The number of tensor multiplets is $T=h_{11}(B_2)-1$ and 
for the clean sheets models $V=(k-1)$, $H=h_{21}(M_3)+1$ 
as well as $h_{11}(M_3)=h_{11}(B_2)+ (k+1)$. In 
particular for $k=0$ one has~\cite{Morrison:1996pp}  
${\rm c}(\mathfrak{g})=30$, which taking into account that 
by Riemann-Roch for rational surfaces  $c_1^2(B_2)=10-h_{11}(B_2)$,   
gives back (\ref{euler3}). Similar matchings hold for 
general $k$.}. If one wants to globalize $k$ more sections 
one has to use elliptic fibrations with $k$ torsion 
points on which the red and the green sections can be 
anchored.  Such points occur in  fibres with more symmetries namely the 
$E_7,E_6,\ldots E_0$ fibers for $k=1,2,\ldots,8$. These fibre where 
considered in~\cite{Klemm:1996hh} in order to study how the $E_8$ 
gauge group realized on the half K3 is broken to $E_8\times U(1)^k$. In particular 
to  represent the additional  sections torically the ambient space of the fibre has 
to be blown up. Of particular interest is how the $E_8$ BPS states split into 
the $E_{8-k}\times U(1)^k$ representations, which can be seen in the
table for the $E_6$ case~\cite{Klemm:1996hh}. The splitting in the $D_5=(E_5)$ case 
was obtained in \cite{Klemm:2004km}.    
\begin{equation} 
\vbox{\offinterlineskip\tabskip=0pt
\halign{\strut\vrule#
&\hfil~$#$
&\vrule#&~
\hfil ~$#$~
&\hfil ~$#$~
&\hfil $#$~
&\hfil $#$~
&\hfil $#$~
&\hfil $#$~
&\hfil $#$~
&\hfil $#$~
&\vrule#
&\hfil ~$#$~
&\vrule#\cr
\noalign{\hrule}
& && U(1)_1\times U(1)_2\times E_6  \quad d_{W_1} 
       & 0    &  1  &  2   &  3    &     4 &  5    &   6     &&U(1)_1\times E_7 \quad  \sum &\cr
\noalign{\hrule}
&d_{W_2}
    &&  &     &      &      &       &       &       &         && &\cr
& 0 &&  &    1&      &      &       &       &       &         &&1&\cr
& 1 &&  &    1&    27&   27 &      1&       &       &         &&56&\cr
& 2 &&  &     &      &   27 &     84&     27&       &         &&138&\cr
& 3 &&  &     &      &      &      1&     27& 27    &     1   &&56&\cr
& 4 &&  &     &      &      &       &       &       &     1   &&1 &\cr
\noalign{\hrule}
&   &&  &     &      &      &   E_8 &       &       &        &&\sum=252 &\cr
\noalign{\hrule}}
\hrule}
\label{instanton}
\end{equation}
\vskip-7pt
\noindent
In the global model again all classes, but the sections  that give rise to the 
$U(1)$, are lost so that the generic global gauge group is  $U(1)^k$. A 
similar interpretation has been given  for the $SU(n)$ groups by the so 
called split spectral cover construction~\cite{Grimm2010c} and has been much 
discussed from the mathematical and phenomenological perspective, 
see e.g. ~\cite{Morrison:2012ei}\cite{Cvetic:2013uta}.  Since the various fibre 
types are connected by toric embeddings~\cite{Berglund:1998va} one can turn on 
flux superpotentials to interpolate between them.   In the following section 
we use the fibre types with $U(1)$ and impose further non-abelian gauge 
symmetries. Like in the $E_8$ case one can work with a restricted set of tops. 
Restricted because in the $E_k$ fibre type the $E_{n>k}$ gauge groups can 
not be realized. Or one can make a local analysis, extending Tates work.

\subsubsection{The $E_7\times U(1)$ case} 
\label{E7u1}
The $E_7$ fibre polyhedron is $\Delta^{(4)}_2$ with $n_F^*=(-1,-2)$. 
As indicated by the data on this point in figure \ref{fig1}, 
there will be one twisted $(1,1)$ class. This is due 
to the fact that the restriction  of the toric divisor 
$z=0$ to the hypersurface splits in two components which 
represent the double sections that come with this fibre type. 
In order to get two toric sections instead one has 
to blow the ambient space of the fibre. This leads to  the 
fibre polyhedron $\Delta^{(6)}$ with $n_F^*=(-1,-2)$, 
which was used in~\cite{Klemm:1996hh} to obtain explicitly the 
$E_7\times U(1)$ representations in table (\ref{instanton}). 
Similar as with the $E_8$ fibre, the  Calabi-Yau manifold with $E_7$  fibre over $\mathbb{F}_1$ 
has two phases and can flop the two sections out of the half K3 to obtain the shrinkable degree 
two del Pezzo $E_7$. The details of the Mori cone and 
the identification of the  Wilson loops parameters with the K\"ahler 
class can be found in~\cite{Klemm:1996hh}.  According to (\ref{WDelta}) the mirror has in the coordinate ring  
$(z,x,y,F,m,\underline{u})$  the form  
\begin{equation} 
F^2 y^2 + x^2 y + F a_3 x y z + a_2 x^2 z^2 + F^2 b_2 y z^2 + F a_1 x z^3 + F^2 a_0 z^4=0.
\label{e7form} 
\end{equation}
Using  the definition of the Stanley Reisner ideal~\ref{complexmoduli} 
and the fact that $l^{*(1)}=(1,1,0,0,-2),l^{*(2)}=(1,0,1,-1,0),l^{*(3)}=(-1,0,0,1,1)$ 
define equivalence classes under the $\mathbb{C}^*$-actions  
$X_i\sim X_i (\mu^{(k)})^{l^{(k)}}$ one finds the single holomorphic section 
at $h_1=(z=0,x=1,y=-1,F=1)$.\footnote{Note that $z=0$ and 
$W_{\Delta}=0$ implies $F^2 y^2 + x^2 y=0$. The solutions of 
this equation could not be mapped to one point on the fibre 
without using the $\mathbb{C}^*$-scaling that involves the 
coordinate associated to $\nu_3$.  However this coordinate does 
not appear in $W_{\Delta}=0$, because it can be eliminated 
by an automorphism of $\mathbb{P}_{\Delta^*}$.} The 
second section is at  $r_1=(z=1,x=1,y=-a_2,F=0)$. This is a 
rational section as the values of $y$ depend of $u$  
via $a_2({\underline u})$. Note that $(z=1,x=0,y=1,F=0)$ is 
in the Stanley-Reisner ideal.

The discriminant on the $E_7\times U(1)$ fiber $\Delta=f^3 -27 g^2$ is determined in terms of the functions $f$ and
$g$ given by
\begin{eqnarray}
f&=&\frac{1}{12} (-24 a_1 a_3 + a_3^4 + 8 a_3^2 (a_2 - b_2) + 
   16 (3 a_0 + a_2^2 + a_2 b_2 + b_2^2)),\\
   g&=&\frac{1}{216} (-36 a_1 a_3^3 + a_3^6 - 144 a_1 a_3 (a_2 - b_2) + 12 a_3^4 (a_2 - b_2) + 
  24 a_3^2 (3 a_0 + 2 a_2^2 - a_2 b_2 + 2 b_2^2) \nonumber \\
  &+& 8 (27 a_1^2 + 8 a_2^3 - 36 a_0 b_2 + 12 a_2^2 b_2 - 8 b_2^3 - 
     12 a_2 (6 a_0 + b_2^2))). \nonumber
\label{E7fg}
\end{eqnarray}

For a 4-fold compactification, it is possible to chose $B_3$ from 4319 different cases.
To exemplify we will  choose the base $B_3=\mathbb{P}^3$ with coordinates $u_i,\, i=1,2,3,4$, to write the codimension 1 enhancements appearing at a given vanishing order of one variable $u_i$.  The results are given in Appendix \ref{E7U1}. Those tables also contain the codimension 1 singularities appearing in a 3-fold with base $B_2=\mathbb{P}^2$ with coordinates $u_i,\, i=1,2,3$, at the point were a certain
$u_i$ vanishes. Let us  fix an $A_4$ singularity at $u_1=0$, then for the $A_4$ singularity the coefficients have the dependence
\begin{eqnarray}
a_3&=&\tilde{a}_3(u_i) + O_{a_3}(u_1)f_{a_3}(u_i), \,  a_2=\tilde{a}_2u_1^2 +O_{a_2}(u_1^2)f_{a_2}(u_i),\,  b_2=\tilde{b}_2u_1 +O_{b_2}(u_1)f_{b_2}(u_i),\,  \\
a_1&=&\tilde{a}_1u_1^2 +O_{a_1}(u_1^2)f_{a_1}(u_i),\, a_0=\tilde{a}_0u_1^3 +O_{a_0}(u_1^3)f_{a_0}(u_i).\,  \nonumber
\end{eqnarray}
The $\tilde{a}_k$ and $b_2$ represent polynomials of $u_i$ which are different from zero when $u_i=0$,
the $\tilde{f}$ represent polynomials on $u_i$ and the $\tilde{O}(u_1^j)$ represent functions on higher order than
$j$ on $u_1$.  With the given vanishing order of the coefficients at $u_1=0$
the 3-fold has the properties: $\chi_3= -204, h^3_{11}=7(0), h^3_{21}=109(0)$ and
for the 4-fold $\chi^4=5898$, $h^4_{11}=7(0)$,  $h^4_{31}=968(0)$ and $h^4_{21}=0.$
As mentioned, other enhancements for the bases $B_3=\mathbb{P}^3$ and $B_2=\mathbb{P}^2$ are summarized in appendix \ref{E7U1}.

In our toric global construction it is possible to write models with different enhancements at different divisors.
Keeping for the 3-fold (4-fold)  the base $\mathbb{P}^2$ ($\mathbb{P}^3$), one can set an $SU(2)$ at $u_1=0$ 
and an $SU(3)$ at $u_2=0$, with the following dependence of the coefficients
\begin{eqnarray}
a_3&=&\hat{a}_3(u_i) + \hat{O}_{a_3}(u_1,u_2)\hat{f}_{a_3}(u_i), \,  a_2=\hat{a}_2(u_i)u_1u_2 +\hat{O}_{a_2}(u_1,u_2)\hat{f}_{a_2}(u_i),\,  \\ 
b_2&=&\hat{b}_2(u_i)u_2 +\hat{O}_{b_2}(u_1^0,u_2)\hat{f}_{b_2}(u_i),\,  a_1=\hat{a}_1(u_i)u_1u_2 +\hat{O}_{a_1}(u_1,u_2)\hat{f}_{a_1}(u_i),\,  \nonumber \\ 
a_0&=&\hat{a}_0(u_i)u_1u_2^2 +\hat{O}_{a_0}(u_1^2,u_1)\hat{f}_{a_0}(u_i).\,  \nonumber
\end{eqnarray}
The vanishing order can be checked in Table \ref{E7U1vanish} from appendix \ref{E7U1}. The $\hat{a}_k$,
$\hat{b}_2$ represent polynomials of $u_i$ which are different from zero when $u_i=0$, the  $\hat{f}$ represent polynomials on $u_i$ and the $\hat{O}(u_1^l u_2^m)$ represent functions on higher order than
$l$ on $u_1$ and higher order of $m$ on $u_2$. For the manifolds  with bases $\mathbb{P}^3$ ($\mathbb{P}^2$) 
which accommodate those singularities  we obtain the properties $\chi_4=5910$, $h^4_{11}=6(0)$, $h^4_{31}=971(0)$ and $h^4_{21}=0$ 
($\chi_3=-210$, $h^3_{11}=165(0)$ and $h^3_{21}=3(0)$).

Let us consider one example with a  different construction. Consider models with heterotic duals, in which
the base $B_3$ is obtained as fibration of $\mathbb{P}^1$ over certain $\tilde{B}_2$. We consider
the case in which $\tilde{B}_2$ is generated by the polytope 7  in Figure \ref{poly}, and there is no twisting thus 
$B_3=\mathbb{P}^1\times \tilde{B}_2$. In addition the 3-fold with heterotic dual is determined by the base $B_2=\mathbb{P}^1\times \mathbb{P}^1$.  Let us localize an $A_4$ singularity at $u_1=0$, the vanishing order for the coefficients is $0, 2, 1, 2, 3$. In the 3-fold
with base $B_2$ the properties are $\chi_3= -196$, $h^3_{11}=8(0)$ and $h^3_{21}=106(0)$. For the 4-fold
with base $B_3$ the properties are $\chi_4= 4092$, $h^4_{11}=11(0)$, $h^4_{31}=663(0)$ and $h^4_{21}=0$.

Finally let us consider cases with heterotic duals  in which the fibration is given by the vectors $\nu^{(i)*}\in \Delta^{(i) *B}_3 $ in (\ref{polyhedrabase}), as well for polytope 7, and with twist $m_1=-1$, $m_i=0, i=2,...,6$. The properties of the 4-fold are given by $\chi^4=3834$, $h^4_{11}=11(0)$,      $h^4_{31}=620(0)$, $h^4_{21}=0$ .

\subsubsection{The $E_6\times U(1)^2$ case} 
\label{E6u12} 
The fibre polyhedron with the triple section 
is  $\Delta^{(1)}_2$  with $n_F^*=(-1,-1)$ and two 
twisted $(1,1)$ class, i.e. a triple section as indicated in figure \ref{fig1}.
In order to get three toric sections and the corresponding 
splitting of the instantons in $E_6\times U(1)^2$ the polyhedron 
$\Delta^{(5)}$  has been considered  in~\cite{Klemm:1996hh}, 
which leads by  (\ref{WDelta}) in the coordinate ring $(z,x,y,F,G,{\underline u})$ to  
\begin{equation} 
G x^2 y + F x y^2 + G^2 a_1 x^2 z + F G b_1 x y z + 
F^2 c_1 y^2 z + F G^2 a_2 x z^2 + F^2 G b_2 y z^2 + F^2 G^2 a_3 z^3=0\ .
\label{e6form}
\end{equation}
Again the holomorphic section is at $h_1=(z=0,x=1,y=1,F=-G)$, 
while the rational ones are at $r_1=(z=1,x=1,y=-a_1 G ,F=0)$ and 
$r_2=(z=1,x=-b_1 F ,y=1 ,G=0)$. Note that the coefficients of 
$x^2 y$ and $ x y^2$ transform in ${\cal O}$ of $B$. It has 
been noted in~\cite{Cvetic:2013uta} that at the expense of 
making all sections rational, one can allow the coefficients 
$a_0$ and $b_0$ instead to transform a restricted set of ample 
line bundles over $B_n$.  

On the $E_6\times U(1)^2$ fiber the functions which determine the discriminant $\Delta=f^3 -27 g^2$ are given by
\begin{eqnarray}
f&=&\frac{1}{12} (b_1^4 - 8 b_1^2 (a_2 + b_2 + a_1 c_1) + 24 b_1 (a_3 + a_1 b_2 + a_2 c_1) + \\
   &&16 (a_2^2 - a_2 b_2 + b_2^2 - 3 a_3 c_1 + a_1^2 c_1^2 - 
      a_1 (3 a_3 + (a_2 + b_2) c_1)))\, ,\nonumber \\
 \end{eqnarray}  
\begin{eqnarray}
      g&=& \frac{1}{216} (b_1^6 - 12 b_1^4 (a_2 + b_2 + a_1 c_1) + 
   36 b_1^3 (a_3 + a_1 b_2 + a_2 c_1) + \nonumber \\
   && 
   24 b_1^2 (2 a_2^2 + a_2 b_2 + 2 b_2^2 - 3 a_3 c_1 + 2 a_1^2 c_1^2 + 
      a1 (-3 a_3 + (a_2 + b_2) c_1)) -  \nonumber \\
 &&  144 b_1 (a_1^2 b_2 c_1 + (a_2 + b_2) (a_3 + a_2 c_1) + 
      a1 (b_2 (a_2 + b_2) - 5 a_3 c_1 + a_2 c_1^2))\nonumber \\
      && - 
   8 (8 a_2^3 - 27 a_3^2 - 12 a_2^2 b_2 - 12 a_2 b_2^2 + 8 b_2^3 + 
      18 a_3 (a_2 - 2 b_2) c1 - 27 a_2^2 c_1^2 + 8 a_1^3 c_1^3 + \nonumber \\
    &&  6 a1 (3 a_3 (-2 a_2 + b_2) - (2 a_2^2 + a_2 b_2 + 2 b_2^2) c1 + 
         12 a_3 c1^2) \nonumber\\
         && - 3 a_1^2 (9 b_2^2 - 24 a_3 c_1 + 4 (a_2 + b_2) c_1^2))).  \nonumber
\label{E6fg}
         \end{eqnarray}        
         
Let us now see the realization of codimension 1 singularities in the $E_6\times U(1)^2$ fiber.
In table \ref{E6U1vanish} 
we have given a list of singularities 
realized in the fiber with a
given vanishing order for the coefficients. In the table are also written the properties
of a toric 3-fold (4-fold) constructed using a base $\tilde{B}_2=\mathbb{P}^2$ ($B_3=\mathbb{P}^3$)
in which the codimension 1 singularities are realized for a given vanishing order of the
coordinates $u_i, i=1,...,3$ ( $i=1,...,4$). As an example let us look at the $A_4$ singularity,
which is one of the rows. The vanishing order for the variable $u_1$ is given by $1, 0, 2, 1, 2, 2$, the properties of the
3-fold  and 4-fold are respectively $\chi_3=-134$, $h^3_{11}=8(0)$, $h^3_{31}=75(0)$, and $\chi_4=3168$,
$h^4_{11}=8(0)$, $h^4_{31}=512(0)$, $h^4_{21}=0$.

Keeping the same basis for the 3-fold and 4-fold,  let us impose an $SU(2)$ singularity 
 at $u_1=0$ and an $SU(3)$ singularity at $u_2=0$. Consider the vanishing order for the coefficients $0,0,1,0,0,1$ at $u_1=0$ and
$1,0,1,1,1, 1$  at $u_2=0$. The properties of the 3-fold and 4-fold that embed the mentioned singularities  are given respectively by
$\chi_3= -140$, $h^3_{11}=7(0)$, $h^3_{21}=77(0)$ and $\chi_4=3180$, $h^4_{11}=7(0)$, $h^4_{31}=515(0)$ and $h^4_{21}= 0$.

Let us look at an example where the 3-fold  and 4-fold are constructed to accommodate heterotic duals, $B_2$ given by the polytope 7 of Figure \ref{poly}, then $B_3=\mathbb{P}^1 \times B_2$. The $\tilde{B}_2$ defining the 3-fold is given by $\tilde{B}_2=\mathbb{P}^1\times \mathbb{P}^1$. The 4-fold (3-fold) will have an $SU(5)$ singularity with vanishing order of  the $\mathbb{P}^1$ variable $u_1$ given by $1, 0, 2, 1, 2, 2$. The properties of the 3-fold will be 
$\chi_3=-128$, $h^3_{11}=9(0)$, $h^3_{21}=73(0)$, and the properties of the
fourfold $\chi_4=2226$, $h^4_{11}=12(0)$, $h^4_{31}=351(0)$ and $h^4_{21}=0$.

Let us look now at the case keeping $B_2$ and constructing the base of the 4-fold as a fibration, with a defined twist $B_3=\mathbb{P}^1\rightarrow B_2$.  The fibration is chosen to be the same as in previous section with $m_1=-1, \, m_i=0, i=2,...,6$. The properties of the 4-fold in this case are given by
$\chi_4=2094$ $h^4_{11}=12(0)$, $h^4_{31}=329(0)$ and $h^4_{21}=0$.

\subsubsection{The extremal heterotic bundle configurations} 
\label{extremalbundle}
So far we have chosen the topology of the heterotic 
gauge bundle in section \ref{heteroticbasis} so that 
the heterotic gauge bundle can be generic enough to 
break both $E_8$'s  i.e. $n=0,1,2$ in (\ref{defn}) 
for the three  dimensional case.

The point $n_F^*$ is defined in (\ref{polyhedrafrombaseandfibre}), 
i.e, for the $E_8$-fibre $n_F^*=(-2,-3)$. 
The embedded base $\mathbb{F}_n$ is bounded by
points $\{u_1=(n_F^*,0,1),u_2=(n_F^*,0,-1),w_1=(n_F^*,1,0),
w_2=(n_F^*,-1,-n)\}$ and the gauge symmetry enhancement occurs  
at the divisor $u_2=0$ in the base, with the 
expected values of the Tate forms. For the extreme 
value $n=12$ one gets the manifold with largest 
known absolute value of the Euler number for $M_3$($W_3$), 
i.e. the point in the extremal north west of the Hodge 
plots.  It has $\chi=-960$, $h_{11}=11$ and 
$h_{21}=491$. This model is the one with the most 
generic bundle $c(V_1)=24$ and $c_2(V)=0$ in the other $E_8$ which can be 
enhances at $u_1=0$ explaining the extremal number 
of deformations in $H^1(M_3,T_{M_3})\sim H_{2,1}(M_3)$. 
Only the cases $n=3,4,5,6,9,12$ can  be compactified torically  
in the way described in~\cite{Morrison:1996pp}.    

We can make the interesting question:
what is the analogue of the maximal 
twisting $n=12$  in the four dimensional case? Here we can 
give the answer for an arbitrary $\mathbb{P}^1$ twisted by $m$  
over $\mathbb{F}_n$, called $\mathbb{F}_{m,n}$. We get an embedding of 
the points of $\mathbb{F}_{m,n}$ by
$\{u_1=(n_F^*,0,0,1),u_2=(n_F^*,0,0,-1),v_1=(n_F^*,0,1,0),v_2=(n_F^*,0,-1,0), 
w_1=(n_F^*,1,0,0), w_2=(n_F^*,-1,-m,-n)\}$. The maximal twisting 
values again for the $E_8$ fibre $n_F^*=(-2,-3)$ fibre are 
$(m=84,n=516)$. Then we have trivial bundle $V_2$ and a $E_8$  
gauge symmetry enhancement at $u_2=0$. This Calabi-Yau 
fourfold is the one with the  largest known absolute 
value of the Euler number for $M_4$($W_4$) $\chi=1820448$ 
and $h_{31}= 303148$, $h_{21}=0$ and remarkably 
$h_{11}=252$. Given the fact that we have three $h_{11}$ 
classes from the base and one from the elliptic fibre, 
it suggest that all the $248$ curves that one finds 
in the $E_8$ fibre, cff~\ref{supokw} have now become 
isolated rational curves in the toric Mori cone of this very 
particular  fourfold.   

\begin{table}[hbt]
\begin{center}
\begin{tabular} {c|ccccc|ccccc|}
& \multicolumn{10}{c}{$E_8$}\\
\hline
&\multicolumn{5}{c}{$A$}& \multicolumn{5}{c}{$B$}  \\ 
\hline 
d& $\chi$ &$h_{11}$& $h_{21}$& $h_{31}$& $h_{22}$& $\chi$ &$h_{11}$& $h_{21}$& $h_{31}$& $h_{22}$ \\
\hline 
3& $-960$ &$11(0)$& $491(0)$& $-$& $-$& $0$ &$251$& $251$& $-$& $-$ \\
4& $1820448$ &$252$& $0$& $303148$& $1213644$& $1820448$ &$151700$& $0$& $151700$& $1213644$ \\
\hline
\end{tabular}
\caption{F-theory duals for the extremal heterotic bundle configurations for the $E8$ fibre}
\label{e8extremal}
\end{center}
\end{table}

\begin{table}[hbt]
\begin{center}
\begin{tabular} {c|ccccc|ccccc|}
& \multicolumn{10}{c}{$E_7$}\\
\hline
&\multicolumn{5}{c}{$A$}& \multicolumn{5}{c}{$B$}  \\ 
\hline 
d& $\chi$ &$h_{11}$& $h_{21}$& $h_{31}$& $h_{22}$& $\chi$ &$h_{11}$& $h_{21}$& $h_{31}$& $h_{22}$ \\
\hline 
3& $-432$ &$11(4)$& $227(0)$& $-$& $-$& $0$ &$119(40)$& $119(0)$& $-$& $-$ \\
4& $201600$ &$120$& $0$& $33472$& $134412$& $201600$ &$16796$& $0$& $16796$& $134412$ \\
\hline
\end{tabular}
\caption{F-theory duals for the extremal heterotic bundle configurations for the $E7$ fibre}
\label{e8extremal}
\end{center}
\end{table}

\begin{table}[hbt]
\begin{center}
\begin{tabular} {c|ccccc|ccccc|}
& \multicolumn{10}{c}{$E_6$}\\
\hline
&\multicolumn{5}{c}{$A$}& \multicolumn{5}{c}{$B$}  \\ 
\hline 
d& $\chi$ &$h_{11}$& $h_{21}$& $h_{31}$& $h_{22}$& $\chi$ &$h_{11}$& $h_{21}$& $h_{31}$& $h_{22}$ \\
\hline 
3& $-240$ &$11(6)$& $131(0)$& $-$& $-$& $0$ &$71(36)$& $71(0)$& $-$& $-$ \\
4& $44928$ &$72$& $0$& $7408$& $29964$& $44928$ &$3740$& $0$& $3740$& $29964$ \\
\hline
\end{tabular}
\caption{F-theory duals for the extremal heterotic bundle configurations for the $E6$ fibre}
\label{e8extremal}
\end{center}
\end{table}

These extremal Calabi-Yau n-folds constitute an infinite series 
of  Calabi-Yau Fermat hypersurfaces $M^A_n$, $n=1,\ldots$, which  
are related to another series of Calabi-Yau Fermat hypersurfaces 
$M^B_n$. A member of the latter is given by the degree $d=m_{n+2}$ constraint  
$\sum_{i=1}^{n+2} x_i^{m_i}=0$ in weighted projective space $\mathbb{P}^{n+1}(d/m_1,\ldots d/m_{n+1},1)$. 
For the series $M^B_n$ we can give a simple induction law over the 
dimension in that $M^B_{n+1}$ is defined by the exponents  
$\{m_1,\ldots,m_{n+1},(m_{n+2}+1),lcm(m_1,\ldots,m_{n+1},(m_{n+2}+1))\}$ 
with the starting manifold $M_1^B$ given for the $E_8$ fibre type 
as $x_1^2+x_2^3+x_3^6=0$, i.e. just an elliptic curve. 
The next case is an K3 given by $x_1^2+x_2^3+x_3^7+x_4^{42}=0$, 
which contains Arnold's self-dual strange singularity 
$E_{12}$ at $x_4=0$ and leads to an K3 hypersurface in 
$\mathbb{P}_{\Delta}$~\cite{MR0429876,MR2426805,MR2278769}. 

The series $M^A_n$ is then obtained by fibering  $M^B_{n-1}$ over $\mathbb{P}^1$.
Their Fermat polynomial is therefore written in terms of the exponents of 
$M^B_{n-1}$ as $\sum_{i_1}^{n} x_i^{m_i} + x_{n+1}^{2 m_{n+2}}+ x_{n+2}^{2 m_{n+2}}=0$. 
E.g. $M_3^A=\{x_1^2+x_2^3+x_3^7+ x_4^{84} + x_5^{84}=0| {\underline x}
\in \mathbb{P}^4(42,28,12,1,1)\}$. The mirror of it sets other remarkable 
records. It has the biggest known gauge group in six dimensions $E_8^{17}\times F_4^{16}
\times G_2^{32}\times SU(2)^{32}$  of rank  $296$ as well as with $193$ the highest known number of 
tensor multiplets, i.e. blow-ups in the base, for a 6d theory~\cite{CPR}. The K\"ahler classes corresponding to them  
as well as the two classes K\"ahler classes from the overall volume of the base and the class of 
the fibre account for the $491$ K\"ahler classes of the mirror.  

Similar series exist for the $E_7$ 
and $E_6$ fibre types, by the same induction starting however with $M_1$ given by the 
elliptic curves $x_1^2+x_2^4+x_3^4=0$ and $x_1^3+x_2^3+x_3^3=0$ respectively.    

The members of the series $M_n^B$ have\footnote{For the K3 he have  
$h^{B\ ver}_{1,1} = h^{B\ hor}_{1,1}=10$, $h^{A \ ver}_{1,1}=2$ and $h^{A\ hor}_{1,1}=18$, i.e. ${\cal S}_A$ .}
\be 
h^B_{1,1} = h^B_{n-1,1}=\frac{h^A_{1,1}+h^A_{n-1,1}}{2}
\ee 
and for the $E_8$ fibre type the highest know total sum 
$\sum_{n,m} h^B_{n,m}$ of Hodge numbers for dimension $d>2$. 
The first members of the series have $\chi(M^B_{2n+1})=0$.

\section{$SU(5)$ models with an $E_8$ Yukawa Point}\label{sec-E8yukawa}
\label{SU5andE8yukawa}  

In this section we analyze models with an $SU(5)$ gauge group which contain
a so called $E_8$ Yukawa point, given by the vanishing orders $ord(f)\geq 4$, $ord(g)=5$ and $ord(\Delta)=10$. 
We want to follow the route of \cite{Heckman:2009mn} to sensible F-theory phenomenology on compact 
Calabi-Yau spaces. There will be a wide class realized by modifying $k,n,m$ 
or ${\Delta^{(i)}}$  and $n_1,\ldots, n_m$ as a starting 
point to achieve matter curves and an $E_8$ point at 
codimension three in a compact case, in particular the examples in Appendix \ref{tate} for 
fourfolds with $B_3=\mathbb{P}^1\times\mathbb{P}^2$ with an $SU(5)$ gauge 
symmetry. Given the models summarized  in table \ref{table:kodaira}, 
there are two strategies either restricting  a model 
with $D_G$ $G\in E_8$ or deforming a model with  $D_{E_8}$. 
We chose the first one,
taking as departure a model with an $SU(5)$ singularity.        

First we describe how  a codimension 3 singularity with vanishing 
order of the coefficients to give an $E_8$ is obtained in the 
global construction described so far. Then we recall the $SU(5)$ F-theory  
GUT models using the notation  which is present in the literature. 
Further, we describe the method of the spectral cover used to parametrize 
the $E_8$ singularities we plan to study. Then we come to our focus which 
is the resolution of two models which posses a potentially $E_8$ singularity: 
Case 1 and Case 2 in order to analyze the consequences of the non-applicability 
of Kodaira classification to singularities with codimension higher than one 
\cite{Esole:2011sm}. For each of this cases we obtain fully resolved spaces, 
and study the the codimension 2 and 3 loci and the gauge 
group charges of the different matter curves.

\subsection{A codimension three $E_8$ point from the global construction}
\label{embedE8}

Consider the global toric construction employed for the tables in appendix \ref{tate}. It has polytope 1 in figure \ref{poly} and no twisting, giving $B_3=\mathbb{P}^1\times \mathbb{P}^2$. In this base it is possible to impose the existence of certain codimension 3 singularities by fixing some moduli.
We have search for a codimension 3 $E_8$ singularity, such that the vanishing of the coefficients in the vicinity
of the singular locus are the ones in Table \ref{table:tatenonsplit}.

Consider the variables of the base  given by $v,w,u_1,u_2,u_3$. Let us choose the 4-fold with
a codimension 1 singularity $SU(5)$ at $v=0$.   This means that we obtain in (\ref{WDelta}) the following dependence of the coefficients
\begin{eqnarray}
a_1&=f_1(u_i,w,c_j)+O(v)g_1(u_i,w,v,c_j),\, &a_2=v f_2(u_i,w,c_j)+O(v^2)g_2(u_i,w,v,c_j),\,  \\
a_3&=v^2f_3(u_i,w,c_j)+O(v^3)g_3(u_i,w,v,c_j),\,& a_4=v^ 3f_4(u_i,w,c_j)+O(v^4)g_4(u_i,w,v,c_j),\,  \nonumber \\
a_6&=v^5 f_6(u_i,w,c_j)+O(v^5)g_6(u_i,w,v,c_j). \nonumber 
\end{eqnarray}

We searched for a singularity in codimension-3 with the vanishing order for an $E_8$
in the locus $v\rightarrow t, u_1+u_2\rightarrow t,u_3\rightarrow t$ (parametrized via $u_1\rightarrow  p + t, u_2 \rightarrow -p$), with $t\rightarrow 0$. This  translates in having
\begin{eqnarray}
a_1&=t l_1(p,w,c_j)+O(t^2)m_1(p,w,c_j),\, &a_2=t^2 l_2(p,w,c_i)+O(t^3)m_2(p,w,c_j),\,  \\
a_3&=t^3l_3(p,w,c_j)+O(t^4)m_3(p,w,c_j),\,& a_4=t^ 4l_4(p,w,c_j)+O(t^5)m_4(p,w,c_j),\,  \nonumber \\
a_6&=t^5 l_6(p,w,c_j)+O(t^6)m_6(p,w,c_j). \nonumber 
\end{eqnarray}
This is achieved by looking at the pieces of the polynomial involving different vanishing order for $t$. Let us look at  the terms of order $t^0$ in $a_1$. The moduli that have to be fixed are
the coefficients of equation
\begin{eqnarray}
(w^2 \alpha_3 u_1^3 + w^2 \alpha_5 u_1^2 u_2 + w^2 \alpha_7 u_1 u_2^2 + w^2 \alpha_{10} u_2^3)
\rightarrow   p^3 w^2 (-\alpha_3+ \alpha_5- \alpha_7+ \alpha_{10})\, ,
\end{eqnarray}
so we need to fix $\alpha_3\rightarrow  \alpha_5- \alpha_7+ \alpha_{10}$. For the coefficient $a_2$
the terms of order $t$ will be
\begin{eqnarray}
&&(v w^3 \beta_4 u_1^6 + v w^3 \beta_8u_1^5 u_2 + 
 v w^3 \beta_{12} u_1^4 u_2^2 + v w^3 \beta_{16} u_1^3 u_2^3 + \\
&& v w^3 \beta_{20} u_1^2 u_2^4 + v w^3 \beta_{24} u_1 u_2^5 + \nonumber
 v w^3 \beta_{28} u_2^6) \\
 && \rightarrow  t p^6 w^3 (\beta_4- \beta_8 + \beta_{12} - \beta_{16} + 
\beta_{20} -  \beta_{24} +  \beta_{28})\, ,\nonumber
 \end{eqnarray}
so we need to fix $\beta_4\rightarrow \beta_8-\beta_{12} + \beta_{16} - \beta_{20} + \beta_{24} -\beta_{28}$. For the coefficient $a_3$ the terms of order $t^2$ are given by
\begin{eqnarray}
&&v^2 w^4\gamma_5u_1^9 + v^2 w^4 \gamma_{10} u_1^8 u_2 + 
 v^2 w^4 \gamma_{15} u_1^7 u_2^2 + v^2 w^4 \gamma_{20} u_1^6 u_2^3 + 
 v^2 w^4 \gamma_{25}u_1^5 u_2^4 +  \\
&&v^2 w^4 \gamma_{30} u_1^4 u_2^5 + v^2 w^4 \gamma_{35} u_1^3 u_2^6 + v^2 w^4 \gamma_{40} u_1^2 u_2^7 + 
 v^2 w^4 \gamma_{45} u_1 u_2^8 + v^2 w^4 \gamma_{50} u_2^9  \nonumber\\
 &&\rightarrow t^2 p^9 w^4( \gamma_5- \gamma_{10} +  \gamma_{15} - \gamma_{20} + 
  \gamma_{25} -  \gamma_{30} + \gamma_{35} - \gamma_{40} + 
\gamma_{45} - \gamma_{50}). \nonumber
\end{eqnarray}
And the moduli should be fixed to make this last term vanish. For the coefficient $a_4$ the following terms contribute to the order $t^3$ to give
\begin{eqnarray}
&& v^3 w^5 (\delta_6 u_1^{12} + \delta_{12} u_1^{11} u_2 + 
   \delta_{18} u_1^{10} u_2^2 + \delta_{24} u_1^9 u_2^3 + 
   \delta_{30} u_1^8 u_2^4 + \delta_{36} u_1^7 u_2^5 + \\
 &&  \delta_{42} u_1^6 u_2^6 + \delta_{48} u_1^5 u_2^7 + 
   \delta_{54} u_1^4 u_2^8 + \delta_{60} u_1^3 u_2^9 + 
   \delta_{66} u_1^2 u_2^{10} + \delta_{72} u_1 u_2^{11} + 
   \delta_{78} u_2^{12}) \nonumber \\
&& \rightarrow t^3p^{12} w^5 (\delta_6 - \delta_{12} + \delta_{18} - \delta_{24} + \delta_{30}- \delta_{36} + \delta_{42} - 
   \delta_{48} + \delta_{54} -\delta_{60} + \delta_{66} -\delta_{72} + \delta_{78}), \nonumber
\end{eqnarray}
which has to vanish by setting the sum of the coefficients $\delta$ to zero. 
By fixing the moduli we have arrived then to a 4-fold with a 
codimension 3 singularity which has the vanishing order of
the so called $E_8$.

However in the given example we have fixed the moduli in excess. 
Let us argue that a more satisfying solution with less relations 
constraining the moduli exists. We could impose the vanishing of two 
coefficients at the required order, for example we can require $f_1=f_2=0$ 
and search for solutions in terms of $u_1$ and $u_2$ (setting $u_3=1$, 
using the fact that $(u_1, u_2, u_3)\in \mathbb{P}^2$), we obtain 
the dependence $u_1(c_{1j},c_{2j})$ and $u_2(c_{1j},c_{2j})$. Having 
these solutions it will be sufficient to achieve the required vanishing 
order to impose the following two relations among the moduli 
\begin{equation}
f_3(u_1(c_{1j},c_{2j}), u_2(c_{1j},c_{2j}),\omega,c_{3j})=f_4(u_1(c_{1j},c_{2j}), u_2(c_{1j},c_{2j}),\omega,c_{4j})=0.
\end{equation} 
From the previous equations it is possible to eliminate the dependence on $w$, having two relations which
depend only on the moduli $c_{1j}, c_{2j}, c_{3j}$ and $c_{4j}$. The vanishing order of the coefficients at $v=0$ and $u_{1,2}$ fixed like $u_{1,2}(c_{1j},c_{2j})$ are given by $1,2,3,4,5$, ensuring a so called $E_8$ at codimension 3. General solutions of this kind can be found in the Calabi-Yau fourfolds we constructed. 


\subsubsection*{Codimension 3 $E_8$ singularities from the spectral cover}

In this section we use the spectral cover construction, and in particular the breaking of the $E_8$ coming
from the decomposition in $SU(5)\times SU(5)_{\perp}$ by turning the values
of the Higgs field $t_i$ to be different from zero. When the
$t_i \rightarrow 0$ for all $i$ the $E_8$ is recovered. Let us employ the relations
for the coefficients of the fibration in terms of the $t_i$s given in (\ref{betas})
and the parametrization for the $t_i$ in terms of the variables $p, q$ given in Case 1
from equation (\ref{E8model-case1}), one obtains that
\begin{eqnarray}\label{case1}
&&a_1= \beta_0  (-4 p^3q^2 -4 q^4 p-10p^2 q^3), a_2=\beta_0  (2 p^3q + 11 p^2 q^2 + 10 p q^3 + 4 q^4) w,\\
&&a_3=\beta_0 (2p^3 + 2p^2 q -2 p q^2) w^2, a_4=\beta_0 (-3 p^2 -4 p q - 5 q^2) w^3. \nonumber\\
&&a_6=\beta_0 w^5.\nonumber
\end{eqnarray}
At the codimension 3 locus determined by $w=p=q=0$ the vanishing order of the coefficients
$a_1,a_2,a_3,a_4,a_6$ is given by $5,5,5,5,5$. This a case for 
the $E_8$ singularity, different to the previously discussed with the vanishing order 1,2,3,4,5. With
the values given in (\ref{case1}) one can check that $ord(f)=5\geq 4$, $ord(g)=5$ and $ord(\Delta)=10$ which according to Table \ref{table:kodaira} and \cite{Esole:2011sm,Bershadsky:1996nh} constitutes an
$E_8$.

We have explored various of the Calabi-Yau constructions
presented and we found examples where in a given patch,
with a fine tuning of the moduli this type of singularity appears. One of this cases
is to have a model with an heterotic dual where $B_3$ is constructed as a fibration of
$\mathbb{P}^1$ over $\mathbb{P}^2$ (polytope 1), with twist parameter  $m_1=-2$
and other twists vanishing $m_k=0$. In this case the coordinates of the $\mathbb{P}^1$ are denoted
by $v,w$ and the coordinates of the $\mathbb{P}^2$ by $u_1,u_2,u_3$.
Using the notation $u_1=p$, $u_2=q$ and setting many of the coefficients
$c_{kj}$ to zero, one finds
\begin{eqnarray}\label{sfix}
&&a_1=v^2 (c_{11} p^3 q^2+c_{12}p^2 q^3+c_{13}p q^4),a_2=v^3u_3^4(c_{21}p^3 q+c_{22}p^2 q^2+c_{23} p q^3+c_{24}q^4)w,\\
&&a_3=v^4u_3^8 (c_{31}p^3+c_{32} p^2 q+c_{33} p q^2)w^2,a_4=v^5u_3^{12}(c_{41} p^2+c_{42} p q+c_{43} q^2)w^3,\nonumber\\
&&a_6=v^7 u_3^{20}c_{61}w^5.\nonumber
\end{eqnarray}
We have omitted the higher order terms. For a patch in which $v=u_3=1$ (\ref{sfix})  coincides with (\ref{case1}). 
One could explore other possibilities, for example with $B_3=\mathbb{P}^1\rightarrow \mathbb{P}^2$, 
twist $m_1=-3$ and $m_k=0$ this singularity kind can be obtained. 
However the polyhedra obtained from this tuning are not reflexive, and we have lost the
global description. 

We have also explored the singularity type of  Case 2 discussed in section
\ref{case2}. This case it is also realized with $B_3=\mathbb{P}^1\rightarrow \mathbb{P}^2$, for 
$m_1=-2$ and other twists $m_k=0$.  But the same difficulty of the lost of reflexivity is
present. We plan to consider in future work a detailed exploration of the global models
we have constructed to achieve a satisfactory global embedding of the codimension 3 
$E_8$ singularity and other singularities relevant to a 4d effective theory.


\subsection{The $SU(5)$ gauge group models}
\label{su5models}

The equation for the elliptic fibration with coordinates $x, y, z$ of a $\mathbb{P}^2$ bundle, in the patch $z=1$, and the base divisor $w=0$ defining a codimension 1 $A_4$ singularity is given by
\begin{equation}\label{tateform2}
 - y^2  + x^3 + \beta_0  w^5 + \beta_2 x w^3 
 + \beta_3 y w^2 + \beta_4 x^2  w + \beta_5  x y = 0 ,\footnote{Here we change the notation with respect to the one
in the equation of an $E_8$ fibre (\ref{tateform}). We do that in order to ease the comparison with the literature \cite{Esole:2011sm,Marsano:2011hv}.}
\end{equation}
in the Tate form (\ref{tateform}), where the $a_i$ coefficients in (\ref{ai}) have been replaced by $\beta_{6-i}w^{i-1}$.  Comparing to the Weierstrass form \ref{Weierstrass},  we can identify the discriminant $\Delta$ of the elliptic fibre and the functions $f$ and $g$.
\begin{align}\label{Deltafg}
\Delta \equiv & -16 (- f^3 + 27g^2) = w^5 \Delta' , \nn \\
f = &\frac{1}{48\times  ^3\sqrt{4}} (-\beta_5^4 - 8 \beta_4 \beta_5^2 w - 16 \beta_4^2 w^2 + 24 \beta_3 \beta_5 w^2 + 48 \beta_2 w^3) ,   \\
g = &\frac{-1}{864} (\beta_5^6 + 12 \beta_4 \beta_5^4 w + 48 \beta_4^2 \beta_5^2 w^2 - 36 \beta_3 \beta_5^3 w^2 +  \nn \\
  &  + 64 \beta_4^3 w^3 - 144 \beta_3 \beta_4 \beta_5 w^3 - 72 \beta_2 \beta_5^2 w^3 + 216 \beta_3^2 w^4 - 
   288 \beta_2 \beta_4 w^4 + 864 \beta_0 w^5)   \nn ,
\end{align}


with 
\begin{equation}\label{Delta'}
 \Delta' = \beta_5 P + w \beta_5 (8 \beta_4 P + \beta_5 R) + w^2 (16 \beta_3^ 2 \beta_4^3 + \beta_5 Q ) 
+ w^3 S + w^4 T + w^5 U ,
\end{equation}
where $P,R,Q,S,T$ and $U$ are polynomials related to  $\beta_i$ and $w$ as 
\begin{align}
P &=\beta^2_3 \beta_4 -\beta_2 \beta_3 \beta_5 + \beta_0 \beta_5^2, \quad 
R=4 \beta_0 \beta_4\beta_5-\beta^3_3 -\beta^2_2 \beta_5
,     \\
Q &=-2(18 \beta _3^3 \beta _4+8 \beta _2 \beta _3 \beta _4^2-15 \beta _2 \beta _3^2 \beta _5+4 \beta _2^2 \beta _4 \beta _5-24 \beta _0 \beta _4^2 \beta _5+18 \beta _0 \beta _3 \beta _5^2),  \nonumber \\
S &=27 \beta _3^4-72 \beta _2 \beta _3^2 \beta _4-16 \beta _2^2 \beta _4^2+64 \beta _0 \beta _4^3
  +96 \beta _2^2 \beta _3 \beta _5-144 \beta _0 \beta _3 \beta _4 \beta _5-72 \beta _0 \beta _2 \beta _5^2,
  \nonumber   \\
T &=8 \left(8 \beta _2^3+27 \beta _0 \beta _3^2-36 \beta _0 \beta _2 \beta _4\right),\quad 
U =432 \beta _0^2.   \nonumber 
\end{align}

So along GUT divisor $w=0$, for general values of $\beta_i$s
and as long as $\Delta'$ remains non-vanishing, 
the vanishing orders of $\Delta$, $f$ and $g$
as we approach $w \ra 0$ are
\begin{equation}
 \ord(\Delta) = 5 , \qquad \ord(f) = 0 , \qquad \ord(g) = 0 .
\end{equation}
and correspond  to a singular curve of type $I_5$, 
thus an $A_4$ (or $SU(5)$) singularity.

Notice also that if we just fix all the parameters $\beta_i$ to zero except $\beta_0$, 
we obtain an $E_8$ singularity.
In fact, such an $SU(5)$ model described by \eqref{tateform2} can be seen as
a higgsing of an $E_8$ down to $SU(5)$,
\begin{equation}
 E_8 \rightarrow SU(5) \times SU(5)_\perp ,
\end{equation}
and the vevs for the Higgs are related to the sections $\beta_i$.
In the Type IIB picture, this can be interpreted locally as a stack of five
D7-branes separated by some distance (encoded by the vev of a Higgs field),
to other branes. As we move them close together the gauge group enhances.
However, from the pure perturbative description we cannot reproduce 
an $E_8$-brane. If however we allow ourselves to deepen in the 
strong coupling regime, we need to include (p,q)-7-branes, and open strings
that attach to 3 or more branes. These unusual
open strings configurations can reproduce 
an $E_8$ algebra \cite{Gaberdiel:1997ud, DeWolfe:1998zf}.


In the following, we proceed to resolve the curve explicitly. 
First notice that (\ref{tateform2}) is singular when $w=x=y=0$. 
The $\beta_i$s depend on the coordinates 
on the base of the elliptic fibration $B_3$. 
They define holomorphic sections of the
 bundle $\cO([6-i]K_{B_3} - [5-i]S_{\text{GUT}})$,
where $K_{B_3}$ is the canonical bundle on the base $B_3$. 
We perform the first blow up $\lambda_1 : [y,x,w]$, 
that is, introducing a coordinate $\lambda_1$ as
\begin{equation}
 y = \lambda_1 \hat y \, , 
 \qquad x = \lambda_1 \hat x \, , \qquad w = \lambda_1 \hat w \, .
\end{equation}
together with the projective relations $[\hat y :\hat  x :\hat  w]$ .
The defining equation \eqref{tateform2} becomes
\begin{equation}
 \lambda_1^{ 2} (- \hat y^{ 2} + \lambda_1 \hat x^{ 3} + \beta_0 \lambda_1^{ 3} \hat w^{ 5} 
+ \beta_2 \lambda_1^{ 2} \hat x\hat  w^{ 3}
+ \beta_3 \lambda_1 \hat y \hat w^{ 2} + \beta_4 \lambda_1 \hat x^{ 2} \hat w^{} + \beta_5 \hat x \hat y) = 0 .
\end{equation}
The expression in brackets is the proper transform that 
defines $\hat Y_4$ (we reserve the symbol $\tilde Y_4$ to the 
fully resolved space), and the new homology class is given by
\begin{equation}
 Y_4 \ra \hat Y_4 + 2 E_1 ,
\end{equation}
where $E_1 = \{ \hat \lambda_1 \} $.  $\hat Y_4$ is still singular in $\hat y=\hat x=\hat \lambda_1$, and we blow up (we will reuse hatted coordinates to denote all the intermediate blow ups, 
to avoid adding too many symbols and because we are mainly interested in 
the original space and the final resolved one),
\begin{equation}
- \hat y^{ 2} + \lambda_2 \hat \lambda_1 \hat x^{ 3} 
+ \beta_0 \lambda_2 \hat \lambda_1^{ 3} \hat w^{ 5} 
+ \beta_2 \lambda_2 \hat \lambda_1^{ 2} \hat x \hat w^{ 3}
+ \beta_3 \hat \lambda_1 \hat y \hat w^{ 2} + \beta_4 \lambda_2 \hat \lambda_1 \hat x^{ 2}\hat  w + \beta_5 \hat x \hat y = 0 ,
\end{equation}
or, rearranging,
\begin{equation}
- \hat y \left( \hat y+ \beta_3 \hat \lambda_1 \hat w^{ 2} + \beta_5 \hat x \right)
+ \lambda_2 \hat \lambda_1 \left(\hat  x^{ 3} + \beta_0 \hat \lambda_1^{ 2}\hat  w^{ 5} 
+ \beta_2 \hat \lambda_1 \hat x \hat w^{ 3}
 + \beta_4 \hat x^{ 2} \hat w \right) = 0 .
\end{equation}

The space is smooth in general points of the GUT divisor, 
but it acquires further 
singularities when we approach 
particular values for the $\beta_i$s, 
corresponding to subregions on $S_{\text{GUT}}$. 
The resolution of these singularity 
enhancements were performed via small resolutions in the work of 
Esole and Yau \cite{Esole:2011sm} and reviewed in a series
of other papers \cite{Grimm:2011tb,Braun:2011zm,Marsano:2011hv}.
These enhancements lead to fiber structures that reproduce the 
fundamental and the antisymmetric representations $\mathbf{5}$ and 
$\mathbf{10}$ of the $SU(5)$ gauge group, the
two representations required to construct matter in realistic
$SU(5)$ models.  Furthermore, there are points on the base 
(codimension three) where the singularities enhance even further,
corresponding to places where matter curves intersect. These points
reproduce the couplings $\mathbf{10 \, 10 \, 5}$ and $\mathbf{10 \, \bar 5 \, \bar 5}$,
required to reproduce the Yukawa couplings between quarks and the 
Higgs field.


In particular, at the point on the base defined by $w=\beta_4=\beta_5=0$,
when these coordinates are treated as non-factorizable holomorphic variables,
 the explicit resolution does not give for the blown-up 
 $\IP^1$s the geometric structure
corresponding to the Dynkin diagram of an $E_6$ group, as expected from 
the counting of the vanishing order as in table \ref{table:kodaira}. Further 
analysis \cite{Marsano:2011hv} showed that although the resolution does 
not reproduce the exact diagram that would be naively expected from 
the Tate's algorithm, it still reproduces the \textbf{10 10 5} coupling, 
necessary to give mass to the top quark in an $SU(5)$ GUTs, the main 
reason for why one considers $E_6$ enhancements in the first place.

However, Esole-Yau resolution \cite{Esole:2011sm} 
does not contemplate further singularities, 
that can arise in particular regions on the moduli space of the base where 
the parameters $\beta_i$ factorize. Such possibility of factorization 
was studied in other contexts 
for example in \cite{Braun:2011zm, Krause:2011xj}.

We also propose a factorization to reproduce a codimension 3 (a point 
on the base) enhancement to an $E_8$ singularity.
Our construction of such splitting follows from the established connections 
between F-theory and heterotic string theory, where the coefficients $\beta_i$ 
are related to the higgsings used to break $E_8$ down to $SU(5)$, as
\begin{equation}
 E_8 \longrightarrow SU(5) \times SU(5)_{\perp} .
\end{equation}
We will briefly review how are the coefficients in the elliptic fiber
related to the Higgs vevs in the next section, following \cite{Donagi:2009ra,Dudas:2010zb}.

\subsection{The $\beta_i$ Coefficients from the Spectral Cover}
\label{co3spectral}

The rough idea of the spectral cover construction 
is to incorporate in a geometrical
description the higgsing of a gauge group.
Our starting point is the $E_8$ group, that we
break down to an SU(5),
\begin{equation}
 E_8 \rightarrow SU(5) \times SU(5)_{\perp} \rightarrow SU(5) \times U(1)^4 .
\end{equation}
There is a Higgs field responsible 
for the breaking, which can be locally described 
as a section of the canonical bundle over the 
SU(5) divisor $S$ with values on the adjoint of $E_8$,
\begin{equation}
 K_S \otimes Adj(E_8) .
\end{equation}
In standard geometrical engineering, 
the gauge groups are identified with 
singularities as the standard ADE 
classification, obtained as a blow-down
of the resolved geometry. They are then broken to smaller subgroups 
by giving non-vanishing volume to some $\IP^1$'s. This corresponds to 
giving a vev to the Cartans of $SU(5)_\perp$. Thus, being a Cartan root, 
the Higgs field we are interested in obeys $[\Phi,\Phi^\dagger]=0$. These solutions
are also relevant since they usually leave $\cN=1$ supersymmetry unbroken.

We next want to describe the Higgs field in terms of its eigenvalues 
and eigenvectors, that is, its spectral data. We introduce a section 
$s$ of the canonical bundle over $S$, $K_S$, and we can write the 
eigenvalue equation
\begin{equation}\label{spectraleigen}
 \det (sI - \Phi) = 0 .
\end{equation}
Since we restrict to the Higgs field 
that leaves the SU(5) but breaks $SU(5)_\perp$, we can expand \eqref{spectraleigen}
in the 5 eigenvalues $t_i$ for the 
fundamental representation of $SU(5)_\perp$ as
\begin{equation}
 \prod_i (s - t_i) = 0 .
\end{equation}
Expanding, we find
\begin{equation}\label{spectral cover1}
 \beta_0 s^5 + \beta_2 s^3 + \beta_3 s^2 + \beta_4 s + \beta_5 =0 .
\end{equation}
The $\beta_1$ is not present since $\beta_1 = t_1 + ... + t_5 = 0$, from the 
tracelessness condition of the roots in $SU(5)$.
Although it is not yet clear, the $\beta_i$s in \eqref{spectral cover1} are 
the same as the elliptic fiber equation in the Tate form, when in the vicinity 
of the GUT divisor. To see that, we first define the ``Tate divisor'' \cite{Marsano:2011hv},
as the equation
\begin{equation}
 \cC_{\text{Tate}} \quad : \quad \beta_0 w^5 + \beta_2 x w^3 + \beta_3 y w^2 + \beta_4 x^2 w + \beta_5 x y = 0 .
\end{equation}
The equation for the elliptic fiber \eqref{tateform2},
\begin{equation}
 -y^2 + x^3 + \left(\beta_0 w^5 + \beta_2 x w^3 + \beta_3 y w^2 + \beta_4 x^2 w + \beta_5 x y\right) = 0 ,
\end{equation}
when restricted to the Tate divisor implies $\frac{y^2}{x^3} = 1$. Also 
we can define the holomorphic section $u=y/x$ on the Tate divisor.
This allows us to write the Tate divisor as
\begin{equation}
 \cC_{\text{Tate}} \quad : \quad \beta_0 w^5 + \beta_2 w^3 u^2 + \beta_3 w^2 u^3 + \beta_4 w u^4 + \beta_5 u^5 = 0 .\label{CTate}
\end{equation}
We then restrict to the vicinity of $w \ra 0$ and close to the singularity on 
the elliptic fiber $y \ra 0$, $x \ra 0$. This in turn implies $u \ra 0$.
We consider the section  $s=w/u$ in (\ref{CTate}), to arrive precisely at the
spectral curve \eqref{spectral cover1}.
The coefficients $\beta_i$ are given in terms of the eigenvalues $t_i$ as
\begin{eqnarray}\label{betas}
\beta_1&=&-\beta_0\sum_i t_i=0 , \qquad \beta_2=\beta_0 \sum_{i\neq j}t_i t_j , \\
\beta_3&=&-\beta_0 \sum_{i\neq j\neq k} t_i t_j t_k , \qquad
\beta_4=\beta_0\sum_{i\neq j\neq k\neq l}t_i t_j t_k t_l , \qquad
\beta_5=-\beta_0 t_1 t_2 t_3 t_4 t_5 . \nonumber
\end{eqnarray}

Recall also that the first enhancements that we encounter 
in the $SU(5)$ model happens when $\Delta'$ in $\Delta= w^5 \Delta'$
has a first order zero. This corresponds to the vanishing of
\begin{eqnarray}\label{delta'neq0}
\beta_5&=&-\beta_0 t_1 t_2 t_3 t_4 t_5\, , \quad \text{or}\\
P_5&=&-\beta_0^3 \prod_{i\neq j}(-t_i-t_j)=\beta_3^2\beta_4-\beta_2\beta_3\beta_5+\beta_0\beta_5^2 . \nn
\end{eqnarray}

The expansion of the $f,g$ and $\Delta$ in terms of the variables on the singularity $w$ and $\{t_i\}$ is given by
\begin{eqnarray}
\Delta=-432 \beta_0^2 w^{10}+\ldots \quad   f=\beta_0 \sum_{i\neq j} t_i t_j w^3+\ldots,\quad g=\beta_0 w^5+\ldots,
\end{eqnarray}
where the dots indicate higher order terms in $w$.
As we want an enhancement to an $E_8$, it follows from the Kodaira classification \ref{table:kodaira}
that we should have
\begin{equation}
 \text{ord}(f) \geq 4 \, , \qquad 
 \text{ord}(g)=5 \, , \qquad 
 \text{ord}(\Delta)=10 \, .
\end{equation}

In order to achieve $\text{ord}(f)\geq 4$ we require 
every term  $t_i t_j$ to vanish, that in turn imposes
at least 4 of the $t_i$s to be zero. 
But from the tracelessness condition, it implies that all should be zero.
We could have advanced this, since the $t_i$s are related to 
higgsings of the underlying $E_8$ group, so setting all $t_i$s
to zero would un-break the $E_8$.

Since we want to describe this enhancement as a codimension 3 locus,
we introduce sections $p$ and $q$, that will be normal sections
to curves on the $SU(5)$ divisor. We could interpret this as the 
normal sections of divisors on the base $B_3$, but as we will 
argue later, having started from the Tate model does not allow this
interpretation.
Looking at $\beta_2$ relation with $\beta_0$ this means that $\sum_{i\neq j}t_i t_j$ 
must be a section of the $K_{B_3}^{-2} \otimes \mathcal{L}^{2}_{SU(5)}$.

We impose now that the $t_i$'s can be written as
\begin{eqnarray}\label{E8model}
t_i=t_i^{p}p+t_i^{q}q.
\end{eqnarray}
The $t_i^p$ and $t_i^q$ could also be sections
of some bundle on the base, 
but for simplicity of the model we consider them to be
constant integer numbers. This implies that $p$ and $q$ 
are sections of $K_{B_3}^{-1} \otimes \mathcal{L}_{SU(5)}$,
and could be homologically equivalent to each other.
The tracelessness condition $\beta_1=0$ implies
\begin{eqnarray}
p(t_1^{p}+t_2^{p}+t_3^{p}+t_4^{p}+t_5^{p})+q(t_1^{q}+t_2^{q}+t_3^{q}+t_4^{q}+t_5^{q})=0\nonumber .
\end{eqnarray}
We also want no trivial solution to \eqref{delta'neq0},
thus $t_i \neq 0$ and $t_i + t_j \neq 0$. In the next sections
we will consider two different cases that satisfy the above 
requirements and reproduce a point of $E_8$ singularity.

\section{Case 1: $E_8$ model with the 5 and the 10 matter representations}
\label{case1}

One choice that satisfies all the requirements 
for an $E_8$ point is
\begin{eqnarray}\label{E8model-case1}
t_1=p,\ \  t_2=q,\ \ t_3= p+2q,\ \ t_4=-2p-q,\ \ t_5=-2q.
\end{eqnarray}

The $\beta_i$s after the replacements \eqref{E8model} become
\begin{equation} \label{betasE8-case1} 
\begin{tabular}{l l}
$\beta_2 = -\beta_0 (3 p^2+4 p q+5 q^2), \quad $&$
\beta_3 = 2 \beta_0 p (p^2+p q-q^2),$ \nn \\ 
$\beta_4 = \beta_0 q (2 p^3+11 p^2 q+10 p q^2+4 q^3), \quad 
$&$\beta_5 = -2 \beta_0 p q^2 (2 p+q) (p+2 q) $. 
\end{tabular}
\end{equation}
It is also convenient, for reference, to write the polynomials $P$ and $R$,
\begin{align}
 P = & 4 \beta_0^3 p^2 (p-2 q) (p-q) q (p+q)^3 (p+3 q) (2 p+3 q) , \\
 R = &-2 \beta_0^3 p (p+q)^2 (4 p^6+4 p^5 q-30 p^4 q^2-41 p^3 q^3+10 p^2 q^4-9 p q^5-18 q^6). \nn
\end{align}

Replacing the values for the $\beta_i$'s, equation \eqref{betasE8-case1},
\begin{eqnarray}
 y^2 - 2 \beta_0 p q^2 (2 p+q) (p+2 q) x y w + 2 \beta_0 p\left(p^2+p q-q^2)\right) y w^2 + x^3 + \beta_0 w^5 + & &  \\
 - \beta_0 \left(3 p^2+4 p q+5 q^2\right) x w^3 + 2 \beta_0 q (2 p^3+11 p^2 q+10 p q^2+4 q^3) x^2 w & = & 0. \nn
\end{eqnarray}

We next calculate explicitly $f$, $g$ and $\Delta$ 
as depending on $w$, $p$ and $q$. There will be codimension 
two and three loci on the base where the vanishing order
of these functions will increase. If we extrapolate the 
result of the Kodaira classification (table \ref{table:kodaira})
to codimension higher than one,
we can extract the information on the ``expected'' gauge group
over each locus.
Away from the enhancement loci and close to $w=0$, we find
\begin{eqnarray}
 f &=& -\frac{1}{3} b0^4 p^4 q^8 (2 p + q)^4 (p + 2 q)^4 + O(w)\, , \nn \\
 g &=& \frac{2}{27} b0^6 p^6 q^12 (2 p + q)^6 (p + 2 q)^6 + O(w)\, , \\
 \Delta &=& -64 (b0^7 p^6 q^9 (p + q)^3 
(2 p^2 + 5 p q + 2 q^2)^4 (2 p^4 
+ 3 p^3 q - 14 p^2 q^2 - 9 p q^3 + 18 q^4)) w^5 + O(w^6) ,\nn
\end{eqnarray}
and therefore $\ord(\Delta)=5$ and $\ord (f) = \ord (g) = 0$.
Over the codimension 2 locus $p=0$ together with $w=0$, 
we obtain $\ord(\Delta)=8$, $\ord (f) = 2$ 
and $\ord (g) = 3$, the vanishing degrees
for an $SO(12)$ singularity, while the curve $q=0$ corresponds to 
an $E_6$ singularity. Notice that in general (non-factorized $\beta_i$s)
$SO(12)$ and $E_6$ singularities appear only at codimension 3, 
reproducing the $10 \bar 5 \bar 5$ and $10 10 5$ couplings, 
respectively. In the next sections we will
discuss what matter representation can be obtained 
after we resolve these singularities.

We can identify other enhancement loci
that we summarize in table \ref{table:E8enhance-case1}.
We will call the $w=q=0$ locus the \textit{$E_6$ matter curve},
even if the explicit resolution lead to something different from an $E_6$.
Similar notation will apply to the other codimension two loci.

At the point $p=q=w=0$ in which we expect 
by construction to get a $E_8$ singularity,
one can check that we get exactly 
$\ord(\Delta)=10$, $\ord (f) = 5$ and $\ord (g) = 5$, the expected
for a $E_8$ singularity. Similarly, we call this codimension
three locus the $E_8$ Yukawa point.

\begin{table}[h]
\centering
 \begin{tabular}{|c|c|c|c|} \hline
  Locus	(in $w=0$) 	& Codim			& $\ord{(\Delta / f / g)}$	& Sing. type 	\\ \hline
  $p=0$			& \multirow{9}{*}{2}	& $8 / 2 / 3 $			& $SO(12)$	\\ \cline{1-1} \cline{3-4}
  $q=0$			&  			& $8 / 3 / 4 $			& $E_6$		\\ \cline{1-1} \cline{3-4}
  $p+q=0$		& 			& $7 / 0 / 0 $ 			& $SU(7)$	\\ \cline{1-1} \cline{3-4}
  $p-2q=0$		&			& \multirow{4}{*}{$6 / 0 / 0 $}	& \multirow{4}{*}{$SU(6)$} \\ \cline{1-1}
  $p-q=0$		&			& 				& 		\\ \cline{1-1}
  $p+3q=0$		& 			&				&		\\ \cline{1-1}
  $2p+3q=0$		&			&				&		\\ \cline{1-1} \cline{3-4} 
  $2p+q=0$		&			& \multirow{2}{*}{$7 / 2 / 3 $}	& \multirow{2}{*}{$SO(10)$} \\ \cline{1-1}
  $p+2q=0$		&			& 				& 		\\ \hline  
  $p=q = 0$		& $3$			& $10 / 5 / 5 $			& $E_8$		\\ \hline
 \end{tabular}
\caption{Codimension 2 and 3 enhancements for the particular model \eqref{E8model}.} \label{table:E8enhance-case1}
\end{table}

\subsection{Resolution} 
\label{case1resol}
After we replace the values of the $\beta_i$'s 
for our chosen factorization \eqref{E8model-case1} the expression for the elliptic curve becomes
\begin{eqnarray}
 y^2 z - 2 \beta_0 p (p^2 + p q - q^2) w^2 y z^2 + 
    2 \beta_0 p q^2 (2 p + q) (p + 2 q) x y z - x^3 - \beta_0 w^5 z^3& & \nn \\ 
    +\beta_0 (3 p^2 + 4 p q + 5 q^2) w^3 z^2 x -  
   \beta_0 q (2 p^3 + 11 p^2 q + 10 p q^2 + 4 q^3) x^2 w z &=& 0 \, . \nn
\end{eqnarray}
After the two first blow ups, 
we again have a binomial variety of the form 
encountered before,
\begin{equation} 
 z y s + \lambda_1 \lambda_2 (\ldots) = 0 ,
\end{equation}
with $s = - y - 2 \beta_0 p q^2 (2 p + q) (p + 2 q) x + 
2 \beta_0 p (p^2 + p q - q^2) w^2 z$.
We resolve the singularities by performing the small resolutions
$\delta_1 : [y,\lambda_1]$ and $\delta_2 : [y,\lambda_2]$,
\begin{eqnarray} \label{b4blowups}
 0 &=& y z\left( \beta_0 p q^2(4 p^2 + 10 p q + 4 q^2) x  
 + \delta_1 \delta_2 y
 + \delta_1 \lambda_1 \beta_0 p (- 2 p^2 - 2 p q + 2 q^2) w^2 z\right) \nn \\ & &
 + \lambda_1 \lambda_2 \left( -\delta_2 \lambda_2 x^3
 - \beta_0 q (2 p^3 + 11 p^2 q + 10 p q^2 + 4 q^3) w x^2 z \right. \\ & &  
 \left. - \beta_0 \delta_1 \lambda_1 (- 3 p^2 - 4 p q - 5 q^2) w^3 x z^2 
 - \beta_0 \delta_1^2 \lambda_1^2 w^5 z^3 \right), \nn
\end{eqnarray}
with the projective relations
\begin{equation}
[\delta_2 \delta_1 \lambda_2 y : \delta_2 \lambda_2 x : w ] , 
\quad [\delta_2 \delta_1 y : x:\delta_1 \lambda_1] , 
\quad [\delta_2 y:\lambda_1] , 
\quad [y:\lambda_2] .  
\end{equation}
The coordinates describe sections of bundles as 
shown in table \ref{table:sections-case1}, in Appendix
\ref{Appendix B: Defining equations}.

At this point, as in Esole-Yau resolution \cite{Esole:2011sm} 
we would have the fully resolved space. The enhancements studied 
do not worsen the singularities, but only split the already existing curves.
In our particular construction, however, 
there are further singularities to be resolved. 
The reason for this is that in our model
the $\beta_i$'s split in a product of the sections $p$ and $q$, 
thus enhancing the vanishing order of the previously smooth terms.

All the further singularities are located on top of $q=0$. After resolving
for generic points in $p$, there are still some singularities on the 
point $q=p=0$.
We can resolve the singularities by many different ways. We choose the
particular resolution

\begin{equation}
\chi_1:[q,\lambda_2,\delta_1] , \quad
\chi_2:[\chi_1,\delta_1], \quad
\chi_3:[\chi_2 ,p, \delta_2], \quad
\chi_4:[\chi_3,\delta_2], \quad
\chi_5:[\chi_4,\delta_2],
\end{equation}
resulting in the following expression for the
resolved fourfold, 
\begin{eqnarray}\label{resolvedCY2}
\left(-\delta_2 \lambda_2 \chi_1 x^3 + b0 q \chi_1 \chi_3^2 \chi_4
 (- 2 p^3 - 
  11 p^2 q \chi_1 \chi_2 - 
  10 p q^2 \chi_1^2 \chi_2^2 - 
  4 q^3 \chi_1^3 \chi_2^3) w x^2 z \right. + \nn & & \\ 
  \left. - \beta_0 \chi_1 \chi_2 \chi_3^2 \chi_4 
  (- 
  3 p^2 - 
  4 p q \chi_1 \chi_2 - 
  5 q^2 \chi_1^2 \chi_2^2) \delta_1 \lambda_1 w^3 x z^2 - 
  \beta_0 \delta_1^2 \lambda_1^2 \chi_1^2 \chi_2^3 \chi_3^2 \chi_4 w^5 z^3 \right) \lambda_1 \lambda_2  \, + & & \\ 
+ \left(\beta_0 p q^2 \chi_1 \chi_3^2 (4 p^2 + 10 p q \chi_1 \chi_2 
+ 4 q^2 \chi_1^2 \chi_2^2 )\chi_4 x \, +\right. & & \nn \\
  \left. + 2 \beta_0  p \chi_3^2 \chi_4 ( -  p^2
    -  p q \chi_1 \chi_2 
    + q^2 \chi_1^2 \chi_2^2 ) \delta_1 \lambda_1 w^2 z
    + \delta_1 \delta_2 y\right) y z  =& & 0\nn
\end{eqnarray}

Notice that some of the blow-ups were
performed by introducing 
$\IP^2$s in the ambient space, 
while in the Esole-Yau small resolutions 
we had only $\IP^1$s. 
As a consequence, we are in fact introducing
a new (three-dimensional) divisor 
on the fourfold, but localized
along codimension $2$ on 
the base (the matter curve). 
That is, the new fiber 
will have to have dimension higher than one.
Up to now, the effect of the resolutions 
was only to modify the one-dimensional fiber of the fourfold,
replacing the singular points by one dimensional curves.
In the resolution we will perform now, we then also modify 
the base of the fibration, 
introducing new submanifolds along the 
enhancement loci.

The difference here to the previous case where we only needed small
resolutions lies on the fact that, in the brane picture, the collisions 
that lead to $SU(6)$ and $SO(10)$ matter curves come from collision 
of the $SU(5)$-brane with an $U(1)$ seven-brane or an O7-plane. 
Both correspond to a non-singular degeneration of the fiber,
and therefore a resolution is not needed. In this case, however,
the collision inducing an $E_6$ enhancement can be understood as
\begin{equation}\label{brokenSU(3)}
 E_6 \ra SU(5) \times SU(3) ,
\end{equation}
and thus the colliding brane would carry with it a singularity 
from the F-theory perspective.
Our local construction however does not allow us to see the 
colliding brane outside the $SU(5)$ locus $\{w=0\}$.

The coordinates in \eqref{resolvedCY2} 
obey the list of projective relations


\begin{eqnarray} \label{E8proj}
& \{\delta_1^2 \delta_2^3 \lambda_1 \lambda_2^2 \chi_1^4 \chi_2^6 \chi_3^9 \chi_4^{12} \chi_5^{15} y: 
    \delta_2^2 \lambda_1 \lambda_2^2 \chi_1 \chi_2 \chi_3^2 \chi_4^3 \chi_5^4 x: z\} \neq \{0: 0: 0\} & \nn \\ \relax
& \{\delta_1 \delta_2^2 \lambda_2 \chi_1^2 \chi_2^3 \chi_3^5 \chi_4^7 \chi_5^9 y: 
\delta_2 \lambda_2 \chi_1 \chi_2 \chi_3^2 \chi_4^3 \chi_5^4 x: w\} \neq \{0: 0: 0\} , \quad \nn  &
     \\ \relax
& \{\delta_1 \delta_2 \chi_1 \chi_2^2 \chi_3^3 \chi_4^4 \chi_5^5 y: x: 
\delta_1 \lambda_1 \chi_1 \chi_2^2 \chi_3^2 \chi_4^2 \chi_5^2\} \neq \{0: 0: 0\}, \quad  & 
\\ \relax
& \{\delta_2 \chi_3 \chi_4^2 \chi_5^3 y:    \lambda_1\} \neq \{0: 0\} , \quad 
 \{y: \lambda_2 \chi_1 \chi_2 \chi_3 \chi_4 \chi_5\} \neq \{0: 0\} , \quad  \nn \\ \relax 
& \{q: \lambda_2: \delta_1 \chi_2 \chi_3 \chi_4 \chi_5\} \neq \{0: 0: 0\} , \quad &
 \nn \\ \relax 
& \{\chi_1: \delta_1\} \neq \{0: 0\} , \quad 
 \{\chi_2: p: \delta_2 \chi_4 \chi_5^2\} \neq \{0: 0: 0\} , \quad \nn \\ \relax 
 &\{\chi_3: \delta_2 \chi_5\} \neq \{0: 0\}: 
 \{\chi_4: \delta_2\} \neq \{0:0\}; & \nn
\end{eqnarray}
and are sections of the bundles from table \ref{table:sections-case1}.

The class of the new fourfold $\tilde Y_4$ 
is given by
\begin{equation}
 [\tilde Y_4] = 3\sigma + 6c_1 -2E_1 -2E_2 
  - E_3 -E_4 -E_5 -E_6 -E_7 -E_8 -E_9 .
\end{equation}

\subsection{Codimension 1 - The GUT Divisor}
\label{case1co1}
At codimension one, when we restrict ourselves to 
$\chi_3\chi_4\chi_5 p\neq 0$, $\chi_1\chi_2\chi_3\chi_4\chi_5 q\neq 0$,
we can simply blow down the five last blow-ups since they 
all sit along
$\chi_3\chi_4\chi_5 p= 0$ or $\chi_1\chi_2\chi_3\chi_4\chi_5 q= 0$. 
Blowing down, we simply return to our space after the 
two small resolutions, \eqref{b4blowups}. Working in the partially or the
full blown-up space the singularity of the fiber is resolved
and we get five curves, ordered as
the picture below,

\begin{figure}[H]
\centering
\begin{picture}(46,11)(-12,0)
 \multiput(3,0)(9,0){4}{\circle{6}}
 \multiput(6,0)(9,0){3}{\line(1,0){3}} 
 \put(16.5,9){\circle{6}}
 \put(3,3){\line(2,1){10}}
 \put(30,3){\line(-2,1){10}}
 \put(1.8,-1){A}
 \put(10.8,-1){B}
\put(19.8,-1){C}
\put(28.8,-1){D}
\put(15,8){X}
\end{picture}
\end{figure}

where the curves are given by the restriction of the ambient divisors
\begin{equation} \label{codim1:divisors-case1}
 A : \delta_1 = 0, \quad B: \delta_2 = 0, \quad C :\lambda_2 = 0, \quad D :\lambda_1 = 0 \, ,
\end{equation}
to the fourfold $\tilde Y_4$. The defining equations are 
exhibited in appendix \ref{Appendix B: Defining equations}.
One can calculate the intersections among the divisors
restricted to the fiber in the same way as was done in 
\cite{Marsano:2011hv,Lawrie:2012gg}, and 
we get precisely the Cartan matrix (times $-1$) 
for an $SU(5)$ group,
\begin{center}
\begin{tabular}{l r r l r r r l} & & & \hspace{-1mm}A & B & C & D  & \hspace{.5mm} X  \\ A & \multicolumn{7}{l}{\multirow{5}{*}{
$\left( 
\begin{tabular}{c c c c : c }
-2 & 1  & 0  & 0  & 1 \\
 1 & -2 & 1  & 0  & 0 \\ 
 0 & 1  & -2 & 1  & 0 \\
 0 & 0  & 1  & -2 & 1 \\ \hdashline
 1 & 0  & 0  & 1  & -2 \\
\end{tabular}
\right)$ }  } \\ 
B & & & & & & & \\ 
C & & & & &  & & \\ 
D & & & & & & &  \\ 
X & & & & & & & \\
\end{tabular}
\end{center}

\subsection{Codimension 2}
\label{case1co2}
Now we look at the codimension 2 loci, where the matter
representations appear.

\subsubsection{10 Matter Curve}
We start with the locus $p+2q=0$ from the singular fourfold, 
that from the vanishing degree corresponded to an $SO(10)$ enhancement,
and from group-theoretical arguments to a matter $\mathbf{10}$ representation. 
After the complete resolution of the fourfold, 
the locus is defined by $p + 2 q \chi_1 \chi_2 \equiv p_0 + 2 q_0 = 0$, and the 
expression for the fourfold reduces to
\begin{eqnarray}\label{matter10}
0&=&-y^2 z \delta_1 \delta_2+x^3 \delta_2 \lambda_1 \lambda_2^2 \chi_1+\beta_0 w^5 z^3 \delta_1^2
\lambda_1^3 \lambda_2 \chi_1^2\chi_2^3 \chi_3^2 \chi_4-4 \beta_0 q^3 w^2 y z^2 \delta_1 \lambda_1 \chi_1^3 
\chi_2^3\chi_3^2 \chi_4\\
&&-9 \beta_0 q^2 w^3 x z^2 \delta_1 \lambda_1^2 \lambda_2 \chi_1^3 \chi_2^3 \chi_3^2 
\chi_4+12 \beta_0 q^4 w x^2 z \lambda_1 \lambda_2\chi_1^4 \chi_2^3 \chi_3^2 \chi_4\, ,\nonumber
\end{eqnarray}
with the projective relations \eqref{E8proj}.
As in \cite{Esole:2011sm,Marsano:2011hv}, 
the divisors living in codimension 1 now rearrange themselves in 
a different way, by splitting into new curves or merging with 
one another. Namely, the restriction to the fourfold of the 
divisor $\delta_1=0$ at the $p - 2 q \chi_1 \chi_2 = 0$ locus becomes
\begin{equation}
\lambda_1 \lambda_2 \chi_1 x^2 (\delta_2 \lambda_2 x + 12 \beta_0 q^4 \chi_1^3 \chi_2^3 \chi_3^2 \chi_4 w z)=0 ,
\end{equation}
that is factorizable in three different curves (obeying the projective relations \eqref{E8proj})
\begin{equation}
 D_{\{p-2q_0\} } : \lambda_1 = 0 , \quad A_{1 \{p-2q_0\} } :\lambda_2 = 0 , 
 \quad A_{2 \{p-2q_0\} }: \delta_2 \lambda_2 x + 12 \beta_0 q^4 \chi_1^3 \chi_2^3 \chi_3^2 \chi_4 w z = 0 .
\end{equation}

A similar splitting happens for the divisor $\lambda_2=0$, as in the fourfold
\begin{equation}
 \delta_1 y z (\delta_2 y + 4 \beta_0 \lambda_1 q^3 \chi_1^3 \chi_2^3 \chi_3^2 \chi_4 w^2 z) = 0 ,
\end{equation}
that factors in
\begin{equation}\label{lambda_2-10-A}
A_{1 \{p-2q_0\} } : \delta_1 = 0 \quad \text{and} \quad 
C_{ \{p-2q_0\} } : \delta_2 y + 4 \beta_0 \lambda_1 q^3 \chi_1^3 \chi_2^3 \chi_3^2 \chi_4 w^2 z = 0 .
\end{equation}
We then  calculate the intersection 
of these curves with the Cartan divisors 
\eqref{codim1:divisors-case1} to get the charges 
of each curve under the Cartan roots, obtaining
(details in appendix \ref{Appendix C})

\begin{table}[h]\centering 
\begin{tabular}{c c c} \hline
Curve & Charges & Weights \\ \hline
$A_{1 \{p-2q_0\} }$ & (-1,1,-1,1,0) &	$\mu_{10}-\alpha_1-\alpha_2-\alpha_3$	\\ \hline
$A_{2 \{p-2q_0\} }$ & (-1,0,0,1,0)  & $-\mu_{10}+\alpha_2+\alpha_3+\alpha_4$ 		\\ \hline
$B_{\{p-2q_0\}}$	& (1,-2,1,0,0)	  & $-\alpha_2$		\\ \hline
$C_{\{p-2q_0\}}$	& (1,0,-1,0,0)	  & $-\mu_{10}+\alpha_1+\alpha_2$		\\ \hline
$D_{\{p-2q_0\}}$ 	& (0,0,1, -2, 1)  & $-\alpha_4$	\\ \hline
$X_{\{p-2q_0\}}$	& (1,0,0,1,-2)	  & $-\alpha_0$		\\ \hline
\end{tabular}
\caption{Curves over the locus $p-2q \chi_1 \chi_2 \equiv p - 2q_0=0$ with their corresponding
charges. The defining equations are in table \ref{table:codim2(p-2q=0)-case2}.}\label{table:matter10}
\end{table}

One can check that the weights obey the charge conservation
\begin{eqnarray}
 A &=& A_{1 \{p-2q_0\} } + A_{2 \{p-2q_0\} } + D_{ \{p-2q_0\} } ,\nn \\
 C &=& C_{\{p-2q_0\}} + A_{1 \{p-2q_0\} } ,\nn
\end{eqnarray}
the curves in table \ref{table:matter10} reproduce 
a complete representation of the matter $\mathbf{10}$,
and they distribute themselves as the Dynkin diagram 
for the $SO(10)$ group.

\subsubsection{Matter 5}
The matter 5 representation is expected to appear at the 
$SU(6)$ enhancement locus, where $P=0$. In our model, 
there are many loci that describe this enhancement.
We will focus on $p-q=0$. An analogous calculation follows
for the other matter $\mathbf{5}$ loci, but with different numerical factors.

At this locus, the curve specified by $\delta_2 = 0$ becomes reducible.
Namely, the defining equation for this curve at the fourfold becomes
(with $\chi_i = 1$)
\begin{equation}
-\beta_0 z \left(-9 q^2 x + \delta_1 \lambda_1 w^2 z\right) \left(-3 \lambda_1 \lambda_2 q^2 w x + 2 q^3 y + 
  \delta_1 \lambda_1^2 \lambda_2 w^3 z\right) ,
\end{equation}
that is factorizable in two curves,
\begin{equation}  
 B_{1 \{p-q_0\}} : \delta_2 = -9 q^2 x + \delta_1 \lambda_1 w^2 z = 0 \quad \text{and}\quad 
  B_{2  \{p-q_0\}} : \delta_2 = -3 \lambda_1 \lambda_2 q^2 w x + 2 q^3 y + 
  \delta_1 \lambda_1^2 \lambda_2 w^3 z = 0
\end{equation}
It is then straightforward to calculate the weights for the new curves.

\begin{table}[h]\centering
\begin{tabular}{c c c} \hline
Curve & Charges & Weights \\ \hline
$A_{1 \{p-q_0\} }$ & (-2,1,0,0,1)		& $- \alpha_1$			\\ \hline
$B_{\{p-q_0\}}$	& (1,-1,0,0,0)		& $-\mu_5 + \alpha_2$	\\ \hline
$B_{\{p-q_0\}}$	& (0,-1,1,0,0)		& $-\mu_5 + \alpha_2$	\\ \hline
$C_{\{p-q_0\}}$	& (0,1,-2,1,0)		& $-\alpha_3$			\\ \hline
$D_{\{p-q_0\}}$ 	& (0,0,1, -2, 1) 	& $-\alpha_4 	$		\\ \hline
$X_{\{p-q_0\}}$	& (1,0,0,1,-2)		& $-\alpha_5 $			\\ \hline
\end{tabular}
\caption{Curves over $p-q_0=0$. Defining equations
in table \ref{table:codim2(p-q=0)-case1}.}.\label{table:matter10-case1} 
\end{table}

Again, the splitting reproduces a full matter $\mathbf{5}$
representation, and arrange themselves as the Dynkin diagram for
the $SU(6)$ group.

\subsubsection{The $p=0$ and the $q=0$ loci}
We now describe two interesting codimension 
2 loci that 
do not appear in previous works.
When $p=0$, the curves split as
\begin{eqnarray}
& &  A: \delta_1 \quad \xrightarrow{p\rightarrow 0} 
\quad 
A_{1 \{p\} } : \delta_2 \lambda_2 x 
+ 4 \beta_0 q^4 \chi_1^3 \chi_2^3 \chi_3^2 \chi_4 w z = 0, \quad
A_{2 \{p\} } : \lambda_2 = 0, \quad
A_{3\{p\}} : \lambda_1 = 0\, , \nn \\
& & B: \delta_2 \quad \xrightarrow{p\rightarrow 0} 
B_{1 \{p\} } : \lambda_2 = 0, \quad
B_{2A \{p\} } : -q^2 \chi_1 x + \delta_1 \lambda_1 w^2 z = 0, \quad
B_{2B \{p\} } : -4 q^2 \chi_1 x + \delta_1 \lambda_1 w^2 z = 0 ,
\nn \\
& &  C: \lambda_2 \quad \xrightarrow{p\rightarrow 0} 
\quad 
A_{3 \{p\} } : \delta_1 = 0, \quad
B_{1 \{p\} } : \delta_2 = 0,
\end{eqnarray}
and rearrange as the $SO(12)$ Dynkin diagram.
The calculation of the intersections between the curves in the 
$p=0$ locus with the Cartan divisors again give the weights of each 
curve under the $SU(5)$ group,

\begin{table}[h]\centering
\begin{tabular}{c c c} \hline
Curve & Charges & Weights\\ \hline
$A_{1 \{p\} }$ & (-1,1,-1,1,0) & $\mu_{10} -\alpha_1-\alpha_2-\alpha_3$		\\ \hline
$A_{2 \{p\} }$ & (-1,0,0,1,0) & $-\mu_{10} +\alpha_2+\alpha_3+\alpha_4$		\\ \hline
$B_{1 \{p\} }$ & (1,0,-1,0,0) & $-\mu_{10}+\alpha_1+\alpha_2$	\\ \hline
$B_{2A\{p\}}$	& (0,-1,1,0,0) & $\mu_5 -\alpha_1 - \alpha_2$  \\ \hline
$B_{2B\{p\}}$	& (0,-1,1,0,0) & $\mu_5 -\alpha_1 - \alpha_2$	\\ \hline
$D_{\{p\}}$ 	& (0,0,1,-2, 1) &$ - \alpha_4$		\\ \hline
$X_{\{p\}}$	& (1,0,0,1,-2)	& $- \alpha_0$		\\ \hline
\end{tabular}
\end{table}

Interestingly, this set of curves reproduces a full representation
$\mathbf{5} + \mathbf{10}$. To check that, notice first that 
$B_{2A\{p\}}+B_{1 \{p\} }=(1,-1,0,0,0) = -\mu_5 + \alpha_1$.
One can then reconstruct the full representation by 
appropriate combinations of the curves above.

Over the $q_0=0$ locus we have to be more careful. The locus 
is described after the blow-up by $q \chi_1 \chi_2 = 0$. Thus, one
has to see the behaviour of the Cartan roots as we approach
$q \rightarrow 0$, $\chi_1 \rightarrow 0$ and $\chi_2 \rightarrow 0$.

The roots as we take $q \rightarrow 0$ are
\begin{eqnarray}
 A: \delta_1 \quad &\xrightarrow{q\rightarrow 0} &
\quad 
A_{ \{q\} } : \delta_2 = 0, \quad
D_{ \{q\} } : \lambda_1 = 0, \nn \\
 B: \delta_2 \quad &\xrightarrow{q\rightarrow 0} &
\quad 
A_{ \{q\} } : \delta_1 = 0, \quad
B_{ \{q\} } : 3 \lambda_1 \lambda_2 p^2 \chi_1 \chi_2 w x 
+ 2 p^3 y 
- \delta_1 \lambda_1^2 \lambda_2 \chi_1^2 \chi_2^3 w^3 z = 0, \nn \\
  C: \lambda_2 \quad &\xrightarrow{q\rightarrow 0}& 
\quad 
C_{ \{q\} } :\delta_2 y 
+ 2 \beta_0 \lambda_1 p^3 \chi_3^2 \chi_4 w^2 z = 0, \nn \\
  D: \lambda_1 \quad &\xrightarrow{q\rightarrow 0} &
\quad 
D_{ \{q\} } :\delta_1 = 0. \nn
 \end{eqnarray}

When we calculate the charges under the $SU(5)$
roots, only the above curves will have non-vanishing
contributions, while all the curves appearing 
as we take $\chi_1 \rightarrow 0$ or 
$\chi_2 \rightarrow 0$ have zero charge\footnote{The 
calculation of the intersections give terms like 
$D_1 S_2 E_i \sigma^2$, that vanish since $D_1$, $S_2$
and $\sigma$ are pullbacks from the blow-down to 
the blow-up space.}.
Thus, the only contributing curves are given in the
table below:

\begin{table}[h]\centering 
\begin{tabular}{c c c} \hline
Curve & Charges & Weights\\ \hline
$A_{\{q\} }$ & (0,-1,1,0,0) & $\mu_{5} -\alpha_1-\alpha_2$		\\ \hline
$B_{\{q\} }$ & (1,-1,0,0,0) & $-\mu_{5} +\alpha_1$		\\ \hline
$C_{\{q\} }$ & (1,0,-1,0,0) & $-\mu_{10}+\alpha_1+\alpha_2$	\\ \hline
$D_{\{q\}}$	& (0,0,1,-2,1) & $\alpha_4$  \\ \hline
$X_{\{q\}}$	& (1,0,0,1,-2) & $\alpha_0$	\\ \hline
\end{tabular}
\caption{Curves of non-vanishing
charges on the $q=0$ locus. Defining equations 
in \ref{table:codim2(q=0)-case1}}\label{table:curvesinq}
\end{table}

There is a breaking of the charge conservation,
as the daughter curves do not reproduce the 
weights of the original (codimension 1) 
ones. Namely, the difference
can be solved by a single curve $M$ of charges $(-1,1,-1,1)$,
\begin{eqnarray}
 A &\rightarrow & 2 M + A_{\{q\}} + D_{\{q\}}  \nn \\
 (-2,1,0,0,1) & = & (-2,2,-2,2,0) + (0,-1,1,0,0) + (0,0,1,-2,1) ,\nn \\
  C &\rightarrow & M + C_{\{q\}}  \nn \\
 (0,1,-2,1,0) & = & (-1,1,-1,1,0) + (1,0,-1,0,0) .\nn
\end{eqnarray}
The curve $M$ has weight in the $\mathbf{10}$ representation,
namely $(\mu_{10} - \alpha_1 - \alpha_2 - \alpha_3)$.

The curves now do not form a recognizable Dynkin diagram.
In fact, if we consider only the curves from table 
\ref{table:curvesinq} the fiber structure is split as 
the picture \ref{fig:sub1} below. If however we include the curves 
not charged under the $SU(5)$, we get a connected 
set of curves, as picture \ref{fig:sub2}.

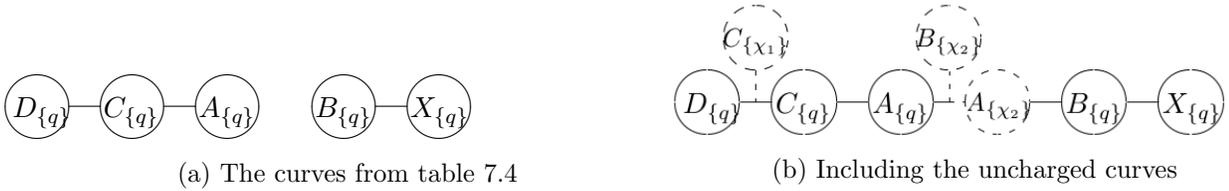
\begin{figure}[H]
\begin{subfigure}{.5\textwidth}
  \begin{picture}(63,20)(0,-5)
\setlength{\unitlength}{4pt}
\put(0,0){\circle{6}}
\put(3,0){\line(1,0){3}}
\put(9,0){\circle{6}}
\put(12,0){\line(1,0){3}}
\put(18,0){\circle{6}}
\put(29,0){\circle{6}}
\put(32,0){\line(1,0){3}}
\put(38,0){\circle{6}}
\put(-2.3,-1){$D_{\{q\}}$}
\put(6.3,-1){$C_{\{q\}}$}
\put(15.3,-1){$A_{\{q\}}$}
\put(26.3,-1){$B_{\{q\}}$}
\put(35.3,-1){$X_{\{q\}}$}
\end{picture}
  \caption{The curves from table \ref{table:curvesinq}}
  \label{fig:sub1}
\end{subfigure} %
\begin{subfigure}{.5\textwidth}
  \includegraphics[scale=0.65]{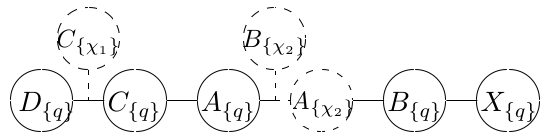}
  \caption{Including the uncharged curves}
  \label{fig:sub2}
\end{subfigure}
\caption{Curves on the $q_0=0$ locus.}
\label{fig:test}
\end{figure}

\subsection{Codimension 3} 
\label{case1co3}

As we move to the point $p_0=q_0=0$, again we have 
more than one vanishing section in the blow-up space, 
since $p_0 = q_0 =0 \rightarrow p \chi_3 \chi_4 \chi_5 = 
q\chi_1 \chi_2 \chi_3 \chi_4 \chi_5 =0$. But just like the $q_0 =0$
locus, only the restriction $p=q=0$ gives non-vanishing 
charge subspaces under the $SU(5)$. These spaces are
given in the table \ref{table:curvesinE8-case1} below,
while the defining equations are given in the 
appendix \ref{Appendix B: Defining equations}.

\begin{table}[h]\centering 
\begin{tabular}{c c c c} \hline
Curve & Equations & Charges & Weights\\ \hline
$A_{\{E8\} }$ & $\delta_1 =\delta_2 = 0$ & (0,-1,1,0,0) & $\mu_{5} -\alpha_1-\alpha_2$	\\ \hline
$C_{\{E8\} }$ & $\delta_2=\lambda_2=0 $&   (1,0,-1,0,0) & $-\mu_{10}+\alpha_1+\alpha_2$	\\ \hline
$D_{\{E8\}}$  & $\lambda_1=\delta_1=0$ &(0,0,1,-2,1) & $\alpha_4$  \\ \hline
$X_{\{E8\}}$ & $w= \tilde Y_4|_{w=0}=0$	& (1,0,0,1,-2) & $\alpha_0$	\\ \hline
\end{tabular}
\caption{Curves of non-vanishing
charges on the $p_0=q_0=0$ point. Equations 
in \ref{table:codim3-case1}.}\label{table:curvesinE8-case1}
\end{table}

We took care not to call these spaces ``curves" since 
unlike the ones appearing in codimension 2, these are
not $\IP^1$s inside the fourfold fiber.
This is a consequence of the fact that by 
blowing up we introduced new exceptional divisors over the 
point of the base $p_0 = q_0$.
The fiber then is not one-complex-dimensional. 

In the next section we will explore another resolution in 
which we have a better control on these higher 
dimensional subspaces appearing after the blow-up.

\section{Case 2: $E_8$ model with $p_0$ and $q_0$ symmetric}
\label{case2}

We now choose for the higgsings \eqref{E8model}
\begin{eqnarray}\label{E8model-case2}
t_1=p,\ \  t_2=q,\ \ t_3= p+q,\ \ t_4=-2q,\ \ t_5=-2p.
\end{eqnarray}
This particular selection is symmetric under the exchange 
$p \leftrightarrow q$. Physically, this means the matter curves 
represented by $p=w=0$ and $q=w=0$ could be exchanged. 
The role of exchange symmetries for the curves 
in phenomenological F-theory models was
explored for example in \cite{Heckman:2009mn}.

The $\beta_i$s after the replacements \eqref{E8model-case2} become
\begin{eqnarray}\label{betasE8-case2}
\beta_2 &= -\beta_0 (3 p^2 + p q + 3 q^2), \quad &\beta_3 = \beta_0 (p + q) (2 p^2 - 3 p q + 2 q^2), \\ 
\beta_4 &= 2 \beta_0 p q (p^2 + 4 p q + q^2), \quad &\beta_5 = -4 \beta_0 p^2 q^2 (p + q). \nn 
\end{eqnarray} 
It is also convenient, for reference, 
to write the polynomials $P$ and $R$,
\begin{align}
 P = &2 \beta_0^3 p (p - 2 q) (p - q)^2 
 (2 p - q) q (p + q)^2 (2 p + q) (p + 2 q) , \\
 R = &-\beta_0^3 (p - q)^2 (p + q) 
 (8 p^6 - 4 p^5 q - 38 p^4 q^2 - 43 p^3 q^3 - 
   38 p^2 q^4 - 4 p q^5 + 8 q^6). \nn
\end{align}

Replacing the values for the $\beta_i$'s, equation \eqref{betasE8-case2},
\begin{eqnarray}
 -y^2 z - 4\beta_0 p^2 q^2 (p+q) x y z + \beta_0 \left[2(p^3 + q^3) - p q (p+q)\right] y w^2z^2 + x^3 + \beta_0 w^5 z^3 + & &  \\
 + \beta_0 \left[-3(p^2 + q^2)-pq\right] x w^3z^2 + 2 \beta_0 p q (p^2 + q^2 + 4 p q) x^2 w z & = & 0. \nn
\end{eqnarray}

We again calculate explicitly $f$, $g$ and $\Delta$ 
with respect to $w$, $p$ and $q$. 
On the codimension 2 locus $p=0$ together with $w=0$, we get
\begin{eqnarray}
 \Delta &=& -432 q^{12} w^8 + \ldots , \nn \\ 
 f &=& - 3 q^2 w^3 + \ldots , \\
 g &=& q^6 w^4 + \ldots ,\nn
\end{eqnarray}
where the $\ldots$ are terms of higher vanishing order. Therefore 
$\ord(\Delta)=8$, $\ord (f) = 3$ and $\ord (g) = 4$, the vanishing degrees
for a $E_6$ singularity. 
There are also other codimension 2 enhancements 
appearing in $p \pm q = 0$, $p \pm 2q = 0$ and $2p \pm q = 0$,
that we summarize in table \ref{tableE8enhance-case2}.
We will call the $w_0=p_0=0$ locus the \textit{$E_6$ matter curve},
even if the explicit resolution leads to something different from an $E_6$.
Similar notation will apply to the other codimension two loci.

At the point $p=q=w=0$ in which we expect 
by construction to get an $E_8$ singularity,
\begin{eqnarray}
 \Delta &=& -432 w^{10} + \ldots , \nn \\ 
 f &=& - 3 p^2 w^3 - p q w^3 - 3 q^2 w^3 + \ldots , \\
 g &=& w^5 + \ldots ,\nn
\end{eqnarray}
Thus $\ord(\Delta)=10$, $\ord (f) = 5$ and $\ord (g) = 5$, the expected
for a $E_8$ singularity. Similarly, we call this codimension
three locus the $E_8$ Yukawa point.

\begin{table}[h]
\centering
 \begin{tabular}{|c|c|c|c|} \hline
  Curve	(in $w=0$) 	& Codim			& $\ord{(\Delta / f / g)}$	& Sing. type 	\\ \hline
  $p=0, q=0$			& \multirow{4}{*}{2}	& $8 / 3 / 4 $			& $E_6$		\\
  $p+q=0$		&  			& $8 / 2 / 3 $			& $SO(12)$	\\
  $p-q=0$		& 			& $7 / 0 / 0 $ 			& $SU(7)$	\\
  $p \pm 2q = 0, q \pm 2p = 0$	&			& $6 / 0 / 0 $			& $SU(6)$	\\  \hline
  $p = q = 0$		& $3$			& $10 / 5 / 5 $			& $E_8$		\\ \hline
 \end{tabular}
\caption{Codimension 2 and 3 enhancements for the particular model \eqref{E8model-case2}.} \label{tableE8enhance-case2}
\end{table}

\subsection{Resolution}
\label{case2resol}

Replacing the expressions for the $\beta_i$s
for the case \eqref{betasE8-case2} into the
expression for the elliptic fibration \ref{tateform} and performing the two blow-ups
$\lambda_1:[y,x,w]$ and $\lambda_2:[y,x,\lambda_1]$ we get,
\begin{eqnarray}
 z y (- y - 4\beta_0 p^2 q^2 (p+q) x + \beta_0 \left[2(p^3 + q^3) - p q (p+q)\right]  \lambda_1 z w^2)+ & & \\
+ \lambda_2 \lambda_1 (\lambda_2 x^3 + \lambda_1^2 w^5 z^3 \beta_0
+ \beta_0 \left[-3(p^2 + q^2)-pq\right] \lambda_1 w^3 x z^2 
+ 2 \beta_0 p q (p^2 + q^2 + 4 p q) x^2 w z)  &=&  0 \, . \nn
\end{eqnarray}

We have a binomial variety of the form 
encountered in \cite{Esole:2011sm,Marsano:2011hv},
\begin{equation}
 z y s + \lambda_1 \lambda_2 (\ldots) = 0 ,
\end{equation}
with $s = - y - 4\beta_0 p^2 q^2 (p+q) x 
+ \beta_0 \left[2(p^3 + q^3) - p q (p+q)\right]  \lambda_1 w^2 z$.
We perform the small resolutions
$\delta_1 : [y,\lambda_1]$ and $\delta_2 : [s,\lambda_2]$,
\begin{equation}
 \begin{cases} 0 & = z y s+ \lambda_2 \lambda_1 (\delta_2 \lambda_2 x^3 + \delta_1^2 \lambda_1^2 w^5 z^3
+ \beta_0 \left[-3(p^2 + q^2)-pq\right] \delta_1 \lambda_1 w^2 x + 2 \beta_0 p q (p^2 + q^2 + 4 p q) x^2 w z)    \\
0 & = \delta_2 s + \delta_1 y + 4\beta_0 p^2 q^2 (p+q) x 
- \beta_0 \left[2(p^3 + q^3) - p q (p+q)\right]  \delta_1 \lambda_1 w^2 z\, , 
\end{cases}
\end{equation}\label{systemE8}
with the projective relations
\begin{equation}
[\delta_2 \delta_1 \lambda_2 y : \delta_2 \lambda_2 x : w ] , 
\quad [\delta_1 y : x:\delta_1 \lambda_1] , 
\quad [y:\lambda_1] , 
\quad [s:\lambda_2] .  
\end{equation}
The coordinates describe sections of the bundles as follows
\begin{table}[h]
\begin{center}
\begin{tabular}{c | c}
 $x$ 		& $\sigma + 2c_1 -E_1 - E_2$	 \\
 $y$		& $\sigma + 3c_1-E_1-E_2 - E_3$	 \\
 $s$		& $\sigma + 3c_1-E_1-E_2 - E_4$	 \\
 $w$		& $S_2 - E_1$ 			 \\
 $z$		& $\sigma$ 			 \\
 $\lambda_1$	& $E_1 - E_3$			 \\
 $\lambda_2$ 	& $E_2 - E_4$			 \\
 $\delta_1$	& $E_3$ 			 \\
 $\delta_2$	& $E_4$				 \\
 $p,q$ 		& $S_2 - c_1$			 \\
 $\beta_0$	& $6c_1 - 5S_2$			 
\end{tabular}
\caption{Coordinates and their respective bundles.}\label{table:sections0-case2}
\end{center}
\end{table}

At this point, as in Esole-Yau resolution \cite{Esole:2011sm} 
we would have the fully resolved space. The enhancements studied 
there do not worsen the singularities, 
but only split the already existing curves.
In our particular construction, however, 
there are further singularities to be resolved. 
The reason for this is that in our model
the $\beta_i$'s split in a product of the sections $p$ and $q$, 
enhancing the vanishing order of the previously smooth terms.

We now proceed to resolve the additional singularities. 
First, we rearrange equation (\ref{systemE8}) as \footnote{We set $z=1$, as we 
are concerned with singularities in the patch $z\neq0$. We reintroduce
it back at the end.}
\begin{equation}\label{b4blowups2}
 \begin{cases}
 0 & = y s+ \lambda_2 \lambda_1 (\delta_2 \lambda_2 x^3 + \delta_1^2 \lambda_1^2 w^5 
+ \beta_0 \left[-3(p^2 + q^2)-pq\right] \delta_1 \lambda_1 w^2 x 
+ 2 \beta_0 p q (p^2 + q^2 + 4 p q) x^2 w)\, ,     \nn \\
0  & =  \delta_2 s + \delta_1 \left( y - \beta_0 \left[2(p^3 + q^3) - p q (p+q)\right]   \lambda_1 w^2 \right) 
+ 4\beta_0 p^2 q^2 (p+q) x \, .
\end{cases}
\end{equation}

We note that all the singularities of the second
equation arise when $\delta_2 = \delta_1 = s = (...) = p = 0$.
or $\delta_2 = \delta_1 = s = (...) = q = 0$.
Similarly as was done to resolve
the $SU(5)$, we introduce an auxiliary
equation $t=y - \left[2(p^3 + q^3) - p q (p+q)\right]\lambda_1 w^2$,
and we work with the system of three equations,
\begin{eqnarray}
 \begin{cases}
 0=&y s+ \lambda_2 \lambda_1 (\delta_2 \lambda_2 x^3 + \delta_1^2 \lambda_1^2 w^5 
+ \beta_0 \left[-3(p^2 + q^2)-pq\right] \delta_1 \lambda_1 w^2 x + 2 \beta_0 p q (p^2 + q^2 + 4 p q) x^2 w)\, ,  \\
 0=&\delta_2 s + \delta_1 t + 4\beta_0 p^2 q^2 (p+q) x\, ,  \\
0=&-t + y - \beta_0 \left[2(p^3 + q^3) - p q (p+q)\right] \lambda_1 w^2 \, .
\end{cases}
\end{eqnarray}

It is now straightforward to resolve this system of equations.
First we resolve the singularity at $q=s=t=\delta_1=\delta_2=0$ 
in the second equation, by
performing the blow up $\chi_1 : [s,t,q]$, that is
\begin{equation}
  s \ra \chi_1 s , \,  t \ra \chi_1 t , \,  q \ra \chi_1 q , \quad \text{with the projective relation}\quad [s:t:q] .
\end{equation}

Notice that this was a $\IP^2$ blow up, not a small resolution
as was done to resolve the curves and Yukawa enhancements in 
the previous section. As a consequence, we are in fact introducing
a new (three-dimensional) divisor on the fourfold, but localized
along codimension $2$ on the base (the matter curve). 
That is, the new fiber will have to have dimension higher than one.
Up to now, the effect of the resolutions 
was only to modify the one-dimensional fiber of the fourfold,
replacing the singular points by one dimensional curves.
In the resolution we will perform now, we then also modify 
the base of the fibration, introducing new submanifolds along the 
enhancement loci.

The difference here to the previous case where we only needed small
resolutions lies on the fact that, in the brane picture, the collisions 
that lead to $SU(6)$ and $SO(10)$ matter curves come from collision 
of the $SU(5)$-brane with an $U(1)$ seven-brane or an O7-plane. 
Both correspond to a non-singular degeneration of the fiber,
and therefore a resolution is not needed. In this case, however,
the collision inducing an $E_6$ enhancement can be understood as
\begin{equation}\label{brokenSU(3)}
 E_6 \ra SU(5) \times SU(3) ,
\end{equation}
and thus the colliding brane would carry with it a singularity 
from the F-theory perspective.
Our local construction however does not allow us to see the 
colliding brane outside the $SU(5)$ locus $\{w=0\}$.

One should also keep in mind that there is a large number of possibilities for the blow-ups,
that might lead to different final resolved manifolds.
Here we perform one of many choices, that leads to a resolved space in few steps. 
There is however no possibility to resolve this singular 
space only via small resolutions. To fully resolve the space, 
we then choose to perform other three blow ups, 
in the following order,
\begin{equation}
 \chi_2 : [s,t,\chi_1] , \, \pi_1 : [s,t,p] , \, \pi_2 : [s,t,\pi_1] ,
\end{equation}
thus introducing four new divisors 
to the ambient fivefold given by $\{\pi_1=0\}$, $\{\pi_2=0\}$,
$\{\chi_1=0\}$ and $\{\chi_2=0\}$. 

The defining equations 
for the elliptic fiber 
consists now on the triple intersection (with $z$ reintroduced)
\begin{eqnarray}
\begin{cases}
0 = & \pi_1 \pi_2^2 \chi_1 \chi_2^2 z s y+  
\lambda_1 \lambda_2 
(-2 \beta_0 p \pi_1 \pi_2 q
\chi_1 \chi_2 (p^2 \pi_1^2 \pi_2^2 + 4 p \pi_1 \pi_2 q \chi_1 \chi_2 
+ q^2 \chi_1^2 \chi_2^2)x^2 w z - \\
& - \beta_0 \delta_1^2 \lambda_1^2 w^5 z^3
    + \beta_0 \delta_1 \lambda_1 (3 p^2 \pi_1^2 \pi_2^2 
+ p \pi_1 \pi_2 q \chi_1 \chi_2 + 
        3 q^2 \chi_1^2 \chi_2^2) w^3 x z^2- 
\delta_2 \lambda_2 x^3)\, , \\
0 = & 
4 \beta_0 p^2 \pi_1 q^2 \chi_1 (p \pi_1 \pi_2 + q \chi_1 \chi_2)x 
+ \delta_2 s 
+ \delta_1 t\, , \\ 
0 = & 
-\pi_1 \pi_2^2 \chi_1 \chi_2^2 t 
- \beta_0 
  \lambda_1 (p \pi_1 \pi_2 + q \chi_1 \chi_2) 
(2 p^2 \pi_1^2 \pi_2^2 - 3 p \pi_1 \pi_2 q \chi_1 \chi_2 + 
     2 q^2 \chi_1^2 \chi_2^2) w^2 z 
+ y \, .
\end{cases}\label{E8eqresy}
\end{eqnarray}
Since the last equation in \eqref{E8eqresy} 
has a simple dependence on $y$, we can use it to eliminate 
$y$ in the other equations, and return to a system of two equations
(with $z$ reintroduced),
\begin{eqnarray}
\begin{cases}
0 = & \pi_1 \pi_2^2 \chi_1 \chi_2^2 z s \left[
\pi_1 \pi_2^2 \chi_1 \chi_2^2 t 
+ \beta_0 
  \lambda_1 (p \pi_1 \pi_2 + q \chi_1 \chi_2) 
(2 p^2 \pi_1^2 \pi_2^2 - 3 p \pi_1 \pi_2 q \chi_1 \chi_2 + 
     2 q^2 \chi_1^2 \chi_2^2) w^2 z \right]+  
\\
 &\lambda_1 \lambda_2 
(-2 \beta_0 p \pi_1 \pi_2 q
\chi_1 \chi_2 (p^2 \pi_1^2 \pi_2^2 + 4 p \pi_1 \pi_2 q \chi_1 \chi_2 
+ q^2 \chi_1^2 \chi_2^2)x^2 w z - \\
& - \beta_0 \delta_1^2 \lambda_1^2 w^5 z^3
    + \beta_0 \delta_1 \lambda_1 (3 p^2 \pi_1^2 \pi_2^2 
+ p \pi_1 \pi_2 q \chi_1 \chi_2 + 
        3 q^2 \chi_1^2 \chi_2^2) w^3 x z^2- 
\delta_2 \lambda_2 x^3)\, , \\
0 = & 
4 \beta_0 p^2 \pi_1 q^2 \chi_1 (p \pi_1 \pi_2 + q \chi_1 \chi_2)x 
+ \delta_2 s 
+ \delta_1 t \, .  
\end{cases} \label{EQns}
\end{eqnarray}
where the coordinates are sections in the bundles described
in table \ref{table:sections-case2},
and they obey the projective relations  
\begin{eqnarray} \label{E8proj-case2}  
&[\delta_1 \delta_2 \lambda_2 y: \delta_2 \lambda_2 x: w] \neq [0: 0: 0] , 
\quad [\delta_1 y: x: \delta_1 \lambda_1] \neq [0: 0: 0] , \quad [y: \lambda_1] \neq [0: 0] ,& \nn \\
&[ \pi_1 \pi_2^2 \chi_1 \chi_2^2 s : \lambda_2 ] \neq [0: 0] , \quad 
[\pi_1 \pi_2^2 \chi_2 s: \pi_1 \pi_2^2 \chi_2 t: q] \neq [0: 0: 0] , & \label{EQScaling}\\ 
& [\pi_1 \pi_2^2 s,\pi_1 \pi_2^2 t,\chi_1]\neq [0:0:0],\quad  [\pi_2s,\pi_2t,p_1]\neq [0:0:0],\quad [s,t,\pi_1]\neq [0:0:0].& \nn
\end{eqnarray}
where $y$ must be replaced by
 \begin{equation}
  y \rightarrow \pi_1 \pi_2^2 
  \chi_1 \chi_2^2 t 
+ \beta_0 
  \lambda_1 (p \pi_1 \pi_2 + q \chi_1 \chi_2) 
(2 p^2 \pi_1^2 \pi_2^2 - 3 p \pi_1 \pi_2 q \chi_1 \chi_2 + 
     2 q^2 \chi_1^2 \chi_2^2) w^2 z .  \label{EQY}
 \end{equation}

The set of equations (\ref{EQns}), (\ref{EQScaling}) and (\ref{EQY}) defines the blown-up fourfold $\tilde Y_4$ 
in the class 
\begin{equation}
[\tilde Y_1] = 3\sigma + 6c_1 -2 E_1 - 2E_2 -E_3 - E_4\, .
\end{equation}

\subsection{Codimension 1 - The GUT Divisor}
\label{case2co1}
The structure here is similar to the previous case. However, the
different order of small resolutions that introduced $\delta_1$
and $\delta_2$ changes the order of the $\IP^1$s on the fiber,
as depicted in the figure below,

\begin{figure}[H]
\centering
\begin{picture}(46,11)(-12,0)
 \multiput(3,0)(9,0){4}{\circle{6}}
 \multiput(6,0)(9,0){3}{\line(1,0){3}} 
 \put(16.5,9){\circle{6}}
 \put(3,3){\line(2,1){10}}
 \put(30,3){\line(-2,1){10}}
 \put(1.8,-1){A}
 \put(10.8,-1){B}
\put(19.8,-1){C}
\put(28.8,-1){D}
\put(15,8){X}
\end{picture}
\end{figure}

where the depicted exceptional divisors are
defined by
\begin{equation}
 A : \delta_1 = 0, \quad B: \lambda_2 = 0, \quad C: \delta_2 = 0, \quad D: \lambda_1 = 0 
\end{equation}
(compare to \eqref{codim1:divisors-case1} ).

\subsection{Codimension 2}
\label{case2co2}
The first loci we look at in codimension 2 
is the one specified by $ p_0 - 2 q_0 = 0$,
that from the Tate's algorithm should correspond
to a locus of $SO(6)$ enhancement.
Computing the intersections with the 
Cartan roots, we get
table \ref{table:charges:codim2(p-2q=0)-case2}.

\begin{table}[h]\centering
\begin{tabular}{c c c} \hline
Curve & Charges & Weights \\ \hline
$A_{\{p-2q\} }$ & (-2,1,0,0,1)		& $- \alpha_1$			\\ \hline
$B_{\{p-2q\}}$	& (1,-2,1,0,0)		& $- \alpha_2$	\\ \hline
$C_{1\, \{p-2q\}}$	& (0,1,-1,0,0)		& $-\mu_5 +\alpha_1 + \alpha_2$	\\ \hline
$C_{2\, \{p-2q\}}$	& (0,0,-1,1,0)		& $\mu_5 -\alpha_1 + \alpha_2 -\alpha_3$			\\ \hline
$D_{\{p-2q\}}$ 	& (0,0,1, -2, 1) 	& $-\alpha_4 	$		\\ \hline
$X_{\{p-2q\}}$	& (1,0,0,1,-2)		& $-\alpha_5 $			\\ \hline
\end{tabular}
\caption{Curves and their respective weights 
at the $p \pi_1 \pi_2 -2 q \chi_1 \chi_2 =0$ locus. 
The defining equations are presented in table \ref{table:codim2(p-2q=0)-case2}.}
\label{table:charges:codim2(p-2q=0)-case2}
\end{table}

This resolution does not contain a curve on the base of $\mathbf{10}$
matter representation. We can however explore other curves where 
some components of the $\mathbf{10}$ representation appear. Take first the 
curve $p_0 + q_0 =0$, that from table 
\eqref{tableE8enhance-case2} corresponds
to an $\SO(12)$ enhancement. 
The curves have the charges under the $SU(5)$ roots presented in
table \ref{table:E8p-q}, 

\begin{table}[h]\centering
\begin{tabular}{c c c} \hline
Curve & Charges & Weights \\ \hline
$A_{1 \,\{p-q\} }$ & (-2,1,0,0,1)		& $- \alpha_1$			\\ \hline
$A_{2 \,\{p-q\} }$ & (0,1,0,-2,1)		& $- \alpha_1+2(\mu_{10}-\alpha_2-\alpha_3-\alpha_4)$			\\ \hline
$A_{3 \,\{p-q\} }$ & (-1,0,0,1,0)		& $-\mu_{10}+\alpha_2+\alpha_3+\alpha_4$\\ \hline
$B_{\{p-q\}}$	& (1,-2,1,0,0)		& $-\alpha_2$	\\ \hline
$C_{1 \,\{p-q\}}$	& (0,-2,1,1,0)		& $- \alpha_3+3(\mu_5-\alpha_1-\alpha_2)$	\\ \hline
$C_{2 \,\{p-q\}}$	& (0,1,-1,0,0)		& $- \mu_5+\alpha_1+\alpha_2$	\\ \hline
$C_{3 \,\{p-q\}}$	& (0,1,-1,0,0)		& $- \mu_5+\alpha_1+\alpha_2$	\\ \hline
$D_{\{p-q\}}$ 	& (0,0,1, -2, 1) 	& $-\alpha_4 	$		\\ \hline
$X_{\{p-q\}}$	& (1,0,0,1,-2)		& $-\alpha_0 $			\\ \hline
\end{tabular}
\caption{Charges and weights of curves at the $p_0-q_0=0$} \label{table:E8p-q}
\end{table}

Similarly as we had in the first resolution \ref{case1}, the
curves at the $p_0 - q_0=0$ locus of the base
reproduce the $\mathbf{5 + 10}$ representation. The charges 
are not being conserved, as one can check that
\begin{eqnarray}
 A & \ra &  A_{1\, \{p-q\}} + A_{2\, \{p-q\}} + A_{3\, \{p-q\}}\, , \\
 C & \ra &  C_{1\, \{p-q\}} + C_{2\, \{p-q\}} + C_{3\, \{p-q\}}\, ,
\end{eqnarray}
while the multiplicities fail to reproduce these splitting.

Another interesting locus is the ``$E_6$'' enhancement curve,
given by $p_0 =0$ (or $q_0=0$). The charges of the fiber curves 
appearing there are given by table \ref{table:E6-case2}.

\begin{table}[b]\centering
\begin{tabular}{c c c} \hline
Curve & Charges & Weights\\ \hline
$A_{1 \,\{p\} }$ & (0,1,0,-2,1)		& $- \alpha_4+(\mu_5-\alpha_1-\alpha_2)$			\\ \hline
$A_{2 \,\{p\} }$ & (-2,1,0,0,1)		& $- \alpha_1$			\\ \hline
$B_{ \{p\}}$	& (1,-2,1,0,0)		& $-\alpha_2$	\\ \hline
$C_{1 \,\{p\}}$	& (0,1,-1,0,0)		& $\mu_5-\alpha_1-\alpha_2$	\\ \hline
$C_{2 \,\{p\}}$	& (0,0,-1,1,0)		& $- \mu_5+\alpha_1+\alpha_2+\alpha_3$	\\ \hline
$C_{3 \,\{p\}}$	& (0,0,-1,1,0)		& $- \mu_5+\alpha_1+\alpha_2+\alpha_3$	\\ \hline
$D_{\{p\}}$ 	& (0,0,1, -2, 1) 	& $-\alpha_4 	$		\\ \hline
$X_{\{p\}}$	& (1,0,0,1,-2)		& $-\alpha_0 $			\\ \hline
\end{tabular}
\caption{Curves of non-vanishing charges along the $p_0=0$ locus.}\label{table:E6-case2}
\end{table}

If we look also at the curves of zero charge,
at this locus some solutions are not simple $\IP^1$s inside the resolved elliptic curve, as before. 
Take as an example the curve $B$ before restricting
to the $p_0$ locus. 
Along the $p_0=\pi_1 \pi_2 q = 0$ 
locus, $\beta_4$ and $\beta_5$ vanish, 
so the defining equation of the curve simplifies to 
\begin{equation}
 \delta_2 =\pi_1 \pi_2^2 \chi_1 \chi_2^2 s (\pi_1 \pi_2^2 \chi_1 \chi_2^2 t + \lambda_1 \beta_3 w^2)+ 
\lambda_1 \lambda_2 (- \delta_1^2 \lambda_1^2 w^5 + \delta_1 \lambda_1 \beta_2 w^2 x) = \delta_1 t =0 .
\end{equation}

However the restriction 
to the $E_6$ curve can be taken to be 
$p=0$, $\pi_1=0$ or $\pi_2=0$. 
When $\pi_1=0$ the curve $B$ reduces to
\begin{equation}
(\pi_1 =0) \quad \delta_2 =
 \delta_1 \lambda_1^2 \lambda_2 (- \delta_1 \lambda_1 w^5 + \beta_2 w^2 x) = \delta_1 t =0 .
\end{equation}
That has as one possible solution
\begin{equation}\label{B*}
 E_{\ast} : \quad \delta_2 =\pi_1 = \delta_1  =0 .
\end{equation}
The rescaling conditions that have to be obeyed for $ E_{\ast}$ are
\begin{equation*}
[0: 0: w], \, [0: x:0]  , \, [\lambda_1 q \chi_1 \chi_2 w z: \lambda_1] , \, 
[0 : \lambda_2 ], \, [0: 0: q], \,  
[0:0:\chi_1] , \, [\pi_2 s:\pi_2 t: p], \,  , [s:t: 0].
\end{equation*}
Using the fact that $\chi_1 \chi_2 q \neq 0$ and the rescaling conditions 
above, we can fix $\chi_1=q=\lambda_2=x=w=\lambda_1=1$, and we are still 
left with the unfixed coordinates $\pi_2$, $s$, $t$ and $q$, 
together with the conditions
\begin{equation}\label{B*proj}
 [\pi_2 s:\pi_2 t:p] , \, [s:t] .
\end{equation}
So, the solution $E_{\ast}$ is actually a 
$\IP^2$ specified by the coordinates and rescaling $[s':t':p]$
blown up at the point $s'=t'=0$ by a $\IP^1$.

There are however minimal solutions that correspond 
to a $\IP^1$ hypersurface inside this $\IP^2$. 
As one concrete example, take the intersection of $B$ with $p=0$,
\begin{equation}
 (p=) \delta_2 =\pi_1 \pi_2^2 \chi_1 \chi_2^2 s (\pi_1 \pi_2^2 \chi_1 \chi_2^2 t + \lambda_1 (q \chi_1 \chi_2)^3 w^2)+ 
\delta_1 \lambda_1^2 \lambda_2 w^2 (- \delta_1 \lambda_1 w^3 + 3 (q \chi_1 \chi_2)^2 x) = \delta_1 t =0 ,
\end{equation}
that has as one possible solution
\begin{equation}
 E_{\ast p} : \quad (p=)\delta_2 = \delta_1 = \pi_1 = 0 .
\end{equation}
We can see from the defining equations of $E_{\ast}$ 
\eqref{B*} and \eqref{B*proj} that
$E_{\ast p}$ is a $\IP^1$ hypersurface inside $E_{\ast}$.

In the M-theory perspective, the four-cycle $\IP^2$
can be wrapped by M5-branes that correspond to a string in the 
remaining dimensions with tension given by the volume of 
the four-cycle \cite{Bershadsky:1996nu}.
When shrunken back to zero volume, the wrapped M5-branes become
tensionless strings. In the effective theory, this corresponds to 
a tensor multiplet becoming massless, leading to a 
breaking of the low-energy effective theory and thus a phase transition. 
In the Type IIB picture, the blow-up introduce a 
one dimensional $\IP^1$ on the fiber and also a $\IP^1$ on the base
along the matter curve (one non-trivially fibered over the other).
This blown-up $\IP^1$ can be wrapped by a D3-brane, that again in the
blown-down limit give rise to a massless string.
Additionally, we might have to worry about string worldsheet instantons
wrapping the vanishing $\IP^1$s.
The $\IP^1$ along the curve might also 
break the Calabi-Yau condition.
A similar blow-up along a curve in Type IIB picture 
was studied in \cite{Grimm:2010gk}.
A more detailed exploration of the role of 
tensionless strings on the theory (or the phase transition) 
arising in this particular setup would be interesting, 
however we do not deal with it in this work.

\subsection{Codimension 3}
\label{case2co3}
We next restrict the elliptic fiber to the Yukawa point, 
that before the blow ups corresponded to the locus
$w_0 = p_0 = q_0 = 0$. Similarly, the restriction to 
the codimension three locus has as solutions some curves that are 
not $\IP^1$s, but again, some of the internal 
$\IP^1$ hypersurfaces appear
as solutions. 

The calculations of the corresponding charges
give the table \ref{table:E8-case2}.

\begin{table}[H]\centering
\begin{tabular}{c c c} \hline
Curve & Charges & \\ \hline
$A_{1 \,\{E8\} }$ & (-2,1,0,0,1)		& $- \alpha_1$			\\ \hline
$A_{2 \,\{E8\} }$ & (0,1,0,-2,1)		& $- \alpha_4 + (-\mu_5+\alpha_1+\alpha_2)$			\\ \hline
$B_{\{E8\}}$	& (1,-2,1,0,0)		& $-\alpha_2$	\\ \hline
$C_{1 \,\{E8\}}$	& (0,1,-1,0,0)		& $\mu_5-\alpha_1-\alpha_2$	\\ \hline
$C_{2 \,\{E8\}}$	& (0,0,-1,1,0)		& $- \mu_5+\alpha_1+\alpha_2+\alpha_3$	\\ \hline
$C_{3 \,\{E8\}}$	& (0,-2,1,1,0)		& $-\alpha_3 + 3(- \mu_5+\alpha_1+\alpha_2)$	\\ \hline
$D_{\{E8\}}$ 	& (0,0,1, -2, 1) 	& $-\alpha_4 	$		\\ \hline
$X_{\{E8\}}$	& (1,0,0,1,-2)		& $-\alpha_0 $			\\ \hline
\end{tabular}
\caption{Curves of non-vanishing charges along the $p_0=q_0=0$ locus.}\label{table:E8-case2}
\end{table}

\subsection{The $p \leftrightarrow q$ symmetry}
\label{case2pqsym}

The monodromies present among the matter curves 
can lead to constrains to the couplings. In the present model 
we see a clear symmetry under exchange of $p \leftrightarrow q$, that 
thus corresponds to the interchange
\begin{equation}
 (t_1, t_2, t_3, t_4, t_5) \rightarrow (t_2, t_1, t_3, t_5, t_4),
\end{equation}
with the $t_i$s specified in \eqref{E8model-case2}. This is similar to the
$\bbZ_2 \times \bbZ_2$ symmetry as the ones described for example 
in \cite{Heckman:2009mn, Antoniadis:2013joa}. We can then restrict
the number of families by their orbits under the symmetry group,
\begin{eqnarray}
 Orb(5_{(1)})&:& p+q \nn \\
 Orb(5_{(2)})&:& p,q \nn \\
 Orb(5_{(3)})&:& p+2q,q+2p \nn \\
 Orb(5_{(4)})&:& p-2q,q-2p \\
 Orb(5_{(5)})&:& p-q,q-p \nn \\
 Orb(10_{(1)})&:& p+q \nn \\
 Orb(10_{(2)})&:& p,q .\nn
\end{eqnarray}
The fact that $10_{(1)}$ and $10_{(2)}$
lie in the same matter curve as $5_{(1)}$ and
$5_{(2)}$, respectively, further restrict the possible 
superpotential couplings we can construct in this model.

\section{Conclusions}
\label{conclusions}

We use the Griffiths-Frobenius structure and homological mirror symmetry to
construct an integer monodromy basis on the primitive horizontal subspace of Calabi-Yau 
fourfolds together with a pairing $\eta$ that determines the K\"ahler potential in the 
complex moduli space.  For all hypersurfaces and complete intersections this yields  a  general method to obtain the 
exact expressions for periods that converge at large radius limits with a radius of 
convergence that is limited by the distance  to the generic conifold divisor.  We 
confirmed this method by comparison with the results of supersymmetric localization 
on the hemi-sphere with supersymmetric boundary condition  by Hori and Romo~\cite{Hori:2013ika}. 
In particular one can specify  boundary data that describe a basis of A-model branes that correspond  
to twisted line bundles. The formalism of~\cite{Hori:2013ika} allows to write down a preferred Mellin-Barnes 
integral representation for the corresponding basis of $B$-model  periods, which relate roughly two regions in the moduli space: 
the large radius and the Landau Ginzburg region.  Homological mirror symmetry and the Hirzebruch 
Riemann Roch formula yield the pairing $\eta$ for these $K$-theory  classes.  This basis  
does not correspond to a minimal integer monodromy basis as can be seen from the rational
entries in $\tilde m$ in (\ref{tildemsixtic}-\ref{tildem}). The relative factor is crucial. e.g. for the 
determination of the actual chiralities in F-theory. 

In conjunction with the ideal of  Picard-Fuchs equations our method of determining the  integer 
monodromy basis allows to give convergent expression  for the periods everywhere in the  moduli
space and  therefore to obtain exact expressions of the superpotential and the  K\"ahler 
potential in all regions of the vev's of the $\mathcal{N}=1$ scalar fields.  Even more basic, but quite 
important  for F-theory phenomenology,  it determines  the correct $\mathcal{N}=1$ coordinates  
in which the effective action is locally parametrized. This is natural for the complex deformations, 
but mirror symmetry extends the method to the complexified K\"ahler parameters. These periods 
have very interesting arithmetic  and modular properties. Because of the occurrence of closed and 
open moduli in the fourfold periods they imply interesting analytic relations between open and  
closed amplitudes.                 

We show in detail that the complex moduli  can be stabilized by the $G_4$-form superpotential  at orbifold points  
and that  different than in the threefold case  for suitable fluxes the scalar potential for 
the complex fields vanishes. This behaviour occurs generically in F-theory vacua 
described by Fermat hypersurfaces. We also analyze   the super potential and 
the K\"ahler potential  close the most generic singularity of a Calabi-Yau fourfold, 
where $S^4$ cycles shrink and a $\mathbb{Z}_2$ singularity appears and show 
that a flux on the vanishing period can drive the theory to the $\mathbb{Z}_2$
orbifold point.  For the generic conifold discriminant of the GKKZ system,  homological mirror 
symmetry maps this vanishing cycle to the structure sheaf of the mirror manifold. 
This information in conjunction with Griffiths transversality allows to see that the $\hat \Gamma$-classes give the right 
modification of the Chern character map to yield an integer monodromy basis and the 
right central charges for the Bridgeland stability conditions~\cite{MR2373143}.

Symmetries  on the Calabi-Yau manifold $M_n$ give rise to symmetries  in the moduli 
space ${\cal M}_{cs}$. Based  on the latter symmetries and the local behaviour 
of the periods  at the invariant sublocus ${\cal S}$ we describe  the conditions 
under which one can find an integral flux that drives the moduli  of the theory 
to the invariant sublocus ${\cal S}$, with and without creating a potential 
on ${\cal S}$. The Hodge bundle  ${\cal H}$ restricted to ${\cal S}$  carries  a 
sub-monodromy problem.  In particular  for Calabi-Yau manifolds embedded as hypersurfaces 
into toric ambient spaces given by a pair of reflexive polyhedra $(\Delta, \Delta^*)$ 
one can characterize certain sub-mondromy problems and the  locus ${\cal S}$ by  
reflexive pairs  $(\ci \Delta, \ci \Delta^*)$  with the embedding  condition 
$\ci \Delta \subset \Delta$.  This yields a restricted ideal of Picard 
Fuchs operators defining the sub-mondromy  problem on ${\cal H}|_{\cal S}$. 
We argue  that one pick  always an integer flux that creates a potential 
with ${\cal S}$ as vacuum manifold.  The two constructions have an overlap, 
because in many cases $\ci \Delta$ can be readily defined by the monomials that are  
invariant under a symmetry in $\mathbb{P}_\Delta$ that acts also on $M_n$.
   
Classical mirror symmetry  is obviously useful  for determining the holomorphic F-theory amplitudes. 
Mirror symmetry  in the $tt^*$ structures could  extend the predictions  beyond these terms.  
Homological mirror symmetry played a small role in our analysis, but there are very interesting 
much further reaching observations. For example for the Hitchin systems with gauge groups $G$  it has been observed  that 
homological mirror symmetry can be identified with Langland's duality exchanging $G$ with the Langland's 
dual group $\hat G$, see \cite{MR2648685} for a review. These Hitchin system arise in the limit of F-theory 
compactifications and for these cases  Langland's duality is implied  by homological mirror symmetry 
on fourfolds. The ubiquitous occurrence of modular  forms in the amplitudes  on fourfolds, when 
expressed in the right coordinates, resonates well with these observations. In particular the 
superpotential and the scalar potential, that we can calculate everywhere in the moduli space inherit 
these modular properties and it would be interesting to make a serious study of axion inflation 
based on these exact  expressions.

In Section \ref{heteroticbasis} we construct families of toric hypersurfaces to give global Calabi-Yau 4-folds with 
elliptic fibrations of type $D_5$, $E_6$, $E_7$ and $E_8$, and 
the most general complex structure, which
can be specialized by turning on $G_4$ fluxes. In this
way  singularities in different co-dimensions are obtained yielding
non-abelian gauge symmetry enhancements, matter curves and Yukawa couplings.
                        
Starting in Section \ref{sec-E8yukawa} we constructed
two models that contained a codimension 
3 locus with an $E_8$ singularity.  First, we discuss
how a codimension 3 $E_8$ singularity can be embedded in
some of the studied global models. 
We imported results from F-theory/Heterotic 
duality, specially the identification 
via the spectral cover of the coefficients 
in the elliptic fiber with the vevs for the Higgs field responsible 
for breaking the $E_8$ gauge group in $SU(5)$ with 
the desired higher codimension enhancements.

To fully resolve the space, we had to 
perform blow-ups that introduced 
two(-complex)-dimensional subspaces 
along the codimension two enhancement loci. After resolving,
we identified the enhancement loci of the two models and 
calculated the charges under the $SU(5)$ group, identifying
the associated weights of $\mathbf{5}$ and $\mathbf{10}$. We argued
that some curves in our particular choice of moduli 
carry a $\mathbf{5} + \mathbf{10}$ representation.

Also, we show that both the models presented do not
give an $E_8$ Dynkin diagram after resolving the $E_8$ point.
Similar feature appears with the codimension 2 locus of
$E_6$ singularity, but is not observed when the singularities
are of $SO(2n)$ or $SU(n)$ types. 
It appears to us that the mismatch between
the expected fiber structure of the resolved space and the 
type of singularity is a feature of the strong coupling 
of the exceptional $E_k$ kinds.

\bigskip
\subsection*{Acknowledgments}
We would like to thank Victor Batyrev, Marcos Marino,  Emanuel Scheidegger and 
Stefan Theisen for correspondences and Sergei Galkin, Daniel Huybrecht, Sheldon Katz,  Duco van Straten and 
Yongbin Ruan for mathematical advises. We benefited from discussion  
with  Pedro Acosta,  Babak Haghighat, Sven Krippendorf, Maximilian Poretschkin, Dustin Ross, and 
Timo Weigand. Special thanks to Kentaro Hori, Hiroshi Irritani,   Hans Jockers 
and Denis Klevers for many useful conversations and explanations of their work.

A.K. and D.V.L. are supported by the DFG grant KL 2271/1-1, D.V.L. wants to thank
the Haussdorff Institut for hospitality and support. D.V.L. was also supported by the PDJ fellowship of 
CNPq, grant number 501107/2012-6. N.C.B. wants to thank for support by a PROMEP postdoctoral grant, 
 the SFB-Tansregio TR33 ``The Dark Universe" (DFG), the EU 7th network program ``Unification in the LHC era" 
(PITN-GA-2009-237920), CEADEN and ``Proyecto de Ciencias B\'asicas: Teor\'{\i}a Cu\'antica de Campos 
en F\'{\i}sica de Part\'{\i}culas y de la Materia Condensada" ICIMAF, CITMA.  

This work was finished while A.K enjoyed the hospitality of 
 MCTP in the Physics Department and the Mathematics Department of the University 
of Michigan in Ann Arbor, supported by the DMS grant DMS1159265.      

\appendix

\section{Gauge symmetries enhancements for $E_8, E_7$ and $E_6$ fiber types.}

All the fiber types $E_6\times U(1)^2, E_7\times U(1)$ and $E_8$ can be brought to the Weierstrass form. In table
\ref{table:kodaira} we have summarized  Kodaira classification of codimension 1 singularities. Using the given information one can obtain the  monodromies using Tate's algorithm. In the following we will describe the obtained enhancement for the different cases.

\begin{table}[hbt]
\begin{center}
\begin{tabular}{|c|c|c|c|c|c|c|c|}
\hline 
 Type 	&$\ord(f)$	& $\ord(g)$	& $\ord(\Delta)$	& $a$ & $j(\tau)$  	&   Group  	& Monodromy	\\ \hline 
$I_0$ 	& $\geq 0$  	& $\geq 0$ 	& $0$ 			&  0 & $\mathbb{R}$ 	&  \textemdash	& 
$\begin{pmatrix}0 & 1\\-1 & 0\end{pmatrix}$ \\ \hline
$I_1$ 	& $0$ 		& $0$ 		& $1$  			&  $\frac{1}{12}$ & $\infty$ 	& $U(1)$	& 
{\small$\begin{pmatrix}1& 1\\0 & 1\end{pmatrix}$}								\\ \hline
$I_n$ 	&$0$ 		&   $0$ 	& $n>1$			& $\frac{n}{12}$ & $\infty$ 	& $A_{n-1}$	& 		
{\small$\begin{pmatrix}1& n\\0 & 1\end{pmatrix}$}								\\ \hline
$II$ 	& $\geq 1$ 	&  $1$ 		& $2$   		& $\frac{1}{6}$ & $0$ 		& \textemdash	& 
{\small $\begin{pmatrix}1& 1\\-1 & 0\end{pmatrix}$}								\\ \hline
$III$ 	& $1$ 		& $\geq 2$  	& $3$ 			& $\frac{1}{4}$ & $1$ 		& $A_1$		& 
{\small $\begin{pmatrix}0 & 1\\-1 & 0\end{pmatrix}$}								\\ \hline
$IV$ 	&  $\geq 2$ 	& $2$		& $4$			& $\frac{1}{3}$ & $0$		& $A_2$		& 
{\small $\begin{pmatrix}0 & 1\\-1 & -1\end{pmatrix}$}								\\ \hline
\multirow{2}{*}{$I^*_n$} & $2$ &$\geq 3$& \multirow{2}{*}{\small $n+6$} & \multirow{2}{*}{\small $\frac{1}{2}+\frac{n}{12}$} & \multirow{2}{*}{\small  $\infty$} 
& \multirow{2}{*}{$D_{n+4}$} & \multirow{2}{*}{$ \begin{pmatrix} -1& -b\\ 0 &- 1 \end{pmatrix}$ }		\\ \cline{2-3}
  	& $\geq 2$	& $3$ 		&  			&  		&  		&		\\ \hline
$IV^*$ 	& $\geq 3$ 	& $4$ 		& $8$ 			& $\frac{2}{3}$  & $0$ 	        & $E_6$		& 
{\small $\begin{pmatrix}-1& -1\\1 & 0\end{pmatrix}$} 								\\ \hline
$III^*$ & $3$		& $\geq 5$	& $9$			& $\frac{3}{4}$ & $1$ 		& $E_7$		& 
{\small  $\begin{pmatrix}0& -1\\1 & 0\end{pmatrix}$}								\\ \hline
$II^*$ & $\geq 4$ 	& $5$ 		& $10$ 			&  $\frac{5}{6}$&  $0$ 		& $E_8$		& 
{\small$\begin{pmatrix}0& -1\\1 & 1\end{pmatrix}$}								\\ \hline
\end{tabular}
\end{center}
\caption{ Kodaira Classification of  singular fibers of an K3 elliptic fibration.
Table extracted from \cite{Esole:2011sm,Bershadsky:1996nh,Katz:2011qp}.}\label{table:kodaira}
\end{table}
\vskip - 1cm

\subsection{The $E_8$ fibre and Tate's algorithm} 
\label{tate} 
In Table \ref{table:tatenonsplit} we have represented the so called non-split cases, which are enhancements that
occur only for a given vanishing order of the coefficients $a_i$ and the discriminant $\Delta$, those
are the more generic cases. In addition, when an extra polynomial factorization occurs one obtains
the split enhancements \cite{Bershadsky:1996nh}. We have checked which of those cases
are realized in the global toric construction with polytop 1 in figure \ref{poly} and no twisting. In this case 
the vectors $\nu^{(i)*}$ in (\ref{polyhedrabase}) are given by $u_1=(1,0,0), u_2=(0,1,0), u_3=(-1,-1,0), w_1=(0,0,1), w_2=(0,0,-1)$ to give
a base $B_3=\mathbb{P}^1\times\mathbb{P}^2$. In Table \ref{tates} we give the occurrence of the explored split enhancements, 
the polynomials which factorization has to be checked are given in Table \ref{spliteqs}. 
Let us explain next how these factorizations occur in some examples.

\begin{table}[hbt]
\begin{center}
\caption{ Tate's Algorithm for the non-split cases of  Table 2 in \cite{Bershadsky:1996nh}\cite{Katz:2011qp}.}
\label{table:tatenonsplit}\footnotesize
\begin{tabular}{|c|c|c|c|c|c|c|c|}
\hline 
 Type 	& group 	& $a_1$	&   $a_2$	&   $a_3$ & $a_4$  	&  $a_6$  	&  $\Delta$\\ \hline 
 $I_0$&-- & 0 & 0 & 0 & 0 & 0& 0\\ \hline
$I_1$ &--&0 & 0 & 1 & 1 & 1 & 1\\ \hline
$I_2$ & $SU(2)$& 0 & 0 & 1 & 1 & 2 &2 \\ \hline
$I_3^{ns}$& $Sp(1)$& 0 & 0 & 2 & 2 & 3 & 3\\ \hline
$I_{2k}^{ns}$& $Sp(k)$& 0 & 0 & $k$ & $k$ & $2k$ & $2k$\\ \hline
$I_{2k+1}^{ns}$& $Sp(k)$ & 0 & 0& $k+1$& $k+1$ & $2k+1$ & $2k+1$\\ \hline
$II$& --& 1 & 1 & 1 & 1 & 1 & 2\\ \hline
$III$& $SU(2)$& 1 & 1 & 1 & 1 &2 & 3 \\ \hline
$IV^{ns}$& $Sp(1)$ & 1 & 1 & 1& 2 & 2 & 4 \\ \hline
$I_0^{*ns}$& $G_2$& 1 & 1 & 2 & 2 & 3 & 6\\ \hline
$I_1^{*ns}$& $SO(9)$ & 1 & 1 & 2 &3 & 4 & 7\\ \hline
$I_2^{*ns}$& $SO(11)$& 1 & 1& 3 & 3 & 5 & 8 \\ \hline
$I_{2k-3}^{*ns}$& $SO(4k+1)$ & 1 & 1 & $k$ & $k+1$& $2k$ & $2k+3$\\ \hline
$I_{2k-2}^{*ns}$ & $SO(4k+3)$& 1 & 1 & $k+1$ & $k+1$& $2k+1$ & $2k+4$\\ \hline
$IV^{*ns}$& $F_4$ & 1 & 2 & 2 & 3 & 4 & 8\\ \hline
$III^*$ & $E_7$ & 1 & 2 & 3 &3 & 5 & 9 \\ \hline
$II^*$ & $E_8$ & 1 & 2 & 3  & 4 & 5 & 10\\ \hline
non-min& -- & 1 & 2 & 3 & 4 & 6 & 12\\ \hline
\end{tabular}
\end{center}
\end{table}
\begin{table}[hbt]
\begin{center}
\caption{ Tate's Algorithm for the split cases of  Table 2 in \cite{Bershadsky:1996nh}\cite{Katz:2011qp}.}
\label{table:tatesplit}\footnotesize
\begin{tabular}{|c|c|c|c|c|c|c|c|}
\hline 
 Type 	& group 	& $a_1$	&   $a_2$	&   $a_3$ & $a_4$  	&  $a_6$  	&  $\Delta$\\ \hline
$I_0^{*ss}$ & $SO(7)$& 1 & 1 & 2 & 2 & 4 &6\\ \hline
$I_0^{*s}$ & $SO(8)^*$& 1 & 1 & 2 & 2 & 4 &6 \\ \hline 
$I_1^{*s}$ & $SO(10)$& 1 & 1 & 2 & 3 & 5 & 7   \\ \hline
$I_2^{*s}$ & $SO(12)^*$& 1 & 1 & 3 & 3 & 5 & 8 \\ \hline
$I_3^s$& $SU(3)$ & 0 & 1 & 1 & 2 & 3& 3 \\ \hline
$I_{2k}^s$ & $SU(2k)$& 0 & 1 & $k$ & $k$ & $2k$ &$2k$ \\ \hline
$I_{2k+1}^s$ & $SU(2k+1)$&0 & 1 & $k$ & $k+1$ & $2k+1$ &$2k+1$\\ \hline
$I_{2k-3}^{*s}$ & $SO(4k+2)$& 1 & 1 & $k$ & $k+1$ & $2k+1$ & $2k+3$ \\ \hline
$I_{2k-2}^{*s}$ & $SO(4k+4)$& 1 & 1 & $k+1$ & $k+1$ & $2k+1$ & $2k+4 $\\ \hline
$IV^{s}$ & $SU(3)$& 1 & 1 & 1 & 2 & 3 &4\\ \hline
$IV^{*s}$ & $E_6$& 1 & 2 & 2 & 3 & 5 & 8 \\ \hline
\end{tabular}
\end{center}
\end{table}
\begin{table}[hbt]
\begin{center}
\caption{ Toric CY realization with no twisting $n_i=0$ in (\ref{polyhedrabase}), of split cases from Table 2 in \cite{Bershadsky:1996nh}. 
($\checkmark$) means that the model  is realized with the splitting, (dns) means that it does not split. (aut) means 
that there is an additional automorphism, which reduces $h_{11}$ and (as) means that there are additional singularities 
due to the global constraints. The latter lead to extra $h_{11}$ forms. $A_{13}$ is the last case in the $A_k$  series 
with $[a_i]=[0,1,6,7,\infty]$ and $D_{13}$ is the last case in the $D_k$  series with $[a_i]=[1,1,6,7,\infty]$. Above this $k$  the polyhedra are not reflexive. }\label{tates}\footnotesize
\begin{tabular}{|c|c|c|c|c|c|c|}
\hline 
 Type 	& group 	& $\chi$  & $h_{11}$ &$ h_{31}$ & $h_{2,1}$& Com. \\ \hline 
$I_4^s$ &$SU(4)$& 14328& 6 & 3277 &0 &  $\checkmark$\\ \hline
$I_5^s$ & $SU(5)$&12978&7& 2148&0   & $\checkmark$ \\ \hline
$I_6^s$ & $SU(6)$&  11682&8& 1931&0  &  $\checkmark$ \\ \hline
$I_7^s$ & $SU(7)$&  10332&9 & 1705&0 & $\checkmark$ \\ \hline
$I_8^s$ &$SU(8)$& 9036& 10 & 1488 &0 & $\checkmark$ \\ \hline
$I_9^s$ & $SU(9)$&7686&11& 1262&0   & $\checkmark$ \\ \hline
$I_{10}^s$ & $SU(10)$&6390    & 12  & 1245 &0  &  $\checkmark$ \\ \hline
$I_{11}^s$ & $SU(11)$& 5040   &  13 &  819 &0 &  $\checkmark$\\ \hline
$I_{12}^s$ & $SU(12)$&3744    &  14 &  602&0  &  $\checkmark$ \\ \hline
$I_{13}^s$ & $SU(13)$& 2232   &  16 &  348 &0 & $\checkmark$ as (1)\\ \hline
$IV^{s}$ & $SU(3)$&    15624     &  5  &   2591&0& $\checkmark$\\ \hline
$I_0^{*ss}$ & $SO(7)$&14328& 6& 2374&0 & dns\\ \hline
$I_0^{*s}$ & $SO(8)^*$& 14328& 6& 2374&0 &dns aut (-1)\\ \hline
$I_1^{*s}$ & $SO(10)$    & 12924 & 8   & 2138        &0 & $\checkmark$ \\ \hline
$I_2^{*s}$ & $SO(12)^*$& 12924 & 8   & 2138(55)& 0&    dns  aut (-1) \\ \hline
$I_3^{*s}$ & $SO(14)$& 10224& 11 & 1695&10& $\checkmark$ \\ \hline
$I_4^{*s}$ & $SO(16)$&  10224& 11 & 1695(55) &10& dns \\ \hline
$I_5^{*s}$ & $SO(18)^*$& 7524 & 14   & 1252&20&  $\checkmark$ (+2)\\ \hline
$I_6^{*s}$ & $SO(20)    $& 7524& 14  & 1252(55) &20& dns (+1) \\ \hline
$I_7^{*s}$ & $SO(22)$&  10224& 17 & 809 &30&  $\checkmark$ (+3) \\ \hline
$I_8^{*s}$ & $SO(24)$&  10224& 17 & 809(55) &30&  dns (+2) \\ \hline
$I_9^{*s}$ & $SO(26)$&  1962& 21 & 338 &40&  $\checkmark$ (+5) \\ \hline
$IV^{*s}$ & $E_6$& 12762& 9& 2110 &0&  $\checkmark$ \\ \hline
\end{tabular}
\end{center}
\end{table}  
 The Kodaira fiber $I_3^s$ \cite{Bershadsky:1996nh} has 
vanishing orders of the coefficients around $w=0$ given by $a_2=a_{2,1}w$, $a_3=a_{3,1}w$, $a_4=a_{4,2}w^2$ and $a_6=a_{6,2}w^2$. Together with the Tate's form expression (\ref{tateform}) for  $z=1$ the required factorization \cite{Bershadsky:1996nh} is
\begin{eqnarray}
y^2+a_1 x y-a_2 x^2&=& (y-r x)(y- s x)\mod w.\label{su3}
\end{eqnarray}
We find that (\ref{tateform}) at $z=1$ can be written as (\ref{su3}) with $r$ and $s$ given by
\begin{eqnarray}
2r&=&-a_1\mp\sqrt{a_1^2+4 a_2}=-a_1\mp\sqrt{\chi^2+ O(w)}=-a_1\mp\chi +\frac{1}{2} O(w), \nonumber\\
2s&=&-a_1\pm\sqrt{a_1^2+4 a_2}=-a_1\pm\sqrt{\chi^2+ O(w)}=-a_1\pm\chi +\frac{1}{2} O(w).\nonumber
\end{eqnarray}
Therefore in a vicinity of $w=0$ a factorization of (\ref{su3}) involving analytic functions $r$ and $s$ occurs $\mod w$ :
\begin{eqnarray}
y^2+a_1 x y-a_2 x^2&=&  \left(y-\frac{1}{2}(-a_1\mp\chi)x\right)\left(y-\frac{1}{2}(-a_1\pm\chi)x\right)\mod w,
\end{eqnarray}
with
\begin{eqnarray}
\chi=w^2 (u_1^3 + u_1^2 u_2 + u_1 u_2^2 + u_2^3 + u_1^2 u_3 + u_1 u_2 u_3 + u_2^2 u_3 + u_1 u_3^2 + u_2 u_3^2 + u_3^3).\label{factsuN}
\end{eqnarray}  
In the sum from (\ref{factsuN}) we have omitted the moduli coefficients, so let us make clear that this factorization occurs without fixing the moduli. This situation is also found for the fibers $I_4^s, I_5^s$, $I_6^s$ and $I_7^s$ represented in Table \ref{tates}, and the factorization occurs exactly in the same way.
The table \ref{spliteqs} gives the list of polynomials on which the non-split, split and semi-split cases occur.
 \begin{table}[hbt]
\begin{center}
\caption{ In the first column the Kodaira fiber type is given, the second column
represents the gauge symmetry and in the last column the polynomials that need to factorize to have the split gauge enhancements 
are given\cite{Bershadsky:1996nh}.}
\label{spliteqs}\footnotesize
\begin{tabular}{|c|c|c|}
\hline 
Kodaira fiber 	& Group 	& Equation\\ \hline 
$I_{2k}^{ns}, I_{2k}^s$ & $Sp(k), SU(2k)$& $Y^2+a_1 X Y-a_2 X^2$\\ \hline
$I_{2k+1}^{ns}, I_{2k+1}^s$ & $\text{unconven.},SU(2k+1)$& $Y^2+a_1 X Y-a_2 X^2$\\ \hline
$I_0^{*ns}, I_0^{*s}, I_0^{*ss}$ & $G_2, SO(8)^*,SO(7)$& $X^3+a_{2,1}X^2+a_{4,2}X+a_{6,3}$ \\ \hline
$I_0^{*s}$ & $SO(8)^*$& $X^2+a_{2,1}X+a_{4,2}$ \\ \hline
$I_{2k-2}^{*ns} , I_{2k-2}^{*s}$ & $SO(4k+3), SO(4k+4)^*,\, k\geq 2$& $a_{2,1}X^2+a_{4,k+1}X+a_{6,2k+1}$ \\ \hline
$I_{2k-3}^{*ns}, I_{2k-3}^{*s}$ & $SO(4k+1), SO(4k+2)$&$Y^2+a_{3,k}Y-a_{6,2 k}$  \\ \hline
$IV^{*ns}, IV^{*s}$ & $F_4, E_6$&$Y^2+a_{3,2}Y-a_{6,4}$  \\ \hline
$IV^{ns}, IV^{s}$ & $\text{unconven.}, SU(3)$&$Y^2+a_{3,1}Y-a_{6,2}$  \\ \hline
\end{tabular}
\end{center}
\end{table} 


Let us see the remaining cases. For $I^{*s}_1$ the factorization that has to occur is the one of the polynomial $Y^2+a_{3,2}Y-a_{6,4}$. Around $v=0$, this can be seen from the roots, that appear when $a_{3,2}^2+4a_{6,4}$ is  a square. Here we obtain
\begin{eqnarray}
a_{3,2}^2+4a_{6,4}&=&w^8 (u_1^9 + u_1^8 u_2 + u_1^7 u_2^2 + u_1^6 u_2^3 + 
  u_1^5 u_2^4 + u_1^4 u_2^5 + u_1^3 u_2^6 + u_1^2 u_2^7 + u_1 u_2^8  \label{aster}\\
  &&   + u_2^9 + u_1^8 u_3 + u_1^7 u_2 u_3 + 
   u_1^6 u_2^2 u_3 + u_1^5 u_2^3 u_3 + u_1^4 u_2^4 u_3 + u_1^3 u_2^5 u_3+ u_1^2 u_2^6 u_3 \nonumber\\
   &&   + u_1 u_2^7 u_3 + 
    u_2^8 u_3 + u_1^7 u_3^2 + u_1^6 u_2 u_3^2 +  u_1^5 u_2^2 u_3^2 + u_1^4 u_2^3 u_3^2 + 
    u_1^3 u_2^4 u_3^2 + u_1^2 u_2^5 u_3^2  \nonumber\\
   && + u_1 u_2^6 u_3^2 +u_2^7 u_3^2 + u_1^6 u_3^3 + u_1^5 u_2 u_3^3 + 
  u_1^4 u_2^2 u_3^3 + u_1^3 u_2^3 u_3^3 + 
    u_1^2 u_2^4 u_3^3 + u_1 u_2^5 u_3^3  \nonumber\\
   &&+ u_2^6 u_3^3 + u_1^5 u_3^4 + u_1^4 u_2 u_3^4 + u_1^3 u_2^2 u_3^4 + 
   u_1^2 u_2^3 u_3^4 + u_1 u_2^4 u_3^4 + u_2^5 u_3^4 +u_1^4 u_3^5  \nonumber\\
   && + u_1^3 u_2 u_3^5 + u_1^2 u_2^2 u_3^5 + 
    u_1 u_2^3 u_3^5 + u_2^4 u_3^5 + u_1^3 u_3^6 +  u_1^2 u_2 u_3^6 + u_1 u_2^2 u_3^6 \nonumber \\
   &&+ u_2^3 u_3^6 + 
    u_1^2 u_3^7 + u_1 u_2 u_3^7 + u_2^2 u_3^7 + u_1 u_3^8 + u_2 u_3^8 + u_3^9)^2\, ,\nonumber
\end{eqnarray}
omitting the moduli coefficients. And the split singularity occurs in our construction. The same occurs
for $I^{*s}_3$, where $Y^2+a_{3,3}Y-a_{6,6}$ factorizes because $a_{3,3}^2+4a_{6,6}$
is a square proportional to  (\ref{aster}). The same occurs for the case 
$I^{*s}_5$ where $a_{3,4}^2+4a_{6,8}$ is a square proportional to (\ref{aster}) 
giving a factorization of $Y^2+a_{3,4}Y-a_{6,8}$.

The other realized split case is the singularity $IV^*$, which appears when $(Y^2 + a_{3,2} Y - a_{6,4} )$ factorizes. This happens when $a_{3,2}^2 + 4 a_{6,4}$ is a square in the vicinity of $v=0$,
and this is the case because is identical to (\ref{aster}).

For $I_0^*$ the polynomial which needs to be analyzed  is
\begin{equation}
X^3+a_{2,1}X^2+a_{4,2} X+a_{6,3}.\label{I0}
\end{equation}
There are the cases non-split, semi-split or split, depending on wether (\ref{I0})
is non-factorizable, factorizable in terms of linear and quadratic terms or factorizable
in three linear terms. The model in this appendix only accommodates the non-split case
generically. Also for the case $I_{0}^{*s}$ with group $SO(8)$, the moduli need specification
to obtain the factorization of $(X^2 + a_{2,1} X + a_{4,2} )$, which occurs when
$a_{2,1}^2 - 4 a_{4,2}$ is a square. The same happens for the singularity $I^{*s}_2$ which can also not been obtained generically. The study presented is partial,  but with our construction we
have the tools to explore systematically by varying the data of the Calabi-Yau, whether the split enhancements not realized here appear generically in other models.

\subsection{Codimension 1 singularities for $E_7\times U(1)$ and the $E_6\times U(1)^2$  fibre .}
\label{E7U1} 
 In this section we give the list of codimension 1 enhancements for the $E_7\times U(1)$ fiber as well as the $E_6\times U(1)^2$. 
In the  tables \ref{E7U1vanish} and  \ref{E7U1vanish}   every column gives the vanishing order 
of the coefficients of the elliptic fibration equation in terms of a coordinate $u_1$, such that at 
$u_1=0$ the given singularity appears. Furthermore for the toric CY construction embedding the singularities, 
we consider the basis $B_2=\mathbb{P}^1$ and $B_3=\mathbb{P}^3$, following the vanishing orders 
of the coefficients the properties of the 3-fold and 4-fold are given.
\begin{table}[htdp]
\begin{center}
\caption{A list of vanishing order of the coefficients in (\ref{e7form}) and the corresponding singularity for 
the $E_7\times U(1)$ fiber.}
\label{E7U1vanish} 
\scriptsize
\begin{tabular}{|c|c|c|c|c|c|c|c|c|c|c|c|c|} \hline
Singularity & $a_3$ &  $a_2$ & $b_2$ & $a_1$ & $a_0$ &  $\chi_3$ & $h^3_{11}$& $h^3_{21}$ & $\chi_4$ & $h^4_{11}$ & $h^4_{31}$ & $h^4_{21}$\\  \hline 
$D_0$& $\geq 1$& $\geq 1$ & $\geq 1$ & $\geq 2$ & 2 &  -228 & 6(0) & 120(4) & 6612 & 6(0) & 1088(15) & 0\\ \hline 
$D_0$& $\geq 1$& 1 & 1 & $\geq 2$ & 3 &  -208 & 7(0) & 111(0) & 6024 & 7(0) & 989(0) & 0\\ \hline 
$D_1$& $\geq 1$& 1 & $\geq 2$ & $\geq 2$ & 3 &  -208 & 7(0) & 111(6) & 6024 & 7(0)  & 989(36) & 0\\ \hline 
$D_1$& $\geq 1$&  $\geq 2$ & 1 & 2 & 3 &  -194 & 8(0) & 105(0) & 5814 & 8(0)  & 953(0)  & 0\\ \hline 
$D_1$& $\geq 1$&  1 & $\geq 2$ & 2 & 4 &  -180 & 8(0) & 98(0) & 5244 & 8(0)  & 858(0)  & 0\\ \hline 
$D_2$& $\geq 1$&  $\geq 2$ & 1 & $\geq 3$ & 3 &  -176 & 9(0) & 97(0) & 5424 & 9(0)   & 887(0)   & 0 \\ \hline 
$D_2$& $\geq 1$&  1 & $\geq 2$ & $\geq 3$ & 4 &  -160 & 13(4) & 93(3) & 4824  & 9(0)  & 802(10)   & 15 \\ \hline 
$D_2$& $\geq 1$&  2 & 1 & $\geq 3$ & 4 &  -156 & 14(4) & 92(0) &  4878 & 10(0)   & 810(0)   & 15 \\ \hline 
$D_2$& $\geq 1$&  1 & 2 & $\geq 3$ & 5 &  -144 & 14(4) & 86(0) & 4392  & 10(0)  & 729(0)  &  15\\ \hline 
$A_1$& $\geq 0$&  1 & 0 & $\geq 1$ & 1 &  -280 & 4(0) & 144(0) & 8064  & 4(0)   &  1332(0) &  0 \\ \hline 
$A_1$& $\geq 0$&  0 & $\geq 1$ & $\geq 1$ & 2 &  -264 & 4(0) & 136(0) & 7338   & 4(0)   & 1211(0)  &  0\\ \hline 
$A_2$& 0 &  $\geq 1$ & $\geq 1$ & 1 & 2 &  -248 & 5(0) & 129(0) &  7074 & 5(0)  & 1166(0)   & 0  \\ \hline 
$A_3$& $\geq 0$&  $\geq 2$ & 0 & $\geq 2$ & 2 &  -240 & 5(0) & 125(2) & 6804 & 5(0)  & 1121(3)   & 0\\ \hline 
$A_3$& 0&  $\geq 2$ & $\geq 1$ & $\geq 2$ & 2 &  -228 & 6(0) & 120(0) &  6612 & 6(0)  & 1088(0)   & 0\\ \hline 
$A_3$& 0&  1 & 1& $\geq 2$ & 3 &  -218 & 6(0) & 115(0) & 6108   & 6(0)    & 1004(0)    & 0 \\ \hline 
$A_3$& $\geq 0$&  0 & $\geq 2$& $\geq 2$ & 4 &  -212 & 5(0) & 111(2) & 5604  &   5(0) & 921(3)   & 0\\ \hline 
$A_4$& 0&  $\geq 2$ & 1& 2 & 3 &  -204 & 7(0) & 109(0) & 5898  &   7(0) & 968(0)   & 0\\ \hline
$A_4$& 0&  1 & $\geq 2$ & 2 & 4 &  -190 & 7(0) & 102(0) & 5328 & 7(0) &  873(0) & 0\\ \hline
$A_5$& $\geq 0$&  $\geq 3$ & 0 & $\geq 3$ & 3&  -204& 6(0) & 108(4) & 5712 & 6(0)  & 938(6)  & 0\\ \hline
$A_5$& 0&  $\geq 3$ & 1 & $\geq 3$ & 3&  -186& 8(0) & 101(0) &  5508 & 8(0) & 902(0)  & 0\\ \hline
$A_5$& 0& 2 & 1 & $\geq 3$ & 4&  -176& 8(0) & 96(0) & 5046 & 8(0) & 825(0) & 0\\ \hline
$A_5$& 0&  $\geq 2$ & $\geq 2$ & 2 & 4&  -176& 8(0) & 96(0) & 5118 & 8(0) & 837(0) & 0\\ \hline
$A_5$& $\geq 0$ &  0 & $\geq 3$ & $\geq 3$ & 6&  -168& 6(0) & 90(4) & 4254 & 6(0) & 695(6)  & 0\\ \hline
$A_5$& 0 &  2 & $\geq 2$ & $\geq 3$ & 4 &  -164& 11(2) & 93(0) & 4878 & 9(0)  & 799(0)  & 3\\ \hline
$A_5$& 0 &  1 & 2 & $\geq 3$ & 5 &  -164& 8(0) & 90(0) & 4560 &  8(0)& 744(0)  & 0\\ \hline
$A_6$& 0 &  $\geq 3$  & 1 & 3 & 4 &  -164& 9(0) & 91(0) & 4884  &  9(0)&  797(0) & 0\\ \hline
$A_6$& 0 &  3  & $\geq 2$ & 3 & 4 &  -152& 12(2) & 88(0) &  4716 & 10(0) & 771(0)  & 3\\ \hline
$A_6$& 0 &  2  & 2 & 3 & 5 &  -144& 12(2) & 84(0) & 4338 & 10(0)& 708(0)  & 3\\ \hline
$A_6$& 0 &  1  & $\geq 3$ & 3 & 6 &  -140& 9(0) & 79(0) & 3966  & 9(0) & 644(0)   & 0\\ \hline
$A_7$& $\geq 0$ &  $\geq 4$  & 0 & $\geq 4$ & 4 &  -172& 7(0) & 93(6) & 4776  & 7(0) & 781(9)& 0\\ \hline
$A_7$& 0 &  $\geq 4$  & 1 & $\geq 4$ & 4 &  -148& 10(0) & 84(0) & 4560   & 10(0) & 742(0)    & 0\\ \hline
$A_7$& 0 &  3  & 1 & $\geq 4$ & 5 &  -138& 10(0) & 79(0) &  4140  & 10(0) & 672(0)   & 0\\ \hline
$A_7$& 0 &  4  & $\geq 2$ & $\geq 4$ & 4 &  -136& 13(2) & 81(0) & 4392   & 11(0)  & 716(0)   & 3\\ \hline
$A_7$& 0 &  3  & 2 & 3 & 5 &  -132& 13(2) & 79(0) &  4176  & 11(0)  &680(0)    & 3\\ \hline
$A_7$& 0 &  0  & 4 & 4 & 8 &  -132& 7(0) & 73(6) &3240  & 7(0)  & 525(9)  & 0\\ \hline
$A_7$& 0 &  3  & 2 & $\geq 4$ & 5 &  -120& 16(4) & 76(0) &  3960 & 12(0)  & 646(0)    & 6\\ \hline
$A_7$& 0 &  2  & $\geq 3$ & 3 & 6 &  -120& 13(2) & 73(0) & 3744  & 11(0)& 608(0)& 3\\ \hline
$A_7$& 0 &  2  & 2 &  $\geq 4$ & 6 &  -120& 13(2) & 73(0) & 3672 & 11(0)& 596(0)& 3\\ \hline
$A_7$& 0 &  1  & 3 & $\geq 4$ & 7 &  -118& 10(0) & 69(0) & 3372   & 10(0)  & 544(0)   & 0\\ \hline
$A_7$& 0 &  2  & $\geq 3$ & $\geq 4$ & 6 &  -108& 16(4) & 70(0) & 3528  & 12(0)  & 574(0)    & 6\\ \hline
$E_6$& $\geq 1$ &  $\geq 2$  & $\geq 2$ & 2 & 3 &  -180& 9(0) & 99(0) &  5604  & 9(0)   & 917(0)     & 0 \\ \hline
$E_6$& $\geq 1$ &  $\geq 2$  & $\geq 2$ & 2 & 4 &  -152& 16(6) & 92(0) & 4824   & 10(0)   & 822(0)     & 36 \\ \hline
$E_7$& $\geq 1$ &  $\geq 2$   &  $\geq 2$  & $\geq 3$ & 3 & -144 & 19(8)  & 91(0) & 4824   & 11(0)   &  851(0)   & 66  \\ \hline
\end{tabular}
\end{center}
\end{table}

\begin{table}[htdp]
\begin{center}
\caption{ A list of the vanishing order of the coefficients in (\ref{e6form}) and the 
corresponding singularity and gauge group for the $E_6\times U(1)^2$ fiber.}
\label{E6U1vanish}\footnotesize
\begin{tabular}{|c|c|c|c|c|c|c|c|c|c|c|c|c|c|} \hline
Singularity & $a_1$ &  $b_1$ & $c_1$ & $a_2$ & $b_2$  &  $a_3$&  $\chi_3$ & $h^3_{11}$& $h^3_{21}$ & $\chi_4$ & $h^4_{11}$ & $h^4_{31}$ & $h^4_{21}$\\  \hline 
$A_1$ & 1 & $\geq 0$   & $\geq 4$  &  $\geq 1$  & 0   & 1 &  -184 &  5(0) & 97(0)   & 4104 & 5(0) & 671(0) & 0   \\ \hline
$A_1$ & 0 & $\geq 0$   & $\geq 1$  &  0  &  $\geq 1$  & 1 &  -184 &  5(0) & 97(0)   & 4104 & 5(0) & 671(0) & 0   \\ \hline
$A_1$ & 0 & $\geq 0$   & 0  &  $\geq 1$  &  $\geq 1$  & 2 &  -168 &  5(0) & 89(0)   & 3546 & 5(0)& 578(0)& 0  \\ \hline
$A_2$ & 1 & 0  & $\geq 1$ & $\geq 1$  & $\geq 1$  & 1 & -168  & 6(0) & 90(0)   & 3840 & 6(0)& 626(0)& 0   \\ \hline
$A_2$ & 0 & 0  & $\geq 1$ & $\geq 1$  & 1  & 2 & -158  & 6(0) & 85(0)   & 3462 & 6(0) & 563(0)& 0   \\ \hline
$A_3$ & 2 &  0 & $\geq 4$ &   2&  0 & 2& -156& 6(0) & 84(2)  & 3444 & 6(0)&560(3) & 0  \\ \hline
$A_3$ & 1 &  0 & $\geq 1$ &   1&  1 & 2& -148& 7(0) & 81(0)  & 3378 & 7(0)&548(0) & 0  \\ \hline  
$A_3$ & 0 &  0 & 2 &   1&  2 & 2& -144& 7(0) & 79(0)  & 3252 & 7(0)&527(0) & 0  \\ \hline 
$A_3$ & 1 &  1 &  1 &   1&  1 & 2& -138& 8(0) & 77(0)  & 3294 & 8(0)&533(0) & 0  \\ \hline
$A_3$ & 0 &  0 &  1 &   1&  $\geq 2$ & 3& -134& 7(0) & 74(0)  & 2916 & 7(0)&471(0) & 0  \\ \hline
$A_3$ & 0 &  $\geq 0$ &  0 &  $\geq 2$&  $\geq 2$ & 4& -128& 6(0) & 70(2)  & 2532 & 6(0)&408(3) & 0  \\ \hline
$A_4$ & 1 &  0 &    $\geq 2$ &   1&  $\geq  2$ & 2& -134& 8(0) & 75(0)  & 3168 & 8(0)&512(0) & 0  \\ \hline
$A_4$ & 2 &  0 &   $\geq 4$ &   $\geq 2$&  1 & 2& -134& 8(0) & 75(0)  & 3168 & 8(0)&512(0) & 0  \\ \hline
$A_4$ &0 & 0 & $\geq 2$ &1 &2 &3 &	-126&  8(0)& 71(0)   & 2862 &8(0) & 461(0) & 0 \\ \hline 
$A_4$ &1 & 0 & $\geq 1$ & $\geq 2$ &1 &3 &	-124&  8(0)& 70(0)   & 2832 & 8(0) & 456(0) & 0 \\ \hline   
$A_4$ &1 & 0 & 1 & $\geq 2$ & $\geq 2$ &3 &	-112&  11(2)& 67(0)   & 2664 & 9(0) & 430(0) & 3 \\ \hline
$A_4$ &0 & 0 & 1 & $\geq 2$ & 2 &4 &	-112&  8(0)& 64(0)   & 2436 & 8(0) & 390(0) & 0 \\ \hline
$A_5$&0 &  $\geq 0$ &    $\geq 3$ &  0 &  $\geq 3$  &  3 & -132 & 7(0) & 73(4) &2904 & 7(0) & 469(6) & 0 \\ \hline
$A_5$&2 &  0 &    $\geq 2$ &  $\geq 2$ &  $\geq 2$  &  2 & -120 & 9(0) & 69(0) &2958 & 9(0) & 476(0) & 0 \\ \hline
$A_5$&1 &  0 &   $\geq 2$ &   1&  2 & 3& -116 & 9(0) & 67(0) & 2778 & 9(0) & 446(0) & 0\\ \hline
$A_5$&0 &  0 &   $\geq 3$ &   1&  $\geq 3$ & 3& -114 & 9(0) & 66(0) & 2700 & 9(0) & 433(0) & 0\\ \hline
$A_5$&1 &  0 &   $\geq 2$ &   $\geq 2$&  2 & 3& -104 & 12(2) & 64(0) & 2610 & 10(0) & 420(0) & 0\\ \hline
$A_5$&0 &  0 &   $\geq 2$ &   $\geq 2$&  $\geq 2$ & 4& -104 & 9(0) & 61(0) & 2382 & 9(0) & 380(0) & 0\\ \hline
$A_5$&0 &  0 &   $\geq 2$ &   1&  $\geq 3$ & 4& -104 & 9(0) & 61(0) & 2406 & 9(0) & 384(0) & 0\\ \hline
$A_5$&1 &  0 &   1 &   2&  2 & 4& -96 & 12(2) & 60(0) & 2340 & 10(0) & 375(0) & 3\\ \hline
$A_5$&0 &  0 &   0 &  3 &  3 & 6& -96 & 7(0) & 55(4) & 1806 & 7(0) & 286(6) & 0\\ \hline
$A_5$&0 &  0 &   2&  2 & 3& 4 &-92& 12(2) & 58(0) & 2238 & 10(0) & 358(0) & 3\\ \hline
$A_5$&0 &  0 &   1 &   2&  3 & 5 & -92 & 9(0) & 55(0) & 2040 & 9(0) & 323(0) & 0\\ \hline$A_6$&3 &  0 &   $\geq 5$ &   3&  1 & 3 & -104 & 10(0) & 62(0) & 2616  & 10(0) & 418(0)    &  0\\ \hline
$A_6$&0 &  0 &   $\geq 3$ &   1&  3 & 4 & -98 & 10(0) & 59(0) &   2376&  10(0) & 378(0)    &  0\\ \hline
$A_6$&2 &  0 &   $\geq 2$ &   2&  2 & 3 & -96 & 13(2) & 61(0) & 2556  & 11(0)  & 410(0)   & 3 \\ \hline
$A_6$&2 &  0 &   $\geq 4$ &   $\geq 3$&  1 & 4 & -94 & 10(0) & 57(0) & 2322 & 10(0)  & 369(0)    & 0  \\ \hline
$A_6$&1 &  0 &   $\geq 3$ &   $\geq 2$&  $\geq 3$ & 3 & -92 & 13(2) & 59(0) & 2448& 11(0)  & 392(0)    & 3  \\ \hline
$A_6$&1 &  0 &   $\geq 2$ &   2&  2 & 4 & -88 & 13(2) & 57(0) & 2286 &  11(0) & 365(0)    & 3  \\ \hline
$A_6$&0 &  0 &   $\geq 3$ &   $\geq 2$&  3 & 4 & -86 & 13(2) & 56(0) & 2208 & 11(0)  & 352(0)   & 3  \\ \hline
$A_6$&0 &  0 &   2 &  2&  3 & 5 & -78 & 13(2) & 52(0) & 1974 & 11(0)   & 313(0)    & 3  \\ \hline
$A_6$&1 &  0 &   2 &  2& $\geq 3$ & 4 & -76 & 16(4) & 54(0) & 2142 & 12(0)   & 343(0)    & 6  \\ \hline
$A_6$&0 &  0 &   1 & $\geq 3$&  3 & 6 & -74 & 10(0) & 47(0) & 1698 &  10(0) & 265(0)   & 0  \\ \hline
$A_6$&1 &  0 &   1 & $\geq 3$&  2 & 5 & -76 & 13(2) & 51(0) & 1944 & 11(0)  & 308(0)    & 3 \\ \hline
$A_6$&1 &  0 &   1 & $\geq 3$&  $\geq 3$ & 5 & -64 & 16(4) & 48(0) & 1800 & 12(0)  & 286(0)    & 6 \\ \hline
$F_4$ &0 & $\geq 0$ & $\geq 4$ & 0 & 4 & 4 & -112 & 8(0) & 64(6) & 2472 & 8(0) & 396(9)& 0\\ \hline 
$A_7$ &0 & 0& $\geq 4$ & 1 &  4 & 4 & -188 & 11(0) & 55(0) & 2256 & 11(0) & 357(0)& 0\\ \hline 
$E_6$ &1 & $\geq 1$&1& 2 & $\geq 2$ &  2 & -96 & 17(6) & 65(0) & 2664 & 11(0) & 461(0)& 36\\ \hline 
$D_0$ &1 & $\geq 1$&$\geq 1$& 1 & 1 &  2 & -138 & 8(0) & 77(0) & 3294 & 8(0) & 533(0)  & 0 \\ \hline
$D_0$ &0 & $\geq 1$&$\geq 1$& $\geq 1$ & $\geq 2$ &  2 & -144 & 7(0) & 79(4) &3252 & 7(0) & 527(15) & 0 \\ \hline
$D_0$ &0 & $\geq 1$& 1 & 1  & $\geq 2$ &  3 & -124 & 8(0) & 70(0)  &2832 & 8(0) & 456(0) & 0 \\ \hline
$D_1$ &1 & $\geq 1$ & $\geq 1$ & $\geq 2$  & 1 &  2 & -124 & 9(0) & 71(0)  &  3084 & 9(0) & 497(0)  & 0 \\ \hline
$D_1$ &0 & $\geq 1$ & $\geq 2$ & 1  & 2 &  3 & -116 & 9(0) & 67(0)  & 2778  & 9(0) & 446(0)  & 0 \\ \hline
$D_1$ &0 & $\geq 1$ & 1 & $\geq 2$   & 2 &  3 & -112 & 11(2) & 67(3)  & 2664  & 9(0) & 430(10) & 3 \\ \hline
$D_1$ & 1 & $\geq 1$ & 1 & $\geq 2$  & 1 &  3 & -104 & 14(4) & 66(0)  & 2664  & 10(0) & 441(0) & 15 \\ \hline
$D_1$ & 0 & $\geq 1$ & 1 & $\geq 2$  & 2 & 4 & -96 & 12(2) & 60(0)  & 2340  & 10(0) & 375(0) & 3 \\ \hline
\end{tabular}
\end{center}
\end{table}

\clearpage

\newpage
\section{Defining Equations for the Resolved Fourfolds}
\label{Appendix B: Defining equations}
In this Appendix we write for reference the
defining equations for the resolved fourfolds.

\subsection{Case 1}
\label{appBcase1}
The first case considered \ref{case1},
\begin{eqnarray}
t_1=p,\ \  t_2=q,\ \ t_3= p+2q,\ \ t_4=-2p-q,\ \ t_5=-2q.\nonumber
\end{eqnarray}
\subsubsection{The resolved elliptic fibration}
\begin{eqnarray}
\left(-\delta_2 \lambda_2 \chi_1 x^3 + \beta_0 q \chi_1 \chi_3^2 \chi_4
 (- 2 p^3 - 
  11 p^2 q \chi_1 \chi_2 - 
  10 p q^2 \chi_1^2 \chi_2^2 - 
  4 q^3 \chi_1^3 \chi_2^3) w x^2 z \right. \nn & & \\ 
  \left. + \beta_0 \chi_1 \chi_2 \chi_3^2 \chi_4 
  (- 
  3 p^2 - 
  4 p q \chi_1 \chi_2 - 
  5 q^2 \chi_1^2 \chi_2^2) \delta_1 \lambda_1 w^3 x z^2 - 
  b0 \delta_1^2 \lambda_1^2 \chi_1^2 \chi_2^3 \chi_3^2 \chi_4 w^5 z^3 \right) \lambda_1 \lambda_2 \nn & &  \\ 
\left(\beta_0 p q^2 \chi_1 \chi_3^2 (4 p^2 + 10 p q \chi_1 \chi_2 
+ 4 q^2 \chi_1^2 \chi_2^2 )\chi_4 x \right. & &  \\
  \left. + 2 \beta_0  p \chi_3^2 \chi_4 ( -  p^2
    -  p q \chi_1 \chi_2 
    + q^2 \chi_1^2 \chi_2^2 ) \delta_1 \lambda_1 w^2 z
    + \delta_1 \delta_2 y\right) y z &=&0 \nn
\end{eqnarray}
with the projective relations
\begin{eqnarray} \label{E8proj-case1}
& \{\delta_1 \delta_2^2 \lambda_1 \lambda_2 \chi_1^2 \chi_2^3 \chi_3^5 \chi_4^7 \chi_5^9 y: 
    \delta_2 \lambda_1 \lambda_2 \chi_1 \chi_2 \chi_3^2 \chi_4^3 \chi_5^4 x: z\} \neq \{0: 0: 0\} & \nn \\ \relax
& \{\delta_1 \delta_2^2 \lambda_2 \chi_1^2 \chi_2^3 \chi_3^5 \chi_4^7 \chi_5^9 y: 
\delta_2 \lambda_2 \chi_1 \chi_2 \chi_3^2 \chi_4^3 \chi_5^4 x: w\} \neq \{0: 0: 0\} , \quad \nn  &
     \\ \relax
& \{\delta_1 \delta_2 \chi_1 \chi_2^2 \chi_3^3 \chi_4^4 \chi_5^5 y: x: 
\delta_1 \lambda_1 \chi_1 \chi_2^2 \chi_3^2 \chi_4^2 \chi_5^2\} \neq \{0: 0: 0\}, \quad  & 
\\ \relax
& \{\delta_2 \chi_3 \chi_4^2 \chi_5^3 y:    \lambda_1\} \neq \{0: 0\} , \quad 
 \{y: \lambda_2 \chi_1 \chi_2 \chi_3 \chi_4 \chi_5\} \neq \{0: 0\} , \quad 
 \{q: \lambda_2: \delta_1 \chi_2 \chi_3 \chi_4 \chi_5\} \neq \{0: 0: 0\} , \quad &
 \nn \\ \relax 
& \{\chi_1: \delta_1\} \neq \{0: 0\} , \quad 
 \{\chi_2: p: \delta_2 \chi_4 \chi_5^2\} \neq \{0: 0: 0\} , \quad 
 \{\chi_3: \delta_2 \chi_5\} \neq \{0: 0\}: 
 \{\chi_4: \delta_2\} \neq \{0:0\}. & \nn
\end{eqnarray}

\subsubsection{Sections}
\begin{table}[H]\centering
\begin{tabular}{c | c}
 $x$ 		& $\cO(\sigma + 2c_1 -E_1 - E_2)$	 \\
 $y$		& $\cO(\sigma + 3c_1-E_1-E_2 - E_3-E_4)$	 \\
 $w$		& $\cO(S_2 - E_1)$ 			 \\
 $z$		& $\cO(\sigma)$ 			 \\
 $\lambda_1$	& $\cO(E_1 - E_2 - E_3)$			 \\
 $\lambda_2$ 	& $\cO(E_2 - E_4-E_5)$			 \\
 $\delta_1$	& $\cO(E_3-E_5-E_6)$ 			 \\
 $\delta_2$	& $\cO(E_4-E_7-E_8-E_9)$				 \\
 $q$ 		& $\cO(Q - E_5)$	 \\
 $\chi_1$ 	& $\cO(E_5 - E_6)$	 		\\
  $p$		& $\cO(P - E_7)$	 \\
 $\chi_2$ 	& $\cO(E_6 - E_7)$		 		\\
 $\chi_3$ 	& $\cO(E_7 - E_8)$		 		\\
 $\chi_4$	& $\cO(E_8 - E_9)$ 			\\
 $\chi_5$	& $\cO(E_9)$					\\
 $\beta_0$	& $\cO(6c_1 - 5S_2)$			\\ 
\end{tabular}
\caption{The coordinates and their corresponding 
bundles for case 1.}\label{table:sections-case1}
\end{table}

\subsubsection{Curves in codimension 1}

\begin{table}[H]\label{table:codim1-case1}
\begin{tabular}{c r r l}
$A :$ & $\delta_1=$ & $-\chi_1 \delta_2 \lambda_1 \lambda_2^2 x^3 + 
 \beta_0 \chi_1 \chi_3^2 \chi_4 q x z(2 p q (2 p + \chi_1 \chi_2 q) (p + 2 \chi_1 \chi_2 q) y -$ \\ 
  &  & $-\lambda_1 \lambda_2 (2 p^3 + 
       11 \chi_1 \chi_2 p^2 q + 10 \chi_1^2 \chi_2^2 p q^2 + 
       4 \chi_1^3 \chi_2^3 q^3) w x  ) $&$=0$  \\ \hline
$B :$ & $\delta_2 = $ & $ -\beta_0 \chi_3^2 \chi_4 z (\chi_1 q x (\lambda_1 \lambda_2 (2 p^3 + 
         11 \chi_1 \chi_2 p^2 q + 10 \chi_1^2 \chi_2^2 p q^2 + 
         4 \chi_1^3 \chi_2^3 q^3) w x - $ \\  & & $
      2 p q (2 p + \chi_1 \chi_2 q) (p + 2 \chi_1 \chi_2 q) y) + 
   \delta_1 \lambda_1 w^2 (\chi_1 \chi_2 \lambda_1 \lambda_2 (3 p^2 + 
         4 \chi_1 \chi_2 p q + 5 \chi_1^2 \chi_2^2 q^2) w x $ \\ & & $ + 2 p (p^2 + \chi_1 \chi_2 p q
         - \chi_1^2 \chi_2^2 q^2) y) z + \chi_1^2 \chi_2^3 \delta_1^2 
      \lambda_1^3 \lambda_2 w^5 z^2)$&$=0$ \\ \hline
$C :$ & $\lambda_2 = $ & $ y z (\delta_1 \delta_2 y + 
   2 \beta_0 \chi_3^2 \chi_4 p (\chi_1 q^2 (2 p + \chi_1 \chi_2 q) (p + 
         2 \chi_1 \chi_2 q) x - $ \\ & & $
      \delta_1 \lambda_1 (p^2 + \chi_1 \chi_2 p q - \chi_1^2 \chi_2^2 q^2) w^2 z)) $&$= 0	$  \\ \hline
$D :$ & $\lambda_1 = $ & $ y z (2 \beta_0 \chi_1 \chi_3^2 \chi_4 p q^2 (2 p + \chi_1 \chi_2 q) (p + 
      2 \chi_1 \chi_2 q) x + \delta_1 \delta_2 y) $&$=0$ \\ \hline
$X :$ & $ w = $ & $ -\chi_1 \delta_2 \lambda_1 \lambda_2^2 x^3 + 
y z(\delta_1 \delta_2 y+2 \beta_0 \chi_1 \chi_3^2 \chi_4 p q^2 (2 p + \chi_1 \chi_2 q) (p + 
      2 \chi_1 \chi_2 q) x ) $&$= 0 $ \\ \hline
\end{tabular}
\caption{\footnotesize Divisors in the codimension 1 locus 
$w=0$ after the complete blow up. Every curve has multiplicity 1.}
\end{table}

\subsubsection{Curves in codimension 2}
\begin{table}[H]
\begin{tabular}{c r r }
Curve & & \multicolumn{1}{c}{Defining equations}  \\ \hline
$A_{\{p-q\}} :$ & $\delta_1=$ & $ 
\delta_2 \lambda_1 \lambda_2^2 \chi_1 x^2 
+ 9 \beta_0 q^4 \chi_1^4 \chi_2^3 \chi_3^2 \chi_4  
 (3 \lambda_1 \lambda_2 w x + 2 q y) z=0$ 
\\ \hline
$B_{1 \{p-q\}} :$ & $\delta_2 = $ & $ -9 q^2 \chi_1 x 
+ \delta_1 \lambda_1 w^2 z  =0 $ \\  \hline
$B_{2 \{p-q\}} :$ & $\delta_2 = $ & 
$-q^2 \chi_1 (3 \lambda_1 \lambda_2 q^2 w x 
+ 2 q y) + \delta_1 \lambda_1^2 \lambda_2 w^3 z =0 $ \\  \hline
$C_{\{p-q\}} :$ & $\lambda_2 = $ & 
$18 \beta_0 q^5 \chi_1^4 \chi_2^3 \chi_3^2 \chi_4 x 
+ \delta_1 \delta_2 y= 0	$  \\ \hline
$D_{\{p-q\}} :$ & $\lambda_1 = $ & $ 
\delta_1 \delta_2 y + 2 \beta_0 q^3 \chi_1^3 \chi_2^3
\chi_3^2 \chi_4
(-9 q^2 \chi_1 x + \delta_1 \lambda_1 w^2 z) =0$ \\ \hline
$X_{\{p-q\}} :$ & $ w = $ & $
-\delta_2 \lambda_1 \lambda_2^2 x^3 
- 18 \beta_0 q^5 x y z + \delta_1 \delta_2 y^2 z= 0 $ \\ \hline
\end{tabular}
\caption{\footnotesize Codimension 2 locus $p_0 - q_0 = 0$.
Exceptional divisors split 
as we approach $p_0 \ra q_0$.
Each curve has multiplicity 1.}\label{table:codim2(p-q=0)-case1}
\end{table}

\begin{table}[H]
\begin{tabular}{c r r c}
Curve & & \multicolumn{1}{c}{Defining equations} & Multipl. \\ \hline
$A_{1 \{p+2q\}} :$ & $\delta_1= $ & $ \lambda_2 =0$ & 2
\\ \hline
$A_{2 \{p+2q\}} :$ & $\delta_1= $ & $\delta_2  \lambda_2 x 
+ 12 \beta_0 q^4 \chi_1^3 \chi_2^3 \chi_3^2 \chi_4  w z =0 $  & 1
\\ \hline
$B_{\{p+2q\}} :$ & $\delta_2 =  $ & $
12 \lambda_2 q^4 \chi_1^2 x^2 
+ \delta_1 q^2 \chi_1 w (9 \lambda_1 \lambda_2 w x 
- 4 q y) z + \delta_1^2 \lambda_1^2 \lambda_2 w^4 z^2 =0 $&1 
\\  \hline
$C_{\{p+2q\}} :$ & $\lambda_2 = $ & $ 
 \delta_2 y
 + 4 \beta_0 \lambda_1 q^3 \chi_1^3 \chi_2^3 
 \chi_3^2 \chi_4 w^2 z=0$ & 1 \\  \hline
$D_{\{p+2q\}} :$ & $\lambda_1 = $ & $
\delta_1 =0$ & 2\\ \hline
$X_{\{p+2q\}} :$ & $ w = $ & $
(-\lambda_1 \lambda_2^2 \chi_1 x^3 + \delta_1 y^2 z)= 0 $&1 \\ \hline
\end{tabular}
\caption{\footnotesize Codimension 2 locus $p_0 + 2 q_0 = 0$.}
\label{table:codim2(p+2q=0)-case1}
\end{table}

\begin{table}[H]
\begin{tabular}{c r r c}
Curve & & \multicolumn{1}{c}{Defining equations} & Multipl. \\ \hline
$A_{1 \{p\}} :$ & $\delta_1= $ & $\lambda_2 =0$ & 2
\\ \hline
$A_{2 \{p\}} :$ & $\delta_1= $ & $\delta_2  \lambda_2 x 
+ 12 \beta_0 q^4 \chi_1^3 \chi_2^3 \chi_3^2 \chi_4  w z =0 $  & 1
\\ \hline
$B_{ 1\{p\}} :$ & $\delta_2 =  $ & $
\lambda_2  =0 $& 2
\\  \hline
$B_{ 2A\{p\}} :$ & $\delta_2 = $ & $
 q^2 \chi_1 x + \delta_1 \lambda_1 w^2 z  =0 $&1 
\\  \hline
$B_{ 2B\{p\}} :$ & $\delta_2 =$ & $
4 q^2 \chi_1 x + \delta_1 \lambda_1 w^2 z =0 $&1 
\\  \hline
$D_{\{p\}} :$ & $\lambda_1 =$ & $
\delta_1 =0$ & 2\\ \hline
$X_{\{p\}} :$ & $ w = $ & $
(-\lambda_1 \lambda_2^2 \chi_1 x^3 + \delta_1 y^2 z)= 0 $&1 \\ \hline 
\end{tabular}
\caption{\footnotesize Codimension 2 locus $p_0 = 0$.}\label{table:codim2(p=0)-case1}
\end{table}

\begin{table}[H]
\begin{tabular}{c r r c}
Curve & & \multicolumn{1}{c}{Defining equations} & Multipl. \\ \hline
$A_{\{q\}} :$ & $\delta_1= $ & $\delta_2 =0$ & 2
\\ \hline
$B_{\{q\}} :$ & $\delta_2= $ & $
3 \lambda_1 \lambda_2 p^2 \chi_1 \chi_2 w x + 2 p^3 y
 + \delta_1 \lambda_1^2 \lambda_2 \chi_1^2 \chi_2^3 
 w^3 z =0 $  & 1
\\ \hline
$C_{\{q\}} :$ & $\lambda_2 = $ & $
 d2 y - 2 \beta_0 \lambda_1 p^3 \chi_3^2 \chi_4 w^2 z =0 $& 2
\\  \hline
$D_{\{q\}} :$ & $\lambda_1 = $ & $
-\lambda_1 \lambda_2^2 \chi_1 x^3 + \delta_1 y^2 z  =0 $&2
\\ \hline
$X_{\{p\}} :$ & $ w = $ & $
(-\lambda_1 \lambda_2^2 \chi_1 x^3 + \delta_1 y^2 z)= 0 $&1 
\\  \hline \hline
$C_{\{\chi_1\}} :$ & $\lambda_2 = $ & $
 \delta_2 y - 2 \beta_0 \lambda_1 
 p^3 \chi_3^2 \chi_4 w^2 z =0 $& 1
\\  \hline
$A_{\{\chi_2\}} :$ & $\delta_1 = $ & $
\delta_2 \lambda_1 \lambda_2^2 x^2 
+ 2 \beta_0 p^3 q (\lambda_1 \lambda_2 w x 
- 2 q y) z = 0 =0 $& 1
\\  \hline
$B_{\{\chi_2\}} :$ & $\delta_2 = $ & $
q \chi_1 x (\lambda_1 \lambda_2 w x 
- 2 q y) + \delta_1 \lambda_1 w^2 y z = 0 $& 1
\\ \hline
$C_{\{\chi_2\}} :$ & $\lambda_2 = $ & $
\delta_1 \delta_2 y 
+ 2 \beta_0 p^3 (2 q^2 \chi_1 x 
- \delta_1 \lambda_1 w^2 z)  =0 $&1
\\  \hline 
$D_{\{\chi_2\}} :$ & $\lambda_1 = $ & $
4 \beta_0 p^3 q^2 \chi_1 x 
+ \delta_1 \delta_2 y =0 $& 1
\\  \hline
\end{tabular}
\caption{\footnotesize Codimension 2 locus $q_0 = 0$. We also
included for reference the curves that fail to intersect
the Cartan divisors.}\label{table:codim2(q=0)-case1}
\end{table}

\subsubsection{``Curves'' in codimension 3}
\begin{table}[H]
\begin{tabular}{c r r r c}
Curve &  \multicolumn{3}{c}{Defining equations}  \\ \hline
$A_{\{``E8"\}} :$ & $p=q=0$ & $\delta_1= $ & $\delta_2 =0$ 
\\ \hline
$C_{\{``E8"\}} :$ & $p=q=0$ &$\delta_2= $ & $\lambda_2 =0$
\\ \hline
$D_{\{``E8"\}} :$ & $p=q=0$ &$\delta_1= $ & $\lambda_1 =0$ 
\\ \hline
$X_{\{``E8"\}} :$ & $p=q=0$ &$w= $ &
 $-\lambda_1 \lambda_2^2 \chi_1 x^3 + \delta_1 y^2 z =0$ 
\\ \hline \hline
  $A_{1\{``E8"\}} :$ & $p=q=0$ &$\delta_2= $ & $\chi_1 =0$ 
\\ \hline
$A_{2\{``E8"\}} :$ & $p=\chi_1=0$ &$\delta_2= $ & =0

  \\ \hline
$A_{3\{``E8"\}} :$ & $p=\chi_1=0$ &$\lambda_2= $ & $\delta_2 =0$ 
\\ \hline
$A_{4\{``E8"\}} :$ & $p=\chi_2=0$ &$\delta_1= $ & $\lambda_1 =0$ 
\\ \hline
$A_{5\{``E8"\}} :$ & $p=\chi_2=0$ &$\delta_1= $ & $\lambda_2 =0$
\\ \hline
$A_{6\{``E8"\}} :$ & $p=\chi_2=0$ &$\delta_1= $ & $\lambda_1 =0$ 
\\ \hline
$A_{7\{``E8"\}} :$ & $\chi_3=0$ &$\delta_1= $ & $\lambda_2 =0$ 
\\ \hline
$A_{8\{``E8"\}} :$ & $\chi_4=0$ &$\delta_1= $ & $\lambda_2 =0$ 
\\ \hline
$A_{9\{``E8"\}} :$ & $\chi_5=0$ &$\delta_1= $ & $\lambda_2 =0$ 
\end{tabular}
\caption{\footnotesize Codimension 3 locus $q_0 = p_0 = 0$.
note that even though 
we call the restrictions of the exceptional 
divisors to the fourfold ``curves'', note that 
some of the restrictions have higher dimensionality, 
as equation \eqref{B*}}.\label{table:codim3-case1}
\end{table}

\subsection{Case 2}

\label{appBcase2}
\subsubsection{The resolved elliptic fibration}
\begin{eqnarray}
\begin{cases}\label{E8eqres}
0=&\pi_1 \pi_2^2 \chi_1 \chi_2^2 s \left[ \pi_1 \pi_2^2 \chi_1 \chi_2^2 t +
  \lambda_1 (p \pi_1 \pi_2 + q \chi_1 \chi_2) (2 p^2 \pi_1^2 \pi_2^2 - 3 p \pi_1 \pi_2 q \chi_1 \chi_2 + 
     2 q^2 \chi_1^2 \chi_2^2)w^2 \right]+  \\ 
& \lambda_1 \lambda_2 \left[-2 p \pi_1 \pi_2 q \chi_1 \chi_2 (p^2 \pi_1^2 \pi_2^2 + 4 p \pi_1 \pi_2 q \chi_1 \chi_2 + 
        q^2 \chi_1^2 \chi_2^2) - \delta_1^2 \lambda_1^2 w^5 +  \right. \\
& \left. \delta_1 \lambda_1 (3 p^2 \pi_1^2 \pi_2^2 + p \pi_1 \pi_2 q \chi_1 \chi_2 + 
        3 q^2 \chi_1^2 \chi_2^2) w^2 x - \delta_2 \lambda_2 x^3\right] \\
 0=&4 p^2 \pi_1 q^2 \chi_1 (p \pi_1 \pi_2 + q \chi_1 \chi_2) + \delta_2 s - 
  \delta_1 t ,
\end{cases}
\end{eqnarray}
where the coordinates are sections in 
the bundles given in table \ref{table:sections-case2},
together with the list of projective relations 

\begin{eqnarray} \label{E8proj-case2}
&[\delta_1 \delta_2 \lambda_2 y: \delta_2 \lambda_2 x: w] \neq [0: 0: 0] , 
\quad [\delta_1 y: x: \delta_1 \lambda_1] \neq [0: 0: 0] , \quad [y: \lambda_1] \neq [0: 0] ,& \nn \\
&[ \pi_1 \pi_2^2 \chi_1 \chi_2^2 s : \lambda_2 ] \neq [0: 0] , \quad 
[\pi_1 \pi_2^2 \chi_2 s: \pi_1 \pi_2^2 \chi_2 t: q] \neq [0: 0: 0] , & \\ 
& [\pi_2 \pi_1 s: \pi_2 \pi_1 t: \chi_1] \neq [0: 0: 0] , \quad 
[\pi_2 s: \pi_2 t: p] \neq [0: 0: 0] , \quad [s: t: \pi_1] \neq [0: 0: 0]. & \nn
\end{eqnarray}
\subsubsection{Sections}
\begin{table}[H]\centering
\begin{tabular}{c | c}
 $x$ 		& $\cO(\sigma + 2c_1 -E_1 - E_2)$	 \\
 $y$		& $\cO(\sigma + 3c_1-E_1-E_2 - E_3)$	 \\
 $t$		& $\cO(\sigma + 3c_1-E_1-E_2 - E_3- E_5 - E_6 - E_7 - E_8)$	 \\
 $s$		& $\cO(\sigma + 3c_1-E_1-E_2 - E_4- E_5 - E_6 - E_7 - E_8)$	 \\
 $w$		& $\cO(S_2 - E_1)$ 			 \\
 $z$		& $\cO(\sigma)$ 			 \\
 $\lambda_1$	& $\cO(E_1 - E_3)$			 \\
 $\lambda_2$ 	& $\cO(E_2 - E_4)$			 \\
 $\delta_1$	& $\cO(E_3)$ 			 \\
 $\delta_2$	& $\cO(E_4)$				 \\
 $p$		& $\cO(S_2 - c_1 - E_7 - E_8)$	 \\
 $q$ 		& $\cO(S_2 - c_1 - E_5 - E_6)$	 \\
 $\chi_1$ 	& $\cO(E_5 - E_6)$	 		\\
 $\chi_2$ 	& $\cO(E_6)$		 		\\
 $\pi_1$ 	& $\cO(E_7 - E_8)$	 		\\
 $\pi_2$ 	& $\cO(E_8)$		 		\\
 $\beta_0$	& $\cO(6c_1 - 5S_2)$			\\ 
\end{tabular}
\caption{Coordinates and their corresponding bundles for case 2.}\label{table:sections-case2}
\end{table}

\subsubsection{Codimension 1}
\begin{table}[H]\label{tableE8codim1}
\begin{tabular}{c r r r }
Curve &  \multicolumn{2}{c}{Defining equations} \\ \hline
$A :$ & $\lambda_1= $ & $ s= 0 $     \\ \hline

$B :$ & $\delta_2 =$ & $ \pi_1 \pi_2^2 \chi_1 \chi_2^2 s 
(\pi_1 \pi_2^2 \chi_1 \chi_2^2 t + \lambda_1 \beta_3 w^2)+ 
\lambda_1 \lambda_2 (\beta_4 - \delta_1^2 \lambda_1^2 w^5 
+ \delta_1 \lambda_1 \beta_2 w^2 x) =0 $   \\ \hline

$C :$ & $\lambda_2= $ & $ \pi_1 \pi_2^2 \chi_1 \chi_2^2 t +
  \lambda_1 \beta_3 w^2 = 0 $   \\ \hline

$D :$ & $\delta_1= $ & $ \pi_1 \pi_2^2 \chi_1 \chi_2^2 s + (\pi_1 \pi_2^2 \chi_1 \chi_2^2 t + 
  \lambda_1 \beta_3 w^2)+ 
\lambda_1 \lambda_2 (\beta_4 - \delta_2 \lambda_2 x^3) = 0
$     \\ \hline

$X :$ & $ w= $ & $ \pi_1^2 \pi_2^4 \chi_1^2 \chi_2^4 s t+ 
\lambda_1 \lambda_2 (\beta_4 - \delta_2 \lambda_2 x^3) =0
$    \\ \hline
\end{tabular}
\caption{\footnotesize Curves in codimension 1 after the complete blow up. To shorten the notation, we have used 
 $\beta_5 =  4 p^2 \pi_1 q^2 \chi_1 (p \pi_1 \pi_2 + q \chi_1 \chi_2)$, 
 $\beta_4 = -2 p \pi_1 \pi_2 q \chi_1 \chi_2 (p^2 \pi_1^2 \pi_2^2 + 4 p \pi_1 \pi_2 q \chi_1 \chi_2 + 
        q^2 \chi_1^2 \chi_2^2)$, 
 $\beta_3 = (p \pi_1 \pi_2 + q \chi_1 \chi_2) (2 p^2 \pi_1^2 \pi_2^2 - 3 p \pi_1 \pi_2 q \chi_1 \chi_2 + 
     2 q^2 \chi_1^2 \chi_2^2)$ and 
$\beta_2 = 3(p \pi_1 \pi_2)^2 + p \pi_1 \pi_2 q \chi_1 \chi_2 + 3(q \chi_1 \chi_2)^2$. All curves 
have multiplicity 1.}
\end{table}

\subsubsection{Curves in codimension 2}
\begin{table}[H]
\begin{tabular}{c r r r}
Curve &  \multicolumn{2}{c}{Defining equations} & Auxiliary Constraint \\ \hline
$A_{\{p_0-2q_0\}} :$ & $\delta_1=$ & 
$\tilde Y_4 |_{(\delta_1=0, 
p_0 - 2 q_0=0)}=0$ & $ (48 \beta_0 q^5 x 
+ \delta_2 s + \delta_1 t = 0) $
\\  \hline
$B_{\{p_0-2q_0\}} :$ 
& $\lambda_2 = $ & 
$\tilde Y_4 |_{(\lambda_2=0,p_0 - 2 q_0=0)} = 0$ 
& $(48 \beta_0 q^5 x 
+ \delta_2 s + \delta_1 t = 0)$ \\  \hline
$C_{1 \{p_0-2q_0\}} :$ & $\delta_2 = $ & 
$t + 12 \beta_0 \lambda_1 q^3 w^2 z = 0$ & $
(48 \beta_0 q^5 x + \delta_1 t
=0) $ \\  \hline
$C_{2 \{p_0-2q_0\}} :$ & $\delta_2 = $ & 
$ 48 \beta_0 (\delta_1^2 \lambda_1^2
\lambda_2 q^3+ 12 q^6 s) + 13 \delta_1^2
\lambda_1 \lambda_2 t=$ & $
(48 \beta_0 q^5 x + \delta_1 t= 0)	$  \\ \hline
$D_{\{p_0-2q_0\}} :$ & $\lambda_1 = $ & $ s = 0$ & $ 
(\delta_1 t + 48 \beta_0 q^5 x = 0)$  \\ \hline
\end{tabular}
\caption{\footnotesize Codimension 
2 locus $p_0 -2 q_0 = 0$.
Exceptional divisors split 
as we approach $p_0 \ra 2 q_0$.
Each curve has multiplicity 1. Since we are 
looking away from the $q_0$ locus, we took the 
blow down limit $\pi_1, \pi_2, \chi_1, \chi_2 
\ra 1$ .}\label{table:codim2(p-2q=0)-case2}
\end{table}

\begin{table}[H]
\begin{tabular}{c r r r}
Curve &  \multicolumn{2}{c}{Defining equations} 
& Additional Constraint \\ \hline
$A_{1 \{p_0 + q_0\}} :$ & $\delta_1=$ 
 &  $ \lambda_1 =0$ & 
$(s = 0)$
\\  \hline
$A_{2 \{p_0 + q_0\}} :$ & $\delta_1=$&  
$ \delta_2 \lambda_2 x 
+ 4 \beta_0 w q^4 z =0$ & 
$(s = 0)$ 
\\  \hline
$A_{3 \{p_0 + q_0\}} :$ & $\delta_1=$ 
 &  $ st+4\beta_0 \lambda_1 \lambda_2
q^4 w x^2 =0$ & 
$(\delta_2 =0) $
\\  \hline
$B_{\{p_0+q_0\}} :$ 
& $\lambda_2 = $ 
& $t=0$ 
& $(\delta_2 =0)$   \\  \hline
$C_{1 \{p_0+q_0\}} :$ & $\delta_2 = $ & $
\lambda_2 =0 $& 
$(t = 0)$  \\  \hline
$C_{2 \{p_0+q_0\}} :$ & $\delta_2 = $  & 
$ 4 q^2 x - \delta_1 \lambda_1 w^2 z= 0	$
& $ (t=0) $  \\ \hline
$C_{3 \{p_0+q_0\}} :$ & $\delta_2 = $ & 
$ q^2 x - \delta_1 \lambda_1 w^2 z= 0 $ 
& $ (t=0) $  \\ \hline
$D_{\{p_0+q_0\}} :$ & $\lambda_1 = $ & $ 
s = 0$ & $
(\delta_1 = 0 )$ \\ \hline
\end{tabular}
\caption{\footnotesize Codimension 
2 locus $p_0 +  q_0 = 0$.
Exceptional divisors split 
as we approach $p_0 \ra 2 q_0$.
Each curve has multiplicity 1. Since we are 
looking away from the $q_0$ locus, we took the 
blow down limit $\pi_1, \pi_2, \chi_1, \chi_2 
\ra 1$ . Notice that even though 
$B$ and $C_1$ have the same equations if we 
take in account the additional constraint, 
they carry different charges.}\label{table:codim2(p+q=0)-case2}
\end{table}

\begin{table}[H]
\begin{tabular}{c r r r}
Curve &  \multicolumn{2}{c}{Defining equations} 
& Additional Constraint \\ \hline
$A_{1 \{p \}} :$ & $\delta_1=$ 
 &  $ \pi_1 \pi_2^2 t
 + 2 \beta_0 \lambda_1 q^3 \chi_1^2 \chi_2
 w^2 z = 0$ & 
$(\delta_2 = 0)$
\\  \hline
$A_{2 \{p \}} :$ & $\delta_1=$&  
$ \lambda_1 =0$ & 
$(s = 0)$ 
\\  \hline
$A_{3 \{p \}} :$ & $\delta_1=$ 
 &  $ st+4\beta_0 \lambda_1 \lambda_2
q^4 w x^2 =0$ & 
$(\delta_2 =0) $
\\  \hline
$B_{\{p \}} :$ 
& $\lambda_2 = $ 
& $\pi_1 \pi_2^2 \chi_1^2 \chi_2^4 
s z (\pi_1 \pi_2^2 t + 2 \beta_0 \lambda_1 
q^3 \chi_1^2 \chi_2 w^2 z)=0$ 
& $(\delta_2s -\delta_1 t =0)$   \\  \hline
$C_{1 \{p \}} :$ & $\delta_2 = $ & $
 \pi_1 \pi_2^2 t + 2 \beta_0 \lambda_1 
q^3 \chi_1^2 \chi_2 w^2 z =0 $& 
$(\delta_1 = 0)$  \\  \hline
$C_{2 \{p \}} :$ & $\delta_2 = $  & 
$ s = 0	$
& $ (\delta_1=0) $  \\ \hline
$C_{3 \{p \}} :$ & $\delta_2 = $ & 
$ 2 s \pi_1 \pi_2^2 q^3 \chi_1^4 \chi_2^5 + \delta_1 \lambda_1 \lambda_2 w (-3 q^2 \chi_1^2 \chi_2^2 x + \delta_1 \lambda_1 w^2 z)= 0 $ 
& $ (t=0) $  \\ \hline
$D_{\{p\}} :$ & $\lambda_1 = $ & $ 
s = 0$ & $
(\delta_1 = 0 )$ \\ \hline
\end{tabular}
\caption{\footnotesize Codimension 
2 locus $p_0 = 0$. Here we have shown 
only the curves with non-vanishing charges,
presented in table \ref{table:E6-case2}.}\label{table:codim2(p=0)-case2}
\end{table}

\subsubsection{Curves in codimension 3}

\begin{table}[H]
\begin{tabular}{c r r r}
Curve &  \multicolumn{2}{c}{Defining equations} 
& Additional Constraint \\ \hline
$A_{1 \{ E_8 \}} :$ & $\delta_1=$ 
 &  $t= 0$ & 
$(\delta_2 = 0)$
\\  \hline
$A_{2 \{E_8 \}} :$ & $\delta_1=$&  
$ \lambda_1 =0$ & 
$(s = 0)$ 
\\  \hline
$B_{\{E_8 \}} :$ 
& $\lambda_2 = $ 
& $t=0$ 
& $(\delta_1 =0)$   \\  \hline
$C_{1 \{E_8 \}} :$ & $\delta_2 = $ & $
 t =0 $& 
$(\delta_1 = 0)$  \\  \hline
$C_{2 \{E_8 \}} :$ & $\delta_2 = $  & 
$ s = 0	$
& $ (\delta_1=0) $  \\ \hline
$D_{\{E_8 \}} :$ & $\lambda_1 = $ & $ 
s = 0$ & $
(\delta_1 = 0 )$ \\ \hline
\end{tabular}
\caption{\footnotesize Codimension 
3 locus $p_0 = q_0 = 0$. Again we present 
only the curves with non-vanishing charges,
from table \ref{table:E8-case2}.}\label{table:codim3-case2}
\end{table}

\section{Calculating the Charges}\label{Appendix C}
We calculate the intersections in the same way
as in \cite{Marsano:2011hv,Lawrie:2012gg}. 
The coordinates are sections 
of the bundles as shown in tables \ref{table:sections-case1}
and \ref{table:sections-case2}, for case 1 and 2, respectively.


In both cases the divisors $P$ and $Q$
are in the same class as $S_2 - c_1$.

We then use the projective relations
\eqref{E8proj-case1} and knowing the corresponding
bundles of each section, we can write 
the vanishing intersections of homology classes.
For example, when we perform the first blow-up
\begin{equation}
 y \ra \lambda_1 y' ,\quad
 x \ra \lambda_1 x' ,\quad 
 w \ra \lambda_1 w' ,
\end{equation}
We introduced a new divisor $E_1$
defined by $\lambda_1=0$ in the ambient 
space, and the relation
\begin{equation}
 (\sigma + 2c_1 -E_1)(\sigma + 3c_1 -E_1)(S_2-E_1)=0 .
\end{equation}
Similarly, we have the relations introduced 
at every blow-up,
\begin{eqnarray}\label{projrel-case1}
 (\sigma + 2c_1 -E_1)(\sigma + 3c_1 -E_1)(S_2-E_1)&=&0 
 \qquad \lambda_1:\, \,(x,y,w) \nn \\ 
 (\sigma + 2c_1 -E_1-E_2)(\sigma + 3c_1 -E_1-E_2)(E_1-E_2)&=&0 
 \qquad \lambda_2:\, \, (x,y,\lambda_1)\nn \\ 
 (\sigma + 3c_1 -E_1-E_2-E_3)(E_1-E_2-E_3)&=&0 
  \qquad \delta_1:\, \,(y,\lambda_1)\nn \\ 
 (\sigma + 3c_1 -E_1-E_2-E_3-E_4)(E_2-E_4)&=&0  
 \qquad \delta_2:\, \,(y,\lambda_2) \\ 
 (Q - E_5)(E_2-E_4-E_5)(E_3-E_5)&=&0  
 \qquad \chi_1:\, \,(q,\lambda_2,\delta_1)\nn \\ 
 (E_5-E_6)(E_3-E_5-E_6)&=&0  
 \qquad \chi_2:\, \,(\chi_1,\delta_1)\nn \\
 (P-E_7)(E_6-E_7)(E_4-E_7)&=&0  
 \qquad \chi_3:\, \,(p,\chi_2,\delta_2)\nn \\
 (E_7-E_8)(E_4-E_7-E_8)&=&0  
 \qquad \chi_4:\, \,(\chi_3,\delta_2)\nn \\
 (E_8-E_9)(E_4-E_7-E_8-E_9)&=&0  
 \qquad \chi_5:\, \,(\chi_4,\delta_2).\nn
\end{eqnarray}

To calculate the Cartan charges
in codimension 1, we need to calculate 
the intersection of each exceptional divisor as restricted
to the fourfold $\tilde Y_4$,
\begin{equation}\label{cartancalculation}
 C_{ij}=D_1 \cdot D_2 \cdot [\tilde Y_4] \cdot \cD_i \cdot \cD_j ,
\end{equation}
Here $\cD_i$ are the exceptional divisors 
in the ambient space and
 $D_1$ and $D_2$ are divisors 
on the base of the ambient space such that
\begin{equation}
 D_1 \cdot_{B_3} D_2 \cdot_{B_3} S_2 = 1 ,
\end{equation}
where the product $\cdot_{B_3}$ is the intersection
restricted to the base $B_3$. This is equivalent
to taking the ambient space product
\begin{equation}
D_1 D_2 S_2 \sigma^2 = 1 ,
\end{equation}
and $\sigma^2$ gives simply a point on the $\IP^2$
fibered over $B_3$. 

We then insert the corresponding classes of the 
exceptional divisors $\cD_i$s, given in table 
\ref{table:sections-case1} for case 1 and 
table \ref{table:sections-case2} for case 2.
The class of the fourfold $\tilde Y_4$ is, for each case,
\begin{eqnarray}
 \text{case 1:}\quad [\tilde Y_4] &=& 3\sigma 
 + 6c_1 -2E_1-2E_2-E_3-E_4-E_5-E_6-E_8-E_9  ,
 \label{fourfoldclass-case1} \\ \relax
 \text{case 2:}\quad [\tilde Y_4] &=& 3\sigma 
+ 6c_1 -2E_1-2E_2-E_3-E_4 .
 \label{fourfoldclass-case2}
\end{eqnarray}

The projective relations \eqref{projrel-case1}
can be expanded to write expressions for the 
highest degree terms like
\begin{equation}\label{E1^3}
 E_1^3 = S_2\sigma^2 + (\ldots)E_1 + (\ldots)E_1^2 
\end{equation}
that when we insert $\cD_i$ and $[\tilde Y_4]$ into 
\eqref{cartancalculation}, 
appear as $D_1 \cdot D_2 \cdot E_i^3$. The only term
of \eqref{E1^3} that contributes to the intersection
calculation is $S_2 \sigma^2$. This follows from 
the fact that terms with lower orders of $\sigma$
fail to give a point fibered over $B_3$ and terms
like $D_1 \cdot D_2 \cdot E_i \cdot \sigma^2$ 
vanish since the $E_i$s are exceptional divisors
while $D_1$ and $D_2$ are pullbacks into the 
original ambient space, and therefore
\begin{equation}
 D_1 \cdot_{B_3} D_2 \cdot_{B_3} E_i=0 .
\end{equation}

To calculate the charges along codimension 
2 and 3 loci, we proceeded in the same way as above,
but looking for terms of the form
\begin{equation}
 P \cdot D_1 \cdot S_2 \cdot \sigma^2 = 0 \, , \qquad
 Q \cdot D_1 \cdot S_2 \cdot \sigma^2 = 0 \, \quad \textsl{or} \quad
 P \cdot Q \cdot S_2 \cdot \sigma^2 = 0,
\end{equation}
where $P$ and $Q$ are the sections corresponding to
$p_0=0$ and $q_0=0$, respectively. 

Notice also that when calculating the charges
for Case 2, the second equation (i.e., the auxiliary
constraint) does not enter the intersection 
calculations.

\section{Alternative interpretation}
Here we mention a somewhat \textit{ad hoc} argument 
to obtain a structure of $\IP^1$s in what we will call the ``F-theory fiber",
the fiber composed simply from the proper transform of the elliptic fiber 
$X$ and a particular subset of the $\IP^1$s.
As we mentioned, the blow-ups that took $q \ra \chi_2 \chi_1 q $ 
introduced two-dimensional spaces located along the matter curve 
that could be interpreted as a mixed resolution of the fiber via 
an one-dimensional space and a resolution of the matter curve on the
base. We assume that the $\IP^1$s 
forming the ``F-theory fiber"
are the ones obtained only by intersection with $p=0$,
even if from our calculations they are not 
charged under the Cartan roots.
One sees that the remaining curves 
intersect precisely as an affine $E_6$ Dynkin diagram, 
with the correct multiplicities (table \ref{tableE6codim2'}.

\begin{table}[h]
\centering
\begin{tabular}{|c c l | c | c |}\hline
\multicolumn{3}{|c|}{Curve} & Mult. & Diagram \\ \hline
$A $&$:$ & $\lambda_1=q=s=\delta_1=0$ & 2 & 
\multirow{9}{*}{\setlength{\unitlength}{0.75mm} 
\begin{picture}(24,42)(2,-5)
\multiput(5,3)(0,9){5}{\circle{6}}   
 \multiput(5,6)(0,9){4}{\line(0,1){3}} 
 \put(3.4,1.7){\small{C}}
 \put(2,10.7){\small{$B_2$}}
 \put(2,19.7){\small{$B_1$}}
 \put(2,28.7){\small{$E_{\ast q}$}}
 \put(2,37.7){\small{$B_3$}}
 \put(8,21){\line(1,0){3}}
 \put(14,21){\circle{6}}
 \put(12,19.7){\small{$A$}}
 \put(17,21){\line(1,0){3}}
 \put(23,21){\circle{6}}
 \put(21.4,19.7){\small{X}}
\end{picture}} \\ \cline{1-4}

$E_{\ast q}$&$ :$ & $\delta_2 =q=\delta_1=\chi_1=0 $& 2  & \\ \cline{1-4}

$B_1$&$ :$ & $\delta_2 =q=s=\delta_1 =0 $ & 3 & \\ \cline{1-4}

$B_2$&$ :$ & $\delta_2 =q= \pi_1 \pi_2^2 \chi_1 \chi_2^2 t + \lambda_1 \beta_3 w^2 = \delta_1 =0 $ & 2 & \\ \cline{1-4}

$B_3$&$ :$ & $\delta_2 = q = 2 \pi_1 \pi_2^2 \chi_2^5 s 
+ \lambda_2 \delta_1 \lambda_1 (-\delta_1 \lambda_1 w^3 + 3 \chi_2^2 x) = t = 0 $ & 1 & \\ \cline{1-4}

$C$&$ :$ & $\lambda_2= q = \pi_1 \pi_2^2 \chi_1 \chi_2^2 t +
  \lambda_1 \beta_3 w^2 = \delta_2 s - \delta_1 t = 0	$ & 1 &  \\ \cline{1-4}

$X$&$ :$ & $w=q=\pi_1^2 \pi_2^4 \chi_1^2 \chi_2^4 s t+ 
\lambda_1 \lambda_2^2 \delta_2  x^3 = \delta_2 s - \delta_1 t = 0 $ & 1 &  \\ \hline
\end{tabular}
\caption{\footnotesize Curves in codimension 2 $w_0=q=0$. The Diagram
is precisely the Dynkin diagram of an affine $E_6$ group.}
\label{tableE6codim2'}
\end{table}

\bibliographystyle{utphys}
\bibliography{ckl}

\def\cprime{$'$} \def\cprime{$'$} \def\cprime{$'$}
\providecommand{\href}[2]{#2}\begingroup\raggedright\begin{thebibliography}{100}

\bibitem{Klemm:1996ts}
A.~Klemm, B.~Lian, S.~Roan, and S.-T. Yau, ``{Calabi-Yau fourfolds for M theory
  and F theory compactifications},''
  \href{http://dx.doi.org/10.1016/S0550-3213(97)00798-0}{{\em Nucl.Phys.}
  {\bfseries B518} (1998) 515--574},
\href{http://arxiv.org/abs/hep-th/9701023}{{\ttfamily arXiv:hep-th/9701023
  [hep-th]}}.

\bibitem{Grassi:2013kha}
A.~Grassi, J.~Halverson, and J.~L. Shaneson, ``{Matter From Geometry Without
  Resolution},'' \href{http://dx.doi.org/10.1007/JHEP10(2013)205}{{\em JHEP}
  {\bfseries 1310} (2013) 205},
\href{http://arxiv.org/abs/1306.1832}{{\ttfamily arXiv:1306.1832 [hep-th]}}.

\bibitem{Cecotti:2009zf}
S.~Cecotti, M.~C. Cheng, J.~J. Heckman, and C.~Vafa, ``{Yukawa Couplings in
  F-theory and Non-Commutative Geometry},''
\href{http://arxiv.org/abs/0910.0477}{{\ttfamily arXiv:0910.0477 [hep-th]}}.

\bibitem{Donagi:1996yf}
R.~Donagi, A.~Grassi, and E.~Witten, ``{A Nonperturbative superpotential with
  E(8) symmetry},'' \href{http://dx.doi.org/10.1142/S0217732396002198}{{\em
  Mod.Phys.Lett.} {\bfseries A11} (1996) 2199--2212},
\href{http://arxiv.org/abs/hep-th/9607091}{{\ttfamily arXiv:hep-th/9607091
  [hep-th]}}.

\bibitem{Grimm:2009ef}
T.~W. Grimm, T.-W. Ha, A.~Klemm, and D.~Klevers, ``{Computing Brane and Flux
  Superpotentials in F-theory Compactifications},''
  \href{http://dx.doi.org/10.1007/JHEP04(2010)015}{{\em JHEP} {\bfseries 1004}
  (2010) 015},
\href{http://arxiv.org/abs/0909.2025}{{\ttfamily arXiv:0909.2025 [hep-th]}}.

\bibitem{Jockers:2009ti}
H.~Jockers, P.~Mayr, and J.~Walcher, ``{On N=1 4d Effective Couplings for
  F-theory and Heterotic Vacua},''
  \href{http://dx.doi.org/10.4310/ATMP.2010.v14.n5.a3}{{\em
  Adv.Theor.Math.Phys.} {\bfseries 14} (2010) 1433--1514},
\href{http://arxiv.org/abs/0912.3265}{{\ttfamily arXiv:0912.3265 [hep-th]}}.

\bibitem{Grimm:2009sy}
T.~W. Grimm, T.-W. Ha, A.~Klemm, and D.~Klevers, ``{Five-Brane Superpotentials
  and Heterotic / F-theory Duality},''
  \href{http://dx.doi.org/10.1016/j.nuclphysb.2010.06.011}{{\em Nucl.Phys.}
  {\bfseries B838} (2010) 458--491},
\href{http://arxiv.org/abs/0912.3250}{{\ttfamily arXiv:0912.3250 [hep-th]}}.

\bibitem{Aganagic:2001nx}
M.~Aganagic, A.~Klemm, and C.~Vafa, ``{Disk instantons, mirror symmetry and the
  duality web},'' {\em Z.Naturforsch.} {\bfseries A57} (2002) 1--28,
\href{http://arxiv.org/abs/hep-th/0105045}{{\ttfamily arXiv:hep-th/0105045
  [hep-th]}}.

\bibitem{Aganagic:2000gs}
M.~Aganagic and C.~Vafa, ``{Mirror symmetry, D-branes and counting holomorphic
  discs},''
\href{http://arxiv.org/abs/hep-th/0012041}{{\ttfamily arXiv:hep-th/0012041
  [hep-th]}}.

\bibitem{Grimm:2008dq}
T.~W. Grimm, T.-W. Ha, A.~Klemm, and D.~Klevers, ``{The D5-brane effective
  action and superpotential in N=1 compactifications},''
  \href{http://dx.doi.org/10.1016/j.nuclphysb.2009.03.008}{{\em Nucl.Phys.}
  {\bfseries B816} (2009) 139--184},
\href{http://arxiv.org/abs/0811.2996}{{\ttfamily arXiv:0811.2996 [hep-th]}}.

\bibitem{Alim:2009bx}
M.~Alim, M.~Hecht, H.~Jockers, P.~Mayr, A.~Mertens, {\em et~al.}, ``{Hints for
  Off-Shell Mirror Symmetry in type II/F-theory Compactifications},''
  \href{http://dx.doi.org/10.1016/j.nuclphysb.2010.06.017}{{\em Nucl.Phys.}
  {\bfseries B841} (2010) 303--338},
\href{http://arxiv.org/abs/0909.1842}{{\ttfamily arXiv:0909.1842 [hep-th]}}.

\bibitem{Alim:2011rp}
M.~Alim, M.~Hecht, H.~Jockers, P.~Mayr, A.~Mertens, {\em et~al.}, ``{Flat
  Connections in Open String Mirror Symmetry},''
  \href{http://dx.doi.org/10.1007/JHEP06(2012)138}{{\em JHEP} {\bfseries 1206}
  (2012) 138},
\href{http://arxiv.org/abs/1110.6522}{{\ttfamily arXiv:1110.6522 [hep-th]}}.

\bibitem{Grimm:2010gk}
T.~W. Grimm, A.~Klemm, and D.~Klevers, ``{Five-Brane Superpotentials, Blow-Up
  Geometries and SU(3) Structure Manifolds},''
  \href{http://dx.doi.org/10.1007/JHEP05(2011)113}{{\em JHEP} {\bfseries 1105}
  (2011) 113},
\href{http://arxiv.org/abs/1011.6375}{{\ttfamily arXiv:1011.6375 [hep-th]}}.

\bibitem{Esole:2011sm}
M.~Esole and S.-T. Yau, ``{Small resolutions of SU(5)-models in F-theory},''
\href{http://arxiv.org/abs/1107.0733}{{\ttfamily arXiv:1107.0733 [hep-th]}}.

\bibitem{Marsano:2011hv}
J.~Marsano and S.~Schafer-Nameki, ``{Yukawas, G-flux, and Spectral Covers from
  Resolved Calabi-Yau's},''
  \href{http://dx.doi.org/10.1007/JHEP11(2011)098}{{\em JHEP} {\bfseries 1111}
  (2011) 098},
\href{http://arxiv.org/abs/1108.1794}{{\ttfamily arXiv:1108.1794 [hep-th]}}.

\bibitem{Heckman:2009mn}
J.~J. Heckman, A.~Tavanfar, and C.~Vafa, ``{The Point of E(8) in F-theory
  GUTs},'' \href{http://dx.doi.org/10.1007/JHEP08(2010)040}{{\em JHEP}
  {\bfseries 1008} (2010) 040},
\href{http://arxiv.org/abs/0906.0581}{{\ttfamily arXiv:0906.0581 [hep-th]}}.

\bibitem{Vafa:1996xn}
C.~Vafa, ``{Evidence for F theory},''
  \href{http://dx.doi.org/10.1016/0550-3213(96)00172-1}{{\em Nucl.Phys.}
  {\bfseries B469} (1996) 403--418},
\href{http://arxiv.org/abs/hep-th/9602022}{{\ttfamily arXiv:hep-th/9602022
  [hep-th]}}.

\bibitem{Morrison:1996pp}
D.~R. Morrison and C.~Vafa, ``{Compactifications of F theory on Calabi-Yau
  threefolds. 2.},'' \href{http://dx.doi.org/10.1016/0550-3213(96)00369-0}{{\em
  Nucl.Phys.} {\bfseries B476} (1996) 437--469},
\href{http://arxiv.org/abs/hep-th/9603161}{{\ttfamily arXiv:hep-th/9603161
  [hep-th]}}.

\bibitem{Aspinwall:1996nk}
P.~S. Aspinwall and M.~Gross, ``{The SO(32) heterotic string on a K3
  surface},'' \href{http://dx.doi.org/10.1016/0370-2693(96)01095-7}{{\em
  Phys.Lett.} {\bfseries B387} (1996) 735--742},
\href{http://arxiv.org/abs/hep-th/9605131}{{\ttfamily arXiv:hep-th/9605131
  [hep-th]}}.

\bibitem{Kachru:1997bz}
S.~Kachru, A.~Klemm, and Y.~Oz, ``{Calabi-Yau duals for CHL strings},''
  \href{http://dx.doi.org/10.1016/S0550-3213(98)00228-4}{{\em Nucl.Phys.}
  {\bfseries B521} (1998) 58--70},
\href{http://arxiv.org/abs/hep-th/9712035}{{\ttfamily arXiv:hep-th/9712035
  [hep-th]}}.

\bibitem{Bershadsky:1998vn}
M.~Bershadsky, T.~Pantev, and V.~Sadov, ``{F theory with quantized fluxes},''
  {\em Adv.Theor.Math.Phys.} {\bfseries 3} (1999) 727--773,
\href{http://arxiv.org/abs/hep-th/9805056}{{\ttfamily arXiv:hep-th/9805056
  [hep-th]}}.

\bibitem{Berglund:1998va}
P.~Berglund, A.~Klemm, P.~Mayr, and S.~Theisen, ``{On type IIB vacua with
  varying coupling constant},''
  \href{http://dx.doi.org/10.1016/S0550-3213(99)00420-4}{{\em Nucl.Phys.}
  {\bfseries B558} (1999) 178--204},
\href{http://arxiv.org/abs/hep-th/9805189}{{\ttfamily arXiv:hep-th/9805189
  [hep-th]}}.

\bibitem{Friedman:1997yq}
R.~Friedman, J.~Morgan, and E.~Witten, ``{Vector bundles and F theory},''
  \href{http://dx.doi.org/10.1007/s002200050154}{{\em Commun.Math.Phys.}
  {\bfseries 187} (1997) 679--743},
\href{http://arxiv.org/abs/hep-th/9701162}{{\ttfamily arXiv:hep-th/9701162
  [hep-th]}}.

\bibitem{Gukov:1999ya}
S.~Gukov, C.~Vafa, and E.~Witten, ``{CFT's from Calabi-Yau four folds},''
  \href{http://dx.doi.org/10.1016/S0550-3213(00)00373-4}{{\em Nucl.Phys.}
  {\bfseries B584} (2000) 69--108},
\href{http://arxiv.org/abs/hep-th/9906070}{{\ttfamily arXiv:hep-th/9906070
  [hep-th]}}.

\bibitem{Witten:1996bn}
E.~Witten, ``{Nonperturbative superpotentials in string theory},''
  \href{http://dx.doi.org/10.1016/0550-3213(96)00283-0}{{\em Nucl.Phys.}
  {\bfseries B474} (1996) 343--360},
\href{http://arxiv.org/abs/hep-th/9604030}{{\ttfamily arXiv:hep-th/9604030
  [hep-th]}}.

\bibitem{Kallosh:2005gs}
R.~Kallosh, A.-K. Kashani-Poor, and A.~Tomasiello, ``{Counting fermionic zero
  modes on M5 with fluxes},''
  \href{http://dx.doi.org/10.1088/1126-6708/2005/06/069}{{\em JHEP} {\bfseries
  0506} (2005) 069},
\href{http://arxiv.org/abs/hep-th/0503138}{{\ttfamily arXiv:hep-th/0503138
  [hep-th]}}.

\bibitem{Katz:1996th}
S.~H. Katz and C.~Vafa, ``{Geometric engineering of N=1 quantum field
  theories},'' \href{http://dx.doi.org/10.1016/S0550-3213(97)00283-6}{{\em
  Nucl.Phys.} {\bfseries B497} (1997) 196--204},
\href{http://arxiv.org/abs/hep-th/9611090}{{\ttfamily arXiv:hep-th/9611090
  [hep-th]}}.

\bibitem{MR2669707}
A.~Klemm, D.~Maulik, R.~Pandharipande, and E.~Scheidegger,
  ``Noether-{L}efschetz theory and the {Y}au-{Z}aslow conjecture,''
  \href{http://dx.doi.org/10.1090/S0894-0347-2010-00672-8}{{\em J. Amer. Math.
  Soc.} {\bfseries 23} no.~4, (2010) 1013--1040}.
  \url{http://dx.doi.org/10.1090/S0894-0347-2010-00672-8}.

\bibitem{greenekantor}
T.-M. Chiang, B.~R. Greene, M.~Gross, and K.~Yakov, ``{Black Hole Condensations
  and the Web of Calabi-Yau Manfolds},''
  \href{http://arxiv.org/abs/9511204}{{\ttfamily arXiv:9511204 [hep-th]}}.

\bibitem{Intriligator:2012ue}
K.~Intriligator, H.~Jockers, P.~Mayr, D.~R. Morrison, and M.~R. Plesser,
  ``{Conifold Transitions in M-theory on Calabi-Yau Fourfolds with Background
  Fluxes},''
\href{http://arxiv.org/abs/1203.6662}{{\ttfamily arXiv:1203.6662 [hep-th]}}.

\bibitem{Curio:2000sc}
G.~Curio, A.~Klemm, D.~Lust, and S.~Theisen, ``{On the vacuum structure of type
  II string compactifications on Calabi-Yau spaces with H fluxes},''
  \href{http://dx.doi.org/10.1016/S0550-3213(01)00285-1}{{\em Nucl.Phys.}
  {\bfseries B609} (2001) 3--45},
\href{http://arxiv.org/abs/hep-th/0012213}{{\ttfamily arXiv:hep-th/0012213
  [hep-th]}}.

\bibitem{Moore:1998pn}
G.~W. Moore, ``{Arithmetic and attractors},''
\href{http://arxiv.org/abs/hep-th/9807087}{{\ttfamily arXiv:hep-th/9807087
  [hep-th]}}.

\bibitem{Beasley:2008dc}
C.~Beasley, J.~J. Heckman, and C.~Vafa, ``{GUTs and Exceptional Branes in
  F-theory - I},'' \href{http://dx.doi.org/10.1088/1126-6708/2009/01/058}{{\em
  JHEP} {\bfseries 0901} (2009) 058},
\href{http://arxiv.org/abs/0802.3391}{{\ttfamily arXiv:0802.3391 [hep-th]}}.

\bibitem{Beasley:2008kw}
C.~Beasley, J.~J. Heckman, and C.~Vafa, ``{GUTs and Exceptional Branes in
  F-theory - II: Experimental Predictions},''
  \href{http://dx.doi.org/10.1088/1126-6708/2009/01/059}{{\em JHEP} {\bfseries
  0901} (2009) 059},
\href{http://arxiv.org/abs/0806.0102}{{\ttfamily arXiv:0806.0102 [hep-th]}}.

\bibitem{Bershadsky:1996nh}
M.~Bershadsky, K.~A. Intriligator, S.~Kachru, D.~R. Morrison, V.~Sadov, {\em
  et~al.}, ``{Geometric singularities and enhanced gauge symmetries},''
  \href{http://dx.doi.org/10.1016/S0550-3213(96)90131-5}{{\em Nucl.Phys.}
  {\bfseries B481} (1996) 215--252},
\href{http://arxiv.org/abs/hep-th/9605200}{{\ttfamily arXiv:hep-th/9605200
  [hep-th]}}.

\bibitem{Sen:1996vd}
A.~Sen, ``{F theory and orientifolds},''
  \href{http://dx.doi.org/10.1016/0550-3213(96)00347-1}{{\em Nucl.Phys.}
  {\bfseries B475} (1996) 562--578},
\href{http://arxiv.org/abs/hep-th/9605150}{{\ttfamily arXiv:hep-th/9605150
  [hep-th]}}.

\bibitem{Minasian:1997mm}
R.~Minasian and G.~W. Moore, ``{K theory and Ramond-Ramond charge},''
  \href{http://dx.doi.org/10.1088/1126-6708/1997/11/002}{{\em JHEP} {\bfseries
  9711} (1997) 002},
\href{http://arxiv.org/abs/hep-th/9710230}{{\ttfamily arXiv:hep-th/9710230
  [hep-th]}}.

\bibitem{Ka1}
Y.~Kawamata, ``{Kodaira dimension of certain algebraic fibre spaces},''
{\em J. Fac. Sci Tokyo University IA} {\bfseries 30} (1983) 1--24.

\bibitem{Grassi1}
A.~Grassi, ``{On Minimal Models of elliptic threefolds},''
{\em Math. Ann.} {\bfseries 290} (1991) 287--301.

\bibitem{MR0206009}
S.~L. Kleiman, ``Toward a numerical theory of ampleness,'' {\em Ann. of Math.
  (2)} {\bfseries 84} (1966) 293--344.

\bibitem{Hirzebruch}
F.~Hirzebruch, {\em {Topological methods in Algebraic Geometry}}.
\newblock {Springer Verlag}, 1966.

\bibitem{MR1288523}
P.~Griffiths and J.~Harris, {\em Principles of algebraic geometry}.
\newblock Wiley Classics Library. John Wiley \& Sons Inc., New York, 1994.
\newblock Reprint of the 1978 original.

\bibitem{MR2103716}
P.~J. Gauntlett, ``Branes, calibrations and supergravity,'' in {\em Strings and
  geometry}, vol.~3 of {\em Clay Math. Proc.}, pp.~79--126.
\newblock Amer. Math. Soc., Providence, RI, 2004.

\bibitem{Louis:1996ya}
J.~Louis and K.~Foerger, ``{Holomorphic couplings in string theory},''
  \href{http://dx.doi.org/10.1016/S0920-5632(97)00071-6}{{\em
  Nucl.Phys.Proc.Suppl.} {\bfseries 55B} (1997) 33--64},
\href{http://arxiv.org/abs/hep-th/9611184}{{\ttfamily arXiv:hep-th/9611184
  [hep-th]}}.

\bibitem{Witten:1996md}
E.~Witten, ``{On flux quantization in M theory and the effective action},''
  \href{http://dx.doi.org/10.1016/S0393-0440(96)00042-3}{{\em J.Geom.Phys.}
  {\bfseries 22} (1997) 1--13},
\href{http://arxiv.org/abs/hep-th/9609122}{{\ttfamily arXiv:hep-th/9609122
  [hep-th]}}.

\bibitem{MR1662447}
J.~H. Conway and N.~J.~A. Sloane, {\em Sphere packings, lattices and groups},
  vol.~290 of {\em Grundlehren der Mathematischen Wissenschaften [Fundamental
  Principles of Mathematical Sciences]}.
\newblock Springer-Verlag, New York, third~ed., 1999.
\newblock With additional contributions by E. Bannai, R. E. Borcherds, J.
  Leech, S. P. Norton, A. M. Odlyzko, R. A. Parker, L. Queen and B. B. Venkov.

\bibitem{Collinucci:2010gz}
A.~Collinucci and R.~Savelli, ``{On Flux Quantization in F-Theory},''
  \href{http://dx.doi.org/10.1007/JHEP02(2012)015}{{\em JHEP} {\bfseries 1202}
  (2012) 015},
\href{http://arxiv.org/abs/1011.6388}{{\ttfamily arXiv:1011.6388 [hep-th]}}.

\bibitem{HS}
M.~Hopkins and S.~I. M., ``{Quadratic functions in geometry, topology and
  M-theory},'' \href{http://dx.doi.org/10.1007/JHEP10(2012)128}{{\em J. Diff.
  Geom} {\bfseries 70} (2012) 329},
  \href{http://arxiv.org/abs/0211216}{{\ttfamily arXiv:0211216 [math]}}.

\bibitem{Batyrev:1994hm}
V.~V. Batyrev, ``{Dual polyhedra and mirror symmetry for Calabi-Yau
  hypersurfaces in toric varieties},'' {\em J.Alg.Geom.} {\bfseries 3} (1994)
  493--545,
\href{http://arxiv.org/abs/alg-geom/9310003}{{\ttfamily arXiv:alg-geom/9310003
  [alg-geom]}}.

\bibitem{Fulton}
D.~Fulton, {\em {Introdcution into Toric Varieties}}.
\newblock {Princeton University Press}, 1997.

\bibitem{CoxKatz}
D.~A. Cox and S.~Katz, {\em {Mirror Symmetry and Algebraic Geometry}}.
\newblock {AMS}, 1999.

\bibitem{MR2810322}
D.~A. Cox, J.~B. Little, and H.~K. Schenck, {\em Toric varieties}, vol.~124 of
  {\em Graduate Studies in Mathematics}.
\newblock American Mathematical Society, Providence, RI, 2011.

\bibitem{Huang:2013yta}
M.-x. Huang, A.~Klemm, and M.~Poretschkin, ``{Refined stable pair invariants
  for E-, M- and [p,q]-strings},''
\href{http://arxiv.org/abs/1308.0619}{{\ttfamily arXiv:1308.0619 [hep-th]}}.

\bibitem{Batyrev:1994pg}
V.~V. Batyrev and L.~A. Borisov, ``{On Calabi-Yau complete intersections in
  toric varieties},''
\href{http://arxiv.org/abs/alg-geom/9412017}{{\ttfamily arXiv:alg-geom/9412017
  [alg-geom]}}.

\bibitem{Batyrev:1994ju}
V.~Batyrev and D.~Dais, ``{Strong McKay correspondence, string theoretic Hodge
  numbers and mirror symmetry},''
\href{http://arxiv.org/abs/alg-geom/9410001}{{\ttfamily arXiv:alg-geom/9410001
  [alg-geom]}}.

\bibitem{Kreuzer:1997zg}
M.~Kreuzer and H.~Skarke, ``{Calabi-Yau four folds and toric fibrations},''
  \href{http://dx.doi.org/10.1016/S0393-0440(97)00059-4}{{\em J.Geom.Phys.}
  {\bfseries 26} (1998) 272--290},
\href{http://arxiv.org/abs/hep-th/9701175}{{\ttfamily arXiv:hep-th/9701175
  [hep-th]}}.

\bibitem{Klemm:2004km}
A.~Klemm, M.~Kreuzer, E.~Riegler, and E.~Scheidegger, ``{Topological string
  amplitudes, complete intersection Calabi-Yau spaces and threshold
  corrections},'' \href{http://dx.doi.org/10.1088/1126-6708/2005/05/023}{{\em
  JHEP} {\bfseries 0505} (2005) 023},
\href{http://arxiv.org/abs/hep-th/0410018}{{\ttfamily arXiv:hep-th/0410018
  [hep-th]}}.

\bibitem{MoriMukai}
S.~Mori and S.~Mukai, ``{On Fano 3-folds with $B2 \ge 2$ in Algebraic Varieties
  and Analytic Varieties, Ed. S. Iitaka},'' {\em {Adv. Studies in Pure Math.}}
  {\bfseries 1} ({1983}) 101.

\bibitem{Cvetic:2013uta}
M.~Cvetic, A.~Grassi, D.~Klevers, and H.~Piragua, ``{Chiral Four-Dimensional
  F-Theory Compactifications With SU(5) and Multiple U(1)-Factors},''
\href{http://arxiv.org/abs/1306.3987}{{\ttfamily arXiv:1306.3987 [hep-th]}}.

\bibitem{Hu:2000pr}
Y.~Hu, C.-H. Liu, and S.-T. Yau, ``{Toric morphisms and fibrations of toric
  Calabi-Yau hypersurfaces},'' {\em Adv.Theor.Math.Phys.} {\bfseries 6} (2003)
  457--505,
\href{http://arxiv.org/abs/math/0010082}{{\ttfamily arXiv:math/0010082
  [math-ag]}}.

\bibitem{Candelas:2000nc}
P.~Candelas, D.-E. Diaconescu, B.~Florea, D.~R. Morrison, and G.~Rajesh,
  ``{Codimension three bundle singularities in F theory},'' {\em JHEP}
  {\bfseries 0206} (2002) 014,
\href{http://arxiv.org/abs/hep-th/0009228}{{\ttfamily arXiv:hep-th/0009228
  [hep-th]}}.

\bibitem{Witten:1995zh}
E.~Witten, ``{Some comments on string dynamics},''
\href{http://arxiv.org/abs/hep-th/9507121}{{\ttfamily arXiv:hep-th/9507121
  [hep-th]}}.

\bibitem{Argyres:1995jj}
P.~C. Argyres and M.~R. Douglas, ``{New phenomena in SU(3) supersymmetric gauge
  theory},'' \href{http://dx.doi.org/10.1016/0550-3213(95)00281-V}{{\em
  Nucl.Phys.} {\bfseries B448} (1995) 93--126},
\href{http://arxiv.org/abs/hep-th/9505062}{{\ttfamily arXiv:hep-th/9505062
  [hep-th]}}.

\bibitem{Maldacena:1997de}
J.~M. Maldacena, A.~Strominger, and E.~Witten, ``{Black hole entropy in M
  theory},'' \href{http://dx.doi.org/10.1088/1126-6708/1997/12/002}{{\em JHEP}
  {\bfseries 9712} (1997) 002},
\href{http://arxiv.org/abs/hep-th/9711053}{{\ttfamily arXiv:hep-th/9711053
  [hep-th]}}.

\bibitem{Klemm:1996hh}
A.~Klemm, P.~Mayr, and C.~Vafa, ``{BPS states of exceptional noncritical
  strings},''
\href{http://arxiv.org/abs/hep-th/9607139}{{\ttfamily arXiv:hep-th/9607139
  [hep-th]}}.

\bibitem{HvS}
J.~Hoffmann and van Straten~D., ``{ Some Monodromy Groups of finite index in
  $SP_4(\mathbb{Z})$},'' \href{http://arxiv.org/abs/1312.3063}{{\ttfamily
  arXiv:1312.3063 [math.AG]}}.

\bibitem{Fuchs:1989yv}
J.~Fuchs, A.~Klemm, C.~Scheich, and M.~G. Schmidt, ``{Spectra and Symmetries of
  Gepner Models Compared to Calabi-yau Compactifications},''
\href{http://dx.doi.org/10.1016/0003-4916(90)90119-9}{{\em Annals Phys.}
  {\bfseries 204} (1990) 1--51}.

\bibitem{Klemm:1990df}
A.~Klemm and M.~G. Schmidt, ``{Orbifolds by Cyclic Permutations of Tensor
  Product Conformal Field Theories},''
\href{http://dx.doi.org/10.1016/0370-2693(90)90164-2}{{\em Phys.Lett.}
  {\bfseries B245} (1990) 53--58}.

\bibitem{Greene:1990ud}
B.~R. Greene and M.~Plesser, ``{Duality in {Calabi-Yau} Moduli Space},''
\href{http://dx.doi.org/10.1016/0550-3213(90)90622-K}{{\em Nucl.Phys.}
  {\bfseries B338} (1990) 15--37}.

\bibitem{MR1798982}
A.~R. Iano-Fletcher, ``Working with weighted complete intersections,'' in {\em
  Explicit birational geometry of 3-folds}, vol.~281 of {\em London Math. Soc.
  Lecture Note Ser.}, pp.~101--173.
\newblock Cambridge Univ. Press, Cambridge, 2000.

\bibitem{Font:1992uk}
A.~Font, ``{Periods and duality symmetries in Calabi-Yau compactifications},''
  \href{http://dx.doi.org/10.1016/0550-3213(93)90152-F}{{\em Nucl.Phys.}
  {\bfseries B391} (1993) 358--388},
\href{http://arxiv.org/abs/hep-th/9203084}{{\ttfamily arXiv:hep-th/9203084
  [hep-th]}}.

\bibitem{Klemm:1992tx}
A.~Klemm and S.~Theisen, ``{Considerations of one modulus Calabi-Yau
  compactifications: Picard-Fuchs equations, Kahler potentials and mirror
  maps},'' \href{http://dx.doi.org/10.1016/0550-3213(93)90289-2}{{\em
  Nucl.Phys.} {\bfseries B389} (1993) 153--180},
\href{http://arxiv.org/abs/hep-th/9205041}{{\ttfamily arXiv:hep-th/9205041
  [hep-th]}}.

\bibitem{Giveon:1990ay}
A.~Giveon and D.-J. Smit, ``{Symmetries on the Moduli Space of (2,2)
  Superstring Vacua},''
\href{http://dx.doi.org/10.1016/0550-3213(91)90193-2}{{\em Nucl.Phys.}
  {\bfseries B349} (1991) 168--206}.

\bibitem{MR909231}
M.~Reid, ``The moduli space of {$3$}-folds with {$K=0$} may nevertheless be
  irreducible,'' \href{http://dx.doi.org/10.1007/BF01458074}{{\em Math. Ann.}
  {\bfseries 278} no.~1-4, (1987) 329--334}.
  \url{http://dx.doi.org/10.1007/BF01458074}.

\bibitem{Berglund:1996uy}
P.~Berglund, S.~H. Katz, A.~Klemm, and P.~Mayr, ``{New Higgs transitions
  between dual N=2 string models},''
  \href{http://dx.doi.org/10.1016/S0550-3213(96)00450-6}{{\em Nucl.Phys.}
  {\bfseries B483} (1997) 209--228},
\href{http://arxiv.org/abs/hep-th/9605154}{{\ttfamily arXiv:hep-th/9605154
  [hep-th]}}.

\bibitem{Klemm:1996kv}
A.~Klemm and P.~Mayr, ``{Strong coupling singularities and nonAbelian gauge
  symmetries in N=2 string theory},''
  \href{http://dx.doi.org/10.1016/0550-3213(96)00108-3}{{\em Nucl.Phys.}
  {\bfseries B469} (1996) 37--50},
\href{http://arxiv.org/abs/hep-th/9601014}{{\ttfamily arXiv:hep-th/9601014
  [hep-th]}}.

\bibitem{BraunSage}
V.~Braun, ``{Toric Geometry and Sage},'' {\em
  {www.stp.dias.ie/~vbraun/ToricLecture.pdf}} .

\bibitem{MR2931227}
E.~Looijenga, ``Trento notes on {H}odge theory,'' {\em Rend. Semin. Mat. Univ.
  Politec. Torino} {\bfseries 69} no.~2, (2011) 113--148.

\bibitem{MR0419438}
M.~Cornalba and P.~A. Griffiths, ``Some transcendental aspects of algebraic
  geometry,'' in {\em Algebraic geometry ({P}roc. {S}ympos. {P}ure {M}ath.,
  {V}ol. 29, {H}umboldt {S}tate {U}niv., {A}rcata, {C}alif., 1974)},
  pp.~3--110.
\newblock Amer. Math. Soc., Providence, R.I., 1975.

\bibitem{MR579742}
Z.~Mebkhout, ``Sur le probl\`eme de {H}ilbert-{R}iemann,'' in {\em Complex
  analysis, microlocal calculus and relativistic quantum theory ({P}roc.
  {I}nternat. {C}olloq., {C}entre {P}hys., {L}es {H}ouches, 1979)}, vol.~126 of
  {\em Lecture Notes in Phys.}, pp.~90--110.
\newblock Springer, Berlin-New York, 1980.

\bibitem{MR743382}
M.~Kashiwara, ``The {R}iemann-{H}ilbert problem for holonomic systems,''
  \href{http://dx.doi.org/10.2977/prims/1195181610}{{\em Publ. Res. Inst. Math.
  Sci.} {\bfseries 20} no.~2, (1984) 319--365}.
  \url{http://dx.doi.org/10.2977/prims/1195181610}.

\bibitem{Kashiwarabook}
K.~Masaki, {\em {D-modules and Microlocal Calculus}}.
\newblock Mathematical Monographs 126. {AMS}, 2000.

\bibitem{Hosono:1993qy}
S.~Hosono, A.~Klemm, S.~Theisen, and S.-T. Yau, ``{Mirror symmetry, mirror map
  and applications to Calabi-Yau hypersurfaces},''
  \href{http://dx.doi.org/10.1007/BF02100589}{{\em Commun.Math.Phys.}
  {\bfseries 167} (1995) 301--350},
\href{http://arxiv.org/abs/hep-th/9308122}{{\ttfamily arXiv:hep-th/9308122
  [hep-th]}}.

\bibitem{MR0344248}
A.~Landman, ``On the {P}icard-{L}efschetz transformation for algebraic
  manifolds acquiring general singularities,'' {\em Trans. Amer. Math. Soc.}
  {\bfseries 181} (1973) 89--126.

\bibitem{MR0417174}
P.~Deligne, {\em \'{E}quations diff\'erentielles \`a points singuliers
  r\'eguliers}.
\newblock Lecture Notes in Mathematics, Vol. 163. Springer-Verlag, Berlin-New
  York, 1970.

\bibitem{MR0382272}
W.~Schmid, ``Variation of {H}odge structure: the singularities of the period
  mapping,'' {\em Invent. Math.} {\bfseries 22} (1973) 211--319.

\bibitem{Donagi:2012ts}
R.~Donagi, S.~Katz, and M.~Wijnholt, ``{Weak Coupling, Degeneration and Log
  Calabi-Yau Spaces},''
\href{http://arxiv.org/abs/1212.0553}{{\ttfamily arXiv:1212.0553 [hep-th]}}.

\bibitem{MR0498552}
P.~Deligne, ``Th\'eorie de {H}odge. {III},'' {\em Inst. Hautes \'Etudes Sci.
  Publ. Math.} no.~44, (1974) 5--77.

\bibitem{Anderson:2013rka}
L.~B. Anderson, J.~J. Heckman, and S.~Katz, ``{T-Branes and Geometry},''
\href{http://arxiv.org/abs/1310.1931}{{\ttfamily arXiv:1310.1931 [hep-th]}}.

\bibitem{MR3011419}
M.~Gross and B.~Siebert, ``Logarithmic {G}romov-{W}itten invariants,''
  \href{http://dx.doi.org/10.1090/S0894-0347-2012-00757-7}{{\em J. Amer. Math.
  Soc.} {\bfseries 26} no.~2, (2013) 451--510}.
  \url{http://dx.doi.org/10.1090/S0894-0347-2012-00757-7}.

\bibitem{MR2248516}
D.~Maulik and R.~Pandharipande, ``A topological view of {G}romov-{W}itten
  theory,'' \href{http://dx.doi.org/10.1016/j.top.2006.06.002}{{\em Topology}
  {\bfseries 45} no.~5, (2006) 887--918}.
  \url{http://dx.doi.org/10.1016/j.top.2006.06.002}.

\bibitem{MR2746343}
D.~Maulik, R.~Pandharipande, and R.~P. Thomas, ``Curves on {$K3$} surfaces and
  modular forms,'' \href{http://dx.doi.org/10.1112/jtopol/jtq030}{{\em J.
  Topol.} {\bfseries 3} no.~4, (2010) 937--996}.
  \url{http://dx.doi.org/10.1112/jtopol/jtq030}. With an appendix by A. Pixton.

\bibitem{MR902936}
I.~M. Gel{\cprime}fand, M.~I. Graev, and A.~V. Zelevinski{\u\i}, ``Holonomic
  systems of equations and series of hypergeometric type,'' {\em Dokl. Akad.
  Nauk SSSR} {\bfseries 295} no.~1, (1987) 14--19.

\bibitem{MR948812}
I.~M. Gel{\cprime}fand, A.~V. Zelevinski{\u\i}, and M.~M. Kapranov, ``Equations
  of hypergeometric type and {N}ewton polyhedra,'' {\em Dokl. Akad. Nauk SSSR}
  {\bfseries 300} no.~3, (1988) 529--534.

\bibitem{Hosono:1994ax}
S.~Hosono, A.~Klemm, S.~Theisen, and S.-T. Yau, ``{Mirror symmetry, mirror map
  and applications to complete intersection Calabi-Yau spaces},''
  \href{http://dx.doi.org/10.1016/0550-3213(94)00440-P}{{\em Nucl.Phys.}
  {\bfseries B433} (1995) 501--554},
\href{http://arxiv.org/abs/hep-th/9406055}{{\ttfamily arXiv:hep-th/9406055
  [hep-th]}}.

\bibitem{MR0188215}
B.~Dwork, ``On the zeta function of a hypersurface. {II},'' {\em Ann. of Math.
  (2)} {\bfseries 80} (1964) 227--299.

\bibitem{MR0242841}
N.~M. Katz, ``On the differential equations satisfied by period matrices,''
  {\em Inst. Hautes \'Etudes Sci. Publ. Math.} no.~35, (1968) 223--258.

\bibitem{Vafa:1989pa}
C.~Vafa, ``{Superstring Vacua},'' in {\em Trieste 1989}, vol.~1989 of {\em
  Proceedings, High energy physics and cosmology* 145-177}, pp.~145--177.
\newblock 1989.

\bibitem{Candelas:1990rm}
{Candelas, Philip and De La Ossa, Xenia C. and Green, Paul S. and Parkes,
  Linda}, ``{A Pair of Calabi-Yau manifolds as an exactly soluble
  superconformal theory},''
\href{http://dx.doi.org/10.1016/0550-3213(91)90292-6}{{\em Nucl.Phys.}
  {\bfseries B359} (1991) 21--74}.

\bibitem{Hori:2013ika}
K.~Hori and M.~Romo, ``{Exact Results In Two-Dimensional (2,2) Supersymmetric
  Gauge Theories With Boundary},''
\href{http://arxiv.org/abs/1308.2438}{{\ttfamily arXiv:1308.2438 [hep-th]}}.

\bibitem{Zagier}
D.~Zagier, ``Integral solutions of ap\'ery-like recurrence equations.,'' in
  {\em Groups and symmetries. From Neolithic Scots to John McKay. Selected
  papers of the conference, Montreal, Canada, April 27–29, 2007, Harnad, John
  (ed.) et al.}, vol.~47 of {\em AMS Proceedings}, pp.~349--366.
\newblock American Mathematical Society, Providence, RI, 2009.

\bibitem{MR2454322}
G.~Almkvist, D.~van Straten, and W.~Zudilin, ``Ap\'ery limits of differential
  equations of order 4 and 5,'' in {\em Modular forms and string duality},
  vol.~54 of {\em Fields Inst. Commun.}, pp.~105--123.
\newblock Amer. Math. Soc., Providence, RI, 2008.

\bibitem{MR915841}
G.~Tian, ``Smoothness of the universal deformation space of compact
  {C}alabi-{Y}au manifolds and its {P}etersson-{W}eil metric,'' in {\em
  Mathematical aspects of string theory ({S}an {D}iego, {C}alif., 1986)},
  vol.~1 of {\em Adv. Ser. Math. Phys.}, pp.~629--646.
\newblock World Sci. Publishing, Singapore, 1987.

\bibitem{MR1027500}
A.~N. Todorov, ``The {W}eil-{P}etersson geometry of the moduli space of {${\rm
  SU}(n\geq 3)$} ({C}alabi-{Y}au) manifolds. {I},'' {\em Comm. Math. Phys.}
  {\bfseries 126} no.~2, (1989) 325--346.
  \url{http://projecteuclid.org/getRecord?id=euclid.cmp/1104179854}.

\bibitem{Witten:1991zz}
E.~Witten, ``{Mirror manifolds and topological field theory},''
\href{http://arxiv.org/abs/hep-th/9112056}{{\ttfamily arXiv:hep-th/9112056
  [hep-th]}}.

\bibitem{Bershadsky:1993cx}
M.~Bershadsky, S.~Cecotti, H.~Ooguri, and C.~Vafa, ``{Kodaira-Spencer theory of
  gravity and exact results for quantum string amplitudes},''
  \href{http://dx.doi.org/10.1007/BF02099774}{{\em Commun.Math.Phys.}
  {\bfseries 165} (1994) 311--428},
\href{http://arxiv.org/abs/hep-th/9309140}{{\ttfamily arXiv:hep-th/9309140
  [hep-th]}}.

\bibitem{Mayr:2001xk}
P.~Mayr, ``{N=1 mirror symmetry and open / closed string duality},'' {\em
  Adv.Theor.Math.Phys.} {\bfseries 5} (2002) 213--242,
\href{http://arxiv.org/abs/hep-th/0108229}{{\ttfamily arXiv:hep-th/0108229
  [hep-th]}}.

\bibitem{Klemm:2007in}
A.~Klemm and R.~Pandharipande, ``{Enumerative geometry of Calabi-Yau
  4-folds},'' \href{http://dx.doi.org/10.1007/s00220-008-0490-9}{{\em
  Commun.Math.Phys.} {\bfseries 281} (2008) 621--653},
\href{http://arxiv.org/abs/math/0702189}{{\ttfamily arXiv:math/0702189
  [math]}}.

\bibitem{BG}
R.~"Bryant and P.~Griffith, ``"some observations on the inﬁnitesimal period
  relations for regular threefolds with trivial canonical bundle",'' in {\em
  Arithmetic and Geometry II}, vol.~36 of {\em Progress in Mathematics},
  pp.~77--102.
\newblock Birkh\"auser, Basel, 1983.

\bibitem{Strominger:1990pd}
A.~Strominger, ``{Special Geometry},''
\href{http://dx.doi.org/10.1007/BF02096559}{{\em Commun.Math.Phys.} {\bfseries
  133} (1990) 163--180}.

\bibitem{Castellani:1990tp}
L.~Castellani, R.~D'Auria, and S.~Ferrara, ``{Special Kahler Geometry: An
  Intrinsic Formulation from N=2 Space-Time Supersymmetry},''
\href{http://dx.doi.org/10.1016/0370-2693(90)91486-U}{{\em Phys.Lett.}
  {\bfseries B241} (1990) 57}.

\bibitem{Greene:1993vm}
B.~R. Greene, D.~R. Morrison, and M.~R. Plesser, ``{Mirror manifolds in higher
  dimension},'' \href{http://dx.doi.org/10.1007/BF02101657}{{\em
  Commun.Math.Phys.} {\bfseries 173} (1995) 559--598},
\href{http://arxiv.org/abs/hep-th/9402119}{{\ttfamily arXiv:hep-th/9402119
  [hep-th]}}.

\bibitem{Mayr:1996sh}
P.~Mayr, ``{Mirror symmetry, N=1 superpotentials and tensionless strings on
  Calabi-Yau four folds},''
  \href{http://dx.doi.org/10.1016/S0550-3213(97)00196-X}{{\em Nucl.Phys.}
  {\bfseries B494} (1997) 489--545},
\href{http://arxiv.org/abs/hep-th/9610162}{{\ttfamily arXiv:hep-th/9610162
  [hep-th]}}.

\bibitem{Doroud:2012xw}
N.~Doroud, J.~Gomis, B.~Le~Floch, and S.~Lee, ``{Exact Results in D=2
  Supersymmetric Gauge Theories},''
  \href{http://dx.doi.org/10.1007/JHEP05(2013)093}{{\em JHEP} {\bfseries 1305}
  (2013) 093},
\href{http://arxiv.org/abs/1206.2606}{{\ttfamily arXiv:1206.2606 [hep-th]}}.

\bibitem{Benini:2012ui}
F.~Benini and S.~Cremonesi, ``{Partition functions of N=(2,2) gauge theories on
  $S^2$ and vortices},''
\href{http://arxiv.org/abs/1206.2356}{{\ttfamily arXiv:1206.2356 [hep-th]}}.

\bibitem{Jockers:2012dk}
H.~Jockers, V.~Kumar, J.~M. Lapan, D.~R. Morrison, and M.~Romo, ``{Two-Sphere
  Partition Functions and Gromov-Witten Invariants},''
\href{http://arxiv.org/abs/1208.6244}{{\ttfamily arXiv:1208.6244 [hep-th]}}.

\bibitem{Honma:2013hma}
Y.~Honma and M.~Manabe, ``{Exact Kahler Potential for Calabi-Yau Fourfolds},''
  \href{http://dx.doi.org/10.1007/JHEP05(2013)102}{{\em JHEP} {\bfseries 1305}
  (2013) 102},
\href{http://arxiv.org/abs/1302.3760}{{\ttfamily arXiv:1302.3760 [hep-th]}}.

\bibitem{MR2003030}
K.~Hori, S.~Katz, A.~Klemm, R.~Pandharipande, R.~Thomas, C.~Vafa, R.~Vakil, and
  E.~Zaslow, {\em Mirror symmetry}, vol.~1 of {\em Clay Mathematics
  Monographs}.
\newblock American Mathematical Society, Providence, RI, 2003.
\newblock With a preface by Vafa.

\bibitem{MR1416353}
P.~Deligne, ``Local behavior of {H}odge structures at infinity,'' in {\em
  Mirror symmetry, {II}}, vol.~1 of {\em AMS/IP Stud. Adv. Math.},
  pp.~683--699.
\newblock Amer. Math. Soc., Providence, RI, 1997.

\bibitem{Ceresole:1992su}
A.~Ceresole, R.~D'Auria, S.~Ferrara, W.~Lerche, and J.~Louis, ``{Picard-Fuchs
  equations and special geometry},''
  \href{http://dx.doi.org/10.1142/S0217751X93000047}{{\em Int.J.Mod.Phys.}
  {\bfseries A8} (1993) 79--114},
\href{http://arxiv.org/abs/hep-th/9204035}{{\ttfamily arXiv:hep-th/9204035
  [hep-th]}}.

\bibitem{Hori:2003ic}
K.~Hori, S.~Katz, A.~Klemm, R.~Pandharipande, R.~Thomas, {\em et~al.},
``{Mirror symmetry},''.

\bibitem{MR2806464}
C.~Hertling and C.~Sabbah, ``Examples of non-commutative {H}odge structures,''
  \href{http://dx.doi.org/10.1017/S147474801100003X}{{\em J. Inst. Math.
  Jussieu} {\bfseries 10} no.~3, (2011) 635--674}.
  \url{http://dx.doi.org/10.1017/S147474801100003X}.

\bibitem{MR2483750}
L.~Katzarkov, M.~Kontsevich, and T.~Pantev, ``Hodge theoretic aspects of mirror
  symmetry,'' in {\em From {H}odge theory to integrability and {TQFT}
  tt*-geometry}, vol.~78 of {\em Proc. Sympos. Pure Math.}, pp.~87--174.
\newblock Amer. Math. Soc., Providence, RI, 2008.

\bibitem{Candelas:1990pi}
P.~Candelas and X.~de~la Ossa, ``{Moduli Space of Calabi-Yau Manifolds},''
\href{http://dx.doi.org/10.1016/0550-3213(91)90122-E}{{\em Nucl.Phys.}
  {\bfseries B355} (1991) 455--481}.

\bibitem{IRITANI1}
H.~Iritani, ``Ruan’s conjecture and integral structures in quantum
  cohomology,'' \href{http://arxiv.org/abs/0809.2749}{{\ttfamily
  arXiv:0809.2749 [math.AG]}}.

\bibitem{Kontsevichgamma}
M.~Kontsevich, ``The gamma genus,'' 2013.
\newblock Talk at the Arbeitstagung Bonn, May 2013.

\bibitem{KlemmYeh}
A.~Klemm and H.-y. Yeh, ``{Periods on Calabi-Yau fourfolds},'' {\em work in
  progress} .

\bibitem{MR1831820}
P.~Seidel and R.~Thomas, ``Braid group actions on derived categories of
  coherent sheaves,''
  \href{http://dx.doi.org/10.1215/S0012-7094-01-10812-0}{{\em Duke Math. J.}
  {\bfseries 108} no.~1, (2001) 37--108}.
  \url{http://dx.doi.org/10.1215/S0012-7094-01-10812-0}.

\bibitem{Hosono:2000eb}
S.~Hosono, ``{Local mirror symmetry and type IIA monodromy of Calabi-Yau
  manifolds},'' {\em Adv.Theor.Math.Phys.} {\bfseries 4} (2000) 335--376,
\href{http://arxiv.org/abs/hep-th/0007071}{{\ttfamily arXiv:hep-th/0007071
  [hep-th]}}.

\bibitem{Jefferson:2013vfa}
R.~A. Jefferson and J.~Walcher, ``{Monodromy of Inhomogeneous Picard-Fuchs
  Equations},''
\href{http://arxiv.org/abs/1309.0490}{{\ttfamily arXiv:1309.0490 [hep-th]}}.

\bibitem{MR2373143}
T.~Bridgeland, ``Stability conditions on triangulated categories,''
  \href{http://dx.doi.org/10.4007/annals.2007.166.317}{{\em Ann. of Math. (2)}
  {\bfseries 166} no.~2, (2007) 317--345}.
  \url{http://dx.doi.org/10.4007/annals.2007.166.317}.

\bibitem{Halverson:2013qc}
J.~Halverson, H.~Jockers, J.~M. Lapan, and D.~R. Morrison, ``{Perturbative
  Corrections to Kahler Moduli Spaces},''
\href{http://arxiv.org/abs/1308.2157}{{\ttfamily arXiv:1308.2157 [hep-th]}}.

\bibitem{Witten:1993yc}
E.~Witten, ``{Phases of N=2 theories in two-dimensions},''
  \href{http://dx.doi.org/10.1016/0550-3213(93)90033-L}{{\em Nucl.Phys.}
  {\bfseries B403} (1993) 159--222},
\href{http://arxiv.org/abs/hep-th/9301042}{{\ttfamily arXiv:hep-th/9301042
  [hep-th]}}.

\bibitem{Herbst:2008jq}
M.~Herbst, K.~Hori, and D.~Page, ``{Phases Of N=2 Theories In 1+1 Dimensions
  With Boundary},''
\href{http://arxiv.org/abs/0803.2045}{{\ttfamily arXiv:0803.2045 [hep-th]}}.

\bibitem{Mayr:2000as}
P.~Mayr, ``{Phases of supersymmetric D-branes on Kahler manifolds and the McKay
  correspondence},''
  \href{http://dx.doi.org/10.1088/1126-6708/2001/01/018}{{\em JHEP} {\bfseries
  0101} (2001) 018},
\href{http://arxiv.org/abs/hep-th/0010223}{{\ttfamily arXiv:hep-th/0010223
  [hep-th]}}.

\bibitem{Candelas:1993dm}
P.~Candelas, X.~De~La~Ossa, A.~Font, S.~H. Katz, and D.~R. Morrison, ``{Mirror
  symmetry for two parameter models. 1.},''
  \href{http://dx.doi.org/10.1016/0550-3213(94)90322-0}{{\em Nucl.Phys.}
  {\bfseries B416} (1994) 481--538},
\href{http://arxiv.org/abs/hep-th/9308083}{{\ttfamily arXiv:hep-th/9308083
  [hep-th]}}.

\bibitem{Candelas:1994hw}
P.~, A.~Font, S.~H. Katz, and D.~R. Morrison, ``{Mirror symmetry for two
  parameter models. 2.},''
  \href{http://dx.doi.org/10.1016/0550-3213(94)90155-4}{{\em Nucl.Phys.}
  {\bfseries B429} (1994) 626--674},
\href{http://arxiv.org/abs/hep-th/9403187}{{\ttfamily arXiv:hep-th/9403187
  [hep-th]}}.

\bibitem{MR1472476}
M.~Passare, A.~K. Tsikh, and A.~A. Cheshel{\cprime}, ``Multiple
  {M}ellin-{B}arnes integrals as periods on {C}alabi-{Y}au manifolds with
  several moduli,'' \href{http://dx.doi.org/10.1007/BF02073871}{{\em Teoret.
  Mat. Fiz.} {\bfseries 109} no.~3, (1996) 381--394}.
  \url{http://dx.doi.org/10.1007/BF02073871}.

\bibitem{MR1619440}
O.~N. Zhdanov and A.~K. Tsikh, ``Computation of multiple {M}ellin-{B}arnes
  integrals by means of multidimensional residues,'' {\em Dokl. Akad. Nauk}
  {\bfseries 358} no.~2, (1998) 154--156.

\bibitem{Haghighat:2008ut}
B.~Haghighat and A.~Klemm, ``{Topological Strings on Grassmannian Calabi-Yau
  manifolds},'' \href{http://dx.doi.org/10.1088/1126-6708/2009/01/029}{{\em
  JHEP} {\bfseries 0901} (2009) 029},
\href{http://arxiv.org/abs/0802.2908}{{\ttfamily arXiv:0802.2908 [hep-th]}}.

\bibitem{Denef:2008wq}
F.~Denef, ``{Les Houches Lectures on Constructing String Vacua},''
\href{http://arxiv.org/abs/0803.1194}{{\ttfamily arXiv:0803.1194 [hep-th]}}.

\bibitem{Bouchard:2008gu}
V.~Bouchard, A.~Klemm, M.~Marino, and S.~Pasquetti, ``{Topological open strings
  on orbifolds},'' \href{http://dx.doi.org/10.1007/s00220-010-1020-0}{{\em
  Commun.Math.Phys.} {\bfseries 296} (2010) 589--623},
\href{http://arxiv.org/abs/0807.0597}{{\ttfamily arXiv:0807.0597 [hep-th]}}.

\bibitem{Seiberg:1994rs}
N.~Seiberg and E.~Witten, ``{Electric - magnetic duality, monopole
  condensation, and confinement in N=2 supersymmetric Yang-Mills theory},''
  \href{http://dx.doi.org/10.1016/0550-3213(94)90124-4}{{\em Nucl.Phys.}
  {\bfseries B426} (1994) 19--52},
\href{http://arxiv.org/abs/hep-th/9407087}{{\ttfamily arXiv:hep-th/9407087
  [hep-th]}}.

\bibitem{MR592569}
K.~Lamotke, ``The topology of complex projective varieties after {S}.
  {L}efschetz,'' \href{http://dx.doi.org/10.1016/0040-9383(81)90013-6}{{\em
  Topology} {\bfseries 20} no.~1, (1981) 15--51}.
  \url{http://dx.doi.org/10.1016/0040-9383(81)90013-6}.

\bibitem{Polchinski:1995sm}
J.~Polchinski and A.~Strominger, ``{New vacua for type II string theory},''
  \href{http://dx.doi.org/10.1016/S0370-2693(96)01219-1}{{\em Phys.Lett.}
  {\bfseries B388} (1996) 736--742},
\href{http://arxiv.org/abs/hep-th/9510227}{{\ttfamily arXiv:hep-th/9510227
  [hep-th]}}.

\bibitem{Greene:1995hu}
B.~R. Greene, D.~R. Morrison, and A.~Strominger, ``{Black hole condensation and
  the unification of string vacua},''
  \href{http://dx.doi.org/10.1016/0550-3213(95)00371-X}{{\em Nucl.Phys.}
  {\bfseries B451} (1995) 109--120},
\href{http://arxiv.org/abs/hep-th/9504145}{{\ttfamily arXiv:hep-th/9504145
  [hep-th]}}.

\bibitem{Giryavets:2003vd}
A.~Giryavets, S.~Kachru, P.~K. Tripathy, and S.~P. Trivedi, ``{Flux
  compactifications on Calabi-Yau threefolds},''
  \href{http://dx.doi.org/10.1088/1126-6708/2004/04/003}{{\em JHEP} {\bfseries
  0404} (2004) 003},
\href{http://arxiv.org/abs/hep-th/0312104}{{\ttfamily arXiv:hep-th/0312104
  [hep-th]}}.

\bibitem{Denef:2004dm}
F.~Denef, M.~R. Douglas, and B.~Florea, ``{Building a better racetrack},''
  \href{http://dx.doi.org/10.1088/1126-6708/2004/06/034}{{\em JHEP} {\bfseries
  0406} (2004) 034},
\href{http://arxiv.org/abs/hep-th/0404257}{{\ttfamily arXiv:hep-th/0404257
  [hep-th]}}.

\bibitem{Louis:2012nb}
J.~Louis, M.~Rummel, R.~Valandro, and A.~Westphal, ``{Building an explicit de
  Sitter},'' \href{http://dx.doi.org/10.1007/JHEP10(2012)163}{{\em JHEP}
  {\bfseries 1210} (2012) 163},
\href{http://arxiv.org/abs/1208.3208}{{\ttfamily arXiv:1208.3208 [hep-th]}}.

\bibitem{Kachru:1995fv}
S.~Kachru, A.~Klemm, W.~Lerche, P.~Mayr, and C.~Vafa, ``{Nonperturbative
  results on the point particle limit of N=2 heterotic string
  compactifications},''
  \href{http://dx.doi.org/10.1016/0550-3213(95)00574-9}{{\em Nucl.Phys.}
  {\bfseries B459} (1996) 537--558},
\href{http://arxiv.org/abs/hep-th/9508155}{{\ttfamily arXiv:hep-th/9508155
  [hep-th]}}.

\bibitem{Kreuzerliste}
M.~Kreuzer and H.~Skarke, ``{Calabi-Yau data web site},''
  \href{http://arxiv.org/abs/http://hep.itp.tuwien.ac.at/~kreuzer/CY/}{{\ttfamily
  http://hep.itp.tuwien.ac.at/~kreuzer/CY/}}.

\bibitem{Morrison:2012js}
D.~R. Morrison and W.~Taylor, ``{Toric bases for 6D F-theory models},''
  \href{http://dx.doi.org/10.1002/prop.201200086}{{\em Fortsch.Phys.}
  {\bfseries 60} (2012) 1187--1216},
\href{http://arxiv.org/abs/1204.0283}{{\ttfamily arXiv:1204.0283 [hep-th]}}.

\bibitem{Sethi:1996es}
S.~Sethi, C.~Vafa, and E.~Witten, ``{Constraints on low dimensional string
  compactifications},''
  \href{http://dx.doi.org/10.1016/S0550-3213(96)00483-X}{{\em Nucl.Phys.}
  {\bfseries B480} (1996) 213--224},
\href{http://arxiv.org/abs/hep-th/9606122}{{\ttfamily arXiv:hep-th/9606122
  [hep-th]}}.

\bibitem{kodaira1}
K.~Kodaira, ``On compact analytic surfaces, iii,'' {\em The Annals of
  Mathematics} {\bfseries 78} no.~1, (1963) 1--40.

\bibitem{kodaira2}
K.~Kodaira, ``On compact analytic surfaces: Ii,'' {\em The Annals of
  Mathematics} {\bfseries 77} no.~3, (1963) 563--626.

\bibitem{Tate}
J.~Tate, ``{Alogarithm for Determining the Type of a Singular Fibre in an
  Elliptic Pencil},'' in {\em Modular Functions of one variable IV}, vol.~476
  of {\em Lecture Notes in mathematics}.
\newblock Springer-Verlag, Berlin, 1975.

\bibitem{Katz:2011qp}
S.~Katz, D.~R. Morrison, S.~Schafer-Nameki, and J.~Sully, ``{Tate's algorithm
  and F-theory},'' \href{http://dx.doi.org/10.1007/JHEP08(2011)094}{{\em JHEP}
  {\bfseries 1108} (2011) 094},
\href{http://arxiv.org/abs/1106.3854}{{\ttfamily arXiv:1106.3854 [hep-th]}}.

\bibitem{Grimm2010c}
T.~W. Grimm and T.~Weigand, ``{On Abelian Gauge Symmetries and Proton Decay in
  Global F-theory GUTs},''. \url{http://arxiv.org/abs/1006.0226}.

\bibitem{Candelas:1996su}
P.~Candelas and A.~Font, ``{Duality between the webs of heterotic and type II
  vacua},'' \href{http://dx.doi.org/10.1016/S0550-3213(96)00410-5}{{\em
  Nucl.Phys.} {\bfseries B511} (1998) 295--325},
\href{http://arxiv.org/abs/hep-th/9603170}{{\ttfamily arXiv:hep-th/9603170
  [hep-th]}}.

\bibitem{Braun:2013nqa}
V.~Braun, T.~W. Grimm, and J.~Keitel, ``{Geometric Engineering in Toric
  F-Theory and GUTs with U(1) Gauge Factors},''
  \href{http://dx.doi.org/10.1007/JHEP12(2013)069}{{\em JHEP} {\bfseries 1312}
  (2013) 069},
\href{http://arxiv.org/abs/1306.0577}{{\ttfamily arXiv:1306.0577 [hep-th]}}.

\bibitem{Berglund:1998ej}
P.~Berglund and P.~Mayr, ``{Heterotic string / F theory duality from mirror
  symmetry},'' {\em Adv.Theor.Math.Phys.} {\bfseries 2} (1999) 1307--1372,
\href{http://arxiv.org/abs/hep-th/9811217}{{\ttfamily arXiv:hep-th/9811217
  [hep-th]}}.

\bibitem{Klemm:1995tj}
A.~Klemm, W.~Lerche, and P.~Mayr, ``{K3 Fibrations and heterotic type II string
  duality},'' \href{http://dx.doi.org/10.1016/0370-2693(95)00937-G}{{\em
  Phys.Lett.} {\bfseries B357} (1995) 313--322},
\href{http://arxiv.org/abs/hep-th/9506112}{{\ttfamily arXiv:hep-th/9506112
  [hep-th]}}.

\bibitem{MR0466134}
E.~Looijenga, ``Root systems and elliptic curves,'' {\em Invent. Math.}
  {\bfseries 38} no.~1, (1976/77) 17--32.

\bibitem{MR515632}
I.~N. Bern{\v{s}}te{\u\i}n and O.~V. {\v{S}}varcman, ``Chevalley's theorem for
  complex crystallographic {C}oxeter groups,'' {\em Funktsional. Anal. i
  Prilozhen.} {\bfseries 12} no.~4, (1978) 79--80.

\bibitem{Katz:1996fh}
S.~H. Katz, A.~Klemm, and C.~Vafa, ``{Geometric engineering of quantum field
  theories},'' \href{http://dx.doi.org/10.1016/S0550-3213(97)00282-4}{{\em
  Nucl.Phys.} {\bfseries B497} (1997) 173--195},
\href{http://arxiv.org/abs/hep-th/9609239}{{\ttfamily arXiv:hep-th/9609239
  [hep-th]}}.

\bibitem{MR0429876}
H.~Pinkham, ``Singularit\'es exceptionnelles, la dualit\'e \'etrange d'{A}rnold
  et les surfaces {$K-3$},'' {\em C. R. Acad. Sci. Paris S\'er. A-B} {\bfseries
  284} no.~11, (1977) A615--A618.

\bibitem{MR1265318}
C.~Voisin, ``Miroirs et involutions sur les surfaces {$K3$},'' {\em
  Ast\'erisque} no.~218, (1993) 273--323. Journ{\'e}es de G{\'e}om{\'e}trie
  Alg{\'e}brique d'Orsay (Orsay, 1992).

\bibitem{MR1420220}
I.~V. Dolgachev, ``Mirror symmetry for lattice polarized {$K3$} surfaces,''
  \href{http://dx.doi.org/10.1007/BF02362332}{{\em J. Math. Sci.} {\bfseries
  81} no.~3, (1996) 2599--2630}. \url{http://dx.doi.org/10.1007/BF02362332}.
  Algebraic geometry, 4.

\bibitem{MR791645}
P.~Slodowy, ``A character approach to {L}ooijenga's invariant theory for
  generalized root systems,'' {\em Compositio Math.} {\bfseries 55} no.~1,
  (1985) 3--32. \url{http://www.numdam.org/item?id=CM_1985__55_1_3_0}.

\bibitem{Klemm:2012sx}
A.~Klemm, J.~Manschot, and T.~Wotschke, ``{Quantum geometry of elliptic
  Calabi-Yau manifolds},''
\href{http://arxiv.org/abs/1205.1795}{{\ttfamily arXiv:1205.1795 [hep-th]}}.

\bibitem{Minahan:1998vr}
J.~Minahan, D.~Nemeschansky, C.~Vafa, and N.~Warner, ``{E strings and N=4
  topological Yang-Mills theories},''
  \href{http://dx.doi.org/10.1016/S0550-3213(98)00426-X}{{\em Nucl.Phys.}
  {\bfseries B527} (1998) 581--623},
\href{http://arxiv.org/abs/hep-th/9802168}{{\ttfamily arXiv:hep-th/9802168
  [hep-th]}}.

\bibitem{Hosono:1999qc}
S.~Hosono, M.~Saito, and A.~Takahashi, ``{Holomorphic anomaly equation and BPS
  state counting of rational elliptic surface},'' {\em Adv.Theor.Math.Phys.}
  {\bfseries 3} (1999) 177--208,
\href{http://arxiv.org/abs/hep-th/9901151}{{\ttfamily arXiv:hep-th/9901151
  [hep-th]}}.

\bibitem{Curio:1997rn}
G.~Curio and D.~Lust, ``{A Class of N=1 dual string pairs and its modular
  superpotential},'' \href{http://dx.doi.org/10.1142/S0217751X97003066}{{\em
  Int.J.Mod.Phys.} {\bfseries A12} (1997) 5847--5866},
\href{http://arxiv.org/abs/hep-th/9703007}{{\ttfamily arXiv:hep-th/9703007
  [hep-th]}}.

\bibitem{Morrison:2012ei}
D.~R. Morrison and D.~S. Park, ``{F-Theory and the Mordell-Weil Group of
  Elliptically-Fibered Calabi-Yau Threefolds},''
  \href{http://dx.doi.org/10.1007/JHEP10(2012)128}{{\em JHEP} {\bfseries 1210}
  (2012) 128},
\href{http://arxiv.org/abs/1208.2695}{{\ttfamily arXiv:1208.2695 [hep-th]}}.

\bibitem{MR2426805}
M.~Kobayashi, ``Duality of weights, mirror symmetry and {A}rnold's strange
  duality,'' {\em Tokyo J. Math.} {\bfseries 31} no.~1, (2008) 225--251.

\bibitem{MR2278769}
W.~Ebeling, ``Mirror symmetry, {K}obayashi's duality, and {S}aito's duality,''
  {\em Kodai Math. J.} {\bfseries 29} no.~3, (2006) 319--336.

\bibitem{CPR}
P.~Candelas, E.~Perevalov, and G.~Rajesh, ``{Matter From Toric Geometry},''
\href{http://arxiv.org/abs/hep-th/9707049}{{\ttfamily arXiv:hep-th/9707049
  [hep-th]}}.

\bibitem{Gaberdiel:1997ud}
M.~R. Gaberdiel and B.~Zwiebach, ``{Exceptional groups from open strings},''
  \href{http://dx.doi.org/10.1016/S0550-3213(97)00841-9}{{\em Nucl.Phys.}
  {\bfseries B518} (1998) 151--172},
\href{http://arxiv.org/abs/hep-th/9709013}{{\ttfamily arXiv:hep-th/9709013
  [hep-th]}}.

\bibitem{DeWolfe:1998zf}
O.~DeWolfe and B.~Zwiebach, ``{String junctions for arbitrary Lie algebra
  representations},''
  \href{http://dx.doi.org/10.1016/S0550-3213(98)00743-3}{{\em Nucl.Phys.}
  {\bfseries B541} (1999) 509--565},
\href{http://arxiv.org/abs/hep-th/9804210}{{\ttfamily arXiv:hep-th/9804210
  [hep-th]}}.

\bibitem{Grimm:2011tb}
T.~W. Grimm, M.~Kerstan, E.~Palti, and T.~Weigand, ``{Massive Abelian Gauge
  Symmetries and Fluxes in F-theory},''
  \href{http://dx.doi.org/10.1007/JHEP12(2011)004}{{\em JHEP} {\bfseries 1112}
  (2011) 004},
\href{http://arxiv.org/abs/1107.3842}{{\ttfamily arXiv:1107.3842 [hep-th]}}.

\bibitem{Braun:2011zm}
A.~P. Braun, A.~Collinucci, and R.~Valandro, ``{G-flux in F-theory and
  algebraic cycles},''
  \href{http://dx.doi.org/10.1016/j.nuclphysb.2011.10.034}{{\em Nucl.Phys.}
  {\bfseries B856} (2012) 129--179},
\href{http://arxiv.org/abs/1107.5337}{{\ttfamily arXiv:1107.5337 [hep-th]}}.

\bibitem{Krause:2011xj}
S.~Krause, C.~Mayrhofer, and T.~Weigand, ``{$G_4$ flux, chiral matter and
  singularity resolution in F-theory compactifications},''
  \href{http://dx.doi.org/10.1016/j.nuclphysb.2011.12.013}{{\em Nucl.Phys.}
  {\bfseries B858} (2012) 1--47},
\href{http://arxiv.org/abs/1109.3454}{{\ttfamily arXiv:1109.3454 [hep-th]}}.

\bibitem{Donagi:2009ra}
R.~Donagi and M.~Wijnholt, ``{Higgs Bundles and UV Completion in F-Theory},''
\href{http://arxiv.org/abs/0904.1218}{{\ttfamily arXiv:0904.1218 [hep-th]}}.

\bibitem{Dudas:2010zb}
E.~Dudas and E.~Palti, ``{On hypercharge flux and exotics in F-theory GUTs},''
  \href{http://dx.doi.org/10.1007/JHEP09(2010)013}{{\em JHEP} {\bfseries 1009}
  (2010) 013},
\href{http://arxiv.org/abs/1007.1297}{{\ttfamily arXiv:1007.1297 [hep-ph]}}.

\bibitem{Lawrie:2012gg}
C.~Lawrie and S.~Schॊfer-Nameki, ``{The Tate Form on Steroids: Resolution and
  Higher Codimension Fibers},''
  \href{http://dx.doi.org/10.1007/JHEP04(2013)061}{{\em JHEP} {\bfseries 1304}
  (2013) 061},
\href{http://arxiv.org/abs/1212.2949}{{\ttfamily arXiv:1212.2949 [hep-th]}}.

\bibitem{Bershadsky:1996nu}
M.~Bershadsky and A.~Johansen, ``{Colliding singularities in F theory and phase
  transitions},'' \href{http://dx.doi.org/10.1016/S0550-3213(97)00027-8}{{\em
  Nucl.Phys.} {\bfseries B489} (1997) 122--138},
\href{http://arxiv.org/abs/hep-th/9610111}{{\ttfamily arXiv:hep-th/9610111
  [hep-th]}}.

\bibitem{Antoniadis:2013joa}
I.~Antoniadis and G.~K. Leontaris, ``Neutrino mass textures from f-theory,''
  {\em Eur.Phys.J.} {\bfseries C73} (2013) 2670,
  \href{http://arxiv.org/abs/1308.1581}{{\ttfamily arXiv:1308.1581 [hep-th]}}.

\bibitem{MR2648685}
E.~Frenkel, ``Gauge theory and {L}anglands duality,'' {\em Ast\'erisque}
  no.~332, (2010) Exp. No. 1010, ix--x, 369--403. S{\'e}minaire Bourbaki.
  Volume 2008/2009. Expos{\'e}s 997--1011.

\end{thebibliography}\endgroup

\end{document}